\documentclass{aastex61}
\received{May 16, 2022}
\revised{Jul 26, 2022}
\accepted{Aug 06, 2022}
\submitjournal{ApJ}

\shorttitle{The Astrophysical Journal}
\shortauthors{Yan et al.}
\usepackage{amssymb,amsmath}
\usepackage{float}
\usepackage{graphicx}
\usepackage{bbding}
\usepackage{threeparttable}
\begin{document}
\title{Modeling H$\alpha$ and He 10830 transmission spectrum of WASP-52b}

\correspondingauthor{Jianheng Guo, guojh@ynao.ac.cn}

\author[0000-0002-8519-0514]{Dongdong Yan}
\affil{Yunnan Observatories, Chinese Academy of Sciences, P.O. Box 110, Kunming 650011, People's Republic of China}
\affiliation{School of Astronomy and Space Science, University of Chinese Academy of Sciences, Beijing, People's Republic of China}
\affiliation{Key Laboratory for the Structure and Evolution of Celestial Objects, CAS, Kunming 650011, People's Republic of China}

\author[0000-0001-9561-8134]{Kwang-il Seon}
\affil{Korea Astronomy and Space Science Institute, 776 Daedeokdae-ro,
Yuseong-gu, Daejeon 34055, Republic of Korea}
\affiliation{Astronomy and Space Science Major, University of Science and
Technology, 217, Gajeong-ro, Yuseong-gu, Daejeon 34113, Republic of Korea}

\author[0000-0002-8869-6510]{Jianheng Guo}
\affil{Yunnan Observatories, Chinese Academy of Sciences, P.O. Box 110, Kunming 650011, People's Republic of China}
\affiliation{School of Astronomy and Space Science, University of Chinese Academy of Sciences, Beijing, People's Republic of China}
\affiliation{Key Laboratory for the Structure and Evolution of Celestial Objects, CAS, Kunming 650011, People's Republic of China}

\author[0000-0003-0740-5433]{Guo Chen}
\affil{Key Laboratory of Planetary Sciences, Purple Mountain Observatory, Chinese Academy of Sciences, Nanjing 210023, China}

\author{Lifang Li}
\affil{Yunnan Observatories, Chinese Academy of Sciences, P.O. Box 110, Kunming 650011, People's Republic of China}
\affiliation{School of Astronomy and Space Science, University of Chinese Academy of Sciences, Beijing, People's Republic of China}
\affiliation{Key Laboratory for the Structure and Evolution of Celestial Objects, CAS, Kunming 650011, People's Republic of China}

%% Note that the \and command from previous versions of AASTeX is now
%% depreciated in this version as it is no longer necessary. AASTeX
%% automatically takes care of all commas and "and"s between authors names.

%% AASTeX 6.1 has the new \collaboration and \nocollaboration commands to
%% provide the collaboration status of a group of authors. These commands
%% can be used either before or after the list of corresponding authors. The
%% argument for \collaboration is the collaboration identifier. Authors are
%% encouraged to surround collaboration identifiers with ()s. The
%% \nocollaboration command takes no argument and exists to indicate that
%% the nearby authors are not part of surrounding collaborations.

%% Mark off the abstract in the ``abstract'' environment.
\begin{abstract}
Escaping atmosphere has been detected by the excess absorption of Ly$\alpha$, H$\alpha$ and He triplet (10830$\rm\AA$) lines. 
Simultaneously modeling the absorption of the H$\alpha$ and He 10830 lines can provide useful constraints about the exoplanetary atmosphere. In this paper, we use a hydrodynamic model combined with a non-local thermodynamic model and a new Monte Carlo 
simulation model to obtain the H(2) and He(2$^3$S) populations. The Monte Carlo simulations of Ly$\alpha$ radiative transfer are performed with assumptions of a spherical stellar Ly$\alpha$ radiation and a spherical planetary atmosphere, for the first time, to calculate the Ly$\alpha$ mean intensity distribution inside the planetary atmosphere, necessary in estimating the H(2) population. We model the transmission spectra of the H$\alpha$ and He 10830 lines simultaneously  in hot Jupiter WASP-52b. We find that models with many different H/He ratios can reproduce the H$\alpha$ observations well if the host star has (1) a high X-ray/extreme ultraviolet (XUV) flux ($F_{\rm XUV}$) and a relatively low X-ray fraction in XUV radiation ($\beta_m$), or (2) a low $F_{\rm XUV}$ and a high $\beta_m$. 
The simulations of He 10830 $\rm\AA$ triplet suggest that a high H/He ratio ($\sim$ 98/2) is required to fit the observation.
%The absorption of He10830 increases with increasing $F_{\rm XUV}$, but its variation with $\beta_m$ appears to be more complicated. 
The models that fit both lines well confine $F_{\rm XUV}$ to be about 0.5 times the fiducial value and $\beta_m$ to have a value around 0.3.
The models also suggest that hydrogen and helium originate from the escaping atmosphere, and the mass-loss rate is about 2.8$\times 10^{11}$ g s$^{-1}$.
%Our work would be a benchmark in studying the exoplanetary hydrogen and helium atmosphere.
\end{abstract}

%% Keywords should appear after the \end{abstract} command.
%% See the online documentation for the full list of available subject
%% keywords and the rules for their use.

\section{Introduction}
Exoplanets orbiting very close to their host stars will receive strong high energy radiations such as XUV and near-ultraviolet (NUV) from the stars. The spectral energy distribution (SED) of the incident radiation can be diverse, depending on the properties of host stars \citep{2011A&A...532A...6S, 2021A&A...649A..96J}. Especially in young stellar systems, the X-ray can occupy a high portion of stellar radiation \citep{2003A&A...397..147P, 2005ApJ...622..680R, 2015A&A...577L...3T}. In some systems, such as A-type stars, the NUV flux can exceed the XUV flux by orders of magnitude \citep{2019ApJ...884L..43G}. In gas-rich exoplanets, the planetary atmosphere can absorb the high-energy photons as a heating source and then expand to overcome the planetary gravitational potential energy \citep{2021ApJ...907L..47Y}. In other words, they are thought to be photo-evaporated and eventually escape the planet. The mass-loss rates of atmospheric escape can
be as high as 10$^9$-10$^{13}$ g s$^{-1}$ \citep{2020MNRAS.491.3435S,2021NatAs.tmp..247B, 2022A&A...657A...6C}. This process can have important ramifications for the planetary composition, evolution and even the population of planets as a whole \citep{2017AJ....154..109F,2018AJ....156..264F}. For instance, the observed valley at 1.5-2.0 Earth radii in the radius distribution of small exoplanets \citep{2020ApJS..247...28H} may be a consequence of atmospheric photo-evaporation of planets \citep{2013ApJ...775..105O}.

With the development of sophisticated telescopes, precise characterization of the exoplanetary atmosphere could be possible. High-resolution transmission spectroscopy of the exoplanetary atmosphere is a prime method to study atmospheric properties. To date, a number of individual chemical species, such as hydrogen, helium, sodium, and oxygen atoms; carbon, magnesium, and iron ions; and water and methane molecules, have been discovered by this method  \citep{2003Nature...422..143,2010A&A...514..72,2012ApJ...751..86,2015Nature...522..459,2018Sci...362.1384A,2019A&A...621A..74A,2019A&A...632A..69Y,2020ApJ...888L..13T,2021ApJ...907L..36G,2021NatAs.tmp..253B,2020A&A...641L...7S,2020MNRAS.493.2215G,2020A&A...641A.123H,2019AJ....158...91S,2007Natur.448..169T,2020MNRAS.496.1638M, 2016ApJ...822L...4E,2020A&A...638A..26S}.

By analyzing the observed transmission spectra of hydrogen  Ly$\alpha$ in the UV band \citep{2003Nature...422..143,2010A&A...514..72,2012ApJ...751..86,2015Nature...522..459, 2021NatAs.tmp..247B}, theoretical models show some hydrogen-rich exoplanets have experienced strong atmosphere escape. While observations of the Ly$\alpha$ line originating from  hydrogens in the ground state H(1s) are highly sensitive to the exoplanetary atmosphere, they are heavily affected by the interstellar absorption and geocoronal Ly$\alpha$ emission, which almost always renders the line core and sometimes parts of the wings unusable. Moreover, such observations are only amenable with spaceborne instrumentation. Unlike Ly$\alpha$, the Balmer lines in the optical band originating from the hydrogen atoms in the first excited state, H(2), are not contaminated by interstellar absorption and can be observed by ground-based instruments. To date, there have been nine exoplanets with detected excess H$\alpha$ absorption \citep{2012ApJ...751..86,2018NatAs...2..714Y,2019AJ....157...69C,2018A&A...616A.151C, 2018AJ....156..154J,2020A&A...635A.171C,2020MNRAS.494..363C,2021A&A...645A..24B,2021A&A...645A..22Y,2022A&A...657A...6C, 2022AJ....163...96B}. The interpretation of the H$\alpha$ signals can help to reveal the atmosphere properties at different altitudes since the population of H(2) atoms is more sensitive to the gas properties than H(1s) \citep{2007Natur.445..511B}. By studying HD 189733b, \cite{{2017ApJ...851..150H}} proposed that the H$\alpha$ abosrption would originate from the same layer where the Na I D absorption is formed.
Therefore, by studying Ly$\alpha$ and H$\alpha$ at the same time, one can obtain more information about the planetary atmosphere. 

The helium triplet in near-infrared is often referred to as He 10830  $\rm\AA$ (hereafter He 10830), and its absorption is caused by the helium atoms at metastable state, He(2$^3$S).
Second only to hydrogen in abundance, with no contamination by the interstellar medium, and approachable by ground-based instruments, all these make the helium triplet emerge as a valuable probe to detect the planetary atmosphere \citep{2000ApJ...537..916S,2018ApJ...855L..11O,2019ApJ...881..133O}.
Up to now, there are several exoplanets with detected excess absorption of this line through transmission spectroscopy; they are WASP-107b, HD 189733b, HD 209458b, GJ 3470b, WASP-69b, HAT-P-11b, HAT-P-32b, and TOI 560.01 \citep{2018Natur.557...68S, 2021AJ....162..284S, 2018A&A...620A..97S, 2020ApJ...894...97N,  2019A&A...629A.110A, 2020AJ....159..115K, 2020A&A...638A..61P, 2018Sci...362.1388N, 2018Sci...362.1384A, 2022A&A...657A...6C,2022AJ....163...67Z}. Tentative detection of He 10830 absorption by the atmosphere of GJ 1214b was also reported recently \citep{2022arXiv220111120O}.
Models have been developed to interpret the observed signals. For example, \cite{2020A&A...636A..13L, 2021A&A...647A.129L} adopted an isothermal Parker wind model to simulate the atmosphere of HD 209458b, HD 189733b, and GJ 3470b and used a radiative transfer of the He 10830 line to reproduce the observed line profiles. They argued that a high H/He number ratio larger than 90/10 is required to fit the observation. Besides, 3D simulations were also performed to explain the He 10830 absorption in WASP-107b, GJ 3470b, and WASP-69b \citep{2021MNRAS.500.1404S,2021MNRAS.503L..23K,2021MNRAS.507.3626K,2021ApJ...914...98W,2021ApJ...914...99W,2022ApJ...927..238R}. A 3D simulation can reproduce the asymmetry features in the transit light curves and transmission spectra well but is time-consuming in parameter exploration.

To date, the absorptions of all three lines (Ly$\alpha$, H$\alpha$, and He 10830) have been detected only in HD 189733b, although the interpretation of the H$\alpha$ signal observed on this planet remains controversial \citep{2016MNRAS.462.1012B,2017AJ....153..185C,2017AJ....153..217C}. In some systems, two of the above three lines have been detected. For instance, Ly$\alpha$ and He 10830 excess absorption lines have  been detected in HD 209458b, GJ 3470b, and HAT-P-11b. The excess absorption of H$\alpha$ and He 10830 lines have been detected in WASP-52b and HAT-P-32b. However, the He 10830 excess absorption in WASP-52b was measured using an ultranarrowband filter rather than spectroscopy.
Simultaneously modeling the transmission spectra of two or more lines can help constrain the physical parameters of the planetary atmosphere. Especially, by fitting the observation of  both the hydrogen and helium lines, one can investigate the atmospheric properties such as composition and dynamics, and its interactions with the host stars. Pioneering work on the H$\alpha$ and He 10830 absorption model was done by \cite{2022A&A...657A...6C}, applying it to HAT-P-32b.
The authors used two independent models to interpret the H$\alpha$ and He 10830 signals separately. The models that can simultaneously explain both the H$\alpha$ and He 10830  $\rm\AA$ absorption of the planetary atmosphere are still rare, which will play a role as a benchmark in studying the exoplanetary atmosphere.

With that in mind, in this paper, we will focus on hot Jupiter WASP-52b, which is observed in both H$\alpha$ and He lines. Using the high resolution spectrograph ESPRESSO at the Very Large Telescope array, \cite{{2020A&A...635A.171C}} detected an excess absorption of about 0.86$\%\pm 0.13\%$ at the H$\alpha$ line center; but there was no modeling effort for this signal.  
With ultranarrowband photometry, \cite{2020AJ....159..278V} placed an upper limit on excess absorption of He 10830 of WASP-52b, indicating the possible presence of helium in the atmosphere. This technique was also used to detect the He 10830 absorption in HAT-P-18b \citep{2021ApJ...909L..10P}.
Due to a lack of spectroscopic observation of the helium triplet, no detailed profiles of the transmission spectra are available. Assuming the same spectral shape as is observed for WASP-69b, WASP-52b was found to have an amplitude of 1.31\% $\pm$ 0.94\% in the deepest line center of He 10830 triplet in the work of \cite{2020AJ....159..278V}. Using high-resolution spectroscopy, \cite{2022arXiv220511579K} recently detected the neutral helium line at 10830 $\rm\AA$ in the atmosphere of WASP-52b. They measured
the spectral shape of the absorption line and found the excess absorption depth of He 10830 to be about $3.44\%\pm 0.31$\% (at $11 \sigma$ significance).
In addition, Na I and K I absorption features were also found in its atmosphere \citep{2016MNRAS.463.2922K, 2017A&A...600L..11C,2017MNRAS.470..742L}. The host star WASP-52 is an active K2 type star, in which some stellar activity features such as chromospheric emission in the Ca II H+K lines, faculae, and occulted star spots were found \citep{2013A&A...549A.134H, 2016MNRAS.463.2922K, 2017MNRAS.465..843M}. \cite{2020MNRAS.491.5361B} studied the effect of star-spot correction on atmospheric retrievals in this planet and concluded that adding the stellar contamination to obtain a ‘good fit’ of a transmission spectrum might hide spectral features that are planetary in nature. This led us to speculate that different stellar radiation levels would affect the observed transmission spectrum.

To model the H$\alpha$ and He 10830 transmission spectra of WASP-52b, we should first calculate the level populations of hydrogen and helium in detail.
The absorption of H$\alpha$ is caused by hydrogens in the first excited state, i.e., the n=2 state, which splits into 2s and 2p sub-states. The number density in the 2p state is predominantly determined by the Ly$\alpha$ radiation field strength inside the atmosphere \citep{2013ApJ...772..144C, 2017ApJ...851..150H, 2021ApJ...907L..47Y}. Thus, a detailed treatment of Ly$\alpha$ radiative transfer is indispensable. Because Ly$\alpha$ is a resonant scattering line, the traditional treatment of the radiative transfer will be very complicated. Monte Carlo simulations, which probabilistically experiment with absorption and scattering processes by using random sampling methods, are often used as an alternative in the radiative transfer calculation \citep{1972ApJ...174..439A,2001ApJ...551..269G,2006MNRAS.367..979H,2011BASI...39..101W, 2013ARA&A..51...63S, 2016A&A...590A..55B,2016ApJ...833..201S, 2020ApJS..250....9S,2021ApJS..254...29X, 2022ApJS..259....3S}. However, the Monte Carlo simulations have not been widely applied in the exoplanetary atmosphere. \cite{2017ApJ...851..150H} used this method to simulate the Ly$\alpha$ resonant scattering in the atmosphere of HD 189733b but by assuming a plane-parallel atmosphere. This motivated us to investigate the influence of different geometries in dealing with stellar illumination and the atmosphere. 

This paper is organized as follows. In Section \ref{sec:Methods}, we introduce the hydrodynamic models to obtain the atmosphere structures and the detailed calculation of the level populations of H(2) and He(2$^3$S). In Section \ref{sec:cal_RT},  we introduce models newly added in LaRT \footnote{LaRT is publicly available via https://doi.org/10.5281/zenodo.5618511 and https://github.com/seoncafe/LaRT.} \citep{2020ApJS..250....9S, 2022ApJS..259....3S}, a Monte Carlo simulation code of Ly$\alpha$ radiative transfer, to deal with the geometrical configuration of the star-planet system. The models assume that the stellar Ly$\alpha$ radiation is a spherical source.
The radiative transfer schemes of H$\alpha$ and He 10830 lines are also described in Section \ref{sec:cal_RT}. In Section \ref{sec: Results}, we present the resluts and compare the models with observations. In Section \ref{sec: Diss}, we discuss
the plane-parallel illumination model, the limb darkening effect, and the influence of the atmosphere size on the transmission spectrum. We also discuss the difference between the isothermal model and our model. In Section \ref{sec: summary}, we conclude by summarizing our findings.

\section{Method}\label{sec:Methods}
\subsection{Hydrodynamic simulation of exoplantary atmosphere}

\begin{figure*}
\gridline{\fig{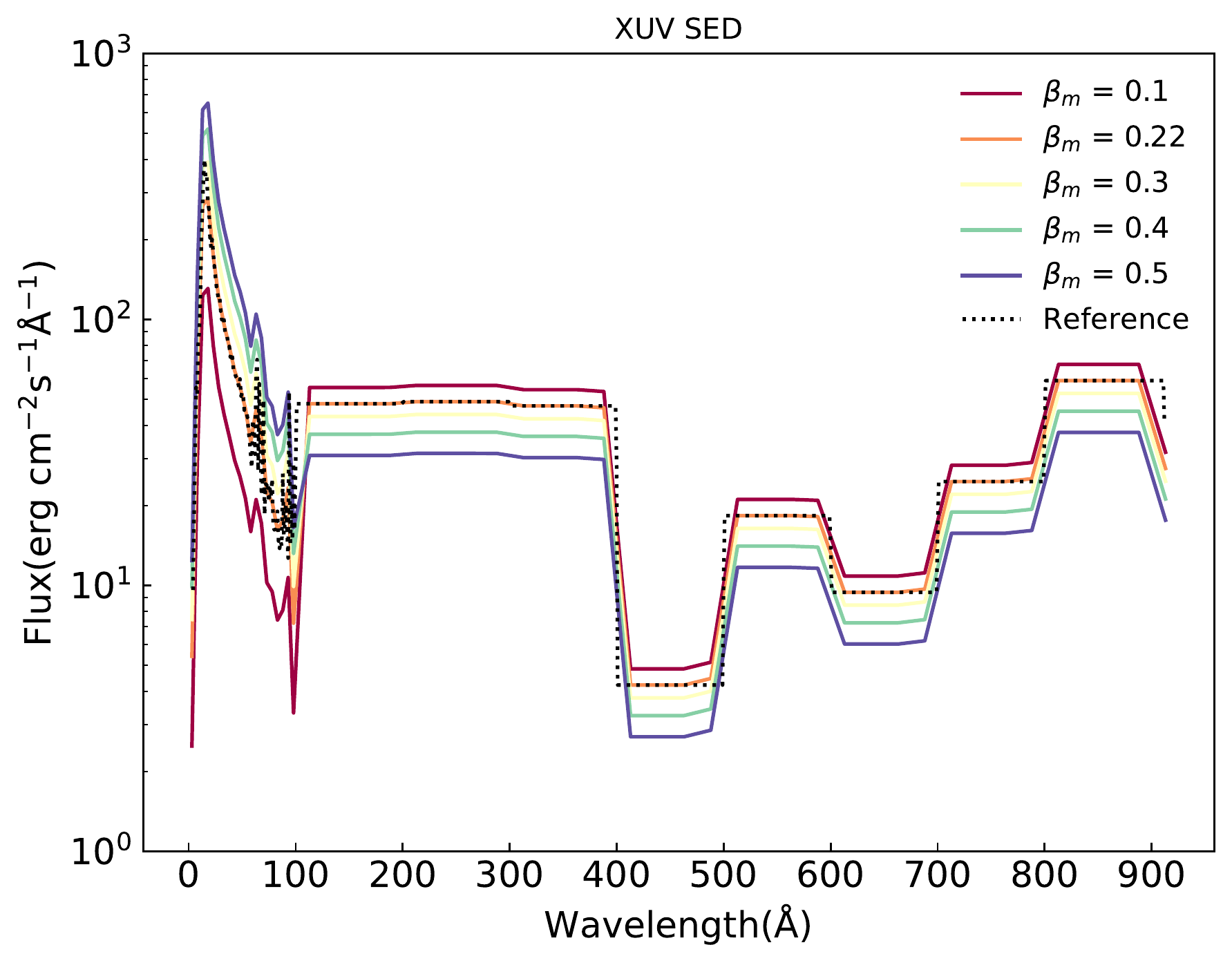}{0.5\textwidth}{(a)}
 \fig{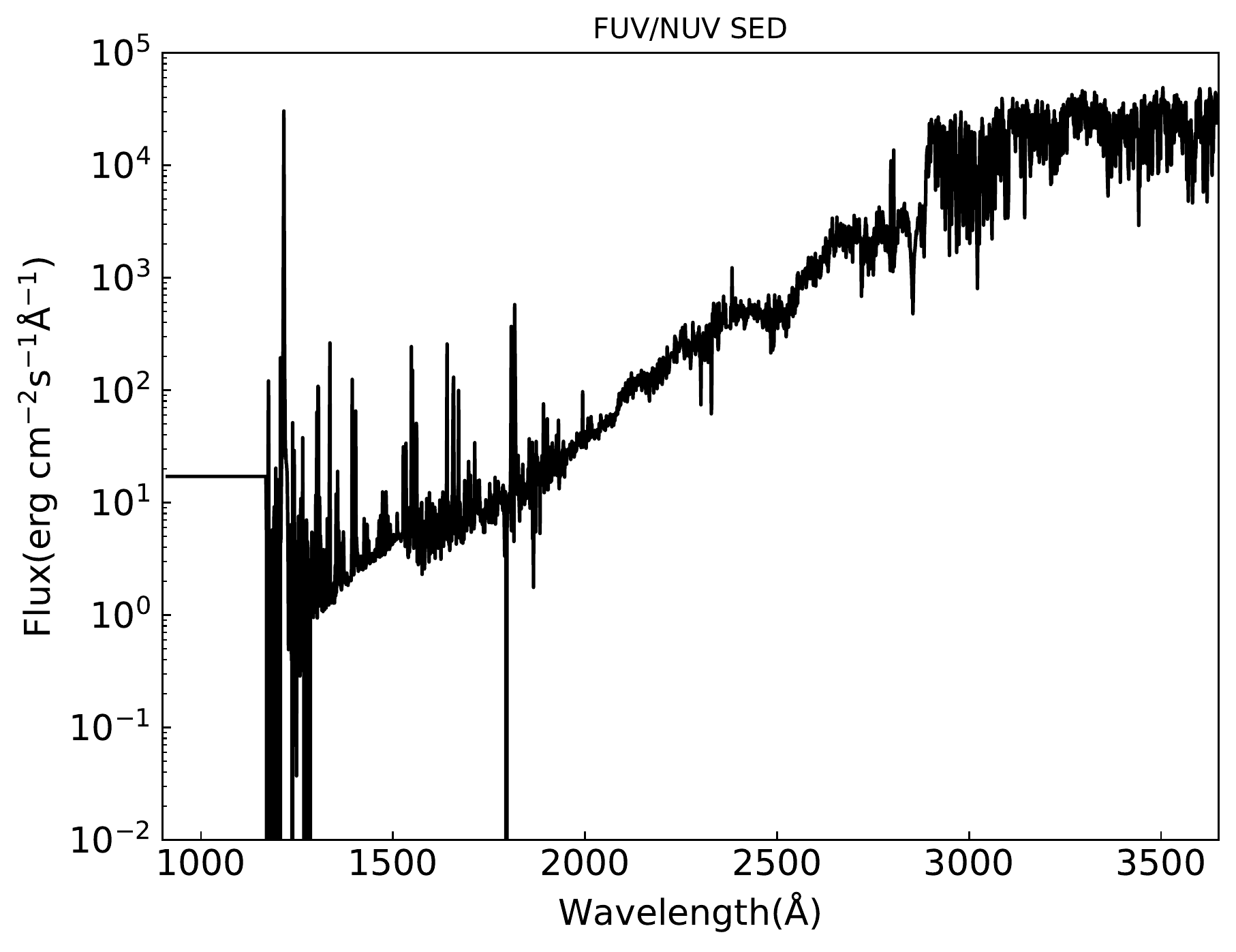}{0.5\textwidth}{(b)}
 }
\caption{The stellar XUV and FUV/NUV spectral energy distribution, which is the spectrum of eps Eri taken from the MUSCLES Treasury Survey \citep{2016ApJ...820...89F} and scaled at the orbital distance of WASP-52b. In panel (a), different solid lines represent the constructed spectra of different $\beta_m$. $\beta_m=0.22$ corresponds to the reference spectrum, which is taken from  MUSCLES Treasury Survey.}\label{XUVSED}
\end{figure*}

We used a 1D hydrodynamic model \citep{2018ChA&A..42...81Y,2021ApJ...907L..47Y} to simulate the atmospheric structure of WASP-52b and obtained the radial profiles of atmospheric temperature, velocity, and particle number densities. The planetary and stellar parameters are based on the reported observations \citep{2013A&A...549A.134H,2020A&A...635A.171C}. WASP-52 is a main-sequence star with $M_\star = 0.87 M_\sun$ and $R_\star = 0.79 R_\sun$. WASP-52b is a hot Jupiter with $M_P = 0.46 M_J$ and $R_P = 1.27 R_J$.
The equilibrium temperature is 1304 K, which is also the temperature at the bottom boundary $R_p$ in our model. The chemical abundance of WASP-52b is assumed to be the same as that of the host star WASP-52 with the solar abundance except for [Fe/H] = 0.03. We first adopt the solar abundance for helium (H/He = 92/8; \cite{2009ARA&A..47..481A}), but later, the effects of the change in the H/He ratio are investigated.
The integrated flux in the XUV band is an important input in the simulations. In the absence of direct observation of the stellar XUV, we use the spectrum of eps Eri from the MUSCLES Treasury Survey \citep{2016ApJ...820...89F} to calculate the $F_{\rm XUV}$ received by the planet. The $F_{\rm XUV}$ is about 34,168 erg cm$^{-2}$ s$^{-1}$ at the orbital distance of about 0.0272 AU. In our calculation, the value is divided by a factor of 4 (hereafter F$_0$), which accounts for the uniform redistribution of the stellar radiation energy around the spherical planet. 
The spectral energy distribution index, $\beta_m$ = F(1-100$\rm\AA$)/F(1-912$\rm\AA$) as defined in \cite{2021ApJ...907L..47Y}, is about 0.22. 
To find the best fit for the observation, we explore parameter space in a range of $F_{\rm XUV}$ = 0.25, 0.5, 1, 2, 4, 6, and 8 $\times$ F$_0$, and $\beta_m$ = 0.1, 0.22, 0.3, 0.4, and 0.5.
Here, $\beta_m$ is the ratio of X-ray to total XUV flux. Note here that the
largest $\beta_m$, 0.5, means that the X-ray flux is half of the total XUV flux,
being the highest value that the late-type stars can have. Works have shown that the stellar EUV flux is higher than the X-ray flux, meaning  $\beta_m \leq 0.5$, in most of the studied systems \citep{2015A&A...577L...3T, 2011A&A...532A...6S, 2021A&A...656A.111S}. Therefore, taking $\beta_m$ = 0.5 as the maximum is reasonable for our purpose.
In our simulation, the XUV spectrum is divided into 53 wavelength bins and we construct the XUV spectra for different $\beta_m$-values. Figure \ref{XUVSED} (a) shows the adopted stellar XUV spectral energy distributions at the orbital distance of WASP-52b. Except for $\beta_m$ = 0.22, other spectra are artificially constructed by changing the high-energy and low-energy parts accordingly.
The pressure at the bottom boundary of the atmosphere is assumed as 1 $\mu$bar. The upper boundary is taken to be 10 $R_P$, which is bigger than the radius of the host star (about 6 $R_P$). 
The Ly$\alpha$ cooling and the stellar tidal force are also considered in the simulations.

%\begin{figure}[t]
%\begin{center}
%  \includegraphics[scale=0.6]{fig1.pdf}
%%\vspace{1cm}
%\caption{The stellar XUV spectral energy distribution, scaled at the orbital distance of WASP-52b. Here, $\beta_m$ denotes the flux ratio $\rm F(1-100\AA)/F(1-912\AA)$. $\beta_m=0.22$ corresponds to the
%spectrum of eps Eri taken from the MUSLES Treasury Survey (France et al. 2016).}\label{XUVSED}
%\end{center}
%\end{figure}

\subsection{Calculation of H(2) populations}\label{method_H2s2p}
To obtain the H$\alpha$ transmission spectrum,
we first calculate the hydrogen population in the n = 2 state by solving the statistical  equation of non-local thermodynamic equilibrium (NLTE). 
Assuming that the atmosphere is in a stationary state, 
the production rates of H(2) are equal to
their loss rates. Therefore, the equation of rate equilibrium \citep{2013ApJ...772..144C,2017ApJ...851..150H} for H(2p) is:
\begin{equation}\label{Eq_2p}
\begin{aligned}
& B_{1s\rightarrow2p}\bar{J}_{Ly\alpha}n_{1s} + C_{1s\rightarrow2p}n_{1s}n_e + C_{(e)2s \rightarrow2p}n_{2s}n_{e} + C_{(H^+)2s \rightarrow2p}n_{2s}n_{H^+}+ \alpha_{2p}n_e n_{H^+} +\beta_{2p}n^2_{e}n_{H^+} \\
& = A_{2p\rightarrow1s}n_{2p} + B_{2p\rightarrow1s}\bar{J}_{Ly\alpha}n_{2p} + C_{2p \rightarrow 1s}n_{2p}n_e + C_{(e)2p\rightarrow2s}n_{2p}n_{e} \\
& +  C_{(H^+)2p\rightarrow2s}n_{2p}n_{H^+} +  C_{2p\rightarrow \infty}n_{2p}n_{e} + \Gamma_{2p}n_{2p} + \frac{1}{r^2}\frac{\partial{}}{\partial{r}}(r^2n_{2p}v)
\end{aligned}
\end{equation}
and the equation of rate equilibrium for H(2s) is:

\begin{equation}\label{Eq_2s}
\begin{aligned}
& C_{1s \rightarrow 2s}n_{1s}n_e +  C_{(e)2p \rightarrow 2s}n_{2p}n_e + C_{(H^+)2p \rightarrow 2s}n_{2p}n_{H^+} + \alpha_{2s}n_e n_{H^+} + \beta_{2s}n^2_{e}n_{H^+}\\
& = C_{2s \rightarrow 1s}n_{2s}n_e +  C_{(e)2s \rightarrow 2p}n_{2s}n_e + C_{(H^+)2s \rightarrow 2p}n_{2s}n_{H^+} + C_{2s\rightarrow \infty}n_{2s}n_{e} + \Gamma_{2s}n_{2s} + A_{2s \rightarrow 1s}n_{2s} + \frac{1}{r^2}\frac{\partial{}}{\partial{r}}(r^2n_{2s}v)
\end{aligned}
\end{equation}
where $n_{1s}$, $n_{2p}$, $n_{2s}$, $n_{H^+}$, and $n_{e}$ are the number densities of H(1s), H(2p), H(2s), H$^+$, and electrons, respectively.
$\bar{J}_{Ly\alpha} = \int J_{\nu} \phi_{\nu} d\nu $ is the Ly$\alpha$ mean intensity, weighted over the Voigt line profile $\phi_{\nu}$ . $C_{(e/H^+)i \rightarrow j}$ is the rate coefficient for the collisional transition from state $i$ to $j$ ($j = \infty$ means collisional ionization by electrons), and the subscript e (or H$^+$) represents the collision with electron (or proton). The equation includes the collisions of $1s \leftrightarrow 2s$, $1s \leftrightarrow 2p$, and $2s \leftrightarrow2p$. $\alpha_{2s}$ and $\alpha_{2p}$ are the effective radiative recombination rate coefficients. $\beta_{2s}$ and $\beta_{2p}$ are the three-body recombination rate coefficients.
$A_{i \rightarrow j}$ and $B_{i \rightarrow j}$ are Einstein coefficients.  $\Gamma_{2s}$ and $\Gamma_{2p}$ represent the photoionization rates :

\begin{equation}\label{Eq_Gamma2s2p}
\Gamma_{2s, 2p}=\int\limits_{\nu_1}^{\nu_0}\frac{4\pi J_{\nu}}{h\nu}
\sigma_{2s, 2p}(\nu)d\nu
\end{equation}
where $\nu_1$ = 3.4eV/$h$ (3646 $\rm\AA$) ($h$ is the planck constant) is the hydrogen ionization threshold for the n = 2 state, and $\nu_0$ = 13.6eV/$h$ (912 $\rm\AA$) is the hydrogen ionization energy from the ground state. The spectrum of the incident ionizing radiation is taken from that of eps Eri in the  MUSCLES Treasury Survey, as shown in Figure \ref{XUVSED} (b). The photoionization cross-sections of H(2p) and H(2s) are cited from ``The Opacity Project" \footnote{http://cdsweb.u-strasbg.fr/topbase/home.html}. Note here the attenuation of the radiation in Equation (\ref{Eq_Gamma2s2p}) is ignored due to the small column densities of H(2s) and H(2p) \citep{2013ApJ...772..144C}. $\frac{1}{r^2}\frac{\partial{}}{\partial{r}}(r^2n_{2p}v)$ and $\frac{1}{r^2}\frac{\partial{}}{\partial{r}}(r^2n_{2s}v)$ are the advection terms of H(2p) and H(2s), repectively, where $v$ is the velocity and $r$ is the altitude of the atmosphere.
The reaction rates and coefficients are listed in Table \ref{tab:reaction_rates}. 

Equations (1) and (2) can be written as :
\begin{eqnarray}
   \begin{split}
a_{11}n_{2p} + a_{12}n_{2s} = A \\
a_{21}n_{2p} + a_{22}n_{2s} = B   
   \end{split}
\end{eqnarray}
Then,
\begin{eqnarray}\label{Eq_n2s2p}
   \begin{split}
n_{2p} = \frac{ A a_{22} - B a_{12} }{a_{11}a_{22} - a_{12}a_{21} }  \\
n_{2s} = \frac{ B a_{11} - A a_{21} }{a_{11}a_{22} - a_{12}a_{21} } \\
   \end{split}
\end{eqnarray}

Here,
\begin{eqnarray}
   \begin{split}
A &=& B_{1s\rightarrow2p}\bar{J}_{Ly\alpha}n_{1s} + C_{1s\rightarrow2p}n_{1s}n_e +  \alpha_{2p}n_e n_{H^+} +\beta_{2p}n^2_{e}n_{H^+} \\
B &=& C_{1s \rightarrow 2s}n_{1s}n_e +  \alpha_{2s}n_e n_{H^+} + \beta_{2s}n^2_{e}n_{H^+} \\
a_{11} &=&  A_{2p\rightarrow1s} + B_{2p\rightarrow1s}\bar{J}_{Ly\alpha} + C_{2p \rightarrow 1s}n_e \\ &\;& + C_{(e)2p\rightarrow2s}n_{e} + 
 C_{(H^+)2p\rightarrow2s}n_{H^+} +  C_{2p\rightarrow \infty}n_{e} + \Gamma_{2p} \\
a_{12} &=& - ( C_{(e)2s \rightarrow 2p}n_e + C_{(H^+)2s \rightarrow 2p}n_{H^+})\\
a_{21} &=& -(C_{(e)2p \rightarrow 2s}n_e + C_{(H^+)2p \rightarrow 2s}n_{H^+})\\
a_{22} &=& C_{2s \rightarrow 1s}n_e +  C_{(e)2s \rightarrow 2p}n_e + C_{(H^+)2s \rightarrow 2p}n_{H^+} + C_{2s\rightarrow \infty}n_{e} + \Gamma_{2s} + A_{2s \rightarrow 1s}
\end{split} 
\end{eqnarray}

By solving the equation of rate equilibrium, we find that n$_{2p}$ is predominantly determined by Ly$\alpha$ pumping. 

\begin{equation}
 n_{2p} \approx \frac{ B_{1s\rightarrow2p}\bar{J}_{Ly\alpha}}{A_{2p\rightarrow1s}}n_{1s} 
\end{equation}
Therefore, a detailed treatment of Ly$\alpha$ radiative transfer in the atmosphere is indispensable. For this purpose, we use LaRT (see \ref{sec:LaRT} for details). 
The incident stellar Ly$\alpha$ and the Ly$\alpha$ generated within the planetary atmospher are the two input sources of the Ly$\alpha$ intensity in this model. Due to a lack of the observation of the stellar Ly$\alpha$ spectrum, we adopt a similar profile to that of \citet{2017ApJ...851..150H}. At the outer boundary of the planetary atmosphere, on which the Ly$\alpha$ photons are incident, a double Gaussian line profile of Ly$\alpha$ with a width of 49 km s$^{-1}$ centered at $\pm$74 km s$^{-1}$ is assumed. The total integrated Ly$\alpha$ flux is assumed to be about 37,300 erg cm$^{-2}$ s$^{-1}$ according to the spectrum of EP Eri (a K2V star, note that it's different from eps Eri which is a K1V star) from \cite{2013ApJ...766...69L}. As we have discussed in a former work \citep{2021ApJ...907L..47Y}, both the input Ly$\alpha$ profile and the total flux can affect the hydrogen population of the n=2 state and thus the H$\alpha$ transmission spectrum. We note that the parameters adopted here could have some uncertainties. Future observation and reconstruction of the Ly$\alpha$ line will provide a more precise input.

The planetary Ly$\alpha$ emission is mainly due to the collisional excitation and recombination. Its emissivity can be approximated as :
\begin{equation}\label{eq_emiss}
S_{\nu}=\frac{(\alpha_B n_{H^+}n_e + C_{(e)1s\rightarrow2}n_{1s}n_e)}{4\pi}
\end{equation} 
where $\alpha_B$ is the case B recombination rate coefficient and $C_{(e)1s\rightarrow2} = C_{1s\rightarrow2s} + C_{1s\rightarrow2p}$.
After obtaining $B_{1s\rightarrow2p}\bar{J}_{Ly\alpha}$ (namely, the scattering rate $P\alpha$ in unit of s$^{-1}$ per atom) inside the atmosphere using LaRT, we then calculate the hydrogen level populations as described in Equation (\ref{Eq_n2s2p}). 

\begin{table}[!htbp]
   \resizebox{\textwidth}{!}   
 { \begin{threeparttable}[t]%
    \centering
    \footnotesize% fontsize
    \setlength{\tabcolsep}{3pt}% column separation
    \renewcommand{\arraystretch}{1.3}%row space 
\caption{Reaction rates related to H(2s) and H(2p).} 
   \begin{tabular}{lllll}
     \hline
  Number & Reaction & Name & Rate (coefficient) & Reference\\
     \hline
     R1 & H$_{1s}$ +  $\gamma$ (Ly$\alpha$)  $\rightarrow$ H$_{2p}$ & $B_{12}\bar{J}_{Ly\alpha}$ & LaRT   &  \cite{2020ApJS..250....9S, 2022ApJS..259....3S}\\       
     R2 & H$_{2p}$ + $\gamma$ (Ly$\alpha$) $\rightarrow$ H$_{1s}$ +  2$\gamma$ (Ly$\alpha$)  & $B_{21}\bar{J}_{Ly\alpha}$ &  $ \frac{1}{3}\times B_{12}\bar{J}_{Ly\alpha}$  &  \\ 
     R3 & e$^- + p \rightarrow H + \gamma$ & $\alpha_B$ & 2.54$\times 10^{-13}(T/10^4K)^{-0.8164-0.0208log(T/10^4K)}$ $\rm cm^3\rm s^{-1}$   & \cite{2011piim.book.....D} \\
     R4 & H$_{1s} + e^- \rightarrow H_{2s} + e^- $  & $C_{1s\rightarrow2s} $ & 1.21$\times 10^{-8}(10^4 K/T)^{0.455}e^{-118400/T}$ $\rm cm^3\rm s^{-1}$  & \cite{2013ApJ...772..144C,Janev2003} \\
     R5 & H$_{2s} + e^- \rightarrow H_{1s} + e^- $  & $C_{2s\rightarrow1s} $ & 1.21$\times 10^{-8}(10^4 K/T)^{0.455}$ $\rm cm^3\rm s^{-1}$ & \\
     R6 & H$_{1s} + e^- \rightarrow H_{2p} + e^- $ & $C_{1s\rightarrow2p}$ & 1.71$\times 10^{-8}(10^4 K/T)^{0.077}e^{-118400/T}$ $\rm cm^3\rm s^{-1}$   &   \cite{2013ApJ...772..144C,Janev2003}\\
     R7 & H$_{2p} + e^- \rightarrow H_{1s} + e^- $ & $C_{2p\rightarrow1s}$ & $\frac{1}{3}\times$1.71$\times 10^{-8}(10^4 K/T)^{0.077}$ $\rm cm^3\rm s^{-1}$  &  \\
     R8 & H$_{2s} + e^- \rightarrow H_{2p} + e^- $  & $C_{(e)2s\rightarrow2p} $ & See Equations (3-4) in the reference &  \cite{1955PPSA...68..457S}\\
     R9 & H$_{2p} + e^- \rightarrow H_{2s} + e^- $  & $C_{(e)2p\rightarrow2s} $ & $\frac{1}{3}\times C_{(e)2s\rightarrow2p}$  &  \\
     R10 &  H$_{2s}$ + H$^+ \rightarrow$ H$_{2p}$ + H$^+ $ & $C_{(H^+)2s\rightarrow2p}$ & See Equations (3-4) in the reference & \cite{1955PPSA...68..457S} \\
     R11 &  H$_{2p}$ + H$^+ \rightarrow$ H$_{2s}$ + H$^+ $ & $C_{(H^+)2p\rightarrow2s}$ &   $\frac{1}{3}\times C_{(H^+)2s\rightarrow2p}$  & \\ 
     R12 & H$_{2s} + \gamma \rightarrow$ H$^+ + e^- $ & $\Gamma_{2s}$&  See Equation (\ref{Eq_Gamma2s2p}) & \\
     R13 &  H$_{2p} + \gamma \rightarrow$ H$^+ + e^- $ & $\Gamma_{2p}$ & See Equation (\ref{Eq_Gamma2s2p})  & \\
     R14 & e$^-$ + H$^+$ $\rightarrow$ H$_{2s}$ + $\gamma $ & $\alpha_{2s}$ & $(0.282+0.047(T/10^4 K)-0.006(T/10^4 K)^2) \alpha_B$ & \cite{2011piim.book.....D} \\
     R15 &  e$^-$ + H$^+$ $\rightarrow$ H$_{2p}$ + $\gamma $ & $\alpha_{2p}$ & $\alpha_{\rm B}$-$\alpha_{2s}$ & \cite{2011piim.book.....D} \\
     R16 & H$^+ $ + 2e$^-$   $\rightarrow$ e$^-$ + H$_{2s}$  &     $\beta_{2s}n^2_{e}n_{H^+}$ & $\beta_{2s} = 1.4\times 10^{-20}\times (2/100)^4\times{(T/10^4)}^{-2}$ $ \rm cm^6 \rm s^{-1}$  & \cite{2011piim.book.....D} \\ 
    R17 &  H$^+ $ + 2e$^-$   $\rightarrow$ e$^-$ + H$_{2p}$    &  $\beta_{2p}n^2_{e}n_{H^+}$ & $\beta_{2p} = \beta_{2s}$ & \cite{2011piim.book.....D} \\ 
    R18 &  H$_{2s}$ + 2e$^- \rightarrow$ e$^-$ + H$^+ $ & $C_{(e)2s\rightarrow \infty}$ &
 $\beta_{2s}/(\frac{h^3}{(2\pi m_e k T)^{\frac{3}{2}}}e^{\frac{3.4\times1.602\times 10^{-12}}{kT}})$     & \cite{2011piim.book.....D}\\
     R19 &  H$_{2p}$ + 2e$^- \rightarrow$ e$^-$ + H$^+ $ & $C_{(e)2p\rightarrow \infty}$ & $\frac{1}{3}C_{(e)2s\rightarrow \infty}$ & \\  
     R20 & H$_{2s}$ $\rightarrow$ H$_{1s}$ + 2$\gamma$ & $A_{2s \rightarrow 1s}$ & 8.26 $\rm s^{-1}$ & \cite{2006agna.book.....O} \\
     R21 & H$_{2p}$ $\rightarrow$ H$_{1s}$ + $\gamma$ & $A_{2p \rightarrow 1s}$ & 6.3$\times$10$^{8}$ $\rm  s^{-1}$  & \cite{2006agna.book.....O} \\
     \hline
  \end{tabular}  
    \end{threeparttable} }  
\label{tab:reaction_rates}
\end{table}

\subsection{Calculation of Helium metastable populations}\label{method_He2s3}
The transitions between the $2{^3}$S and $2{^3}$P$_J$ levels of helium emit a triplet line in the near-infrared band. The wavelengths of
the triplet are 10829.09, 10830.25, and 10830.33$\rm\AA$ in the air, with the last two lines unresolved in practice \citep{1971PhRvA...3..908D}. The oscillator strengths for the three lines are 0.059902, 0.17974, and 0.29958, respectively \footnote{From NIST Atomic Spectra
Database, https://www.nist.gov/pml/atomic-spectra-database.}. 
The absorption of He 10830 is caused by the helium atoms at a metastable state He(2$^3$S). We assume that the atmosphere is in a stationary state, and the helium atoms are composed of He(1s) and He(2$^3$S).
Using the hydrodynamic simulations, we can obtain the number density structure of the total neutal helium $n_{He}$ and ionized helium $n_{He^+}$. To obtain the number density of He(2$^3$S), we solve the following equation of rate equilibrium:

\begin{equation}\label{Eq_He2s3}
\begin{aligned}
n_{He^+}n_{e}\alpha_3 + n_1n_eq_{13a} 
= n_3\Phi_3 e^{-\tau_3}  + n_3A_{31} + n_3n_eq_{31a} + n_3n_eq_{31b} + n_3n_HQ_{31}+ \frac{1}{r^2}\frac{\partial{}}{\partial{r}}(r^2n_{3}v)
\end{aligned}
\end{equation}
where $n_1 = n_{He} - n_3$. $n_1$ and $n_3$ are the number densities of He(1s) and He(2$^3$S). $\alpha_3$ and $\Phi_3$ are, respectively, the recombination rate coefficient and photoionization rate of He(2$^3$S). $\tau_3$ is the optical depth of He(2$^3$S) calculated using flux averaged cross section as described by \cite{2018ApJ...855L..11O}.
In our simulation, the value of $\tau_3$ is so small that $e^{-\tau_3}$ is very close to 1. Therefore, this attenuation factor of the radiation in the photoionization rate is ignored.
Here the incident ultraviolet spectrum (912-2600$\rm\AA$) that ionizes He(2$^3$S) is also taken from the MUSCLES Treasure Survey, as shown in Figure \ref{XUVSED} (b). The integrated flux in this FUV and NUV band is about 2.6$\times 10^5$ erg cm$^{-2}$ s$^{-1}$.
$A_{31}$ is the radiative transition rate between the 2$^3$S and 1s state. $q_{13a}$, $q_{31a}$, and $q_{31b}$ are the collision transition rate coefficients due to electrons between 1s and 2$^3$S, 2$^3$S and 2$^1$S, 2$^3$S and 2$^1$P state, respectively. $Q_{31}$ is the collisional rate coefficient due to hydrogen atoms. 
$\frac{1}{r^2}\frac{\partial{}}{\partial{r}}(r^2n_{3}v)$ is the advection term of He(2$^3$S).
Note that the values of the advection terms can be either positive or negative. In Table \ref{tab:reaction_rates_He}, we list the reaction rates and coefficients related to the helium metastable population. 
By solving the equation, we can obtain the population of helium.

Here, we note that the advection terms of H(2p), H(2s), and He(2$^3$S) in Equations (\ref{Eq_2p}), (\ref{Eq_2s}), and (\ref{Eq_He2s3}) are negligible compared to other
dominant sources and sink terms. We, therefore, ignored them in our calculations. For further discussion, see Sections \ref{sub:H2_pop} and \ref{sub:He_pop}.

\begin{table}[!htbp]
  \begin{threeparttable}[t]%
    \centering
    \footnotesize% fontsize
    \setlength{\tabcolsep}{3pt}% column separation
    \renewcommand{\arraystretch}{1.3}%row space 
\caption{Reaction rates related to He(2$^3$S).}    
   \begin{tabular}{lllll}
     \hline
  Number & Reaction & Name & Rate (coefficient)\tnote{(1)} & Reference\\
     \hline
     R1 &  He$^+$ + e$^-$ $\rightarrow$  He$_{2^3S}$  & $\alpha_3$ & $ 2.10\times10^{-13}(T/10^4)^{-0.778}\rm cm^{3}\rm s^{-1}$   & \cite{1999ApJ...514..307B}  \\  
     R2 &   He$_{2^3S}$ + $h\nu$ $\rightarrow$ He$^+$ + e$^-$  & $\Phi_3e^{-\tau_3}$ & This work   &  \cite{1971JPhB....4..652N} \\   
     R3 &    He$_{1s}$  + e$^-$ $\rightarrow$ He$_{2^3S}$ + e$^-$  & $q_{13a}$ & $2.10\times 10^{-8}\sqrt{\frac{13.6 eV}{kT}}e^{-\frac{-19.81}{kT}} \gamma_{13a} \rm cm^{3}\rm s^{-1}$ &  \cite{2000AAS..146..481B} \\  
     R4 &   He$_{2^3S}$ + e$^-$ $\rightarrow$ He$_{2^1S}$  + e$^-$  & $q_{31a}$ &  $2.10\times 10^{-8}\sqrt{\frac{13.6 eV}{kT}}e^{-\frac{-0.8 eV}{kT}} \frac{ \gamma_{31a}}{3} \rm cm^{3}\rm s^{-1}$  &    \cite{2000AAS..146..481B}  \\   
     R5 &   He$_{2^3S}$ + e$^-$ $\rightarrow$ He$_{2^1P}$  + e$^-$  & $q_{31b}$ &  $2.10\times 10^{-8}\sqrt{\frac{13.6 eV}{kT}}e^{-\frac{-1.4 eV}{kT}} \frac{ \gamma_{31b}}{3} \rm cm^{3}\rm s^{-1}$  &    \cite{2000AAS..146..481B}   \\  
     R6 &   He$_{2^3S}$  $\rightarrow$ He$_{1s}$  + h$\nu$ & $A_{31}$ &  $1.272\times10^{-4}\rm s^{-1}$ & \cite{1971PhRvA...3..908D} \\  
     R7 &   He$_{2^3S}$ + H $\rightarrow$ He$_{1s}$  + H & $Q_{31}$ &  $5.0\times10^{-10}\rm cm^{3}\rm s^{-1}$ &  \cite{1982ApJ...255..489R} \\         
     \hline
  \end{tabular}
    \begin{tablenotes}
    \item[(1)] Note here $\gamma_{13a}$, $\gamma_{31a}$, and $\gamma_{31b}$ are the effective collision strength between  1s and 2$^3$S, 2$^3$S and 2$^1$S, 2$^3$S and 2$^1$P state, respectively \citep{2000AAS..146..481B}.
  \end{tablenotes}
  \end{threeparttable}%  
\label{tab:reaction_rates_He}
\end{table}

\section{radiative transfer } \label{sec:cal_RT}

\subsection{Monte Carlo simulation of Ly$\alpha$ radiative transfer }\label{sec:LaRT}
We use LaRT, a Ly$\alpha$ Monte Carlo radiative transfer code, to simulate the Ly$\alpha$ resonant scattering in the planetary atmosphere. The code has been used to solve Ly$\alpha$ radiative transfer in the diffuse interstellar medium or circumgalactic medium \citep{2020ApJS..250....9S, 2020ApJ...901...41S, 2022ApJS..259....3S}, where the density of hydrgen is relatively low, and the Ly$\alpha$ source is usually inside hydrogen gas. The situation is different for the close-in exoplanet systems because the stellar Ly$\alpha$ is an external source and the planetary atmosphere has a relatively high number density. Therefore, we modified the code to make it applicable in the exoplanet systems.

The previous Ly$\alpha$ Monte Carlo simulations for the exoplanetary atmosphere assumed that both the stellar Ly$\alpha$ source and the planetary atmosphere are plane-parallel \citep{2017ApJ...851..150H}. However, because the distance between the star and planet is only a few stellar radii for close-in planets, the stellar Ly$\alpha$ radiation cannot be considered plane-parallel. In addition, the plane-parallel atmosphere assumption is likely to cause differences from the spherical atmosphere.
Therefore in this work, we use mainly a spherical stellar Ly$\alpha$ source and a spherical atmosphere assumption. This model will be called a spherical illumination (or spherical-spherical) model. 
In Section \ref{sec: Diss}, we will also discuss two additional cases concerning the geometries of the stellar Ly$\alpha$ source and planetary atmosphere. The two cases are (1) plane-parallel stellar Ly$\alpha$ source + plane-parallel atmosphere (plane-plane model), and (2) plane-parallel stellar Ly$\alpha$ source +  spherical atmosphere (plane-spherical model). Table \ref{tab:Lya_models} defines these models.
The main difference in these models is how the incident Ly$\alpha$ photons in LaRT are sampled or generated.

\begin{table}[!hp]
   \begin{threeparttable}[t]
    \centering
    \footnotesize% fontsize
    \setlength{\tabcolsep}{8pt}% column separation
    \renewcommand{\arraystretch}{1.3}%row space 
\caption{Models defined in LaRT}    
   %\begin{tabular}{c|c|c|c|c|c|c|c|c|c}
   \begin{tabular}{l|l|l}
     \hline
 Models\tnote{(1)} & Geometry of the stellar Ly$\alpha$ source & Geometry of the planetary atmosphere \\
 \hline
spherical illumination model\tnote{(2)} & spherical & spherical \\
plane-plane model & plane-parallel & plane-parallel \\
plane-spherical model & plane-parallel & spherical \\
 \hline
   \end{tabular}
  \begin{tablenotes}
    \item[(1)]In this work, the main results are obtained using the spherical illumination model. The plane-plane and plane-spherical models will be discussed in Section \ref{sec: Diss}.
   \item[(2)]This model can also be called as a spherical-spherical model.
  \end{tablenotes}
  \end{threeparttable}%
\label{tab:Lya_models}
\end{table}

In the following, we introduce the spherical illumination model. The model can be described in two parts: the generation of Ly$\alpha$ photons, mainly related to the system geometry, and the radiative transfer of the photons in the medium. In this paper, we will mainly discuss the first part, i.e., how to generate photons efficiently. For the discussion of the second part, one can refer to \cite{2020ApJS..250....9S} and \cite{2022ApJS..259....3S}.

\begin{figure}[t]
\begin{center}
   \includegraphics[scale=0.6]{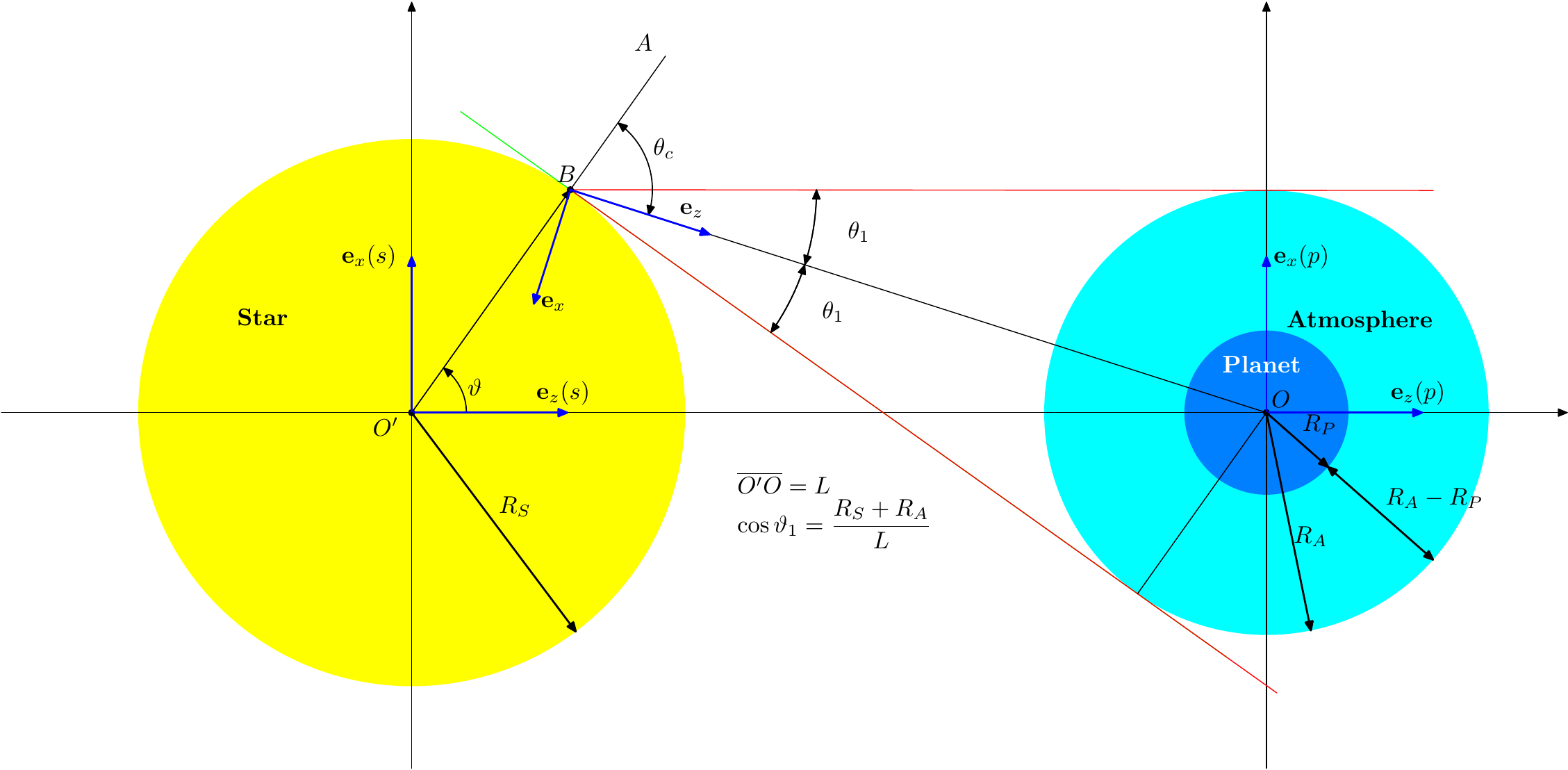}
  \includegraphics[scale=0.6]{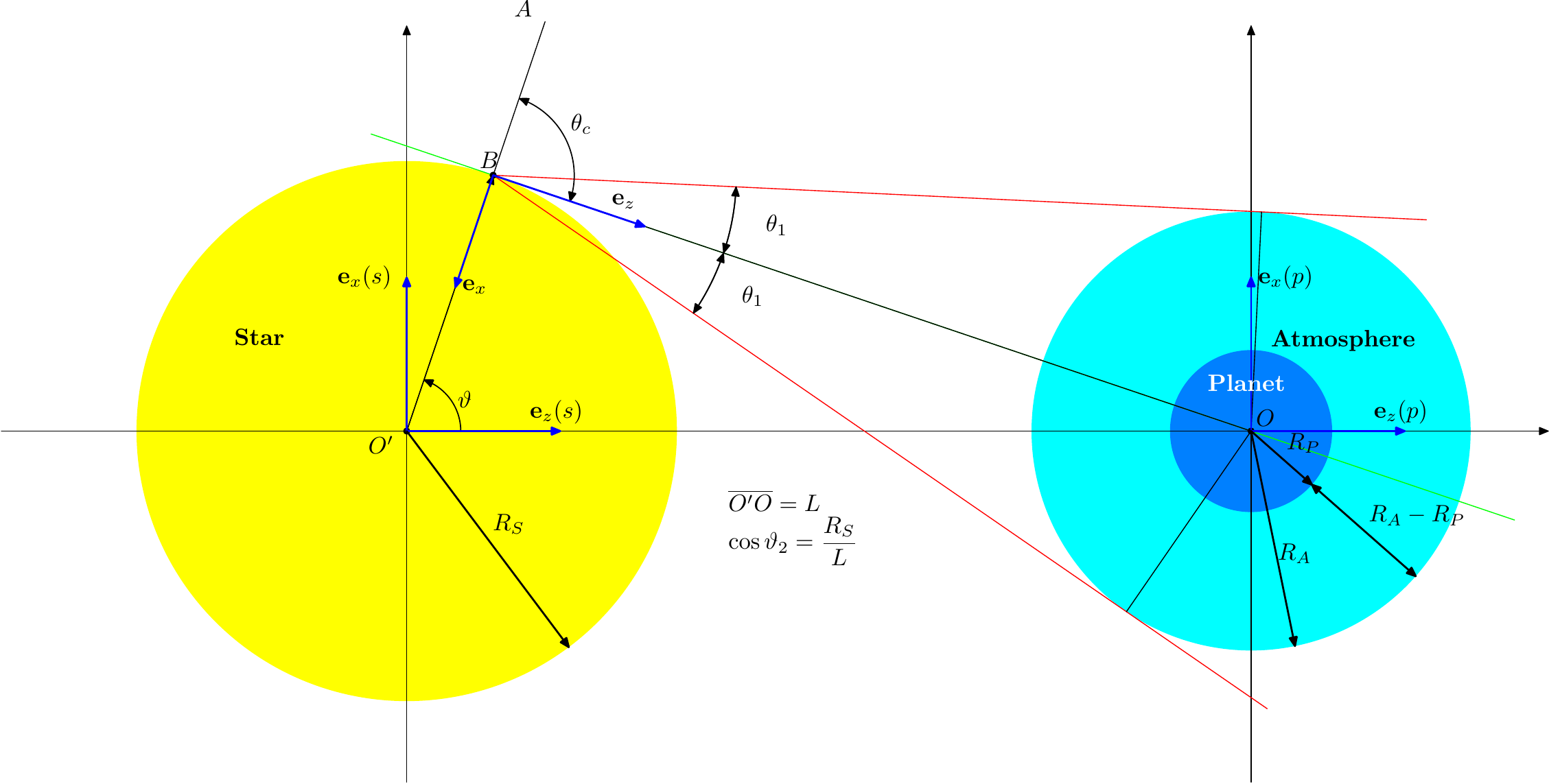}
  \includegraphics[scale=0.6]{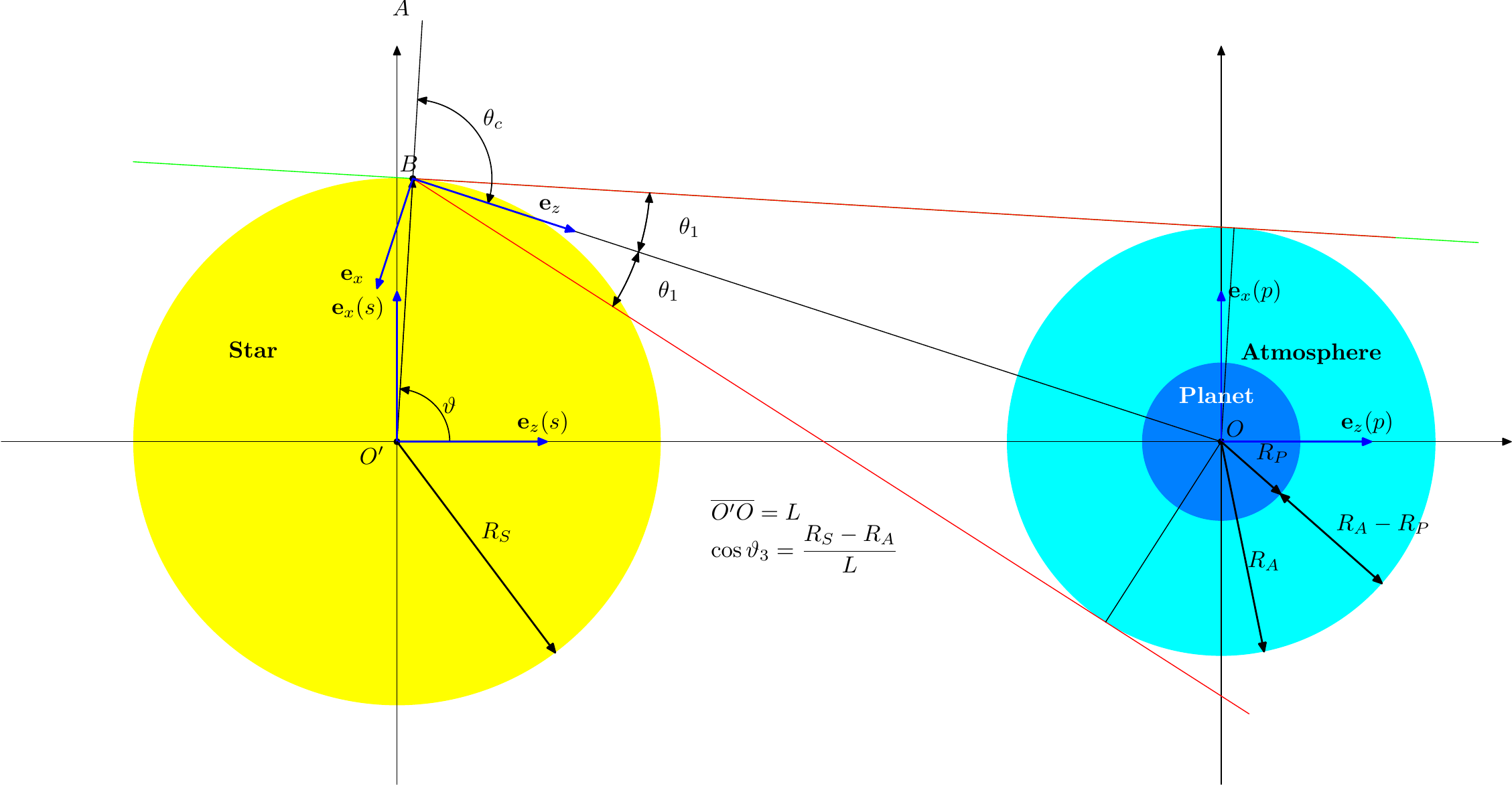} 
%\vspace{1cm}
\caption{Three critical angles $\vartheta_1, \vartheta_2, \vartheta_3$.
Here, the green line represents the tangent plane of the stellar
sphere at the emission point $B$. $R_S$, $R_P$, $R_A$ are the radii of the star, planet, and atmosphere, respectively. $\mathbf{e}_i(s)$ and $\mathbf{e}_i(p)$ are the basis vectors of the rest frames of the star and the planet, respectively. } \label{fig1}
\end{center}
\end{figure}

To sample the initial photons, we first construct the basic reference frames (Figure \ref{fig1}). They are the rest frames of the star and planet, the basis vectors of which are denoted by $\mathbf{e}_i(s)$ and $\mathbf{e}_i(p)$, where $i=x, y, z$. The origins of $\mathbf{e}_i(s)$ and $\mathbf{e}_i(p)$ coincide with the centers of the star and planet, respectively. The basis vectors $\mathbf{e}_z(s)$ and $\mathbf{e}_z(p)$ are both along the line of $\overline{O'O}$ and $\mathbf{e}_i(s)$ is parallel to $ \mathbf{e}_i(p)$. The radii of the star and the planet ($R_P$) plus its atmosphere, and the distance between the centers of the star and planet are denoted by $R_S, R_A$, and $L$, respectively.
In our fiducial model, 
the photons are assumed to be emitted isotropically from the surface of the stellar. The polar and azimuth angles of the photon emission site on the stellar surface are denoted by $(\vartheta, \varphi)$ and those of the initial photon propagation direction are $(\theta, \phi)$. 

Now, we describe how to sample the position and propagation vectors as following. First, we sample the polar angle $\vartheta$ (or $\bar{\mu} = \cos\vartheta$) of the photon emission position on the stellar surface according to the probability distribution function proportional to $\Delta S(\vartheta)$, the solid angle subtended by photon rays toward the planetary atmosphere, as described in Section \ref{appendix_sec1}. Next, the azimuth angle $\varphi$ is obtained by $\varphi = 2\pi\xi$ using a uniform random number $\xi$ between 0 and 1.
Then the position vector of emission site in the frame $\mathbf{e}_i(s)$ can be expressed as
\begin{eqnarray}
   \mathbf{r}_{\text{em}} = R_S\hat{\mathbf{r}}_{\text{em}} = R_S(\sqrt{1-\bar\mu^2}\cos\varphi,
\sqrt{1-\bar\mu^2}\sin\varphi, \bar{\mu}),
\end{eqnarray} 
where $\hat{\mathbf{r}}_{\text{em}}=\mathbf{r}_{\text{em}}/R_S$ is the unit normal vector of the stellar surface at $\mathbf{r}_{\text{em}}$. In the frame $\mathbf{e}_i(p)$, the position vector can be expressed as $\mathbf{r} = \mathbf{r}_{\rm em} - \mathbf{L}$, where $\mathbf{L}=(0, 0, L)$.

Before sampling the initial photon direction, we introduce three critical angles defined by
\begin{eqnarray}
  \cos\vartheta_1=\frac{R_S+R_A}{L}, \,\,\cos\vartheta_2=\frac{R_S}{L}, \,\,\cos\vartheta_3=\frac{R_S-R_A}{L}.
\end{eqnarray} 
The geometrical meaning of the three angles is illustrated in Figure \ref{fig1}.
The first critical angle $\vartheta_1$ is the angle when the tangent plane of the star at emission point $B$ is tangent to the surface of the planetary atmosphere. When $0 \leq\vartheta \leq \vartheta_1$, the photons emitted from point $B$ toward the planetary atmosphere will not be blocked by the star; in other words, all the photons emitted within the cone of $\theta_1$ (the cone defined by the two red lines) can arrive at the planetary atmosphere.
$\vartheta_2$ is the second critical angle when the tangent plane of the star at emission point $B$ goes through the center of the planet. At $\vartheta = \vartheta_2$, only half of the photons in the cone of $\theta_1$ can arrive at the planetary atmosphere.
And the third critical angle $\vartheta_3$ is another angle when the tangent plane of the star at emission point $B$ is tangent to the planetary atmosphere. When $\vartheta \geq \vartheta_3$, no photons emitted from point $B$ can arrive at the planetary atmosphere.

In choosing a random initial propagation direction $\mathbf{p}$ of the photon, it is convenient first to sample the directions for a constant flux $dF = I\cos\theta dA={\rm constant}$, where $\theta$ is the angle between the normal vector of the stellar surface and the direction vector, and $dA$ a small area element on the surface. We then multiply a factor
proportional to $\cos\theta$ to the photon weight to yield an isotropic radiation field ($I={\rm constant}$; this is because $I$ is sampled to be inversely proportional to $\cos\theta$ when $dF ={\rm constant}$ is assumed).
To do so, we should first construct a triad at $\mathbf{r}_{\rm em}$. The basis vectors of this triad is denoted by $\mathbf{e}_i$. A photon direction defined with respect to this triad is denoted by $(\mu, \phi)$, where $\mu=\cos\theta$. The choice of the triad can be arbitrary.
For example, we can simply set $\mathbf{e}_i = \mathbf{e}_i(s)$ and sample the direction by $\mu=\cos\theta=-1+2\xi_1, \,\,\phi=2\pi\xi_2$, where $\xi_1$ and $\xi_2$ are mutually independent, uniformly-distributed random numbers between 0 and 1. Then, we have $\mathbf{p}=p_x\mathbf{e}_x+p_y\mathbf{e}_y+p_z\mathbf{e}_z$, where $p_x = \sqrt{1-\mu^2}\cos\phi, \,p_y = \sqrt{1-\mu^2}\sin\phi$, and $p_z = \mu$. We next accept only the rays that hit the planetary atmosphere.
This procedure is quite simple but has a drawback; the acceptance rate is quite low since a large portion of trajectories of photons generated will either miss the planetary atmosphere or go toward the stellar interior. The acceptance rate equals $\Delta S(\vartheta)/4\pi$, where $\Delta S(\vartheta)$ is the solid angle formed by all effective directions that the photons can arrive at the planetary atmosphere in the $\mu$-$\phi$ plane. In other words, the number $\Delta N(\vartheta)$ of accepted samples at $\vartheta$ is proportional to $\Delta S(\vartheta)/4\pi$.

To overcome this shortage, we adopt an alternative method, in which the basis vectors are constructed as follows: 
$\mathbf{e}_z = -\hat{\mathbf{r}}=\mathbf{r}/|\mathbf{r}|$, $\mathbf{e}_y =
\hat{\mathbf{r}}_{\text{em}}\times\mathbf{e}_z/|\hat{\mathbf{r}}_{\text{em}}\times\mathbf{e}_z|=\hat{\mathbf{r}}_{\text{em}}\times\mathbf{e}_z/\sin\theta_c$, and $\mathbf{e}_x = \mathbf{e}_y\times\mathbf{e}_z$. For any emission site,
the $z$-axis of this triad always points toward the center of the planet. From Figure \ref{fig1}, one can see that the solid angle subtended by the planetary atmosphere with respect to this triad is related to the angle $\theta_1$ given by
\begin{eqnarray}\label{eq_mu1}
 \mu_1 = \cos\theta_1 = \frac{\sqrt{BO^2-R_A^2}}{BO},
\end{eqnarray}
where $BO=\sqrt{L^2+R_S^2-2R_SL\bar{\mu}}$. In the $\mu$-$\phi$ plane of the reference frame $\mathbf{e}_i$, the solid angle that photons could reach the planetary atmosphere will form a rectangle $\rm A=\{(\mu, \phi)|\mu_1\le \mu \le 1,\, -\pi\le \phi\le\pi\}$. On the other hand, the area B formed by the rays that will not be blocked by the star is determined by the condition $\hat{\mathbf{r}}_{\text{em}}\cdot \mathbf{p}_{\text{em}}\ge 0$, and $\hat{\mathbf{r}}_{\text{em}}\cdot \mathbf{p}_{\text{em}} = 0$ gives the boundary. Expanding this equation, we obtain the expression for the boundary curve by solving 
$\sqrt{1-\mu^2}\sin\theta_c\cos\phi-\mu\cos\theta_c=0$:
\begin{eqnarray}\label{exp_mu_phi}
   \begin{split}
   \mu(\phi) = \left\{\begin{array}{lr}
      \displaystyle\frac{\sin\theta_c\cos\phi}{\sqrt{1-\sin^2\theta_c\sin^2\phi}}, &\quad \text{if  } \vartheta<\vartheta_2,\\
      \displaystyle\frac{-\sin\theta_c\cos\phi}{\sqrt{1-\sin^2\theta_c\sin^2\phi}},& \quad \text{if  } \vartheta>\vartheta_2.
     \end{array}\right.
   \end{split}
\end{eqnarray}
where $\theta_c$ is the angle between the vectors $\hat{\mathbf{r}}_{\text{em}}$ and $\mathbf{e}_z$ given by
\begin{eqnarray}
   \begin{split}
   \theta_c=\left\{\begin{array}{llr}
     &\displaystyle \arcsin\left(\frac{L}{BO}\sin\vartheta\right), &\quad \text{if  } \vartheta<\vartheta_2, \\
     &\displaystyle \pi-\arcsin\left(\frac{L}{BO}\sin\vartheta\right), &\quad \text{if  } \vartheta\ge\vartheta_2.
     \end{array}\right.
   \end{split}
\end{eqnarray}

\begin{figure}[t]
\begin{center}
  \includegraphics[scale=0.5]{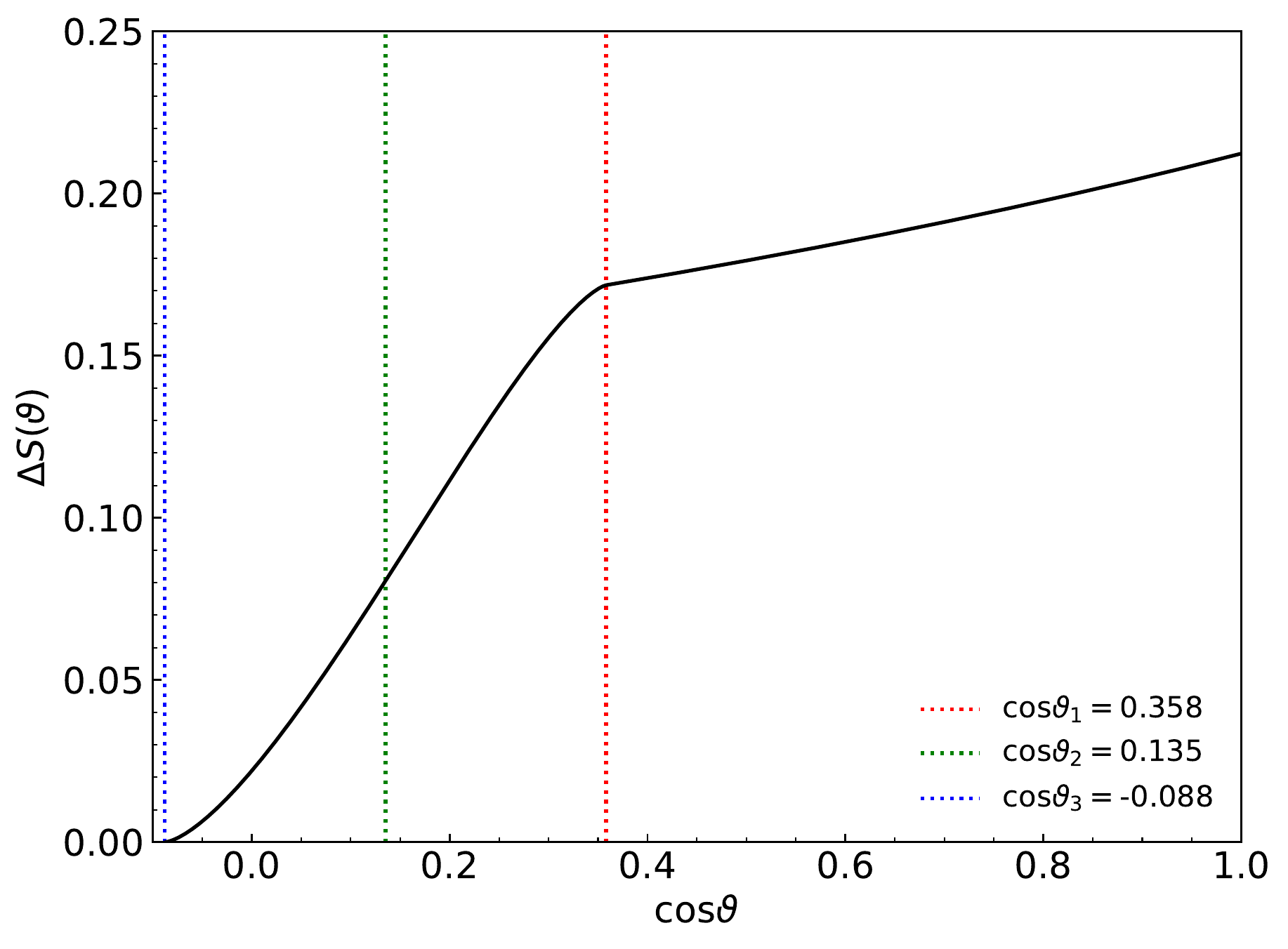}
%\vspace{1cm}
\caption{The solid angle $\Delta S(\vartheta)$ of the planet subtended by a point
located at a polar coordinate $\vartheta$ on the stellar surface. The dotted vertical lines denote the three critical angles defined in the text: cos$\vartheta_1$ (red), cos$\vartheta_2$ (green), and cos$\vartheta_3$ (blue).
Here, $R_S$ = 6.06697 $R_P$ , $R_A$ = 10 $R_P$, and $L$ = 44.8611856 $R_P$ for WASP-52b.} \label{fig3}
\end{center}
\end{figure}

Then, the intersection of A and B, $\rm C =A \cap B$, gives the valid area whose points corresponding to the photon directions that could reach the planet asmosphere eventually. The area C is a useful quantity in our sampling algorithm and is denoted by $\Delta S(\bar{\mu})$, which is dependent on the position of the emission site and hence a function of $\bar{\mu}$ (or $\vartheta$). The functional form of $\Delta S(\bar{\mu})$ is given by:

\begin{eqnarray}\label{deltaSmu1}
   \begin{split} 
     \Delta S(\vartheta) =
        \left\{\begin{array}{lll}
           &\displaystyle 2\pi(1-\mu_1), \quad\quad & 0\le \vartheta <\vartheta_1,\\
           &\displaystyle 2\pi(1-\mu_1)-2 \int_0^{\phi_1}\mu(\phi)d\phi+2\mu_1\phi_1, 
                 \quad\quad & \vartheta_1\le\vartheta < \vartheta_2,\\
           &\displaystyle 2 \int_{\phi_1}^\pi\mu(\phi)d\phi - 2(\pi-\phi_1)\mu_1, 
                 \quad\quad &  \vartheta_2\le\vartheta < \vartheta_3. 
        \end{array}\right.
   \end{split}
\end{eqnarray}
where $\mu_1 = \cos\theta_1$ as given by Equation (\ref{eq_mu1})
and $\phi_1$ is the root of equation $\mu(\phi)=\mu_1$, given by
\begin{eqnarray}\label{eq_phi1}
   \begin{split}
   \phi_1=\left\{\begin{array}{llr}
       &  \displaystyle \arccos\left(\frac{\mu_1\cos\theta_c}{\sqrt{1-\mu_1^2}\sin\theta_c}\right),& \quad \text{if  } \vartheta<\vartheta_2,\\
       &  \displaystyle \pi-\arccos\left(\frac{\mu_1|\cos\theta_c|}{\sqrt{1-\mu_1^2}\sin\theta_c}\right), &\quad \text{if  } \vartheta>\vartheta_2.
     \end{array}\right.
   \end{split}
\end{eqnarray}
Since $\displaystyle\int \mu(\phi)d\phi=\arcsin(\sin\theta_c\sin\phi)$ using the expression of function $\mu(\phi)$ given by Equation \eqref{exp_mu_phi}, we get
\begin{eqnarray}\label{deltaSmu2}
   \begin{split} 
     \Delta S(\vartheta) =
        \left\{\begin{array}{lll}
           &\displaystyle 2\pi(1-\mu_1), \quad\quad & 0\le \vartheta <\vartheta_1,\\
           &\displaystyle 2\pi(1-\mu_1)-2 \arcsin(\sin\theta_c\sin\phi_1)+2\mu_1\phi_1, 
                 \quad\quad & \vartheta_1\le\vartheta < \vartheta_2,\\
           &\displaystyle 2 \arcsin(\sin\theta_c\sin\phi_1) - 2(\pi-\phi_1)\mu_1, 
                 \quad\quad & \vartheta_2\le\vartheta < \vartheta_3.
        \end{array}\right.
   \end{split}
\end{eqnarray} 
Figure \ref{fig3} shows the curve of function $\Delta S(\vartheta)$. One can see that as $\vartheta$ increases, the area decreases monotonically.

For a given location on the stellar surface, determined by $(\vartheta$, $\varphi)$, the angles $(\theta$, $\phi)$ for the photon propagation direction vector are obtained as follows.
The algorithm to sample the initial photon direction $\mathbf{p}$ is divided into three cases:\\

\begin{figure*}[t]
\begin{center}
  \includegraphics[scale=0.7]{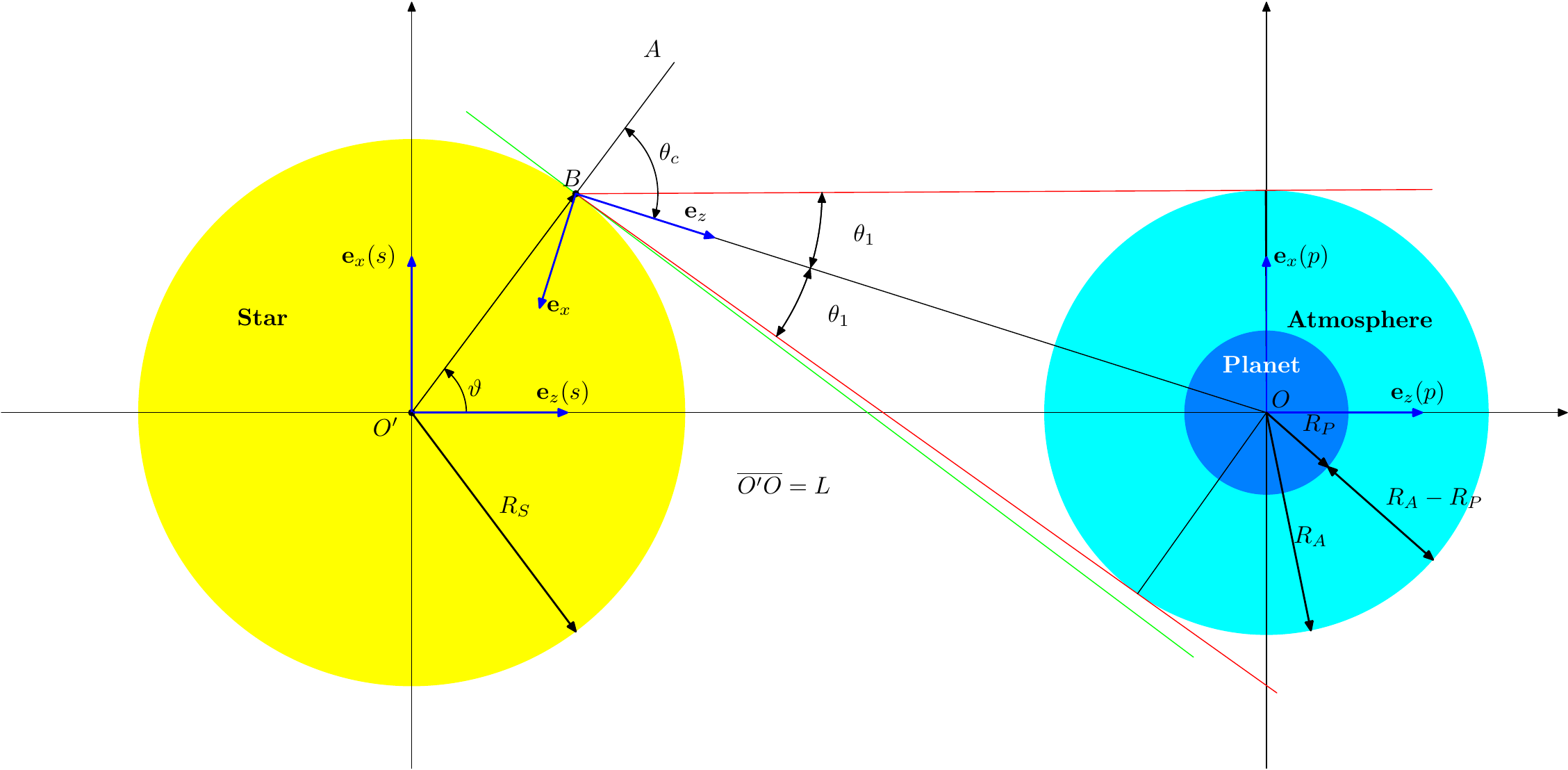}
  \includegraphics[scale=1.2]{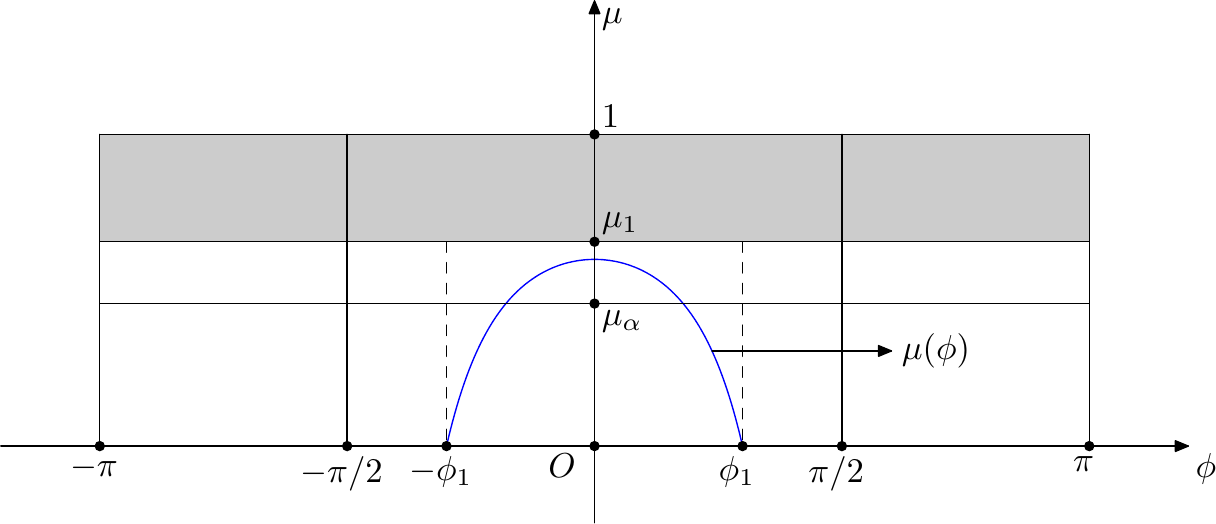}
%\vspace{1cm}
\caption{An illustration of Case 1, as $0\le\vartheta\le \vartheta_1$. The upper panel is a schametic of the photon generation. The lower panel shows the $\mu-\phi$ plane in which the solid angle of the effective photons is denoted by the grey area. The blue line represents the curve $\mu(\phi)$. The gray area denotes the valid region in the $\mu-\phi$ plane. Here, $\mu_{\alpha} = \cos\theta_1$ when $\vartheta = 0$.} \label{appendix1s}
\end{center}
\end{figure*}

{\bf{Case 1} ($0\le\vartheta\le \vartheta_1$) : }In this case, the tangent plane of the star at $\mathbf{r}_{\rm em}$ does not intersect with the planetary atmosphere. The geometrical relationships between the emission site and the planetary atmosphere, and the corresponding valid region in the $\mu$-$\phi$ plane are shown in the upper and lower panels of Figure \ref{appendix1s}, respectively. From this figure, we can sample the photon direction by:
\begin{eqnarray}
   \begin{split}
    \begin{array}{l}
         \mu = \mu_1 + (1-\mu_1)\xi_1,\\
         \phi = - \pi + 2\pi\xi_2.
     \end{array}
   \end{split}
\end{eqnarray}

\begin{figure*}[t]
\begin{center}
  \includegraphics[scale=0.7]{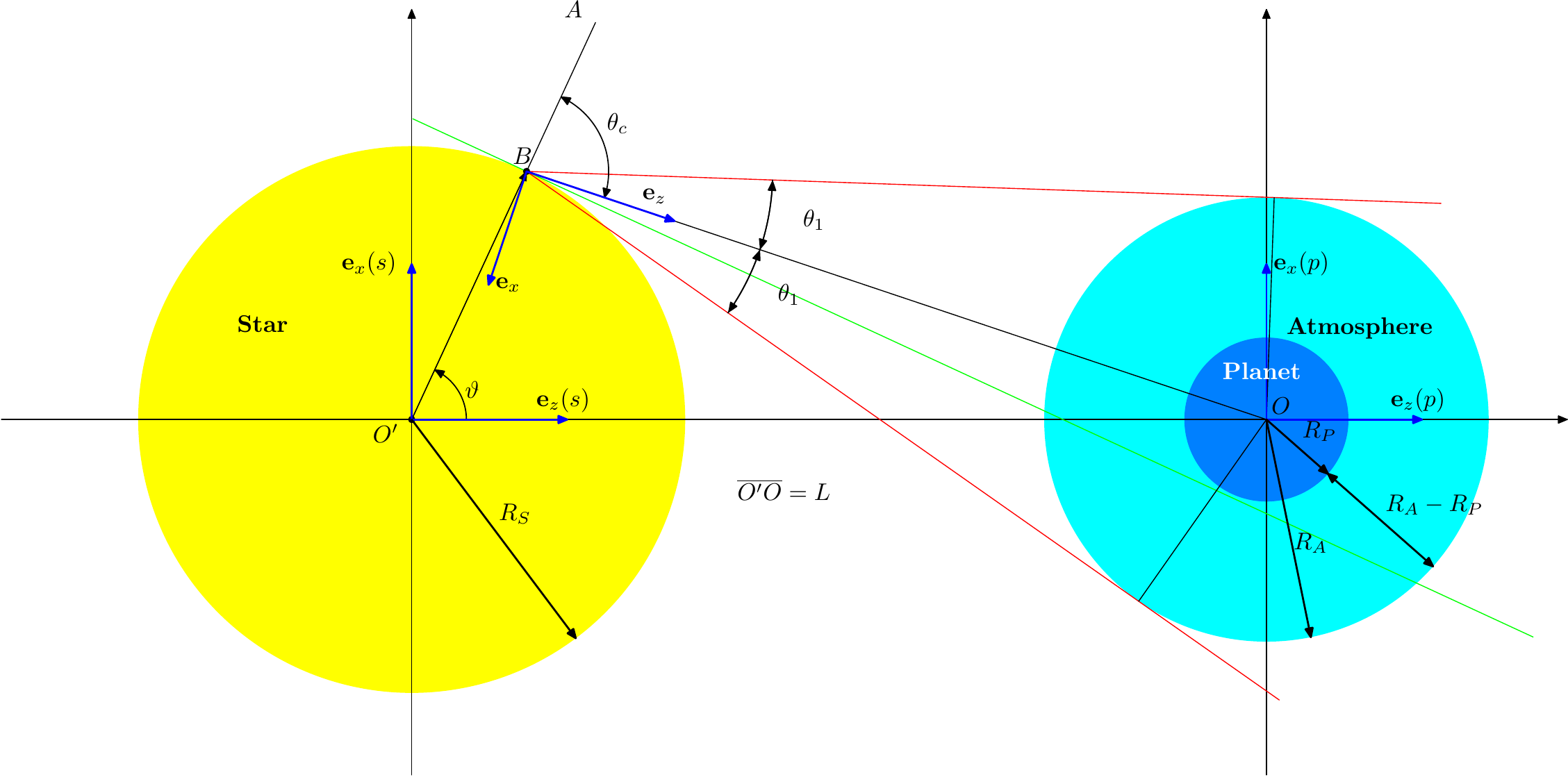}
  \includegraphics[scale=1.2]{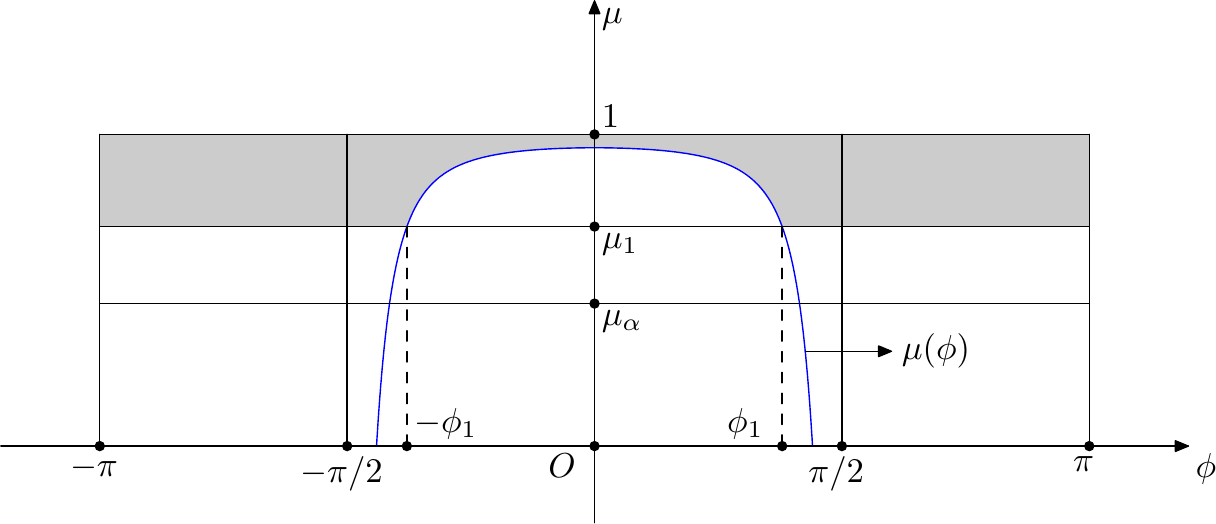}
%\vspace{1cm}
\caption{The same as in Figure \ref{appendix1s}, but for Case 2, as $\vartheta_1\le\vartheta\le \vartheta_2$. } \label{appendix1s2}
\end{center}
\end{figure*}

{\bf{Case 2} ($\vartheta_1\le\vartheta\le \vartheta_2$) :} In this case, the tangent plane of the star at $\mathbf{r}_{\rm em}$ intersects with the `first' half of the planetary atmosphere that is divided equally by the plane $BO$. The geometry and the area of the valid region are shown in Figure \ref{appendix1s2}. The sampling algorithm is:
\begin{eqnarray}
   \begin{split}
 &\text{do}\\
 &\quad\quad\mu = \mu_1 + (1-\mu_1)\xi_1,\\
 &\quad\quad\phi = - \pi + 2\pi\xi_2.\\
 &\quad\quad \text{if not } ((-\phi_1\le\phi\le\phi_1)\text{ and } \mu\le\mu(\phi))\text{ then }\\
% &\quad\quad \text{else}\\
 &\quad\quad \quad \quad\text{exit}\\
 &\quad\quad \text{end if}\\
 &\text{end do}\\
   \end{split}
\end{eqnarray}

\begin{figure*}[t]
\begin{center}
  \includegraphics[scale=0.7]{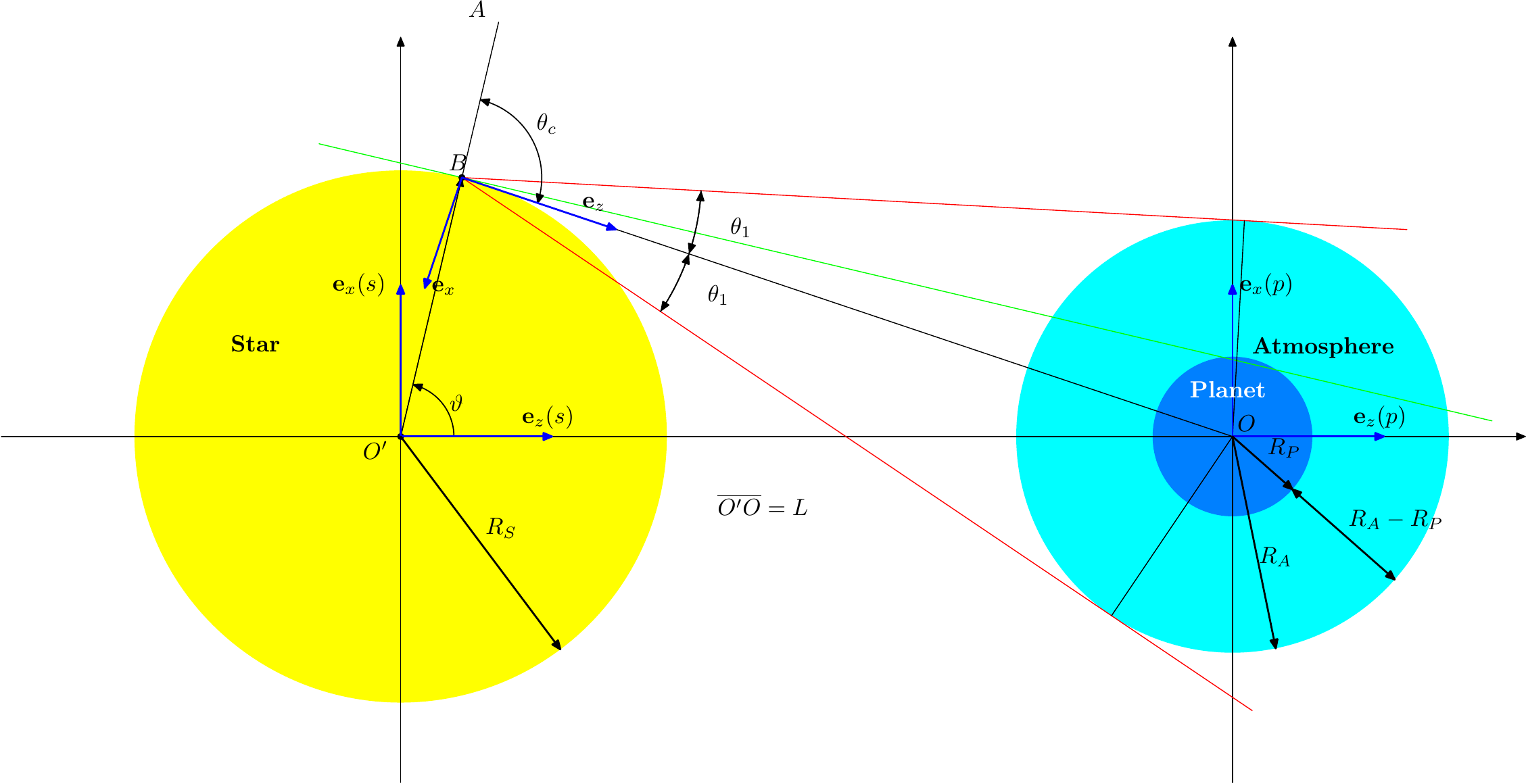}
  \includegraphics[scale=1.2]{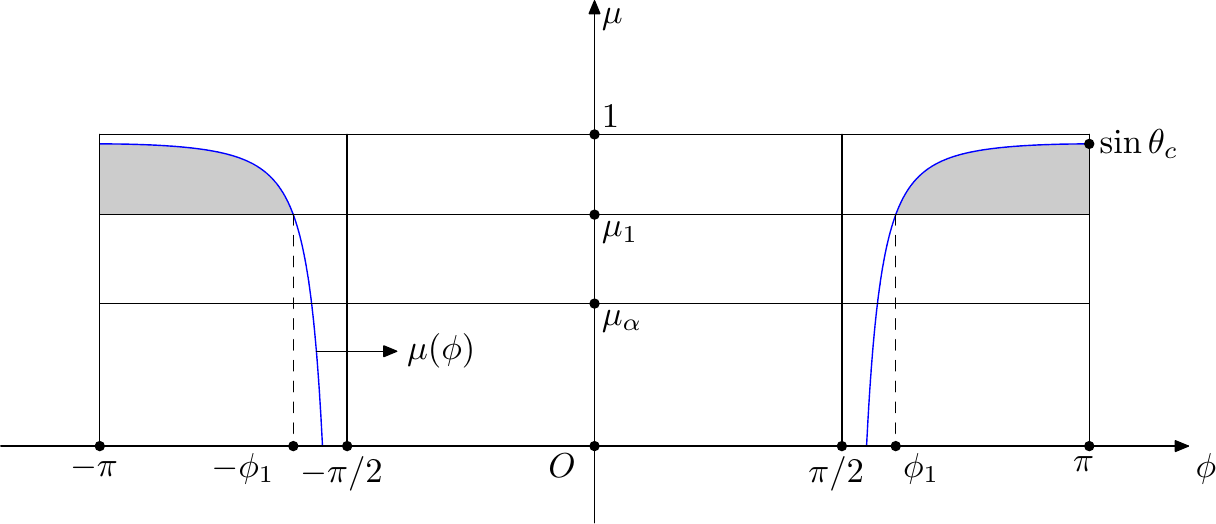}
%\vspace{1cm}
\caption{The same as in Figure \ref{appendix1s}, but for Case 3, as $\vartheta_2\le\vartheta\le \vartheta_3$. } \label{appendix1s3}
\end{center}
\end{figure*}

{\bf{Case 3} ($\vartheta_2\le\vartheta\le \vartheta_3$) :} In this case, the tangent plane of the star at $\mathbf{r}_{\rm em}$ intersects with the `second' half of the planetary atmosphere that is divided equally by the plane $BO$. The geometry and the area of the valid region are shown in Figure \ref{appendix1s3}. Then we have:

\begin{eqnarray}
   \begin{split}
 &\text{do}\\
 &\quad\quad\mu = \mu_1 + (\sin\theta_c-\mu_1)\xi_1,\\
 &\quad\quad\phi = \phi_1 + 2(\pi-\phi_1)\xi_2.\\
 &\quad\quad \text{if }(\phi>\pi) \phi = \phi - 2\pi,\\
 &\quad\quad \text{if } (\mu_1\le\mu\le\mu(\phi))\text{ then }\\
 &\quad\quad \quad \quad\text{exit}\\
 &\quad\quad \text{end if}\\
 &\text{end do}\\
   \end{split}
\end{eqnarray}

As $\mu$ and $\phi$ are obtained, the initial direction of the photon is given by
$\mathbf{p}=p_x\mathbf{e}_x+p_y\mathbf{e}_y+p_z\mathbf{e}_z$, where
$p_x = \sqrt{1-\mu^2}\cos\phi, p_y=\sqrt{1-\mu^2}\sin\phi, p_z=\mu$.

We also implemented a simpler, second method, which samples a photon direction $\mathbf{p}$ as follows:
\begin{eqnarray}    
 \begin{split}
 &\text{do}\\ 
 &\quad\quad\mu = \mu_\alpha + (1-\mu_\alpha)\xi_1,\\
 &\quad\quad\phi = -\pi + 2\pi \xi_2.\\ 
 &\quad\quad \text{if } ((\mathbf{r}\cdot\mathbf{p})^2+R_A^2-\mathbf{r}\cdot\mathbf{r}\ge 0)\text{then}\\
 &\quad\quad \quad \quad\text{exit}\\
 &\quad\quad \text{end if}\\ 
 &\text{end do} 
  \end{split}
\end{eqnarray}
where $\mu_\alpha = \sqrt{1-R_A^2/(L-R_S)^2}$ is the cosine of $\theta_1$ as $\vartheta = 0$. This second method adopts an equal sampling area $2\pi(1-\mu_\alpha)$ for any $\vartheta$, and then uses the rejection method to sample $\vartheta$. Thus, one can obtain the acceptance rate for this method, i.e.,
\begin{eqnarray}
   \begin{split}
      f_{\rm accept} =\left.\ \int_{0}^{\vartheta_3}
          \Delta S(\vartheta)\sin\vartheta d\vartheta  \right/ \int_{0}^{\vartheta_3}
          2\pi(1-\cos\alpha)\sin\vartheta d\vartheta\\
        =\left.\int_{0}^{\vartheta_3}  \Delta S(\vartheta)\sin\vartheta d\vartheta  \right/ [ 2\pi(1-\cos\alpha)(1-\cos\vartheta_3) ].
   \end{split}
\end{eqnarray} 
We have confirmed that the above two different methods give the same results. Here, we stress again that the polar and azimuth angles $(\theta, \phi)$ were sampled by assuming a constant flux on the stellar surface. The isotropic
radiation is obtained by multiplying a factor of $2 (\hat{\mathbf{r}}_{\rm em} \cdot {\mathbf{p}}_{\rm em}$) to the photon weight. 

\begin{figure*}[t]
\begin{center}
  \includegraphics[scale=0.6]{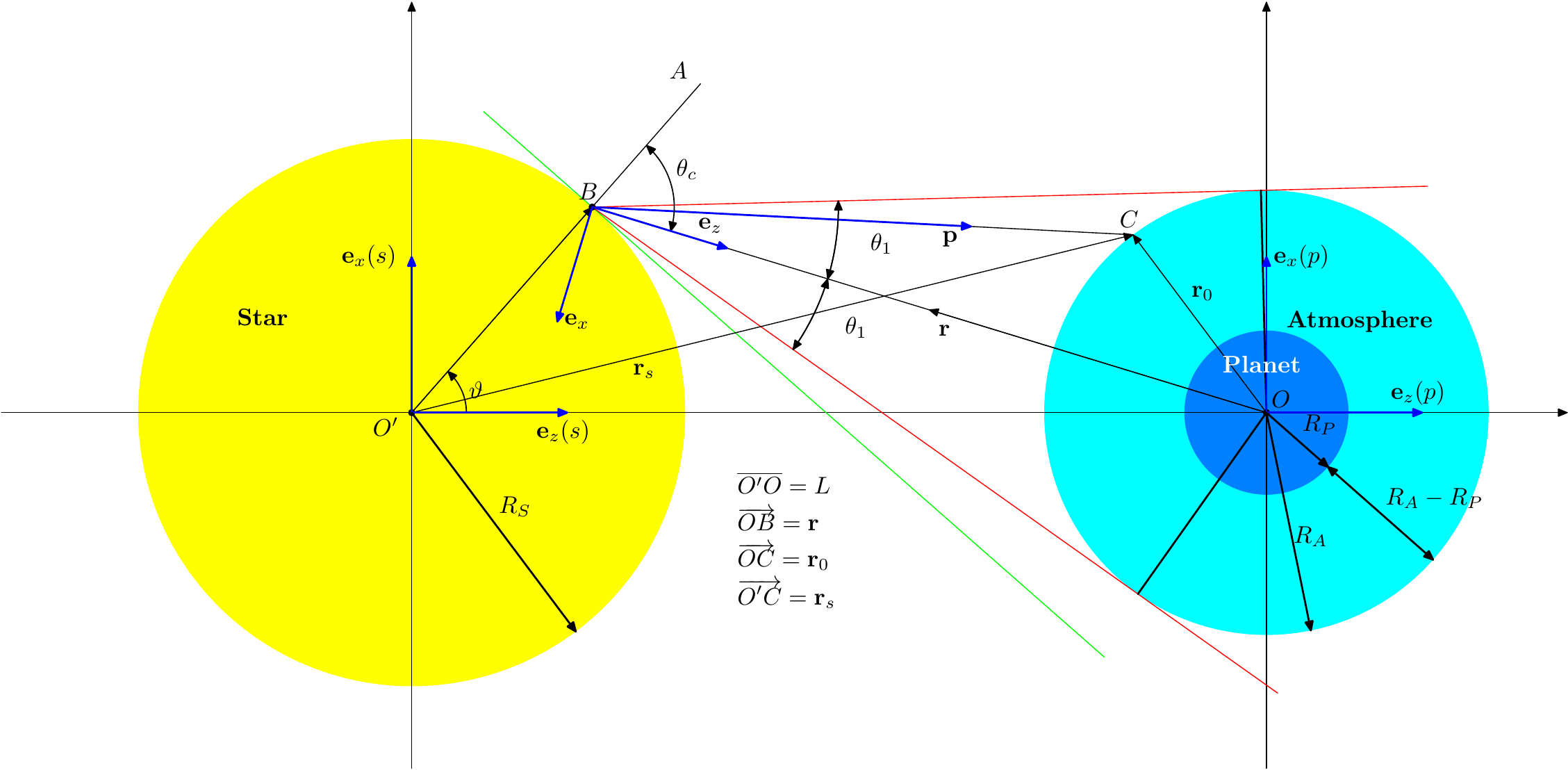}
\caption{The geometrical relationships between vectors $\mathbf{r},
\mathbf{r}_s,\mathbf{r}_0$, and $\mathbf{p}$.} \label{fig2}
\end{center}
\end{figure*}

As shown in Figure \ref{fig2}, once 
$\mathbf{r}_{\text{em}}$ and $\mathbf{p}$ are obtained, we can readily get $\mathbf{r}_0$, the intersection of the photon trajectory with the outer boundary of the atmosphere, in the reference frame $\mathbf{e}_i(p)$.
The intersecting position is given by $\mathbf{r}_0=\mathbf{r}+s\mathbf{p}$, where $s=-\mathbf{r}\cdot\mathbf{p}-\sqrt{(\mathbf{r}\cdot\mathbf{p})^2+R_A^2-\mathbf{r}\cdot\mathbf{r}}$. Here, $s$ is the distance to the point where the photon hits the planetary atmosphere.
The Ly$\alpha$ radiative transfer calculation is performed as described in \cite{2020ApJS..250....9S}, starting from this position. In our radiative transfer simulation, photons entering the planetary radius are assumed to be completely destroyed by the planet. In addition to the stellar Ly$\alpha$ photons, we also simulate the radiative transfer of Ly$\alpha$ photons emitted from the planetary atmosphere. We sample the initial radial coordinates of photons according to the probability distribution function proportional to $r^2 S_{\nu}(r)$. Here, $S_{\nu}(r)$ is the emissivity calculated using Equation ($\ref{eq_emiss}$). Then, the polar and azimuth angles of propagation vector are randomly generated in the planetary
coordinate system. The photon direction vector is sampled to be isotropic.

Now we need to calculate $\mathcal{L}_A / \mathcal{L}_{\rm S}$ (called the flux factor in this work), namely the ratio between the total luminosities
received by the planetary atmosphere and emitted by the star. This factor is required to
convert the total stellar luminosity to the luminosity incident onto the atmosphere. We have that
\begin{eqnarray}
   \begin{split}
      \mathcal{L}_{\rm S}&=\int_0^{2\pi}d\varphi\int_0^{\pi}\sin\vartheta d\vartheta
            \int_0^{2\pi}d\phi \int_0^1 I(\mu)\mu d\mu = 8\pi^2\int_0^1 I(\mu)\mu d\mu,\\
      \mathcal{L}_A&=\int_0^{2\pi}d\varphi\int_0^{\vartheta_3}\sin\vartheta d\vartheta
            \int\int_{\Delta S(\vartheta)} F_0(\mu, \phi) d\mu d\phi, 
   \end{split}
\end{eqnarray}
where $F_0(\mu, \phi)=I(\mu_{\rm em})\mu_{\rm em}$, $\mu_{\rm em}=\hat{\mathbf{r}}_{\rm em}\cdot\mathbf{p}$ and $I(\mu)$ is the intensity on the stellar surface. Obviously, $\mu_{\rm em}$ is a function of $\mu$ and $\phi$. The integral $\mathcal{L}_A$ can be divided into three terms, i.e., $\mathcal{L}_A = \mathcal{L}_1 + \mathcal{L}_2 + \mathcal{L}_3$, where
\begin{eqnarray}
   \begin{split}
      \mathcal{L}_1 &=  2\pi\int_0^{\vartheta_1}\sin\vartheta d\vartheta \int_{-\pi}^\pi d\phi
         \int_{\mu_1}^1 F_0(\mu, \phi) d\mu, \\
      \mathcal{L}_2 &= 2\pi\int_{\vartheta_1}^{\vartheta_2}\sin\vartheta d\vartheta 2\left[
             \int_0^{\phi_1}d\phi\int_{\mu(\phi)}^1  F_0(\mu, \phi)d\mu +
             \int_{\phi_1}^\pi d\phi\int_{\mu_1}^1 F_0(\mu, \phi)d\mu  \right], \\
      \mathcal{L}_3 &= 2\pi\int_{\vartheta_2}^{\vartheta_3}\sin\vartheta d\vartheta 2\left[
             \int_{\phi_1}^\pi d\phi\int_{\mu_1}^{\mu(\phi)}  F_0(\mu, \phi) d\mu  \right].
   \end{split}
\end{eqnarray}
The luminosity ratio $\mathcal{L}_A / \mathcal{L}_{\rm S}$ can be calculated directly from the above integrals or obtained by adding up the photon weights while the Monte Carlo simulation is performed.

\subsection{The inversion and alias methods}\label{appendix_sec1}
In our first method to sample $\bar{\mu}$, the coordinate $\bar{\mu}$ of an initial photon location is drawn from a probability distribution function (PDF) given by

\begin{eqnarray}
   \begin{split}
      p(\bar\mu) = \frac{1}{S}\Delta S(\bar\mu),
   \end{split}
\end{eqnarray}
where $S=\int_0^{\cos\vartheta_3}\Delta S(\bar\mu) d\bar\mu$ is the normalization factor. 
One can sample $p(\bar\mu)$ by the inverse cumulated distribution function (CDF) method \citep{Devroye1986}. From Equation \eqref{deltaSmu2}, one can see that $\Delta S(\bar\mu)$ is a complicated function of $\bar\mu$; thus, we must implement the inverse CDF method in a numerical rather than analytical way. The procedure of our implementation is given as follows.
 
We obtain the cumulative probability distribution function $F(\bar\mu)$ of $p(\bar\mu)$:
\begin{eqnarray}
   \begin{split} 
      F(\bar{\mu}) = \int_{\cos\vartheta_\text{max}}^{\bar{\mu}} p(\bar\mu')d\bar\mu'.
   \end{split}
\end{eqnarray}
We first make a numerical table of $(\bar{\mu}_i, F_i)$ such that $\Delta F \equiv F_{i+1}-F_{i} = 1/N$ for $i=0,\cdots,N$, where $F_i =
F(\bar{\mu_i})$ and N is the number of bins. In a range of $F_j$ and $F_{j+1}$, we approximate $F({\bar\mu})$ to be a linear function
$f(\bar{\mu})$, i.e.,
\begin{eqnarray}
   \begin{split} 
      \frac{f(\bar\mu)-F_j}{F_{j+1}-F_{j}} = \frac{ \bar\mu -\bar\mu_j}{\bar\mu_{j+1}-\bar\mu_{j}}.
   \end{split}
\end{eqnarray}
For a random number $\xi$, we can readily find a subscript $j$ such that $F_j\le\xi\le F_{j+1}$ because $\Delta F = constant$. Then, we get
\begin{eqnarray}
   \begin{split} 
       \mu_\xi = \frac{\xi-F_j}{\Delta F}(\bar\mu_{j+1}-\bar\mu_{j}) + \bar\mu_j,
   \end{split}
\end{eqnarray}
where $j =[\frac{\xi}{\Delta F}]$. Here, [x] is the function of taking the integer part of x.
We use this $\bar\mu_\xi$ as an approximation of $\cos\vartheta$. 
As an alternative method to sample $\bar{\mu}$ efficiently, we also implemented the alias method \citep{Walker1974, Walker1977}. The inversion and alias methods were also used to sample the photon frequency (wavelength) from the input Ly$\alpha$ spectral shape and the initial location of Ly$\alpha$ photon emitted in the planetary atmosphere.

\subsection{H$\alpha$ and He 10830 Radiative transfer} \label{subsec:model-3}
\begin{figure}[t]
\begin{center}
  \includegraphics[scale=0.6]{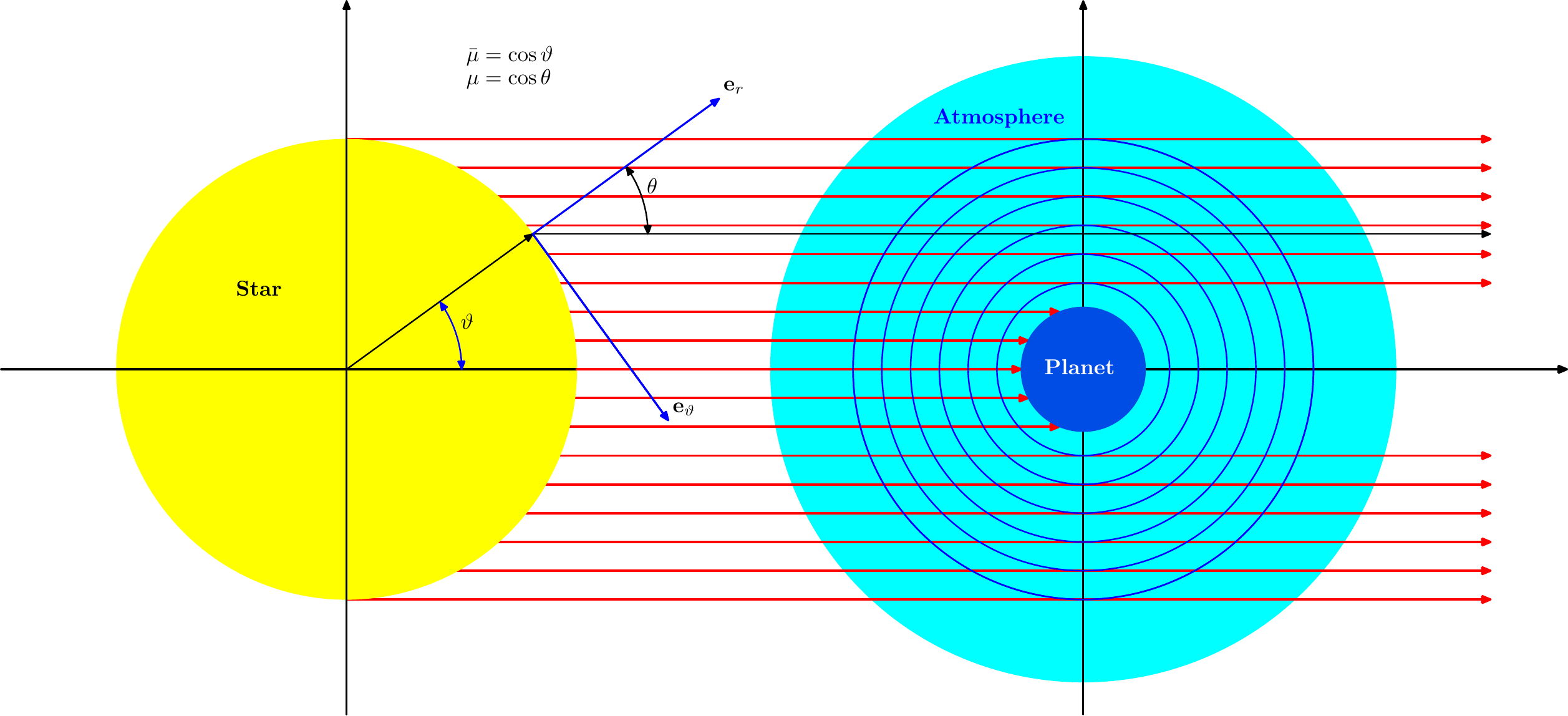}
  \includegraphics[scale=0.6]{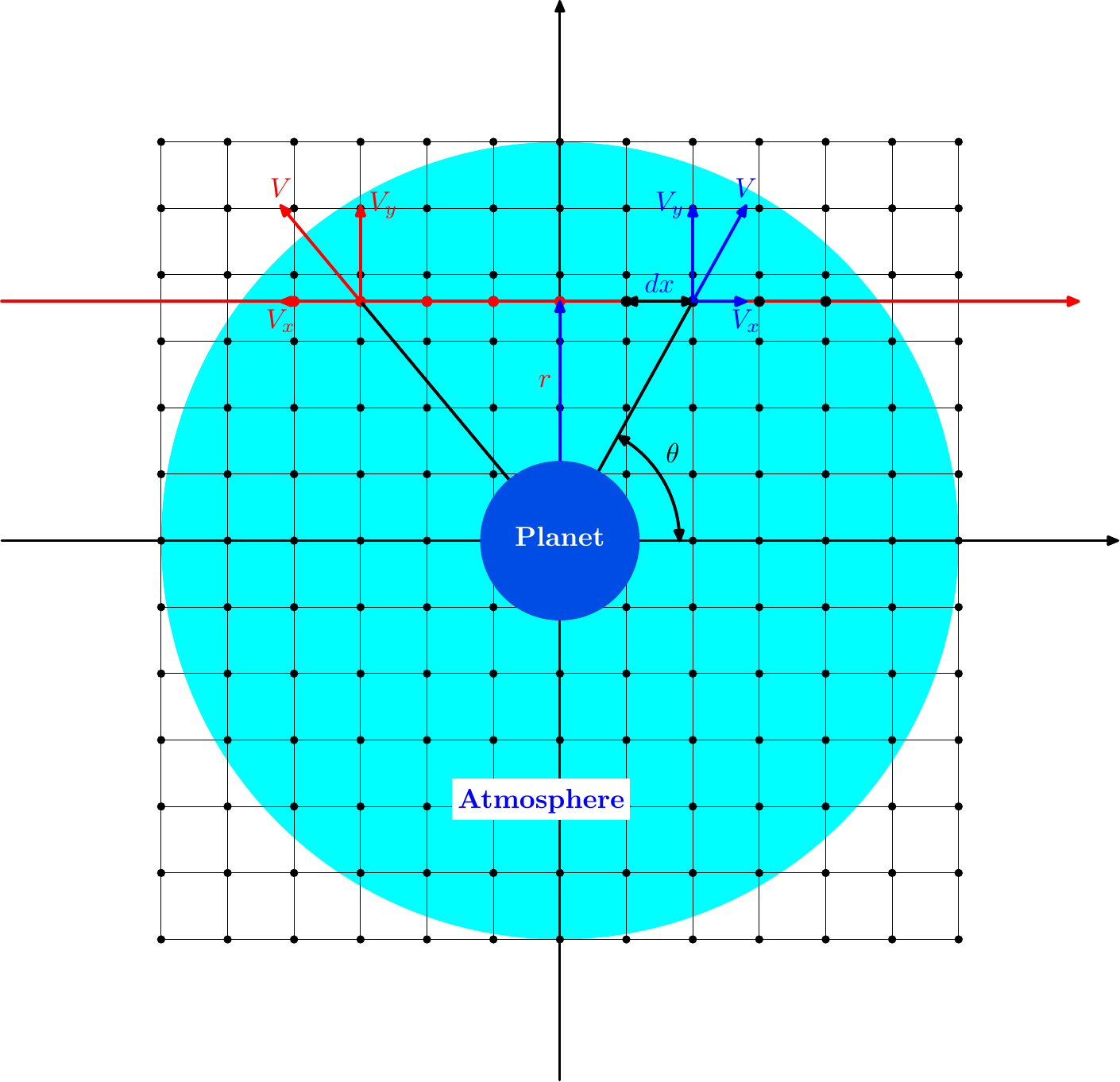}
%\vspace{1cm}
\caption{Schematic of H$\alpha$ and He 10830 radiative transfer. In the bottom panel, the radial velocity is decomposed of two orthogonal components $V_x$ and $V_y$, where $V_x$ causes the Doppler shift along the line of sight.} \label{fig_schematic}
\end{center}
\end{figure}
When calculating the H$\alpha$ and He 10830 transmission spectra, we only consider the true absorption in calculating the radiative transfer. The formula are quite similar for the two lines, except for the physical parameters.
The geometrical relationships and variables for H$\alpha$ and He 10830 radiative transfer are demonstrated in Figure \ref{fig_schematic}. Then, the observed line profiles
at frequency $\nu$ for the cases with and without absorption can be, respectively, obtained as follows:

\begin{eqnarray}\label{eqa10a}
   \begin{split} 
   \left\{ 
       \begin{array}{ll} 
       L'_\nu(obs) & = \displaystyle 2\pi  \int_0^{R_S} I_\nu(r)rdr,\\
       L_\nu(obs) & =  \displaystyle2\pi  \int_0^{R_S} I_\nu(r)
                 \exp[-\tau(r)]rdr,
       \end{array}
   \right.
   \end{split}
\end{eqnarray}
where $L'_\nu(obs)$ and $L_\nu(obs)$ are the observed luminosity during the out-of- and in-transit phases, and $r$ is the projected radius on a
plane perpendicular to the star-planet line. $\tau = n\sigma dl$ is the optical depth along the ray path, where $n$ is the number density of the relevant atomic species, $\sigma$ is the cross-section of absorption, and $dl$ is a finite traveling pathlength of the photons. 
In our radiative transfer simulations, the stellar H$\rm\alpha$ line center is at 6562.8 $\rm\AA$.
For the He 10830 lines, we consider only the two unresolved lines (10830.25, and 10830.33). Because the oscillator strength of 10829.09 line is much lower than the unresolved lines, its contribution is not considered here and this does not affect our conclusion.
The cross-sections of H$\alpha$ and He 10830 absorption lines are evaluated via,
\begin{equation}
\sigma_{23}=f_{23}\frac{\pi e^2}{m_ec}\phi_{\nu}
\end{equation}
in $\rm cm^2$,
where $e$ is the elementary charge of an electron, m$_e$ is the electron mass, and $f_{23}$ is the oscillator strength. $f_{23}$ =  0.64108 at 6562.8 $\rm\AA$ and $f_{23}$ =  0.17974 + 0.29958 =  0.47932 at the He 10830 unresolved lines (from NIST Atomic Spectra
Database). 
$\phi_\nu$ is the line profile given by the normalized Voigt function $H(a,u)/(\sqrt{\pi}\Delta \nu_D)$,
where $u$ is the frequency offset ($u = (\nu-\nu_0)/\Delta\nu_D$) normalized by the Doppler width $\displaystyle\Delta\nu_D = \frac{\nu_0}{c}\sqrt{\frac{2kT}{m}}$ and $\displaystyle a = \frac{\Gamma}{4\pi \Delta \nu_D}$ is the damping parameter \citep{Rybicki 2004}.

Notice that ${L'}_\nu(obs)$ and ${L}_\nu(obs)$ can also be expressed as an integral over $\mu$ (= $\cos\theta$), instead of $r$. 
\begin{eqnarray}\label{eqa10b}
   \begin{split} 
      \left\{ 
       \begin{array}{ll} 
       L'_\nu(obs) &= R_S^2\displaystyle \int_0^{2\pi}d\Phi\int_{0}^1 I_\nu(\bar{\mu})\bar{\mu} d\bar{\mu}\\
        &= 2\pi R_S^2\displaystyle \int_0^1I_\nu(\bar{\mu})\bar{\mu}d\bar{\mu},\\
       L_\nu(obs) & = 2\pi R_S^2\displaystyle \int_0^1I_\nu(\bar{\mu})\bar{\mu}
                 \exp[-\tau(\bar{\mu})]d\bar{\mu}.
   \end{array}
   \right.
   \end{split}
\end{eqnarray}
Equations (\ref{eqa10a}) and (\ref{eqa10b}) are equivalent because $r = R_{S}\cos\vartheta$ and $\sin\vartheta\lvert d\sin\vartheta \rvert = \cos\vartheta\lvert d\cos\vartheta \rvert$.

Finally, the transmission spectrum can be expressed as
\begin{equation}
\displaystyle TS=\frac{L'_\nu(obs)-L_\nu(obs)-L_\nu(abs)}{-L'_\nu(obs)}
\end{equation}
where $L_\nu(abs)$ is the attenuated flux by planet itself. If the integral is taken over the projection disk of the star, then $\frac{L_\nu(abs)}{L'_\nu(obs)} = \frac{R_{P}^2}{R_{S}^2}$.
The absorption depth by the planetary atmosphere is $-TS (\times 100\%$).

Usually, an isotropic source will be assumed for stellar radiation. However, the surface brightness of the stellar disk may change according to the limb angle \citep{2015A&A...574A..94Y,2015A&A...582A..51C}. The brightness profile as a function of $\mu$ is $I_\nu = 1.0 $ for the isotropic radiation case and $I_\nu = \mu + \frac{2}{3}$ for the limb darkening case described by the Eddington approximation \citep{Rybicki 2004}. The attenuated fraction by the planet disk is $L(abs) = 2(\int_{\mu_0}^{1}\mu d\mu) = (1.0 - \mu_0^2)$ for isotropic and  
$L(abs) = \frac{3}{2}(\int_{\mu_0}^{1}(\mu + \frac{2}{3})\mu d\mu) = 1-\frac{\mu_0^2}{2}(1+\mu_0)$ for the Eddington limb darkening case, where $\mu_0 = \sqrt{1.0 - R_{P}^2/R_{S}^2}$. Note here the integrated luminosity is normalized to unity. We will consider the limb darkening case in Section \ref{sec: Results} and discuss the results obtained for the isotropic radiation in Section \ref{sec: Diss}. 
The limb darkening laws of Ly$\alpha$ and H$\alpha$ are considered self-consistently. When the stellar Ly$\alpha$ radiation is assumed to be isotropic (or limb darkened by Eddington approximation), the stellar H$\alpha$ radiation is also supposed to be isotropic (or limb darkened by Eddington approximation) correspondingly. When the stellar Ly$\alpha$ radiation is assumed to follow the Eddington limb darkening law, we multiply the photon weight by a limb darkening factor $\frac{3}{2}(\mu + \frac{2}{3})\mu$.

\section{Results}\label{sec: Results}

We find that the absorption depth at the He 10830 line center for all models with H/He = 92/8 exceeds the 3$\sigma$ upper limit of the observation (see Section \ref{sub:He_pop} for details). This likely suggests that the abundance of helium was assumed to be too high in the models with H/He = 92/8. 
Therefore, to model the He 10830 observation, lower abundances of helium in the atmosphere need to be examined. \cite{2020A&A...636A..13L, 2021A&A...647A.129L}  reproduced the observed absorption depth of He 10830 line in HD 209458b, HD 189733b, and GJ 3470b and found that a high H/He ratio is required. By simulating the metastable He 10830 absorption, \cite{2022ApJ...927..238R} also constrained the helium abundance in the upper atmosphere of HD 189733b and found its H/He ratio is very high.
Therefore, in the following, we investigate additional models for WASP-52b with H/He = 98/2, 99/1, and 99.5/0.5. For each H/He, we calculate the cases of $F_{\rm XUV}$ = 0.25, 0.5, 1, 2, 4, 6, 8 $\times$ F$_0$, and $\beta_m$ = 0.1, 0.22, 0.3, 0.4, 0.5. 
\subsection{Asmosphere structures from the hydrodynamic simulations}

\begin{table}[!htbp]
   \begin{threeparttable}[t]
    \centering
    \footnotesize% fontsize
    \setlength{\tabcolsep}{5pt}% column separation
    \renewcommand{\arraystretch}{1.3}%row space 
\caption{Mass-loss rates of the models (in units of $10^{11}$ g s$^{-1}$).}    
   \begin{tabular}{c|ccccc|ccccc}
     \hline
  &  $\beta_{m}$ = 0.1 & 0.22 & 0.3 & 0.4  & 0.5  &   0.1 & 0.22 & 0.3 & 0.4  & 0.5\\
  \hline
  $F_{\rm XUV}$ &   H/He=  &  92/8  & &   &  &   H/He = &   98/2  & &   & \\
        \hline
     0.25F$_{0}$     & 1.37 & 1.41  & 1.60 & 1.76 &  1.83  & 1.33  &   1.32  &  1.43 & 1.57  & 1.63 \\
 %  \hline
     0.5     & 2.64 &  2.73 & 3.05 & 3.31 & 3.38    & 2.59  & 2.59   & 2.79 & 3.05 & 3.10\\  
     1    & 4.89 &  5.04 & 5.61 & 5.97 &   6.01   &  4.87 & 4.92  & 5.29 & 5.69 & 5.72  \\  
     2     & 8.65 & 9.08 &  9.92 & 10.30 &  10.58  & 8.81  & 9.01  & 9.66 & 10.12 & 10.05 \\  
     4     & 14.63 &  15.76 & 16.72 & 17.37 & 18.40   & 15.31  & 15.86   & 16.75 & 17.10 & 16.95\\  
     6   & 19.51 & 21.25  & 22.29 &  23.54& 25.09  & 20.71  &  21.62 &  22.59 & 22.88 & 22.82\\   
     8  & 23.76 & 25.99 & 27.25 & 29.13  & 30.87  &  25.40  & 26.66  & 27.67 & 27.97 & 28.13 \\ 
    \hline
      &   $\beta_{m}$ = 0.1 & 0.22 & 0.3 & 0.4  & 0.5    &  0.1 & 0.22 & 0.3 & 0.4  & 0.5  \\
      \hline
  $F_{\rm XUV}$  &   H/He = &   99/1  & &   &     &   H/He = &   99.5/0.5  & &   & \\
     \hline
     0.25F$_{0}$  & 1.35 & 1.30  & 1.37  & 1.52  & 1.57 & 1.34 &  1.29 &  1.32 & 1.46  & 1.54   \\
 %  \hline
     0.5  & 2.61 &  2.56  & 2.71 & 2.97 & 3.04    & 2.61 &  2.54  & 2.64 &   2.88  & 2.98    \\  
     1  &  4.92 & 4.89  &  5.18 & 5.59 &  5.64    & 4.90  &   4.86 &  5.08 & 5.49  &  5.57   \\  
     2   & 8.90 & 8.98  & 9.52 & 10.00 &  10.00 & 8.89 & 8.94  & 9.39 & 9.93  & 9.93   \\ 
     4 &  15.44 & 15.84  & 16.65 & 17.09 & 16.92   &  15.51 & 15.79 & 16.50 & 16.99 &  16.86    \\
     6   &  20.94 &  21.63 & 22.54 & 22.89 & 22.73 & 21.06  & 21.58  & 22.40 & 22.81  &  22.64   \\   
     8   &  25.74 &  26.71 &  27.68 & 28.00 &  27.94  &  25.91 &  26.67  &  27.56 & 27.93  &  27.82   \\   
          \hline     
  \end{tabular}
  \end{threeparttable}%
\label{tab:mass_loss_rates}
\end{table}

First, we use the 1D hydrodynamic atmosphere escape code to simulate the atmosphere structures and obtain the temperature, velocity, and densities of hydrogen and helium (the atmosphere structures of helium will be described in Section \ref{sub:He_pop}). The mass-loss rates were also calculated, as shown in Table \ref{tab:mass_loss_rates}, for various models
considered in this paper. The mass-loss rate is proportional to $F_{\rm XUV}$ for low
$F_{\rm XUV}$, while for high $F_{\rm XUV}$,  it appears to increase nonlinearly. This is due to the Ly$\alpha$ cooling effect considered in the hydrodynamic simulations. The mass-loss rate also increases slightly with the increase of spectral index $\beta_m$ for the models with H/He = 92/8, as studied by \cite{2016ApJ...818..107}. It tends to increase in general with increasing $\beta_m$ also in other models with H/He = 98/2, 99/1, and 99.5/0.5. However, when $F_{\rm XUV} \lesssim F_0$ and $\beta_m \lesssim 0.22$, or when $F_{\rm XUV} \gtrsim 4F_0$ and $\beta_m \gtrsim 0.4$, the trend appears to be reversed.
In addition, we find that the mass-loss rate does not change significantly as the H/He ratio varies.

\begin{figure*}
\gridline{\fig{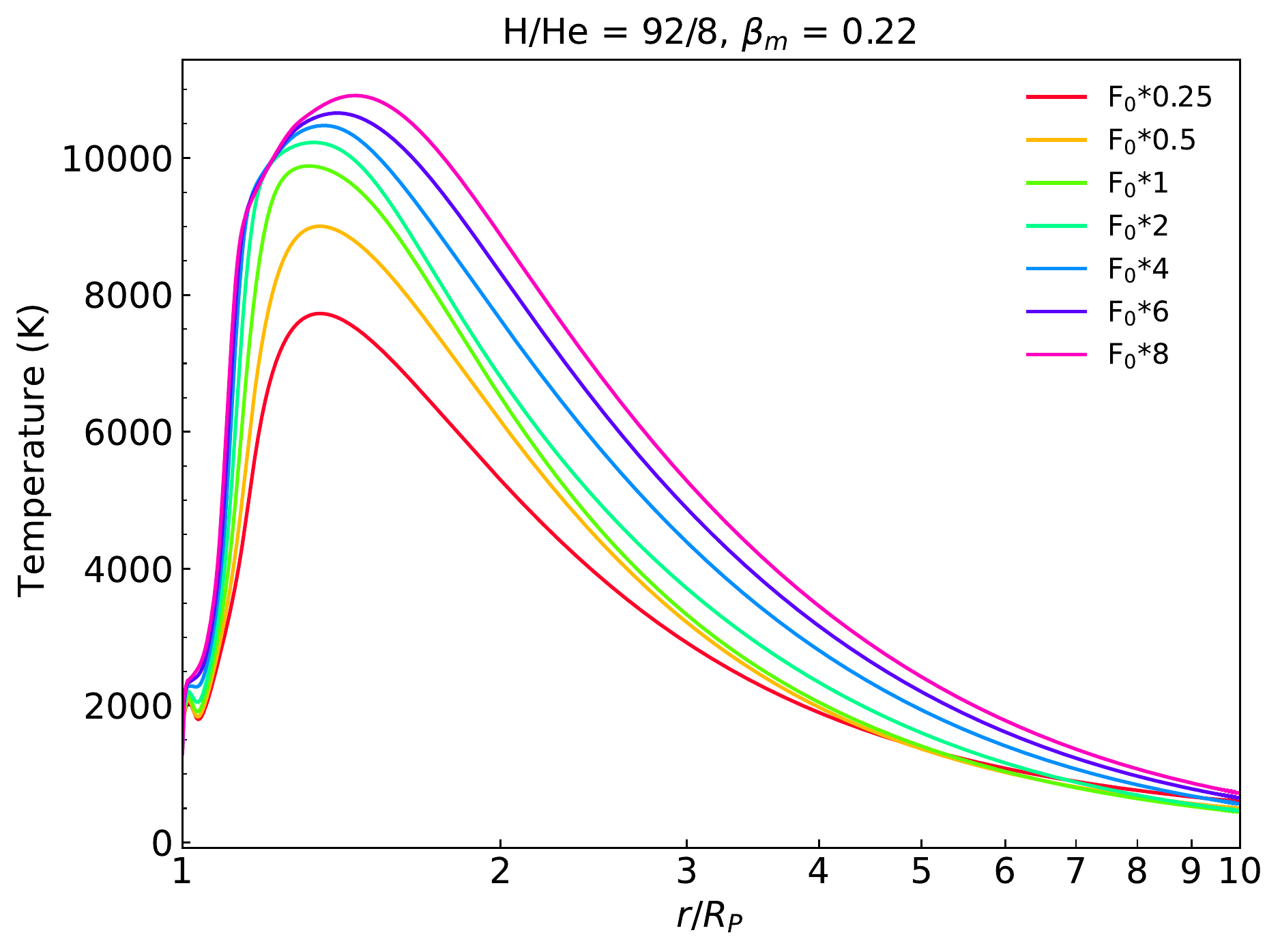}{0.53\textwidth}{(a)}
 \fig{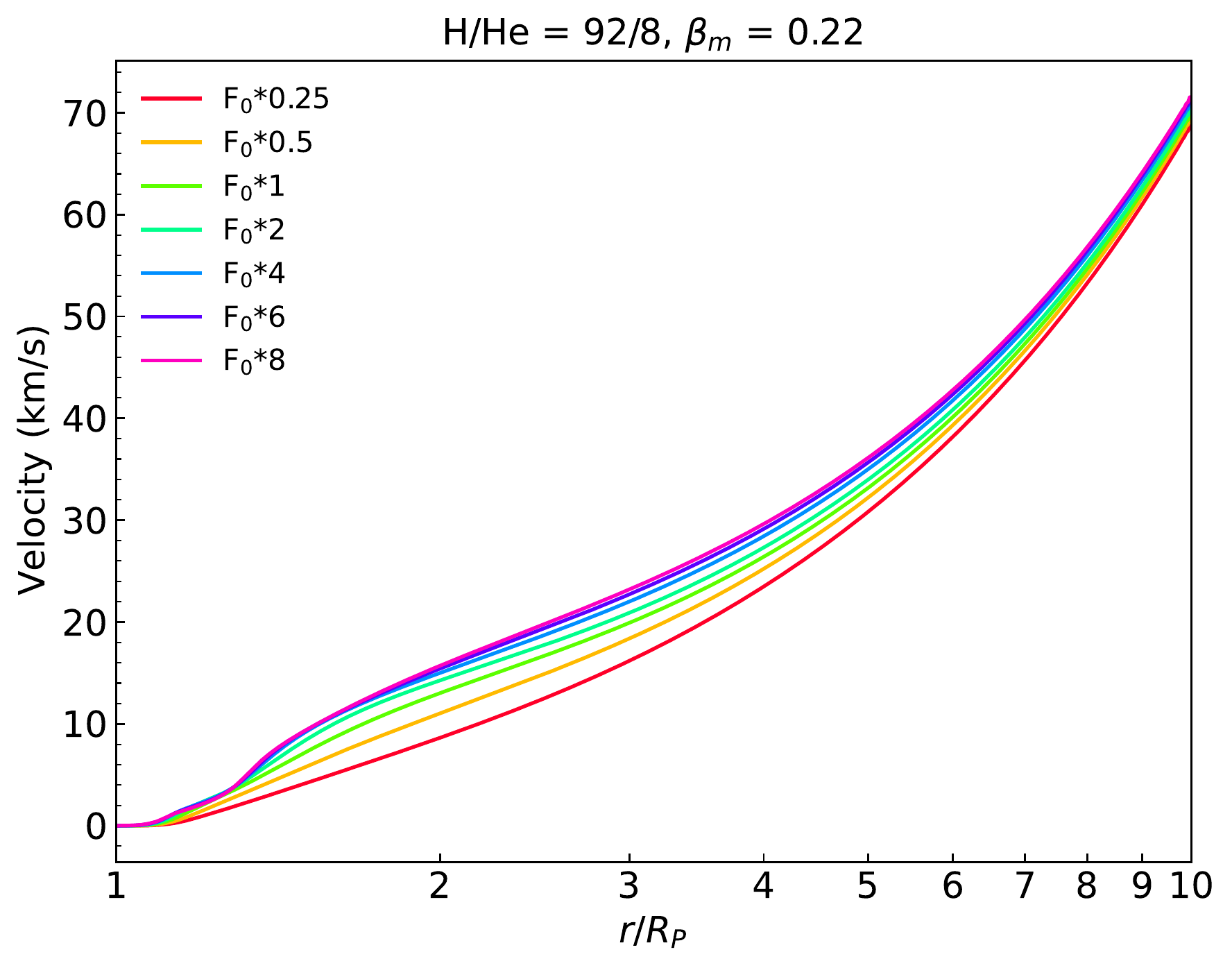}{0.5\textwidth}{(b)}
 }
\gridline{\fig{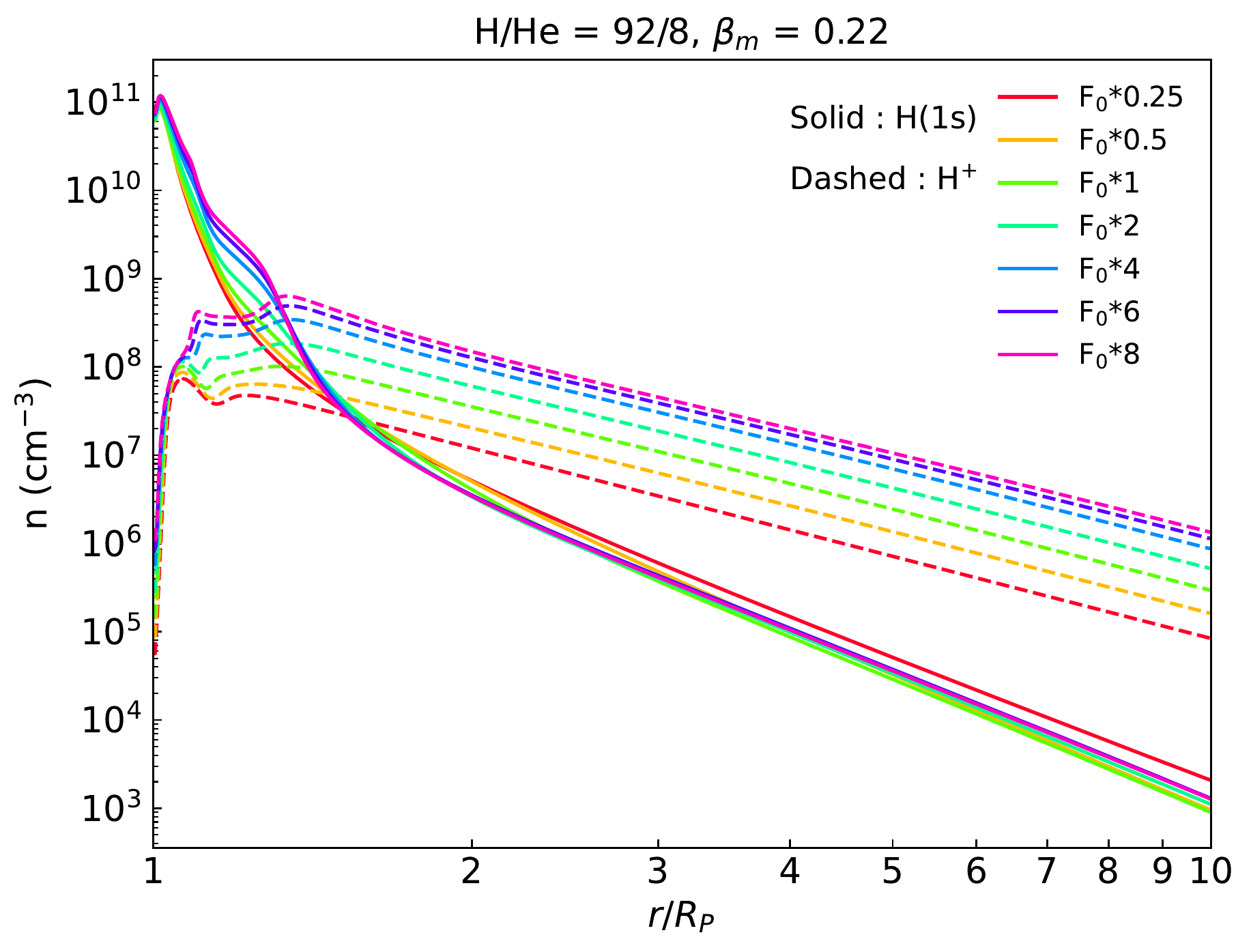}{0.5\textwidth}{(c)}
 \fig{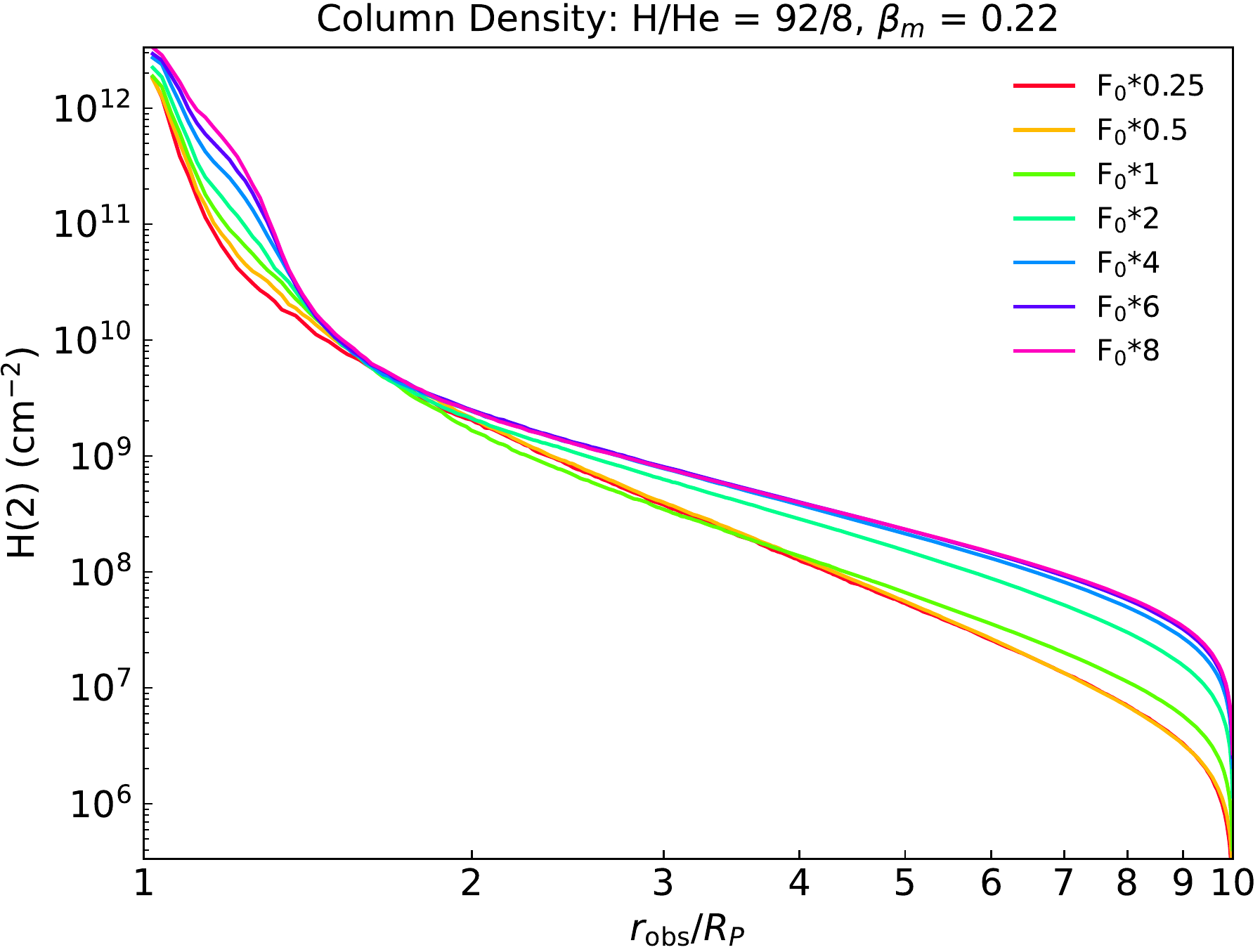}{0.5\textwidth}{(d)}
 }
\gridline{\fig{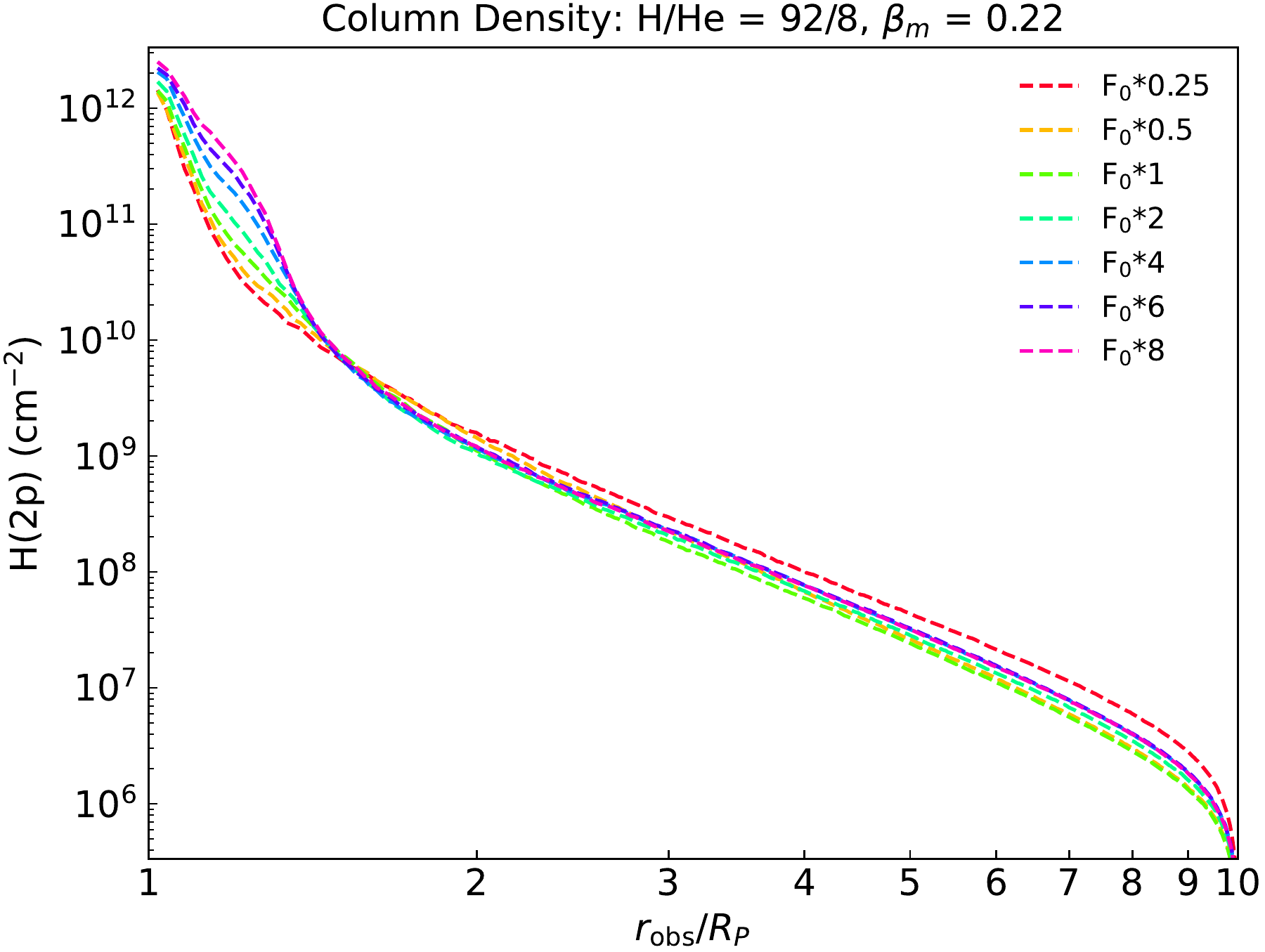}{0.5\textwidth}{(e)}
 \fig{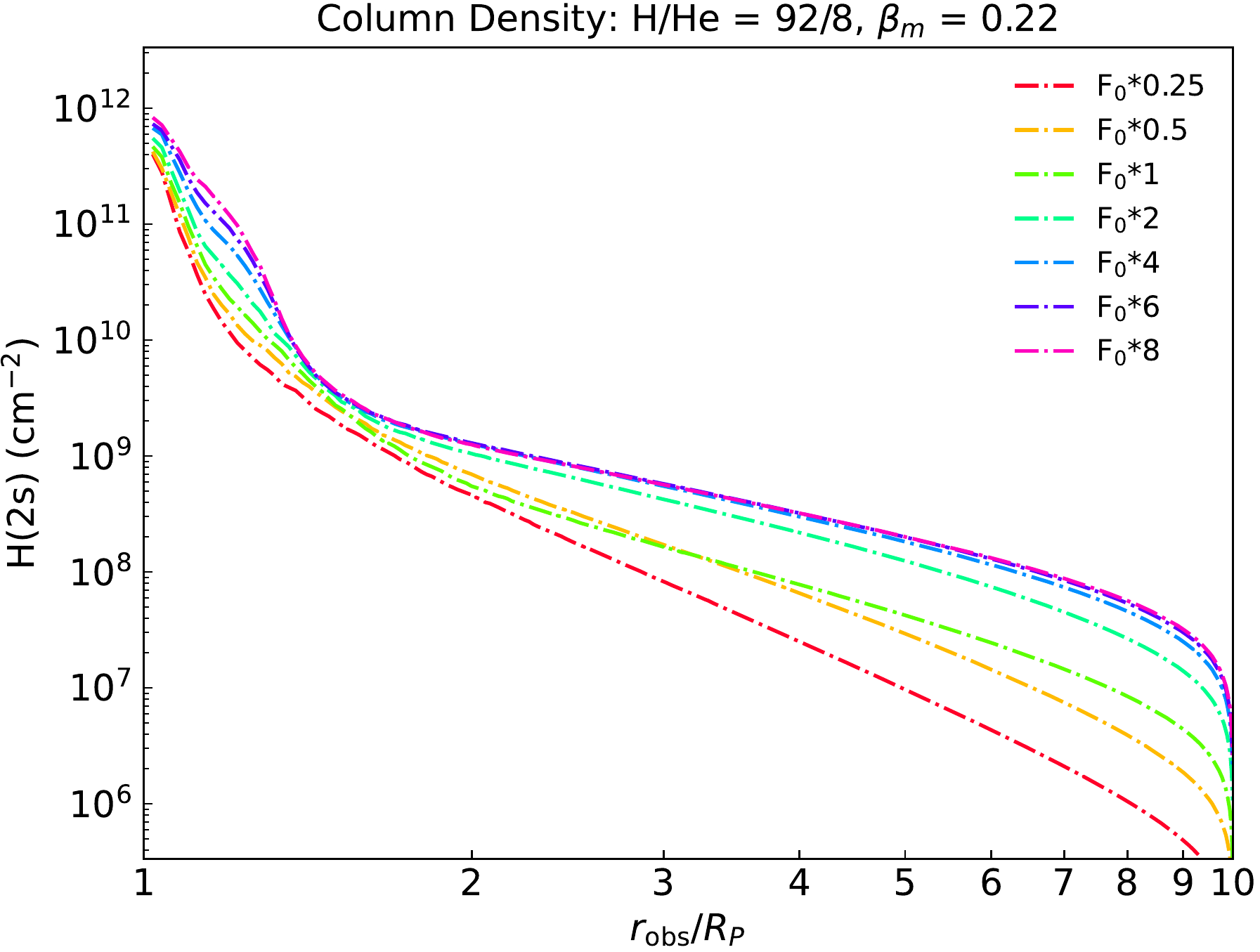}{0.5\textwidth}{(f)}
 } 
\caption{Atmosphere structures for models with various $F_{\rm XUV}$ when H/He = 92/8, $\beta_m$ = 0.22. Here, $r$ denotes the altitude (the radial distance from the planet center) and $r\rm_{obs}$ is the projected
distance onto a plane perpendicular to the z-axis.}\label{atm_90_10_b022}
\end{figure*}

\begin{figure*}
\gridline{\fig{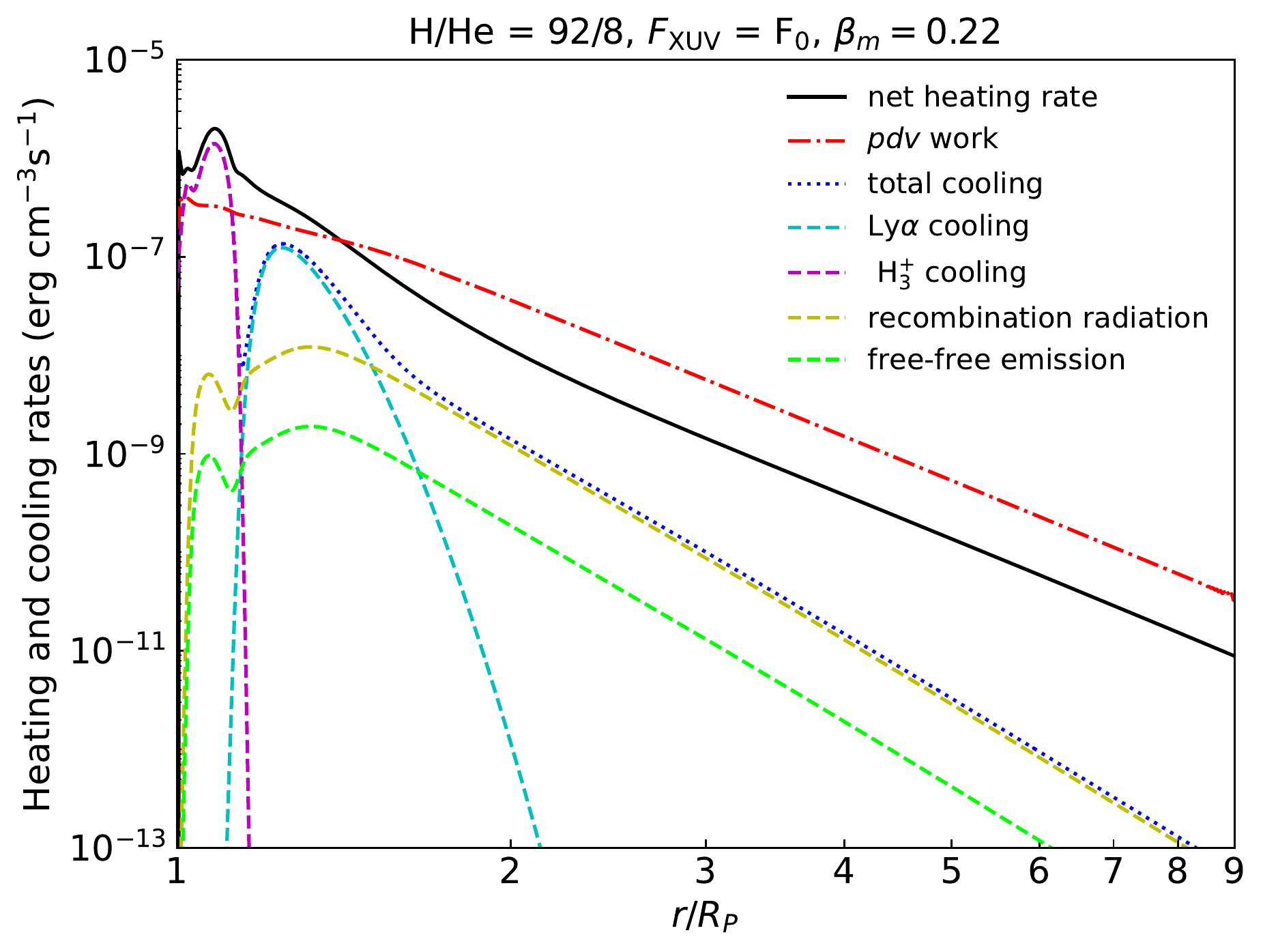}{0.53\textwidth}{}
}
\caption{Heating and cooling rates for the model of H/He = 92/8, $F_{\rm XUV}$ = F$_0$, and $\beta_m$ = 0.22. The total cooling rate plotted in the blue dotted line is the sum of the cooling rate by Ly$\alpha$, H$_3^+$, recombination radiation, and free-free emission.}\label{heating_cooling}
\end{figure*}

Figure \ref{atm_90_10_b022} shows the atmosphere structures as a function of altitude for models with different $F_{\rm XUV}$ but with the same $\beta_m$ = 0.22 and H/He = 92/8. Figure \ref{atm_90_10_b022} (a) shows that the atmospheric temperature increases with the increase of $F_{\rm XUV}$. When $F_{\rm XUV}$ = 0.25F$_0$, the highest temperature is about 6800 K, and it increases to about 14,000 K when $F_{\rm XUV}$ = 8F$_0$. The reason for this trend is that a higher $F_{\rm XUV}$ will lead to more ionization of the atmosphere and thus more heating. As altitude increases, the temperature first rises and then decreases in all models. The temperature increases mainly due to the XUV heating and decreases mainly due to the atmosphere expansion, also called ``$pdv$ work" or ``adiabatic expansion." Figure \ref{heating_cooling} shows the heating and cooling rates for the model of H/He = 92/8, $F_{\rm XUV}$ = F$_0$, and $\beta_m$ = 0.22. The net heating rate is the sum of XUV and chemical heating rates. Note that the chemical heating rate can be negative sometimes. The cooling mechanisms include the cooling by ``$pdv$ work" and other cooling mechanisms, with the latter including Ly$\alpha$ cooling, H$_3^+$ cooling, recombination radiation, and free-free emission \citep{2009ApJ...693...23M,2021ApJ...907L..47Y}. We can see that at $r \lesssim 1.1 R_P$, the cooling is dominated by H$_3^+$ cooling; above this altitude, the ``$pdv$ work" is the dominant cooling term. The Ly$\alpha$ cooling can be effective when the temperature is high; and the cooling by recombination radiation and free-free emission can be ignored throughout the atmosphere.
When $r \gtrsim 1.8 R_P$, the cooling by ``$pdv$ work" exceeds the heating rates, decreasing the temperature substantially. As a result, the temperature at $\sim 4 R_P$ can be as low as $\sim$ 2000 K for the model of H/He = 92/8, $F_{\rm XUV}$ = F$_0$, and $\beta_m$ = 0.22.
Figure \ref{atm_90_10_b022} (b) shows atmosphere expansion velocity as a function of altitude. The difference in velocity at different $F_{\rm XUV}$ becomes smaller with the increase in altitude. A higher $F_{\rm XUV}$ causes a slightly higher velocity because a higher $F_{\rm XUV}$ can provide more energy to overcome the planetary gravitational potential, leaving more kinetic energy for the atmosphere. Figure \ref{atm_90_10_b022} (c) shows the number density of hydrogen atoms in the 1s state H(1s) and that of hydrogen ions H$^+$. Obviously, a higher $F_{\rm XUV}$ causes more H$^+$ ions. The number density of H(1s) increases with the increase of $F_{\rm XUV}$ at low altitudes (radii) of $r \lesssim 1.5$-$2 R_P$, but it doesn't show much difference above 2 $R_P$. This would be because the mean absorption radius of XUV flux is below 2 $R_P$. 
\cite{2019ApJ...880...90Y} performed a statistical analysis of the mean absorption radius of the stellar XUV radiation and proposed that it is in the range of 1.1-1.7 $R_P$ for Jupiter-like planets. This work is consistent with that one. 

\begin{figure*}
\gridline{\fig{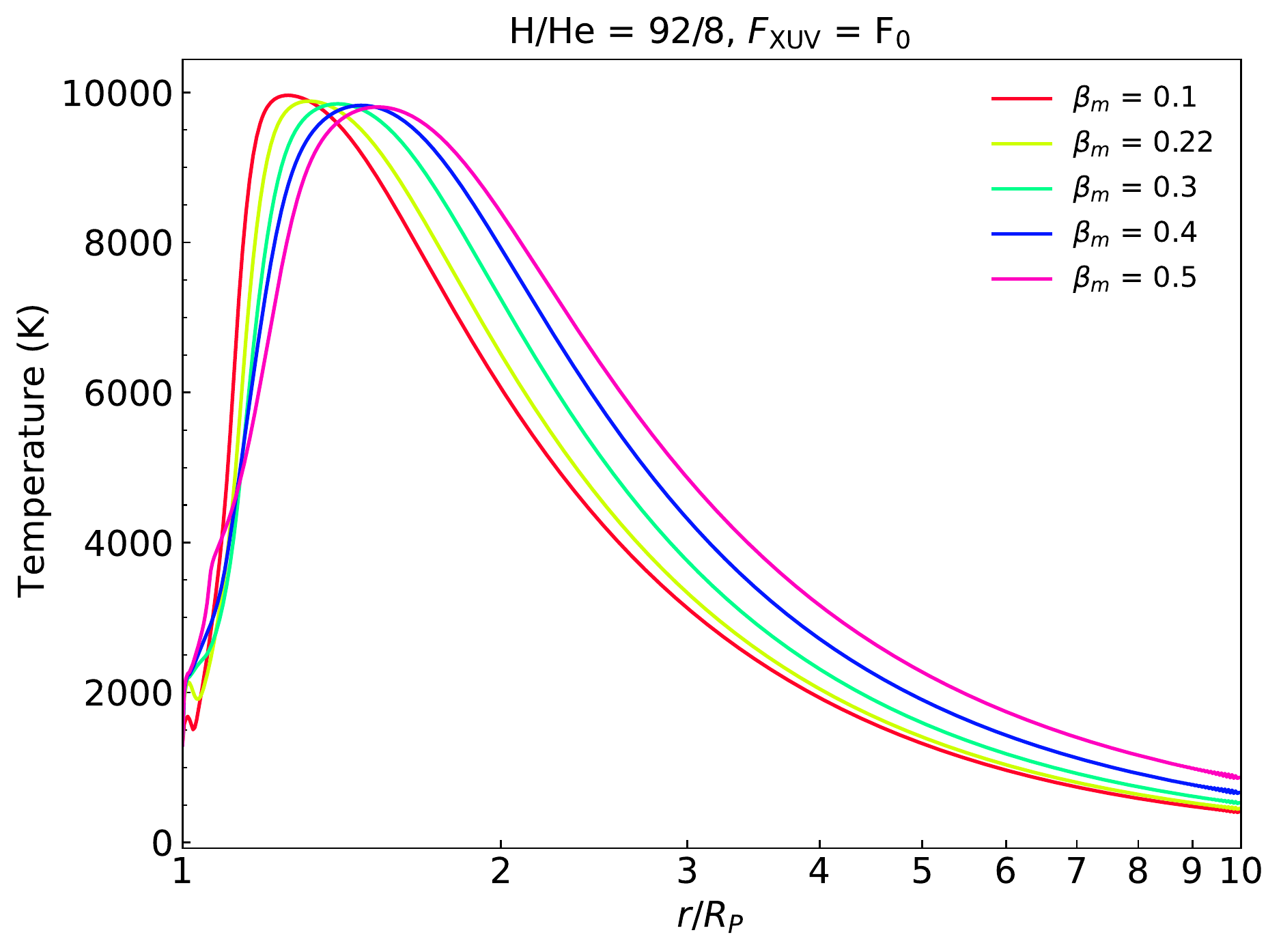}{0.53\textwidth}{(a)}
 \fig{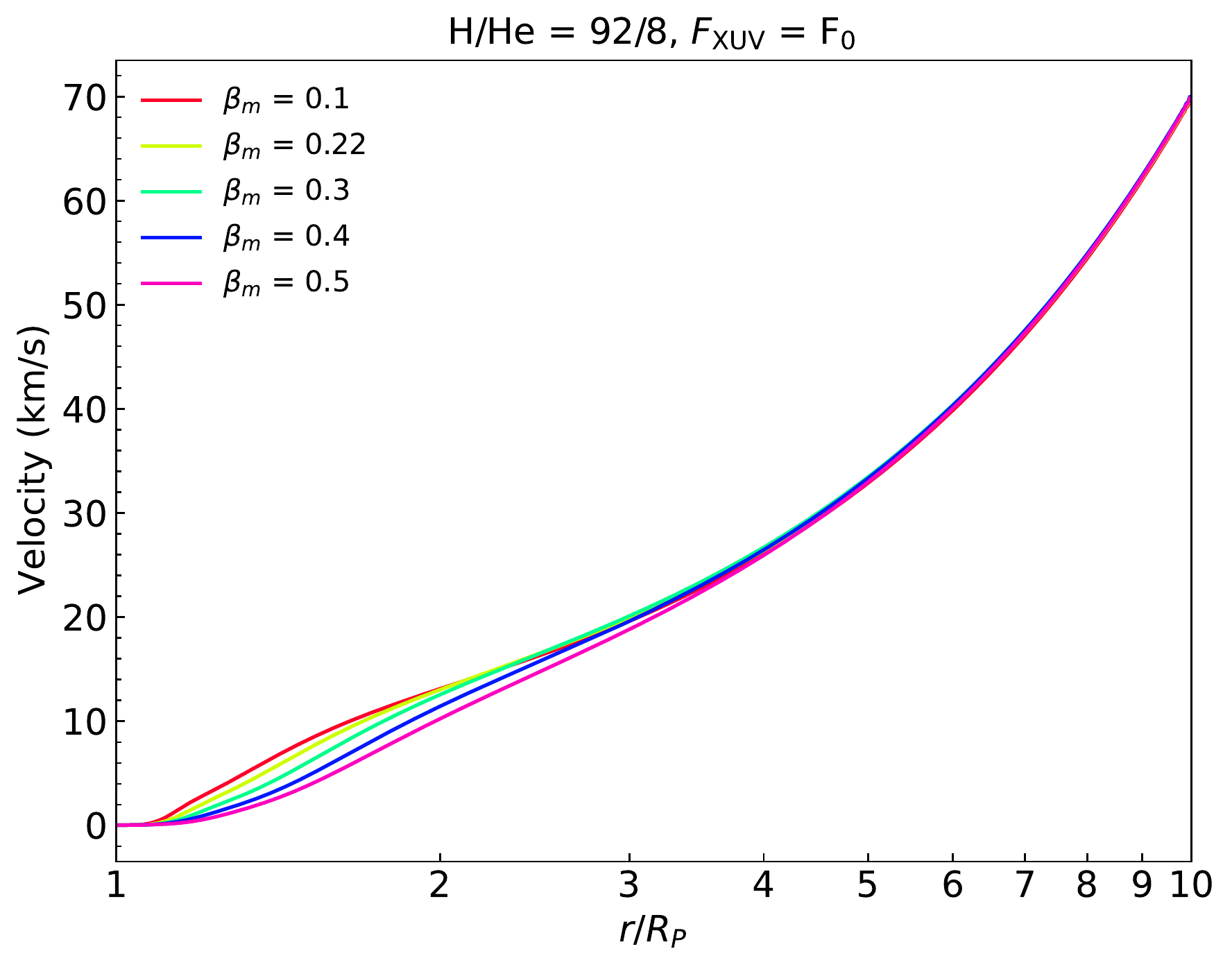}{0.5\textwidth}{(b)}
 }
\gridline{\fig{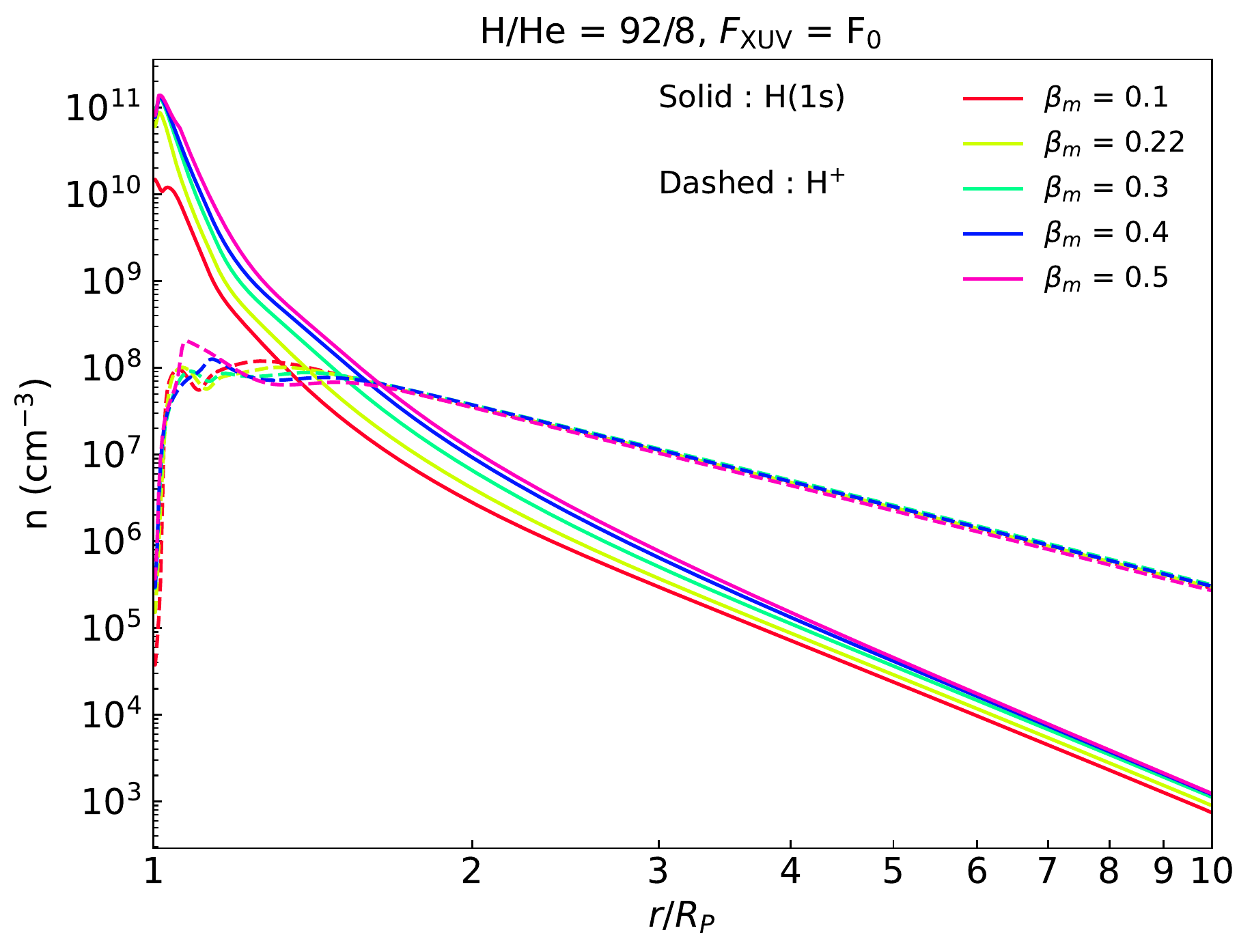}{0.5\textwidth}{(c)}
 \fig{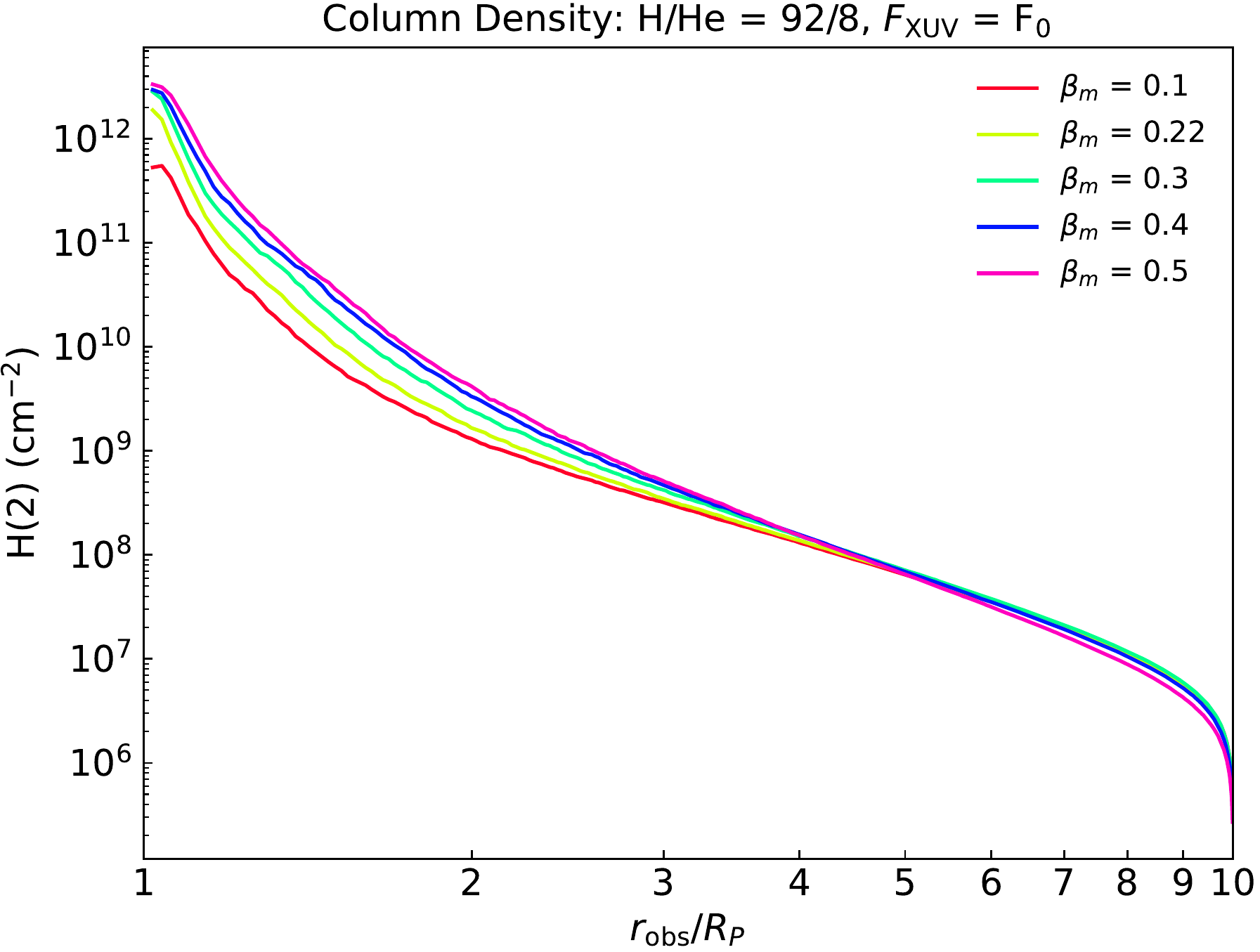}{0.5\textwidth}{(d)}
 }
\gridline{\fig{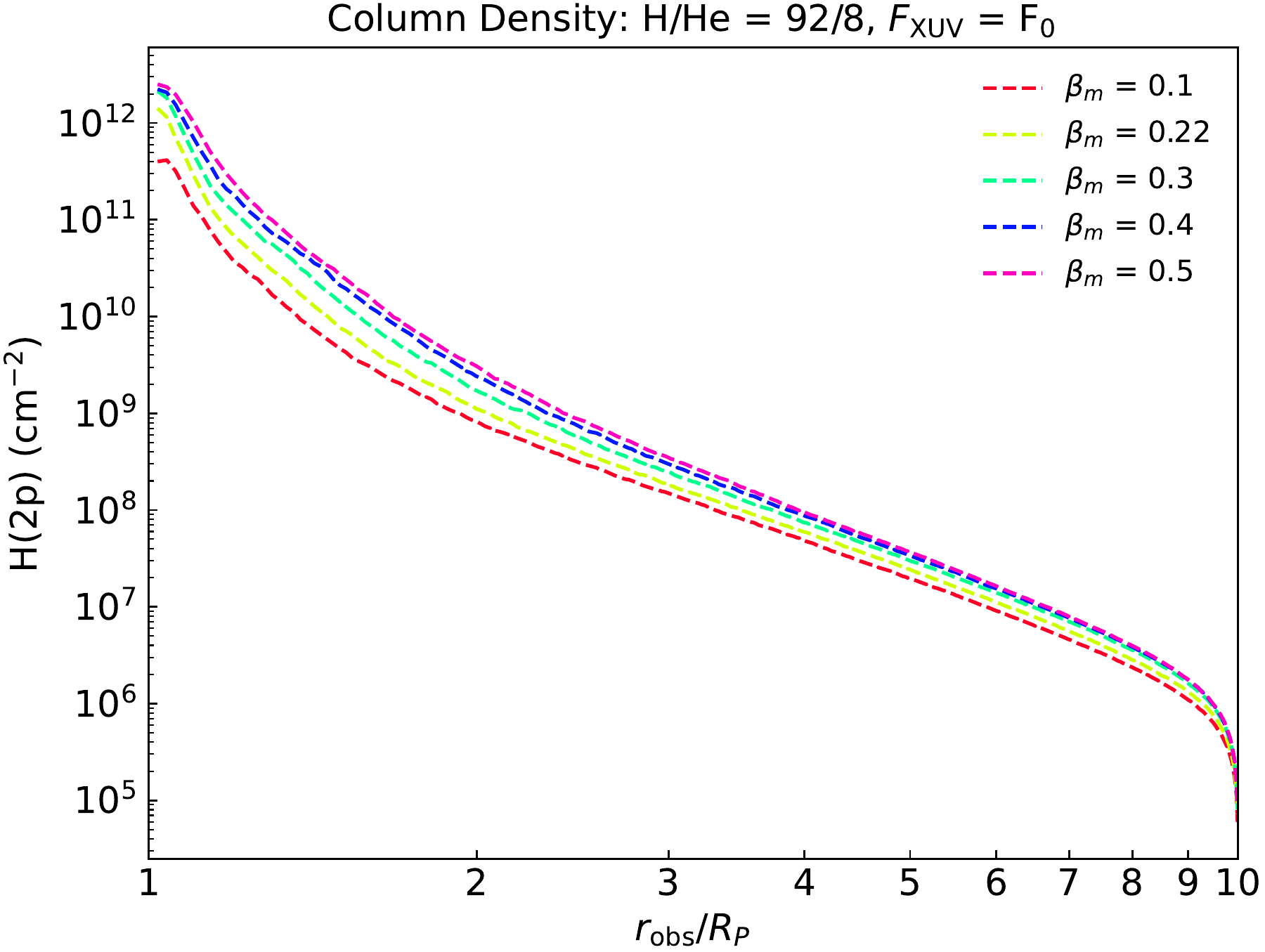}{0.5\textwidth}{(e)}
 \fig{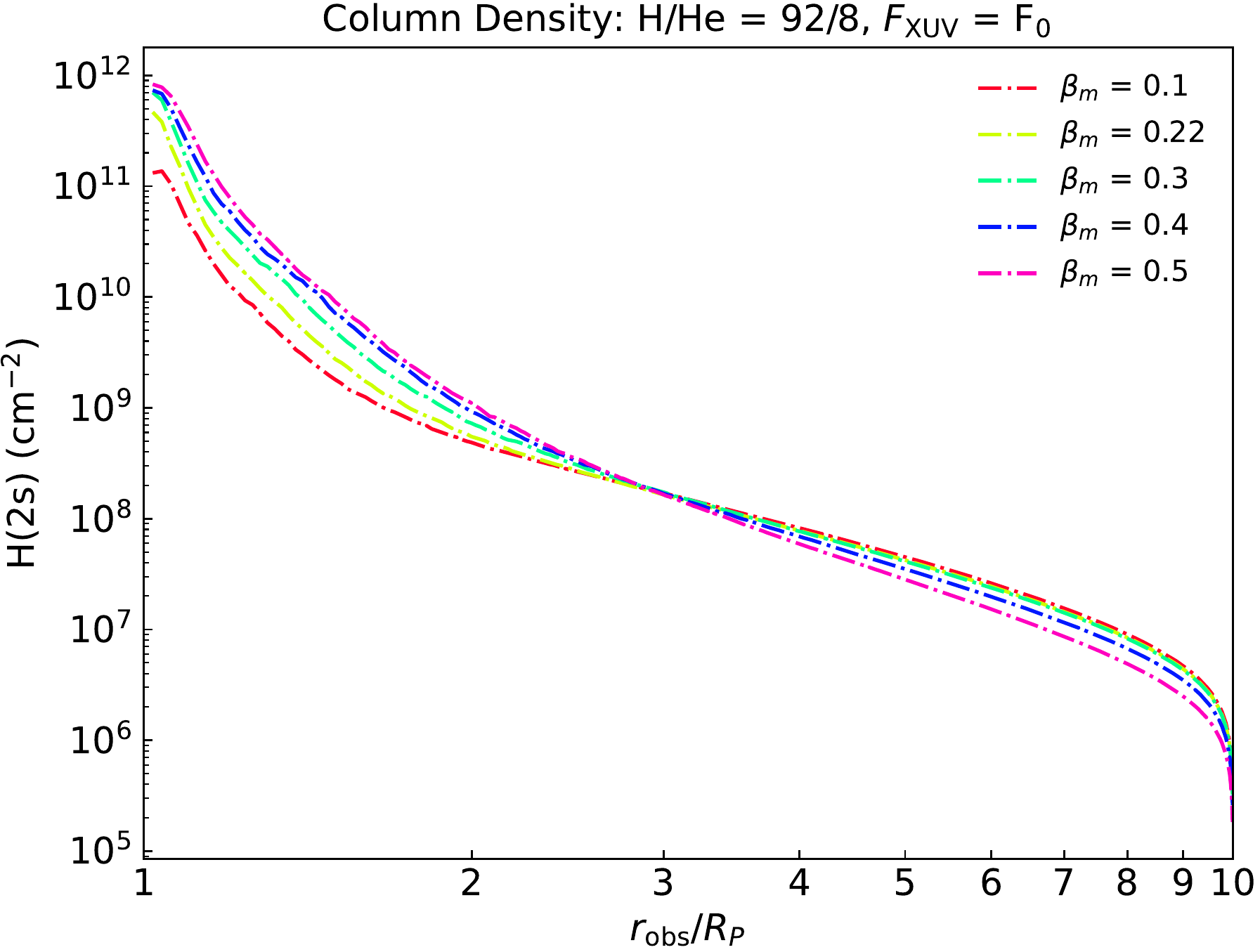}{0.5\textwidth}{(f)}
 } 
\caption{Atmosphere structures for models with various $\beta_m$ when H/He = 92/8, $F_{\rm XUV}$ = F$_0$.}\label{atm_90_10_fxuv1}
\end{figure*}

In Figure \ref{atm_90_10_fxuv1}, we show the atmosphere structures as a function of altitude for models with different $\beta_m$ but with the same $F_{\rm XUV}$ = F$_0$ and H/He = 92/8. Figure \ref{atm_90_10_fxuv1} (a) shows that with the increase of $\beta_m$, the highest temperature of the atmosphere decreases slightly. However, the altitude with the highest temperature is found to increase as $\beta_m$ increases. This can be due to a high XUV mean absorption radius for models with a small EUV portion in the XUV spectrum.
The temperature firstly increases with the altitude until it reaches the highest temperature due to the absorption of XUV energy. Then later, it drops with the altitude because of the atmosphere expansion. Figure \ref{atm_90_10_fxuv1} (b) shows that the velocity decreases with the increase of $\beta_m$ when $r$ $\lesssim$ 5$R_P$. Above the altitude of $\approx 5 R_P$, no significant change in the velocity profile with a variation of $\beta_m$ is found. Figure \ref{atm_90_10_fxuv1} (c) shows the number density of hydrogen atoms and ions. We can see that a larger $\beta_m$ leads to more neutral hydrogen atoms. This is because a larger $\beta_m$ corresponds to a smaller portion of EUV flux that can ionize hydrogen. 

Figures \ref{atm_fxuv1_b022} shows the atmosphere structures for models with different H/He ratios but with the same $F_{\rm XUV}$ = F$_0$ and $\beta_m$ = 0.22. 
Figures \ref{atm_fxuv1_b022} (a-c) show the temperature, velocity, and number density of H(1s) as a function of altitude. We can see no significant differences
in these physical quantities between the models with H/He = 98/2, 99/1, and 99.5/0.5. However, the model with H/He = 92/8 shows
some differences from other models with higher H/He ratios. The model with H/He = 92/8 shows a relatively higher temperature, lower velocity, and lower number density of H(1s). 

Because the hydrodynamic simulations can not give the number densities of H(2s), H(2p), and He(2$^3$S), we calculate the population of these species by solving the NLTE statistical equation as mentioned in Sections \ref{method_H2s2p} and \ref{method_He2s3}.
The column densities of H(2), H(2p), and H(2s) vs. the projected radius, calculated in this way, are shown in Figures \ref{atm_90_10_b022} (d)-(f), \ref{atm_90_10_fxuv1} (d)-(f), and \ref{atm_fxuv1_b022} (d)-(f), which will be discussed in the next section.
\subsection{Ly$\alpha$ mean intensity and H(2) populations}\label{sub:H2_pop}

\begin{figure*}
\gridline{\fig{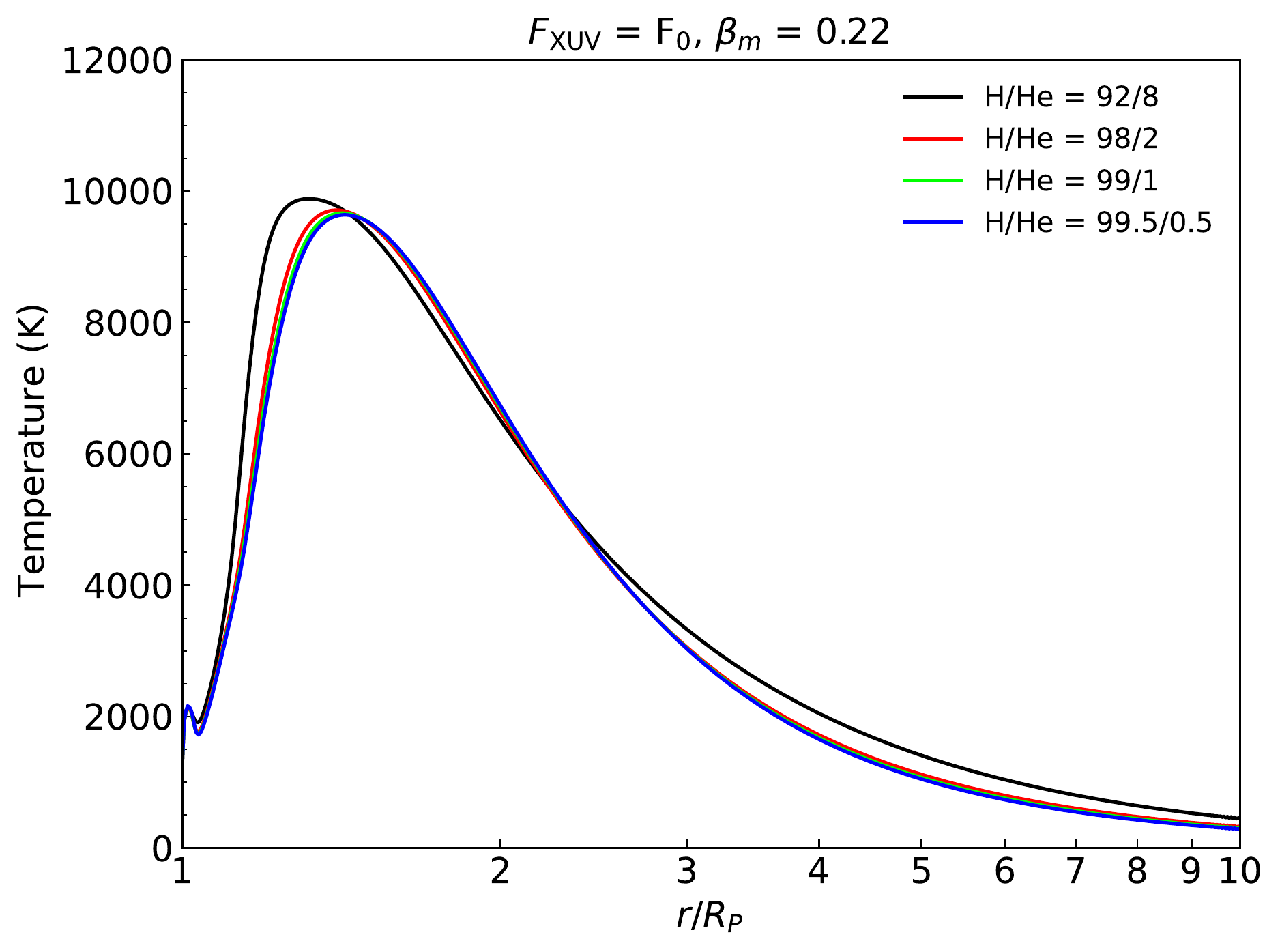}{0.53\textwidth}{(a)}
 \fig{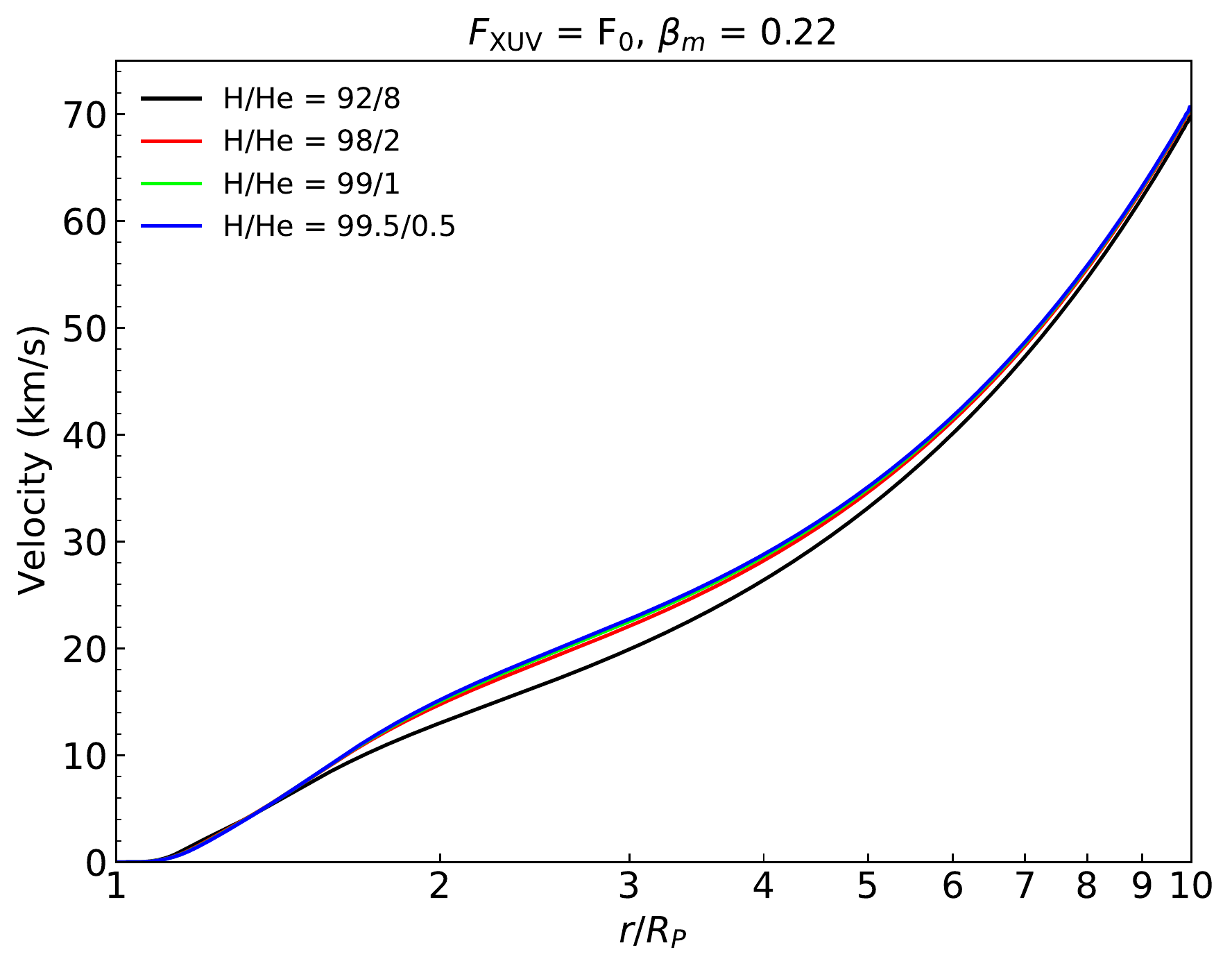}{0.5\textwidth}{(b)}
 }
\gridline{\fig{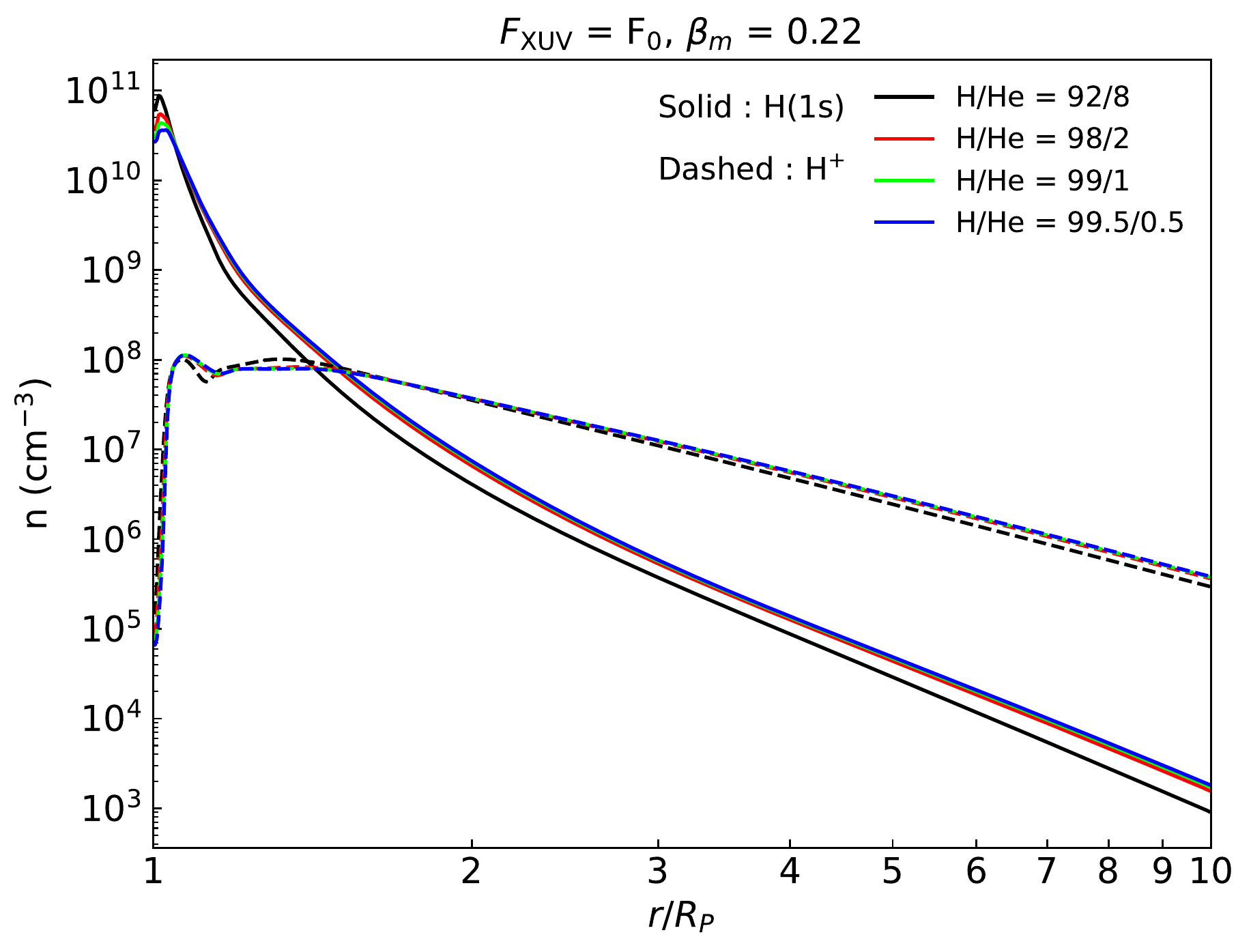}{0.5\textwidth}{(c)}
 \fig{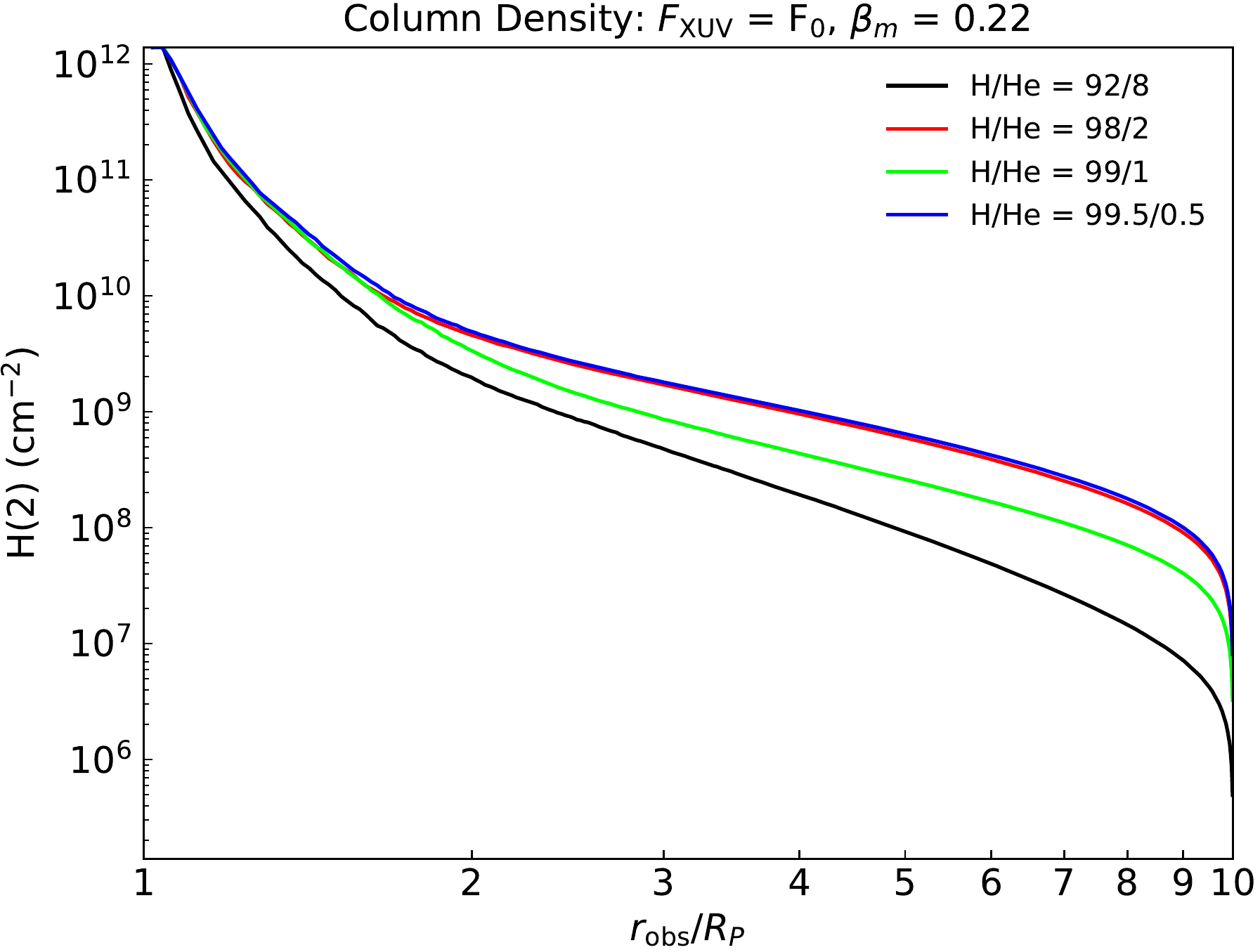}{0.5\textwidth}{(d)}
 }
\gridline{\fig{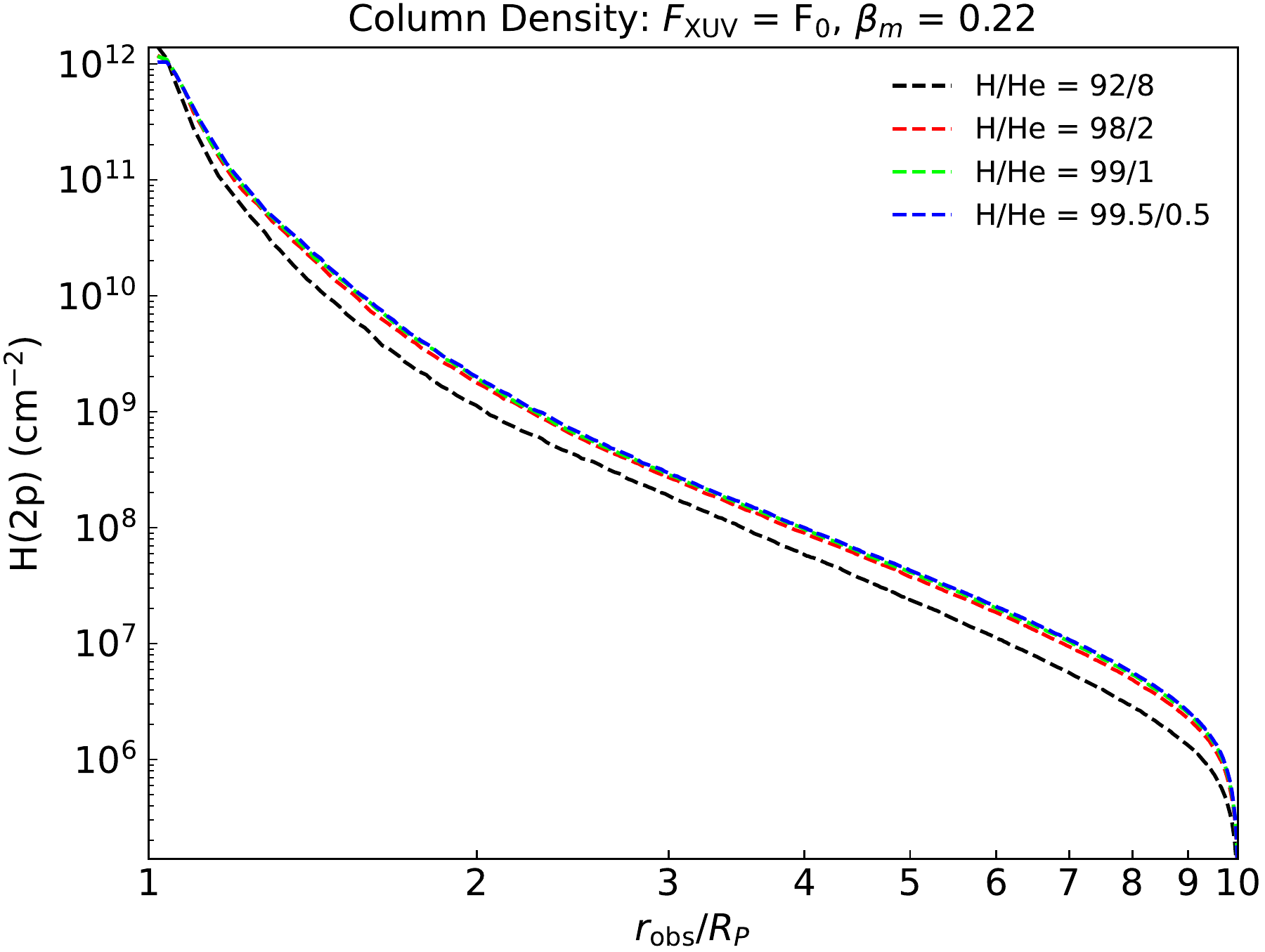}{0.5\textwidth}{(e)}
 \fig{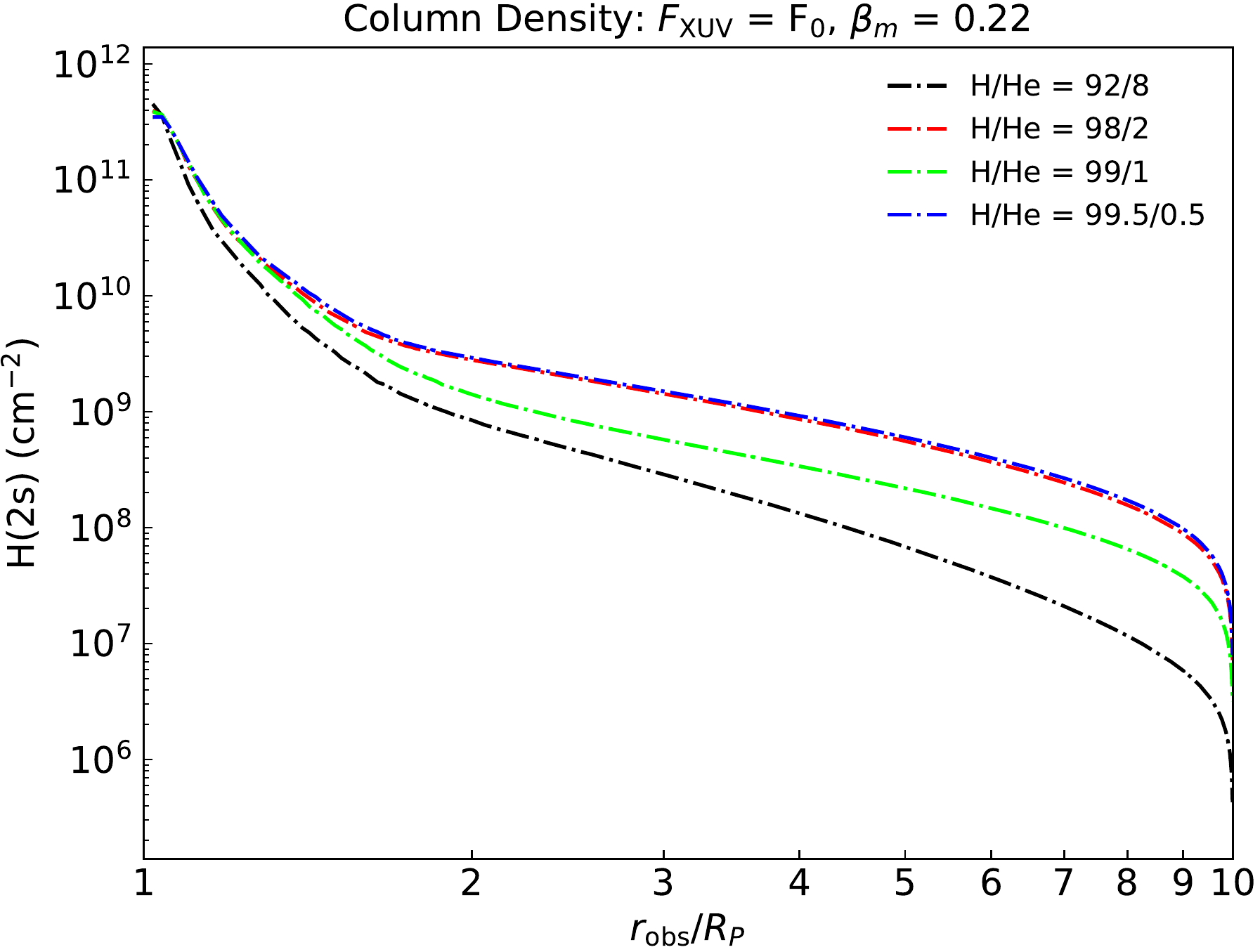}{0.5\textwidth}{(f)}
 }
 
\caption{Variation of the atmosphere structures with H/He ratio when $F_{\rm XUV}$ = F$_0$ and $\beta_m$ = 0.22.}\label{atm_fxuv1_b022}

\end{figure*}
Using LaRT, we simulate the Ly$\alpha$ mean intensity ($\bar{J}_{Ly\alpha}$) inside the atmosphere. Both the stellar and planetary Ly$\alpha$ source contribute to the total Ly$\alpha$ intensity, but they have different symmetries. The spatial distribution of Ly$\alpha$ mean intensity from the stellar part has a cylindrical symmetry, because the atmosphere has a 1D spherical symmetry, but the stellar Ly$\alpha$ photons penetrate the atmosphere in a direction connecting the star and planet. However, the Ly$\alpha$ mean intensity from the planetary part has a 1D spherical symmetry. In the simulations, we treat the stellar and planetary parts separately and then combine the results to obtain the distribution of the total mean intensity. The total Ly$\alpha$ mean intensity and thus the populations of H(2) have a cylindrical symmetry.

\begin{figure*}
\gridline{\fig{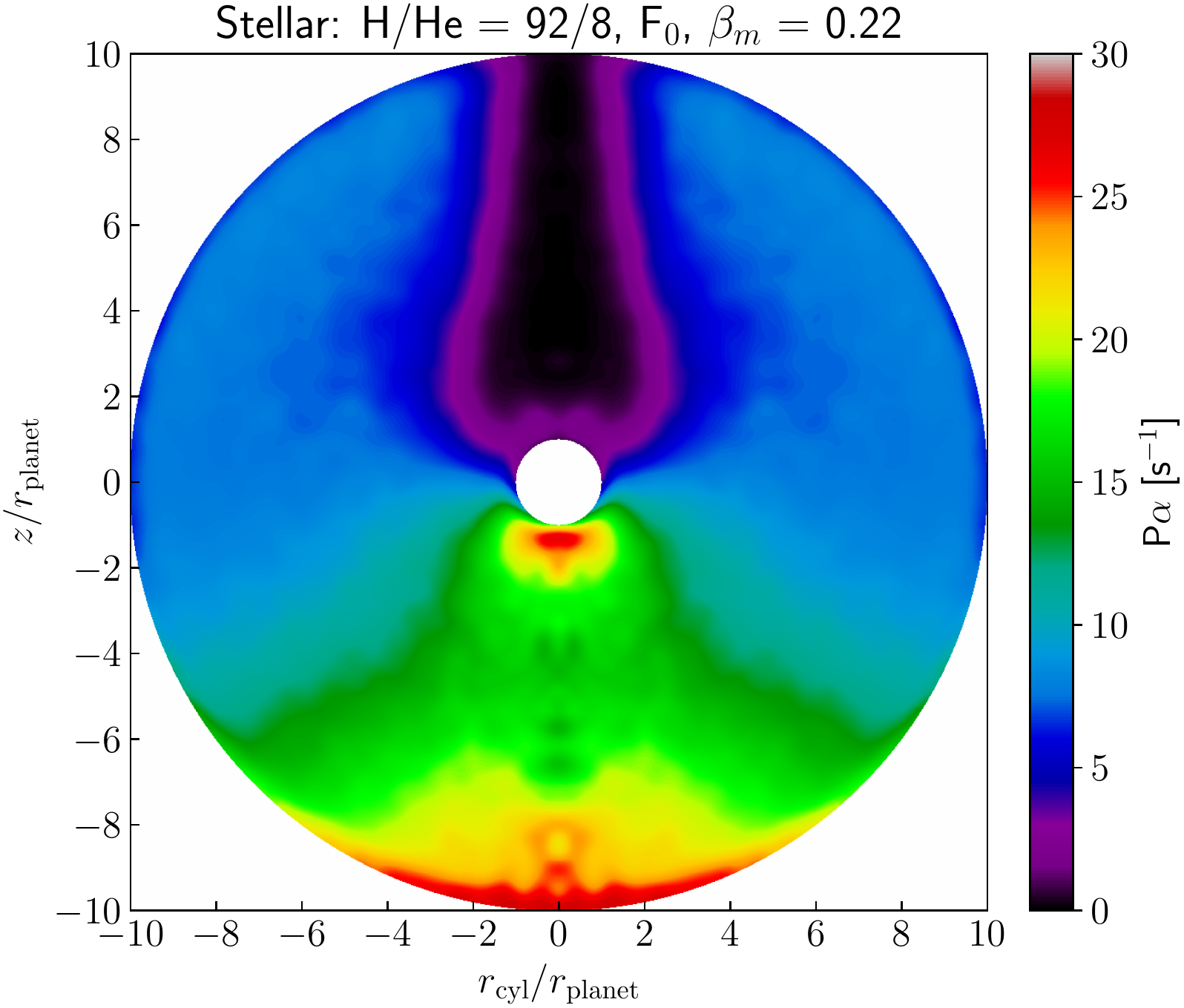}{0.5\textwidth}{(a)}
 \fig{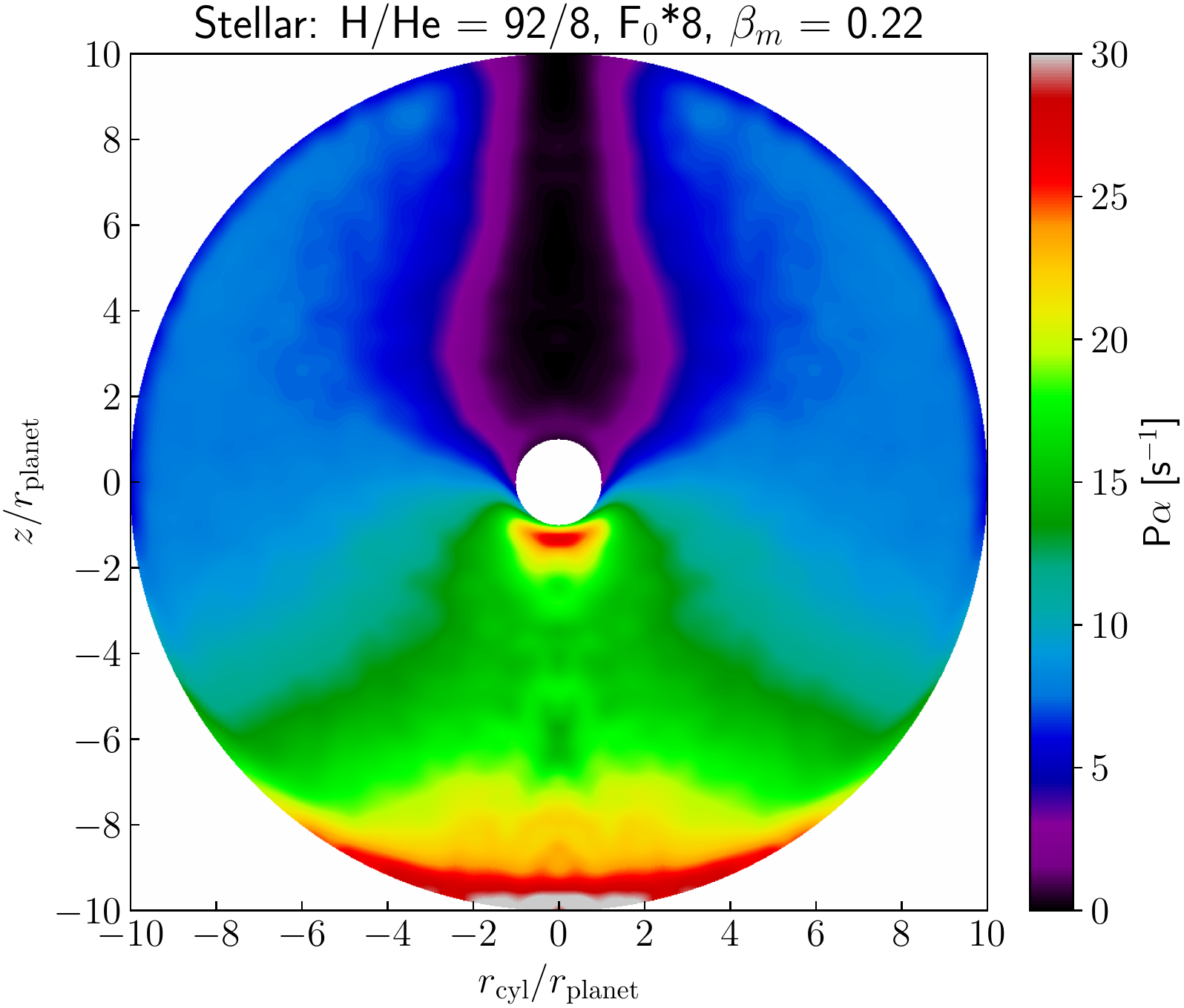}{0.5\textwidth}{(b)}
 }
\gridline{\fig{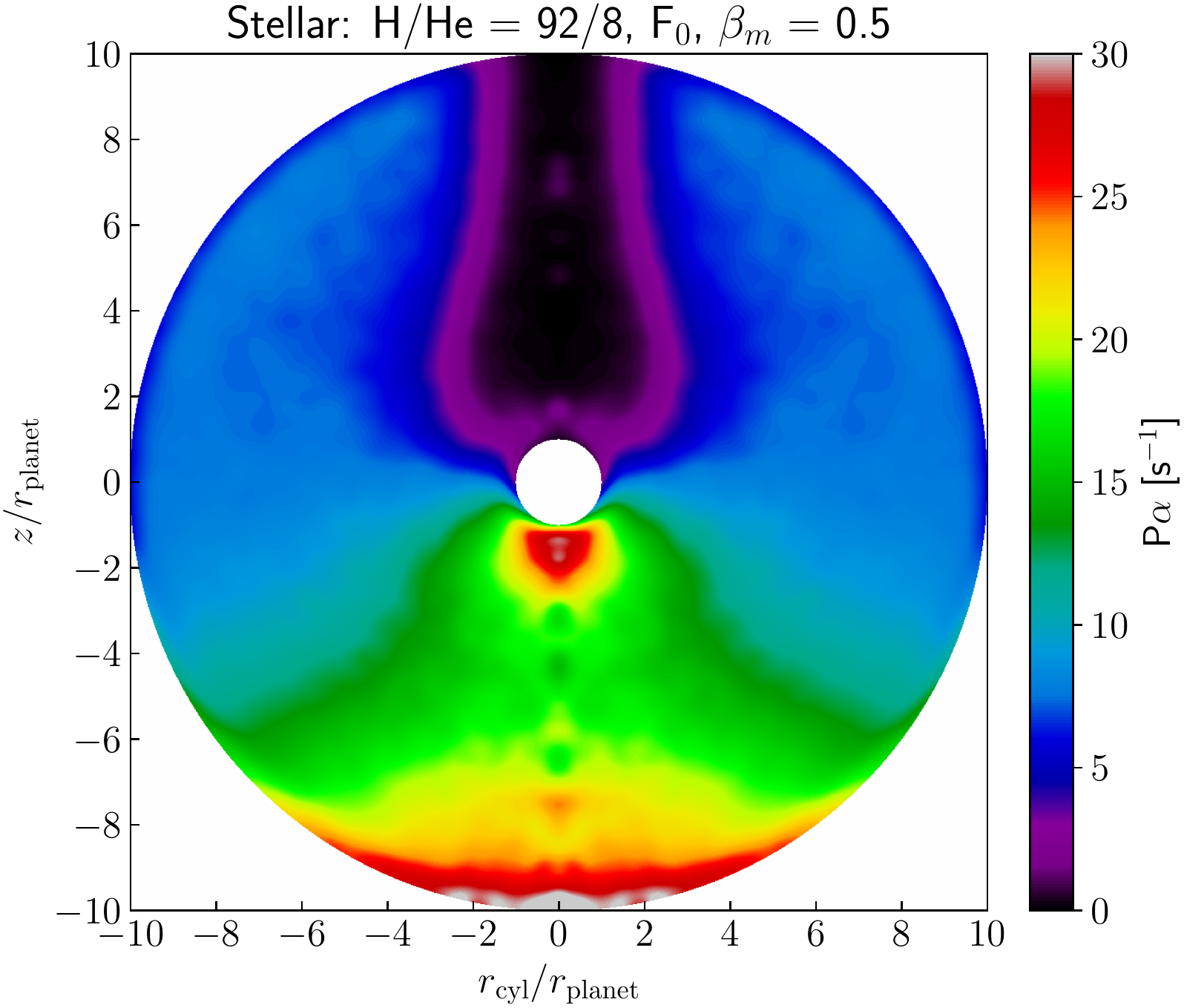}{0.5\textwidth}{(c)}
 \fig{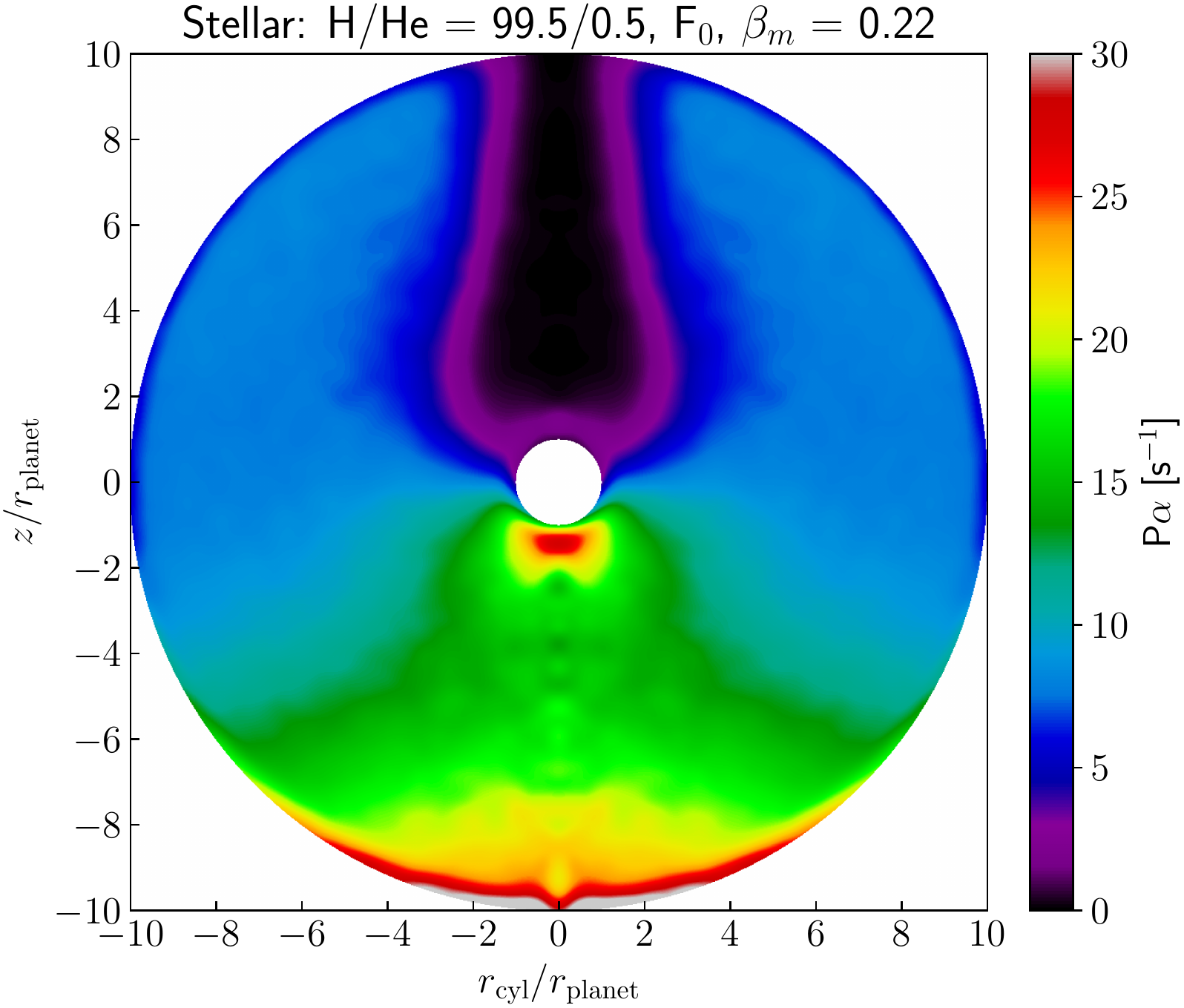}{0.5\textwidth}{(d)}
 }
\caption{$P\alpha$ for various models, in which only the stellar Ly$\alpha$ source is considered. (a) H/He = 92/8, $F_{\rm XUV}$ = F$_0$, $\beta_m$ = 0.22; (b) H/He = 92/8, $F_{\rm XUV}$ = 8F$_0$, $\beta_m$ = 0.22; (c) H/He = 92/8, $F_{\rm XUV}$ = F$_0$, $\beta_m$ = 0.5; (d) H/He = 99.5/0.5, $F_{\rm XUV}$ = F$_0$, $\beta_m$ = 0.22. Note in panels (b)-(d), the value of $P\alpha$ is slightly larger than 30 in the gray region near the edge.}
\label{fig_Pa_stellar}
\end{figure*}

In Figure \ref{fig_Pa_stellar}  we show $P\alpha$ (scattering rate in units of s$^{-1}$ per atom; $P\alpha$ = $B_{12}\bar{J}_{Ly\alpha}$) due to the stellar Ly$\alpha$ in the $z$-$r_{\rm cyl}$ plane calculated for various models. The $z$-axis connects the centers of the star and planet. 
The stellar Ly$\alpha$ comes from the -$z$ axis. In the cylindrical coordinate, $P\alpha$ is independent of the azimuth angle.
The $P\alpha$ images were obtained by azimuthally averaging the 3D cartesian cube data about the $r_{\rm cyl} = 0$ axis and then smoothed out with a Gaussian filter of $\sigma = 10$ pixels, corresponding to 0.2 $R_P$, to reduce the noisy features intrinsic to Monte Carlo calculations. In the original figures, a vertial steak owing to noise was noticeable along the $r_{\rm cyl} = 0$ axis. The steak was primarily caused because (1) most photons incident upon the atmosphere along this axis are reflected
away, and only a small number of photons pass through near the axis, and (2) only 1 or 4 cartesian cells were used to average azimuthally. To reduce the noise, one has to increase the number of photon packets by a factor proportional to the square of the signal-to-noise ratio, which increases the computational cost by the same factor. Instead, one may smooth out the images at the expense of the image resolution. We adopted the latter approach only for the display purpose in the figure; however, the H(2) population calculations were made using the original $P\alpha$ data.
We can see that the strongest $P\alpha$ appears at the edge region of the atmosphere, then drops rapidly and decreases almost by
half in the surrounding regions. This is because the incident stellar Ly$\alpha$ intensity is highest at the atmosphere upper part and the photons incident onto the atmosphere tend to be reflected by scattering.  
It is well known that the Ly$\alpha$ photons are reflected away when they go into a higher density region from a lower density region \citep{1991ApJ...370L..85N, 2006MNRAS.367..979H,2016ApJ...833L..26G}.
$P\alpha$ becomes stronger again at $r\sim 1-2R_P$. This is because some of the Ly$\alpha$ photons (mostly wing photons) will penetrate into deeper regions by chance. Then, they will be trapped there, yielding a huge number of scatterings, which increases $P\alpha$ there. 
In the nightside region, where z$\geq$0, $P\alpha$ drops dramatically. Because of blocking by the planet rocky core, only a small fraction of Ly$\alpha$ photons can reach the shadow region of the planet.

Figure \ref{fig_Pa_stellar} (a) is for the model with H/He = 92/8, $F_{\rm XUV}$ = F$_0$, and $\beta_m$ = 0.22. To investigate the influence of different $F_{\rm XUV}$, we show $P\alpha$ of the model with H/He = 92/8, $F_{\rm XUV}$ = 8F$_0$, and $\beta_m$ = 0.22 in Figure \ref{fig_Pa_stellar} (b). The difference of $P\alpha$ is not very discernible from the two figures.
It shows a similar trend to that in Figure \ref{fig_Pa_stellar} (a), but the extent of the regions with the highest $P\alpha$ (plotted in red color) at the outer boundary is a little larger. To investigate the influence of $\beta_m$, we show the $P\alpha$ of the model with H/He = 92/8, $F_{\rm XUV}$ = F$_0$, and $\beta_m$ = 0.5 in Figure \ref{fig_Pa_stellar} (c).
It also presents a similar trend to Figures \ref{fig_Pa_stellar} (a) and (b) but tends to have higher $P\alpha$ values at the dayside boundary impacted by Ly$\alpha$ and
the deeper regions near the planet.
In Figure \ref{fig_Pa_stellar}(d), we show $P\alpha$ of the model with H/He = 99.5/0.5, $F_{\rm XUV}$ = F$_0$, and $\beta_m$ = 0.22 in order to examine the influence of H/He ratio on $P\alpha$. The highest $P\alpha$ values in this model are only marginally less extended at the outer boundaries and the regions near the
planet than in (b) and (c) models, but seems to be slightly more extended than in (a) model.

\begin{figure*}
\gridline{\fig{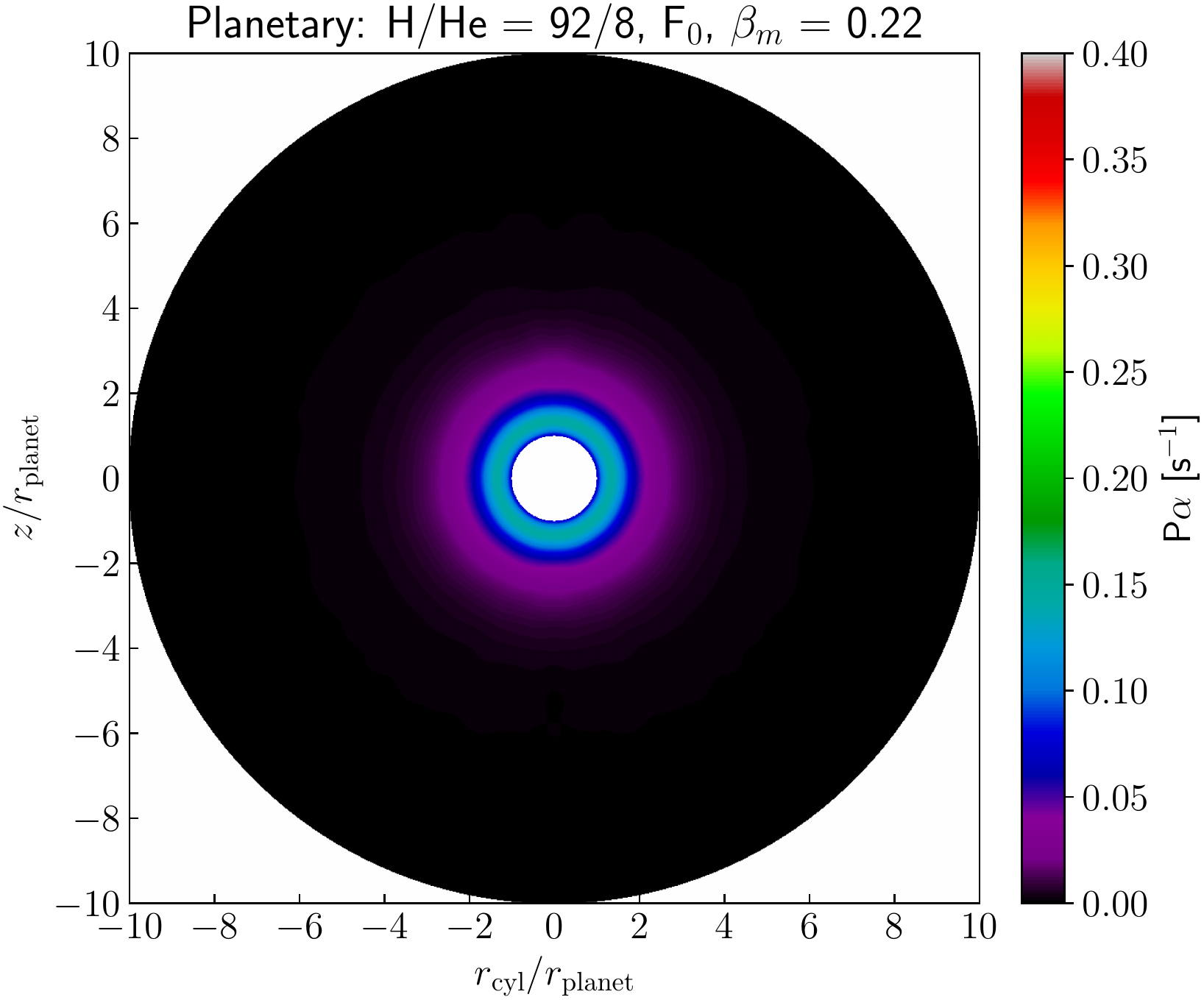}{0.5\textwidth}{(a)}
 \fig{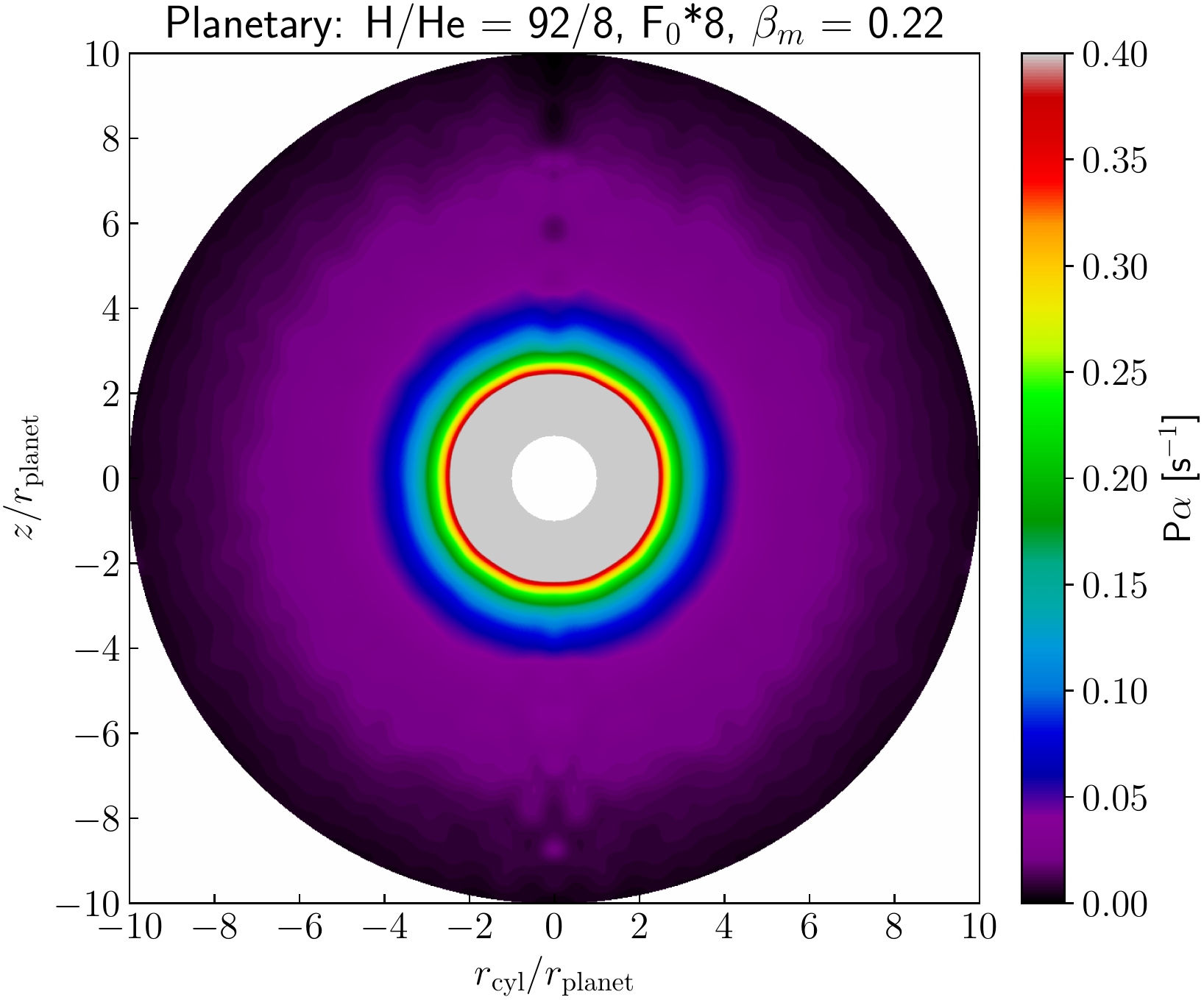}{0.5\textwidth}{(b)}
 }
\gridline{\fig{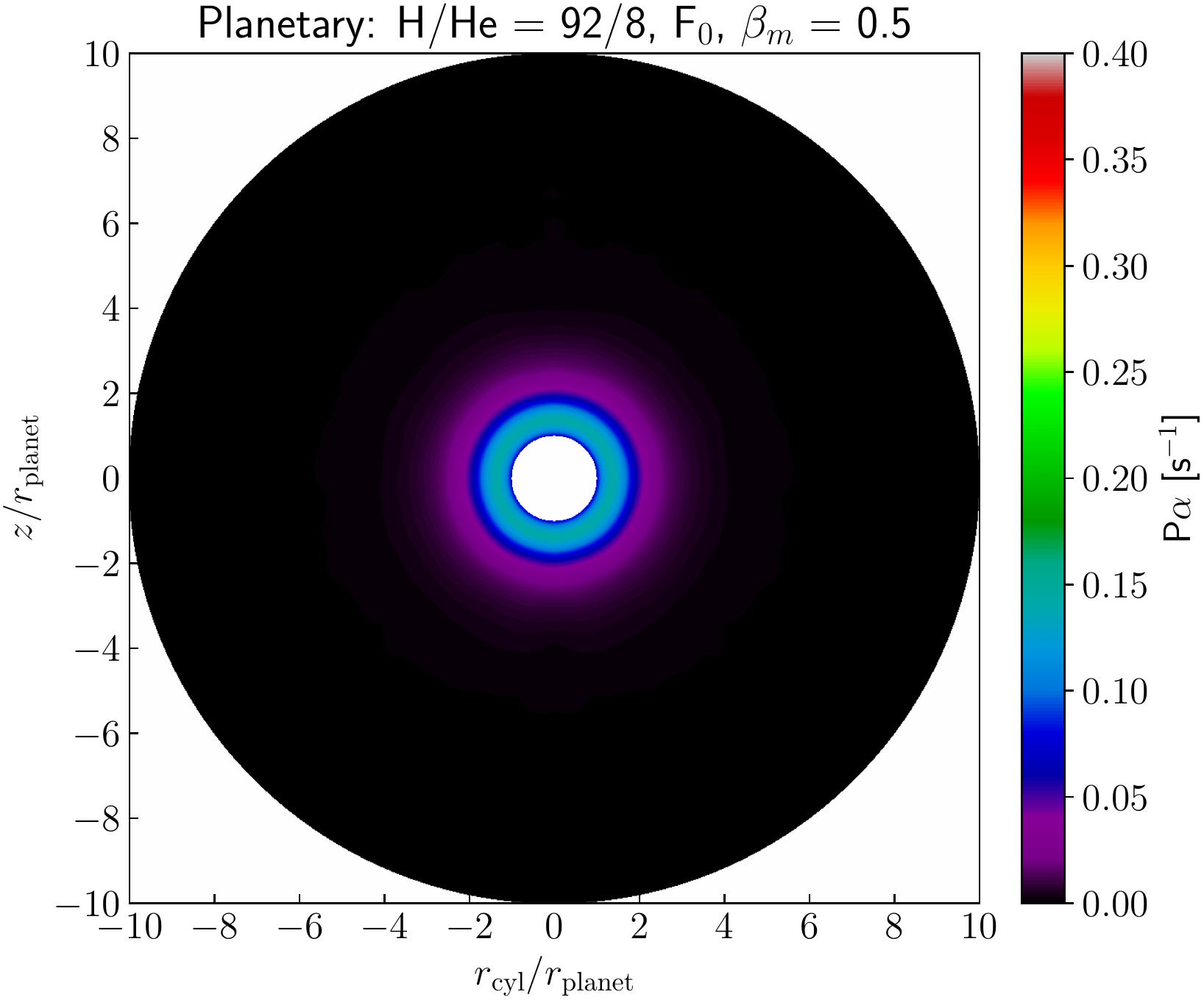}{0.5\textwidth}{(c)}
 \fig{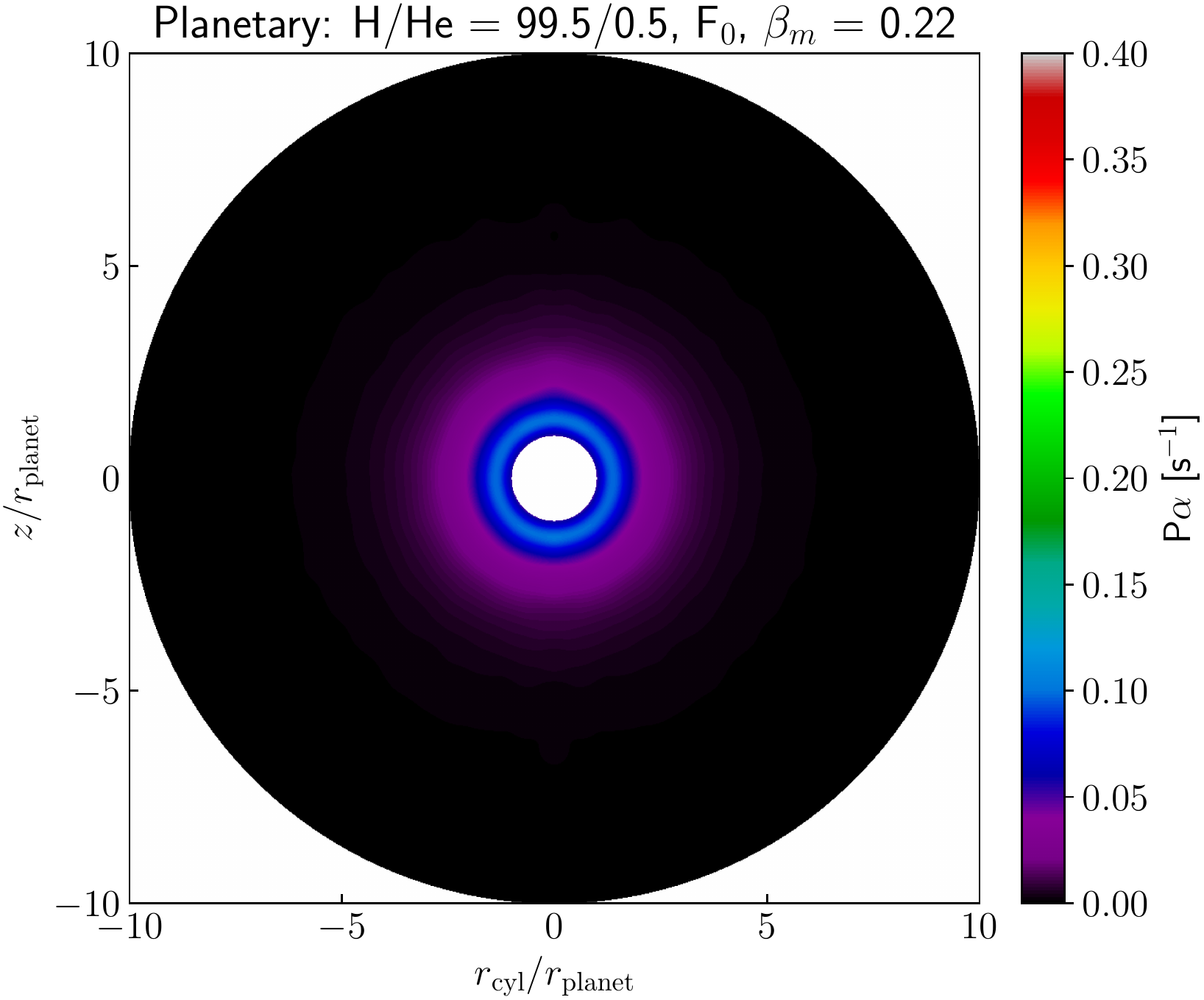}{0.5\textwidth}{(d)}
 }
\caption{$P\alpha$ for various models, in which only the planetary Ly$\alpha$ source is considered. (a) H/He = 92/8, $F_{\rm XUV}$ = F$_0$, $\beta_m$ = 0.22; (b) H/He = 92/8, $F_{\rm XUV}$ = 8F$_0$, $\beta_m$ = 0.22; (c) H/He = 92/8, $F_{\rm XUV}$ = F$_0$, $\beta_m$ = 0.5; (d) H/He = 99.5/0.5, $F_{\rm XUV}$ = F$_0$, $\beta_m$ = 0.22.}
\label{fig_Pa_diffuse}
\end{figure*}

Figures \ref{fig_Pa_diffuse} (a)-(d) show $P\alpha$ caused by the planetary Ly$\alpha$ for the models corresponding to those in Figure \ref{fig_Pa_stellar}, which is only tenths of a few s$^{-1}$.
Compared to that of the stellar Ly$\alpha$, it is almost negligible, especially in the dayside of the atmosphere. The model with $F_{\rm XUV}$ = 8F$_0$ is found to give the highest $P\alpha$ values in Figure \ref{fig_Pa_diffuse}. However, the Ly$\alpha$ luminosity from the planetary atmosphere is negligible, and the standing-out result shown in Figure \ref{fig_Pa_diffuse} (b) does not affect the final result in the population of H(2).
After we obtain $P\alpha$, we then calculate the populations of H(2) by solving the NLTE statistical equation. 

\begin{figure*}
\gridline{\fig{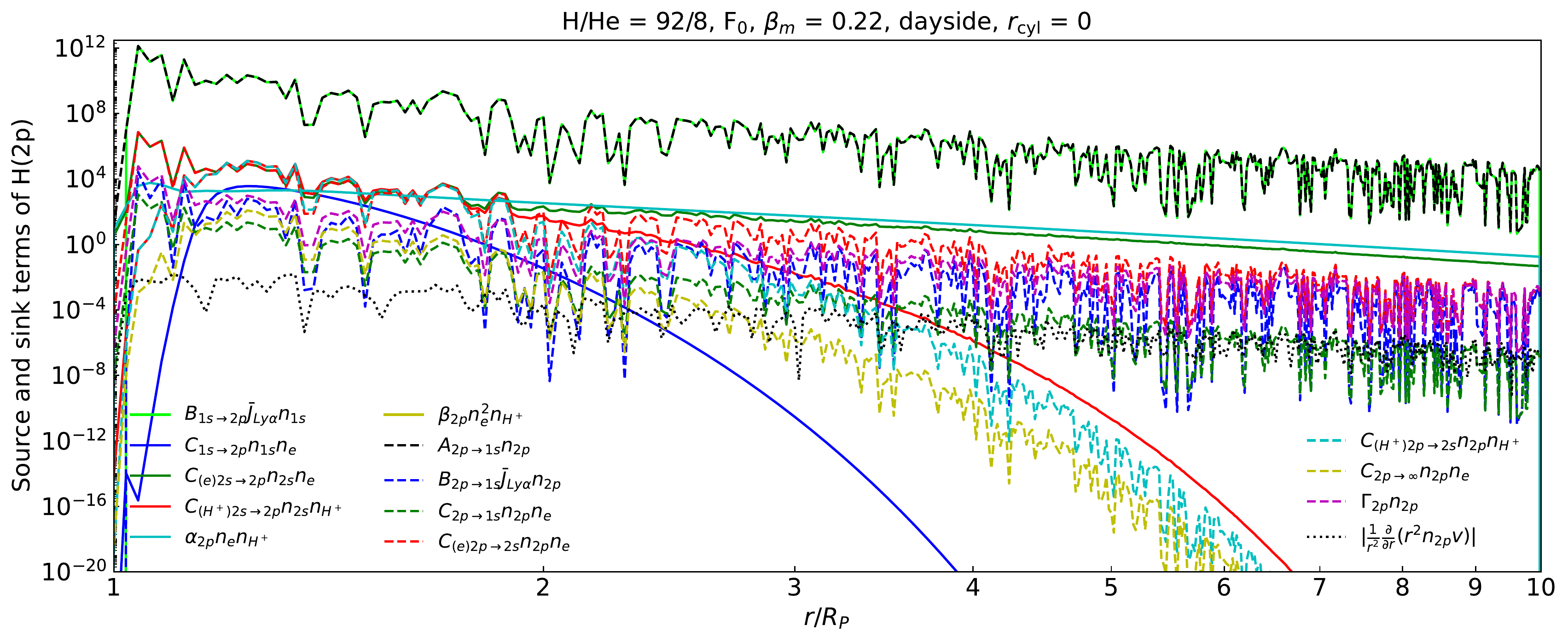}{0.9\textwidth}{(a)}
 }
\gridline{\fig{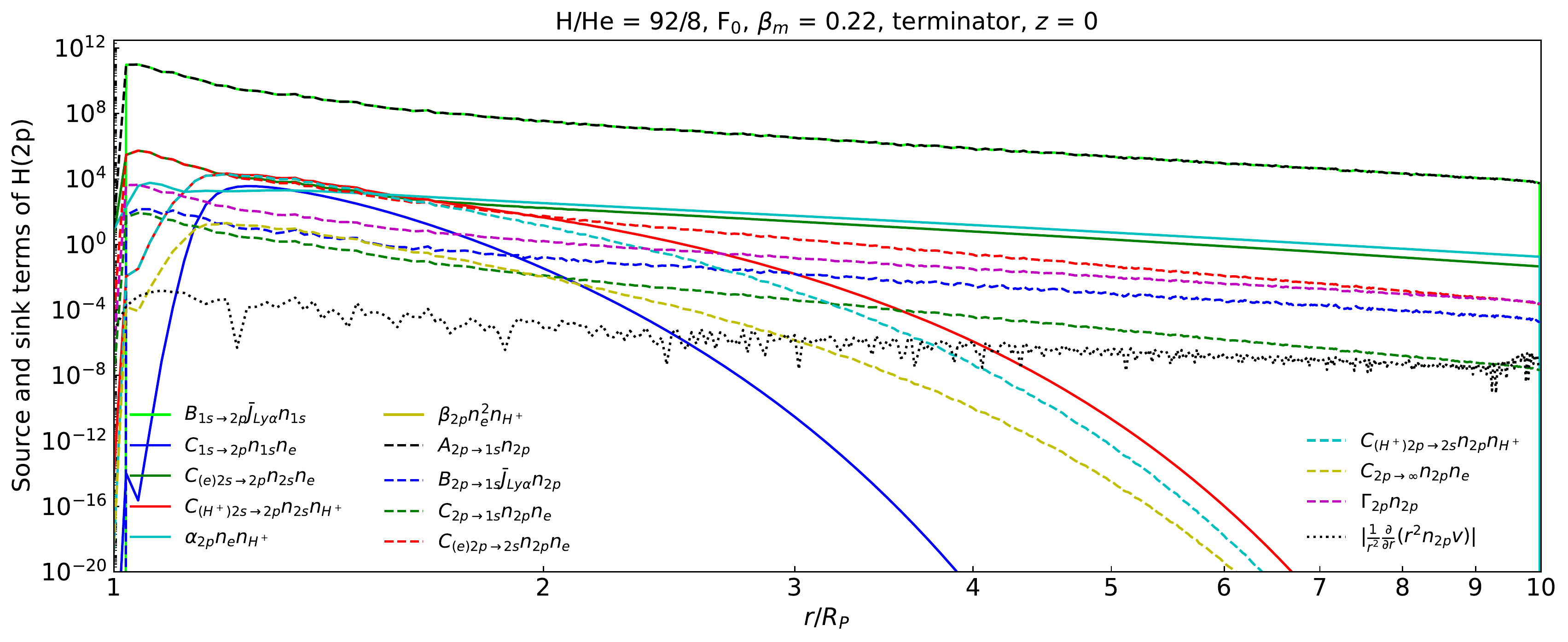}{0.9\textwidth}{(b)}
 }
 \gridline{\fig{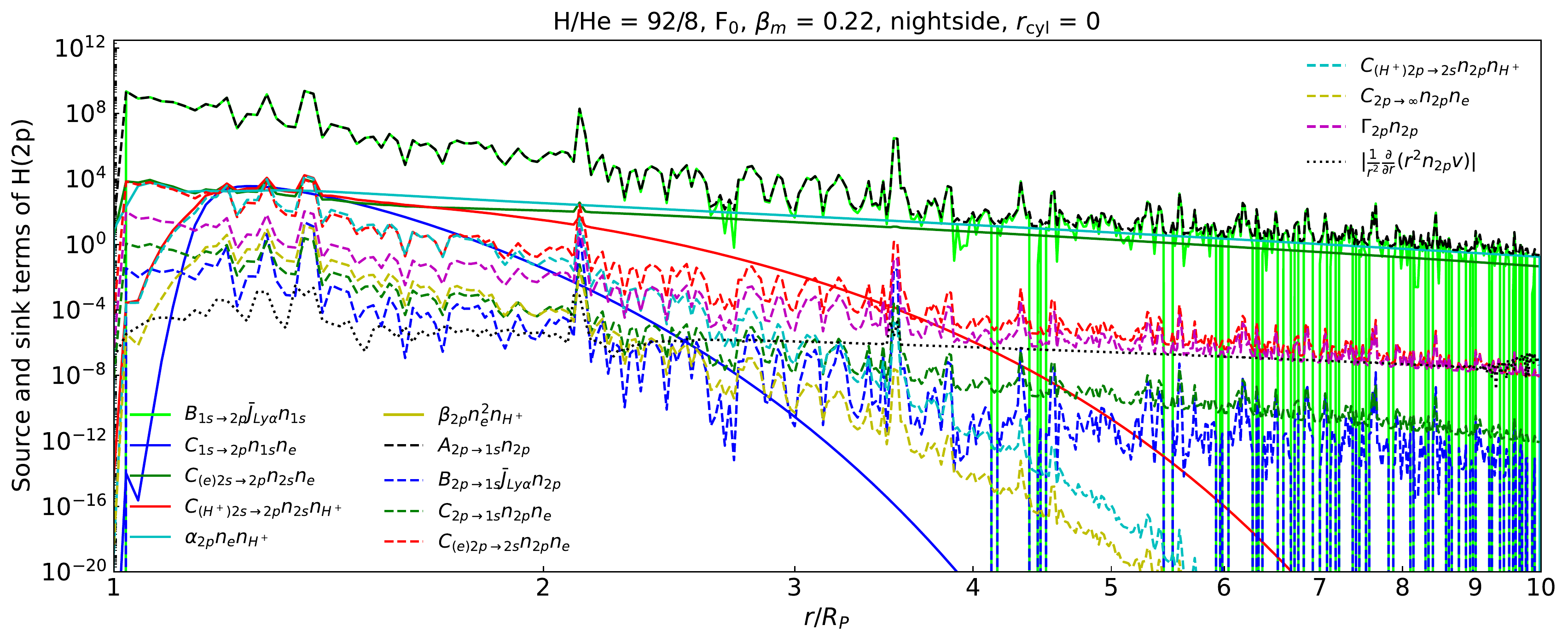}{0.9\textwidth}{(c)}
 }
\caption{The radial profiles of the source (solid lines) and sink (dashed lines) terms for H(2p) in the model of H/He = 92/8, $F_{\rm XUV}$ = F$_0$, and $\beta_m$ = 0.22. The black dotted line represents the absolute value of the advection term $\frac{1}{r^2}\frac{\partial{}}{\partial{r}}(r^2n_{2p}v)$. From top to bottom, the figure shows the cases of the dayside where $r_{\rm cyl}$ = 0, the terminator where $z$ = 0, and the nightside where $r_{\rm cyl}$ = 0, respectively. In each panel, the three-body recombination term $\beta_{2p}n^2_{e}n_{H^+}$ is so small that the line representing this term doesn't appear in the plot.}
\label{fig_H2p_source}
\end{figure*}

To clearly illustrate the mechanisms that influence the production and destruction of H(2p), Figure \ref{fig_H2p_source} shows the radial profiles of the source (solid lines) and sink (dashed lines) terms of H(2p) for the model of H/He = 92/8, $F_{\rm XUV}$ = F$_0$, and $\beta_m$ = 0.22. 
From top to bottom, the figure shows the cases of the dayside where $r_{\rm cyl} = 0$, the terminator where $z=0$, and the nightside
where $r_{\rm cyl}=0$, respectively. We can see that the dominant production mechanism of H(2p) is the photoexcitation by Ly$\alpha$
(Ly$\alpha$ pumping), as shown in the solid green line. In the nightside ($r \gtrsim 4R_p$), where the Ly$\alpha$ intensity
decreases substantially, the terms due to the recombination to H(2p) and collisional excitation from H(2s) are comparable to that
due to the Ly$\alpha$ pumping. The figure also demonstrates that the advection term is negligible. We calculated the advection term $\frac{1}{r^2}\frac{\partial{}}{\partial{r}}(r^2n_{2p}v)$ using the density profile of H(2p) calculated ignoring advection. In the figure, the black dotted line represents the absolute value of the term calculated in this way. Even though this
estimation of the advection term would not be self-consistent, it is clear that the advection term has little effect on the production or destruction of
H(2p). Therefore, we can safely ignore it in our study.

\begin{figure*}
\gridline{\fig{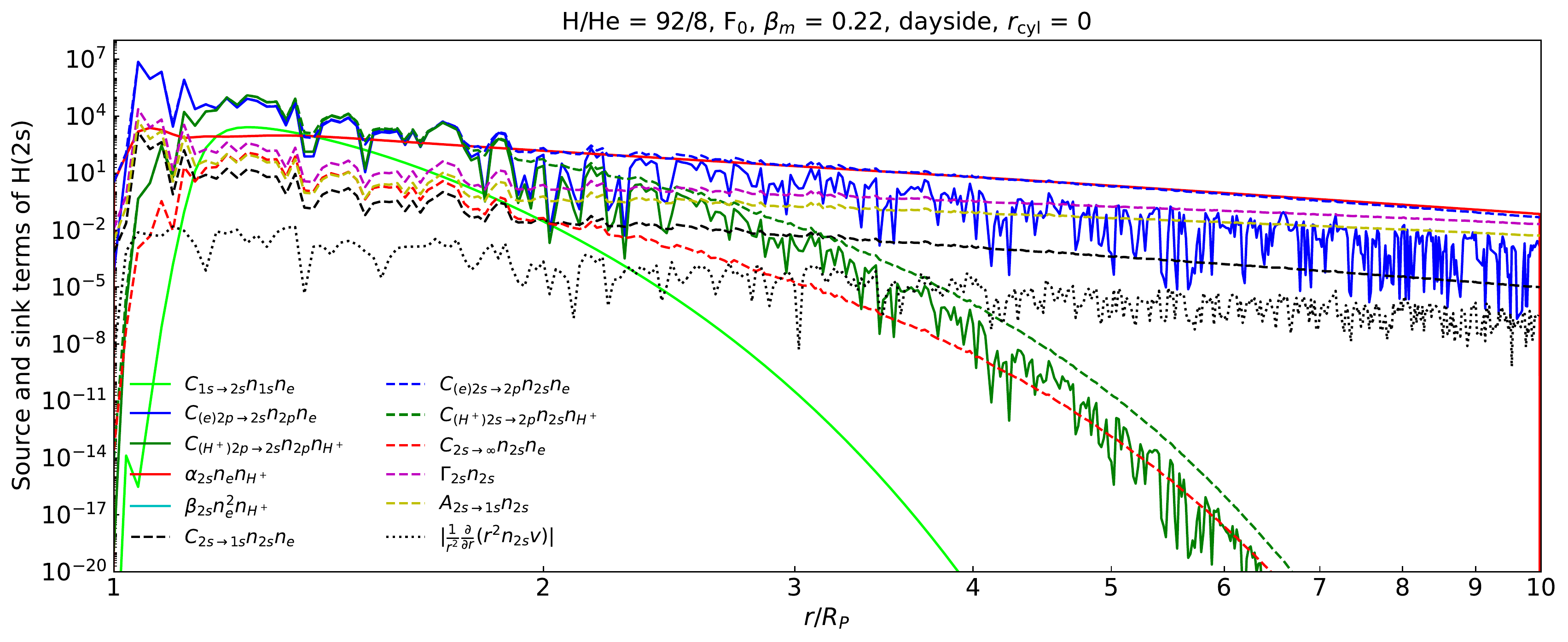}{0.9\textwidth}{(a)}
 }
\gridline{\fig{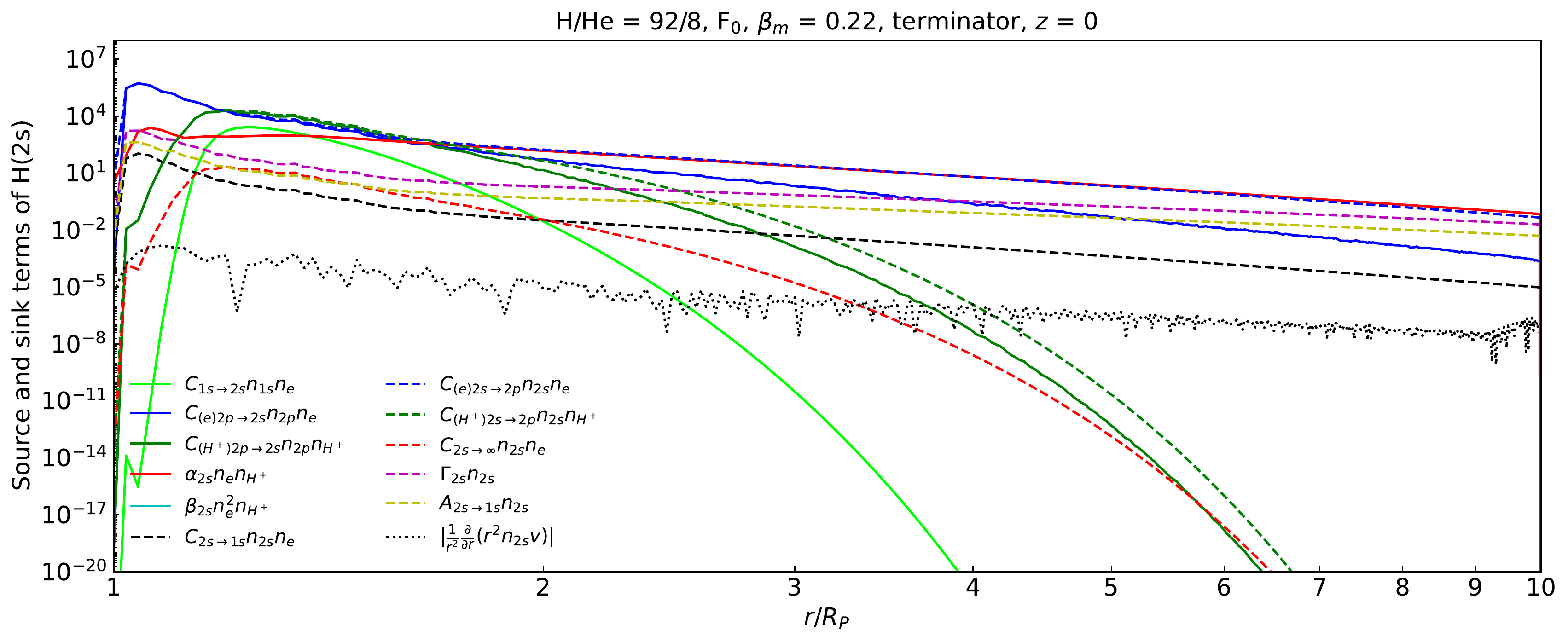}{0.9\textwidth}{(b)}
 }
 \gridline{\fig{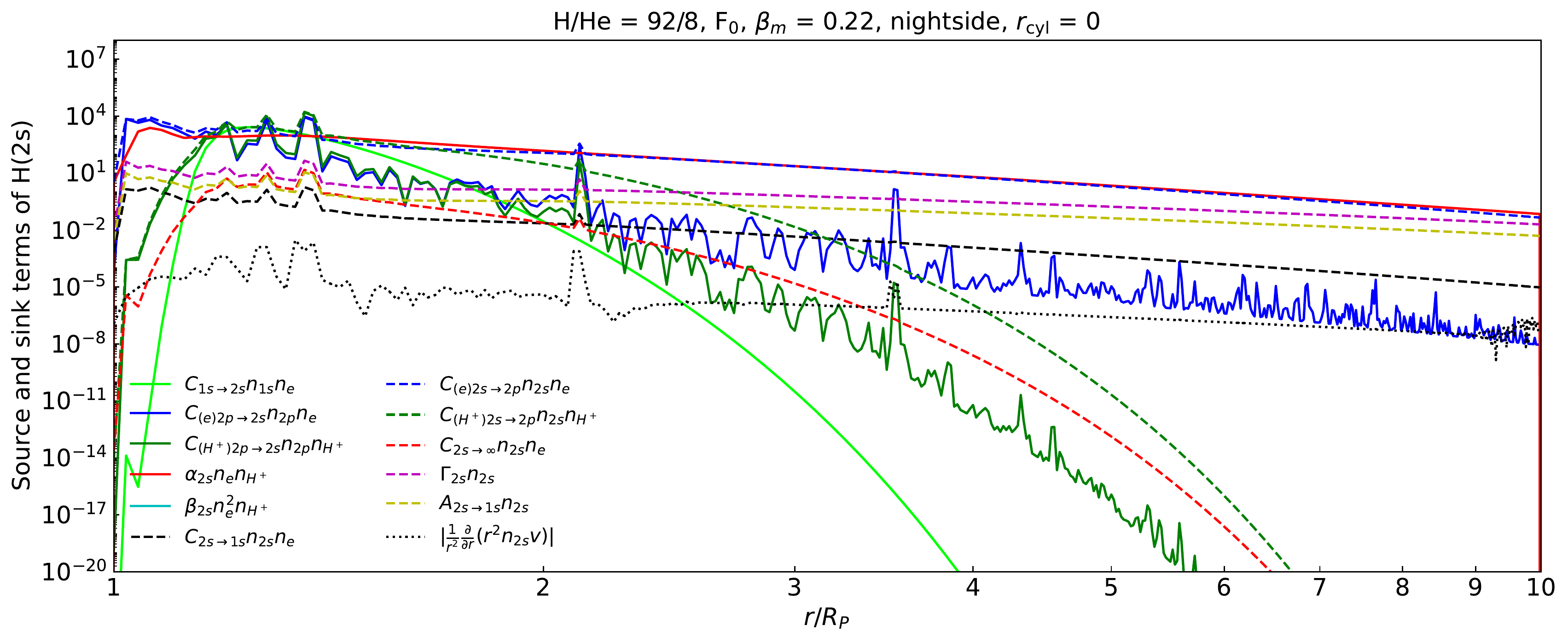}{0.9\textwidth}{(c)} 
 }
\caption{The same as Figure \ref{fig_H2p_source}, but for H(2s).}
\label{fig_H2s_source}
\end{figure*}

Figure \ref{fig_H2s_source} shows the source and sink terms for H(2s). The dominant source of H(2s) is the collisional transition from H(2p) due to electrons and protons at lower altitude ($\lesssim 2 R_P$ for the dayside case), and the recombination becomes dominant at higher altitude. The destruction of H(2s) is dominated by the collisional transition to H(2p) and the photoionization is also effective at some altitudes. Similarly to that for H(2p), the advection term for H(2s) is too small compared to other terms and can be ignored.

\begin{figure*}
\gridline{\fig{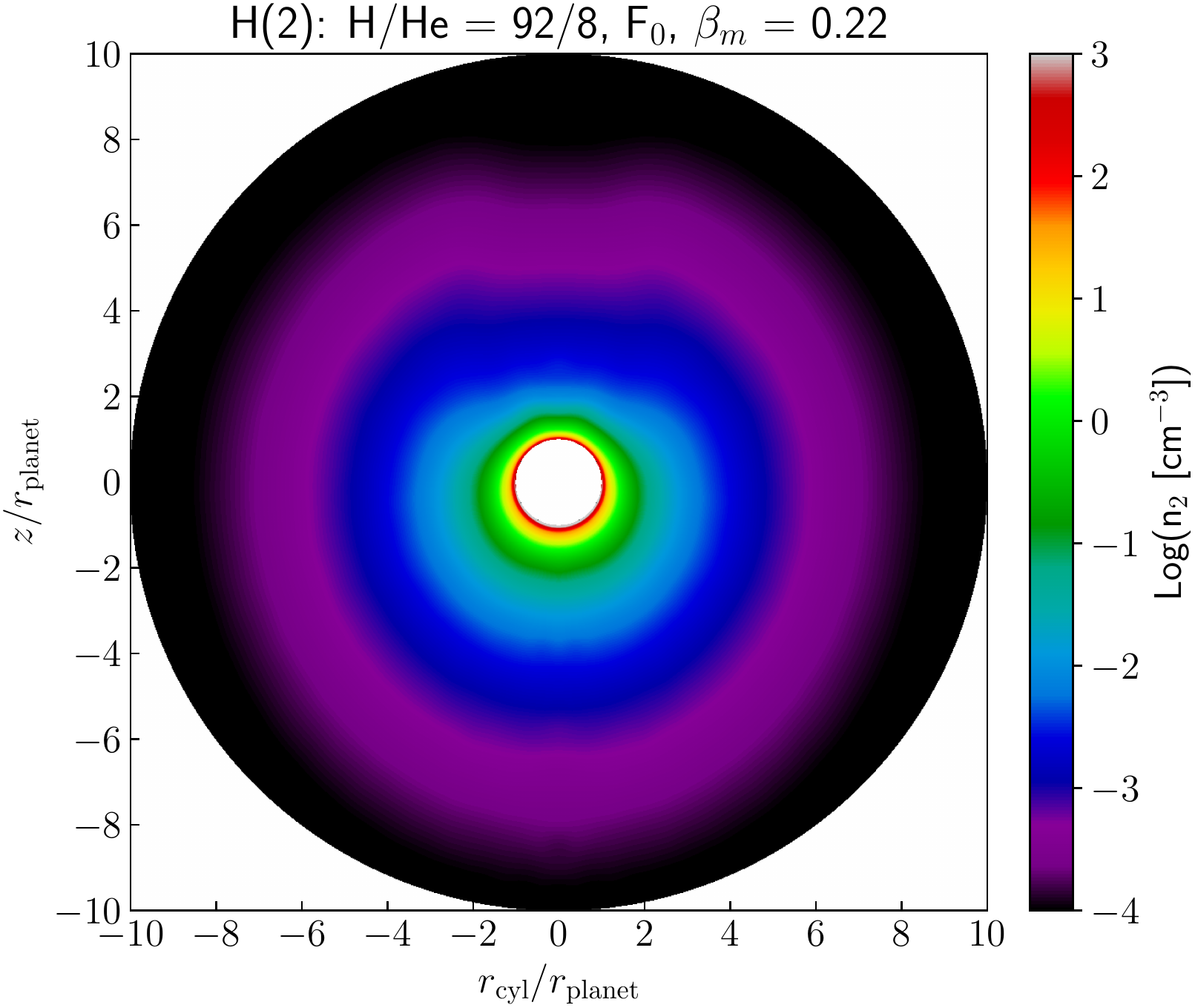}{0.5\textwidth}{(a)}
 \fig{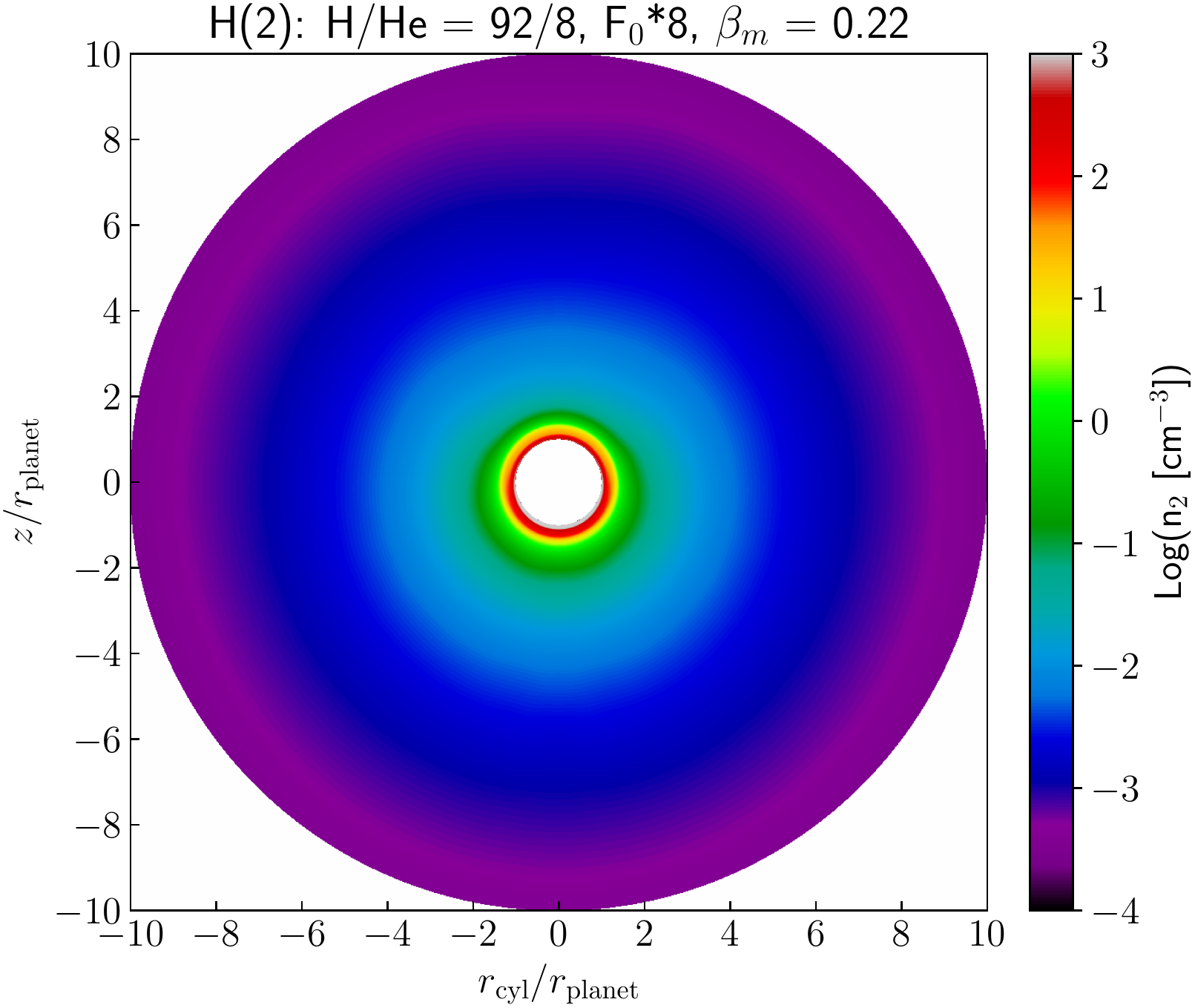}{0.5\textwidth}{(b)}
 }
\gridline{\fig{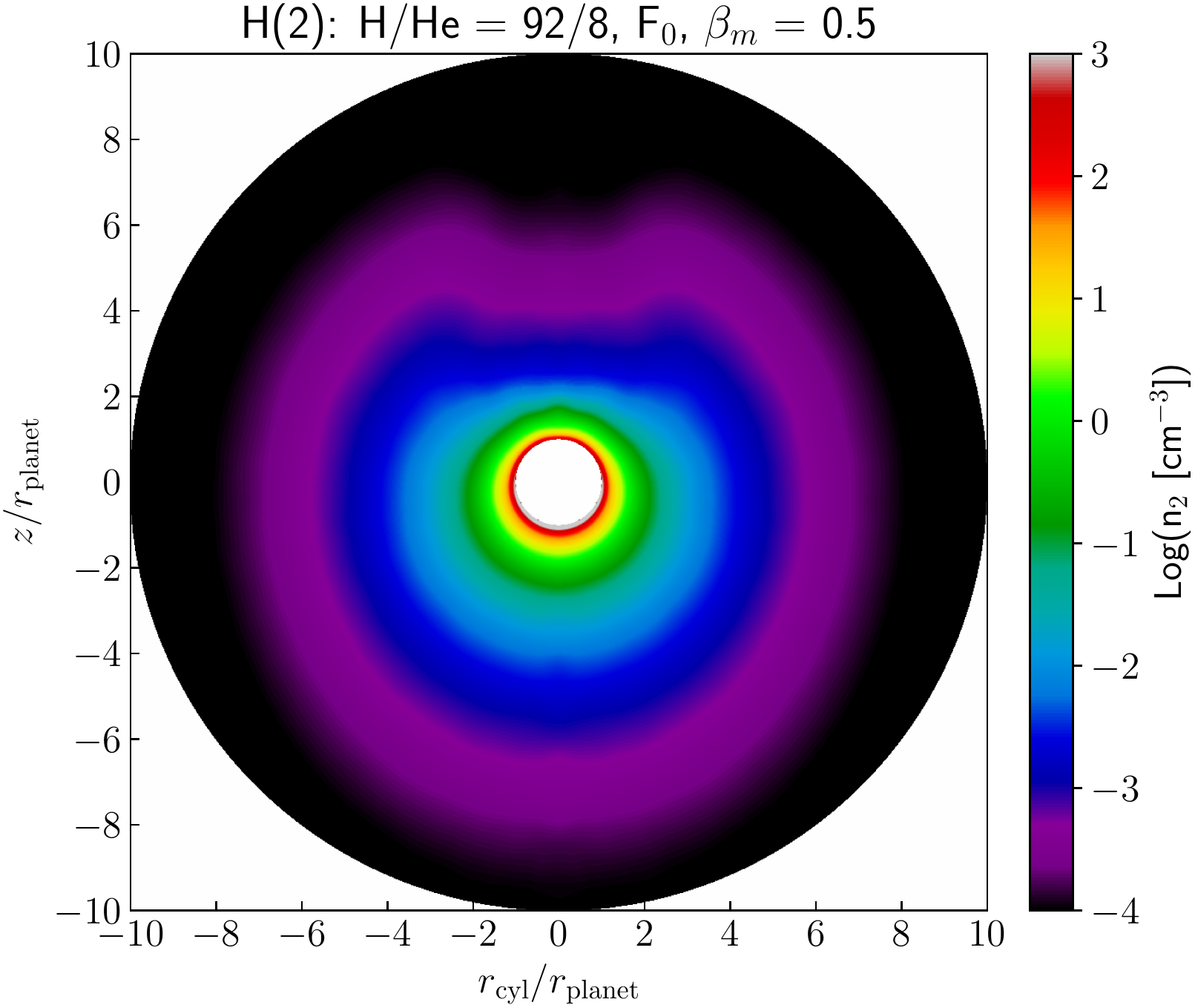}{0.5\textwidth}{(c)}
 \fig{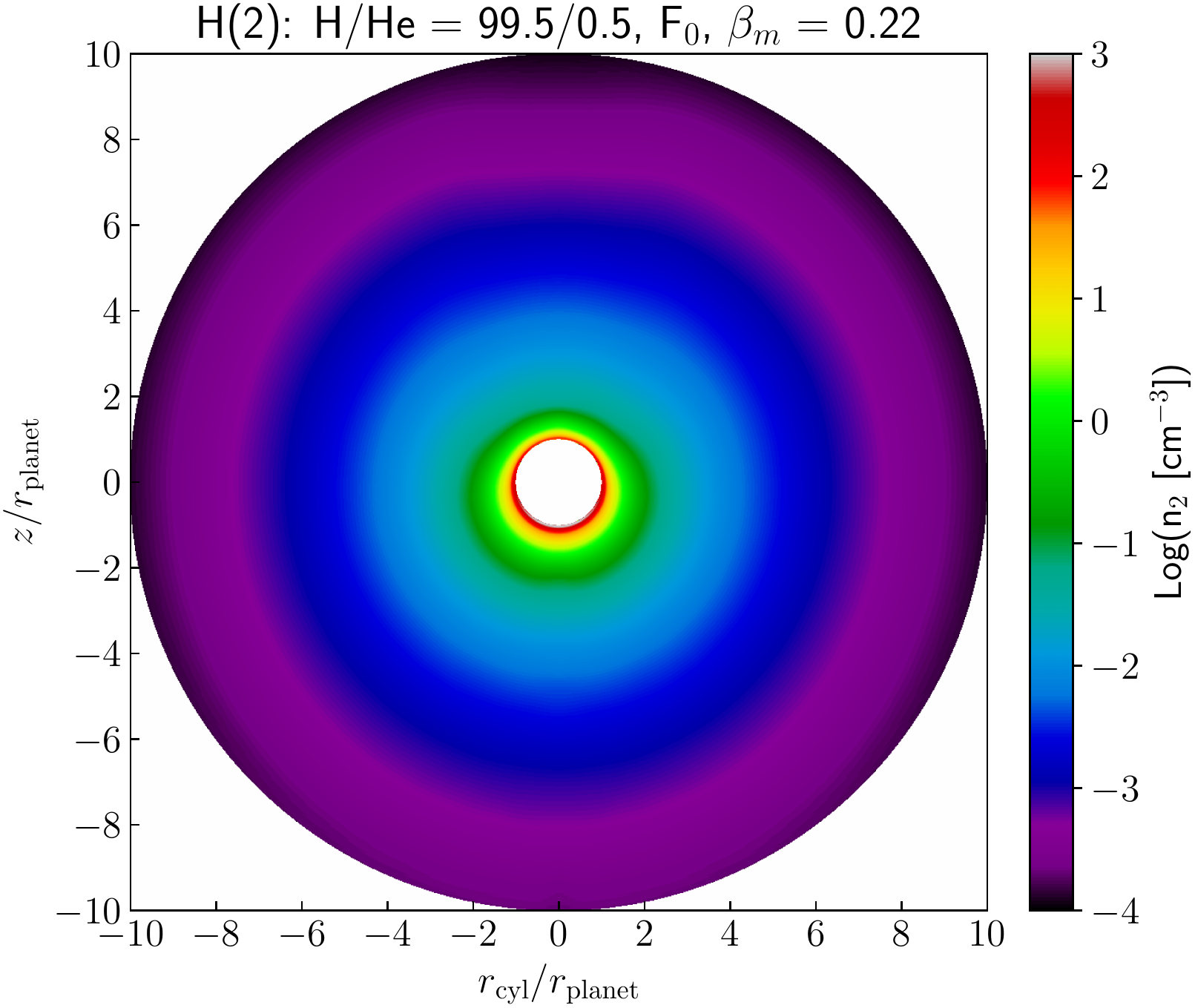}{0.5\textwidth}{(d)}
 }
\caption{H(2) number density for various models. (a) H/He = 92/8, $F_{\rm XUV}$ = F$_0$, $\beta_m$ = 0.22; (b) H/He = 92/8, $F_{\rm XUV}$ = 8F$_0$, $\beta_m$ = 0.22; (c) H/He = 92/8, $F_{XUV}$ = F$_0$, $\beta_m$ = 0.5; (d) H/He = 99.5/0.5, $F_{\rm XUV}$ = F$_0$, $\beta_m$ = 0.22.}
\label{fig_H2_pop}
\end{figure*}
Figures \ref{fig_H2_pop} (a)-(d) show the number density of H(2) of the models corresponding to those in Figure \ref{fig_Pa_stellar}. We can see that most H(2) atoms are present in the dayside. We also found a higher $\beta_m$ leads to a higher number density of H(2), especially near the lower atmosphere. A higher $F_{\rm XUV}$ and H/He ratio are found to increase the number density of H(2) at high altitudes.
Because the differences in $P\alpha$ and the number density of H(2) are not very discernible in Figure \ref{fig_H2_pop}, to demonstrate the influence of $F_{\rm XUV}$ on the density of H(2) more clearly, we examine the column  densities of H(2), H(2p), and H(2s) as a function of the projected radius ($r_{obs}/R_P$) in Figures \ref{atm_90_10_b022} (d)-(f). The column densities of H(2p) and H(2) exhibit similar overall trends to the number density of H(1s). At the relatively low altitudes, the column density of H(2) has almost the same structure as H(2p) because H(2p) is dominant in the population of H(2). On the other hand, at high altitudes, it is strongly affected or even dominated by H(2s). The column density of H(2s) increases with the increase of $F_{\rm XUV}$. 
Likewise, Figures \ref{atm_90_10_fxuv1} (d)-(f) show the column densities of H(2), H(2p), and H(2s) as a function of $r_{obs}$ for the models with different $\beta_m$. We find that the column densities of both H(2p) and H(2s) increase with $\beta_m$. As a result, a higher $\beta_m$ leads to more H(2) atoms. 

Figures \ref{atm_fxuv1_b022} (d)-(f) show the column densities of H(2), H(2p), and H(2s) for models with different H/He ratios. Between the models with H/He = 98/2, 99/1, and 99.5/0.5,
we can see that the column densities of H(2p) have little difference, but the column densities of H(2s) and thus H(2) differ slightly. The model with H/He = 92/8, however, shows relatively lower column densities of H(2), H(2p), and H(2s), compared to those with higher H/He ratios. The lower column density of H(2) in the model with H/He = 92/8 is mainly because the H(1s) density is low, as can be seen in Figure \ref{atm_fxuv1_b022} (c). The scattering rate $P\alpha$ that determines the population of H(2p), however, doesn't show much difference compared to the case of H/He = 99.5/0.5, $F_{\rm XUV}$ = F$_0$, and $\beta_m$ = 0.22, as shown in Figure \ref{fig_Pa_stellar}(d).

\subsection{Modeling H$\alpha$ transmission spectrum}

\begin{figure*}
\gridline{\fig{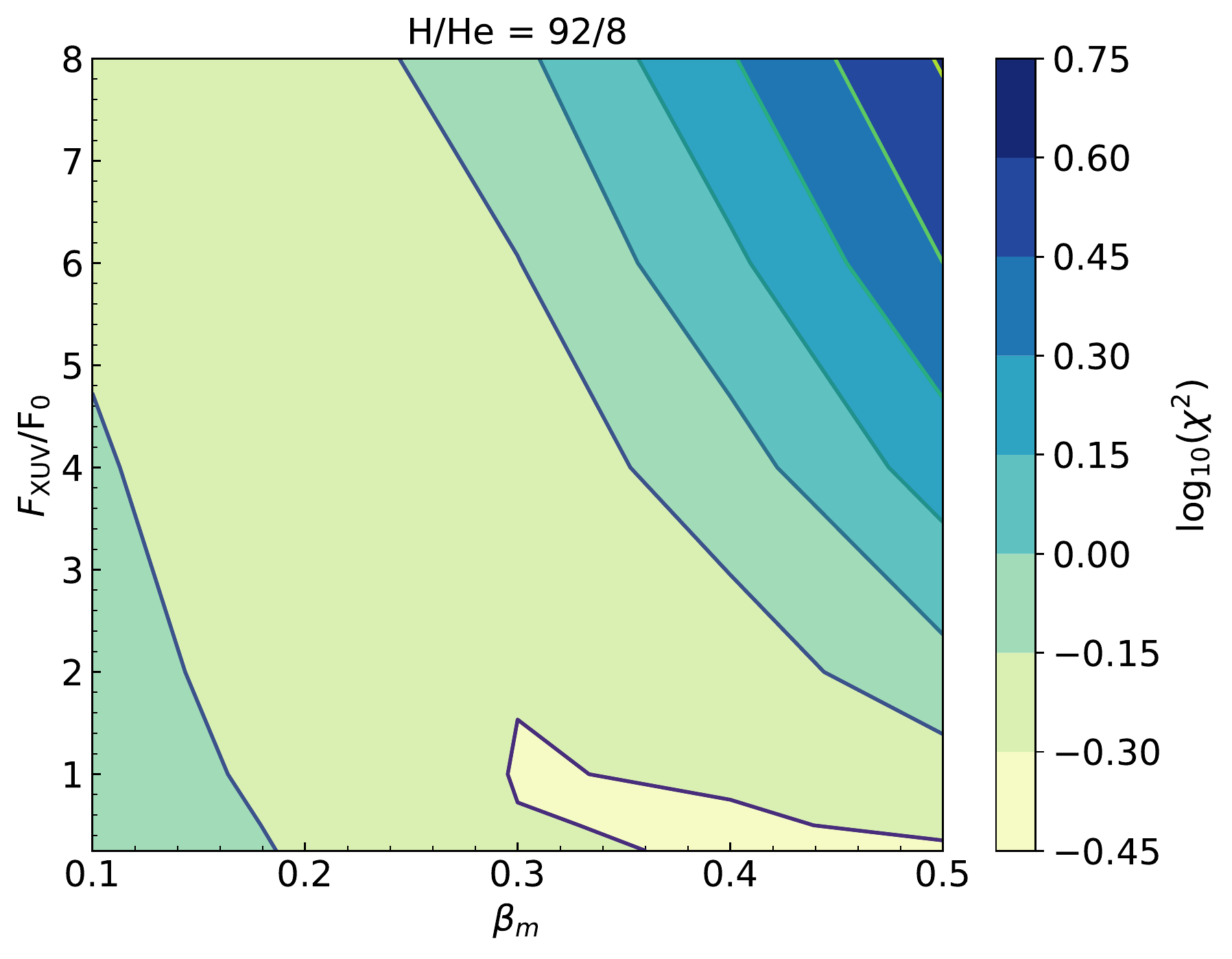}{0.5\textwidth}{(a)}
 \fig{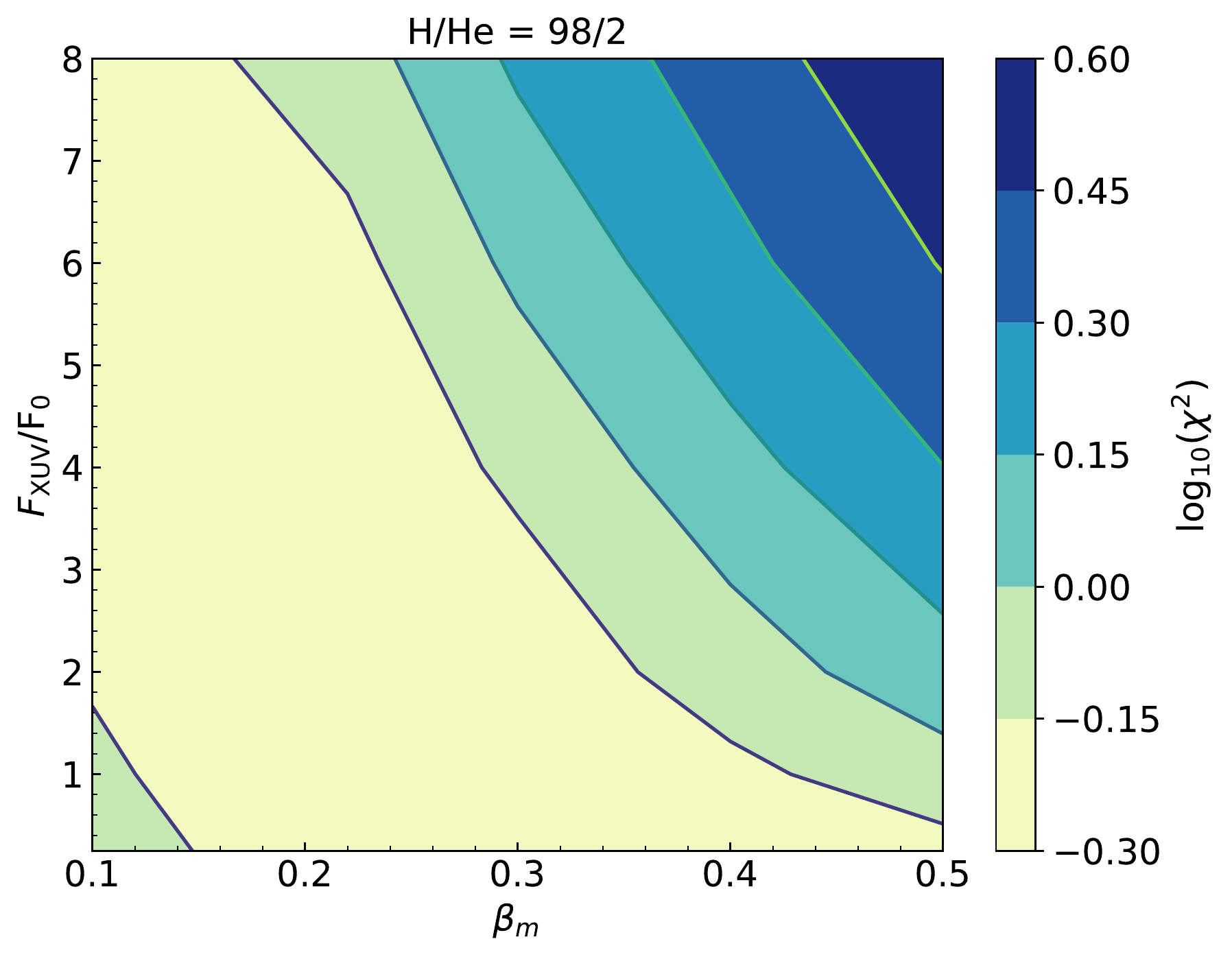}{0.5\textwidth}{(b)}
 }
\gridline{\fig{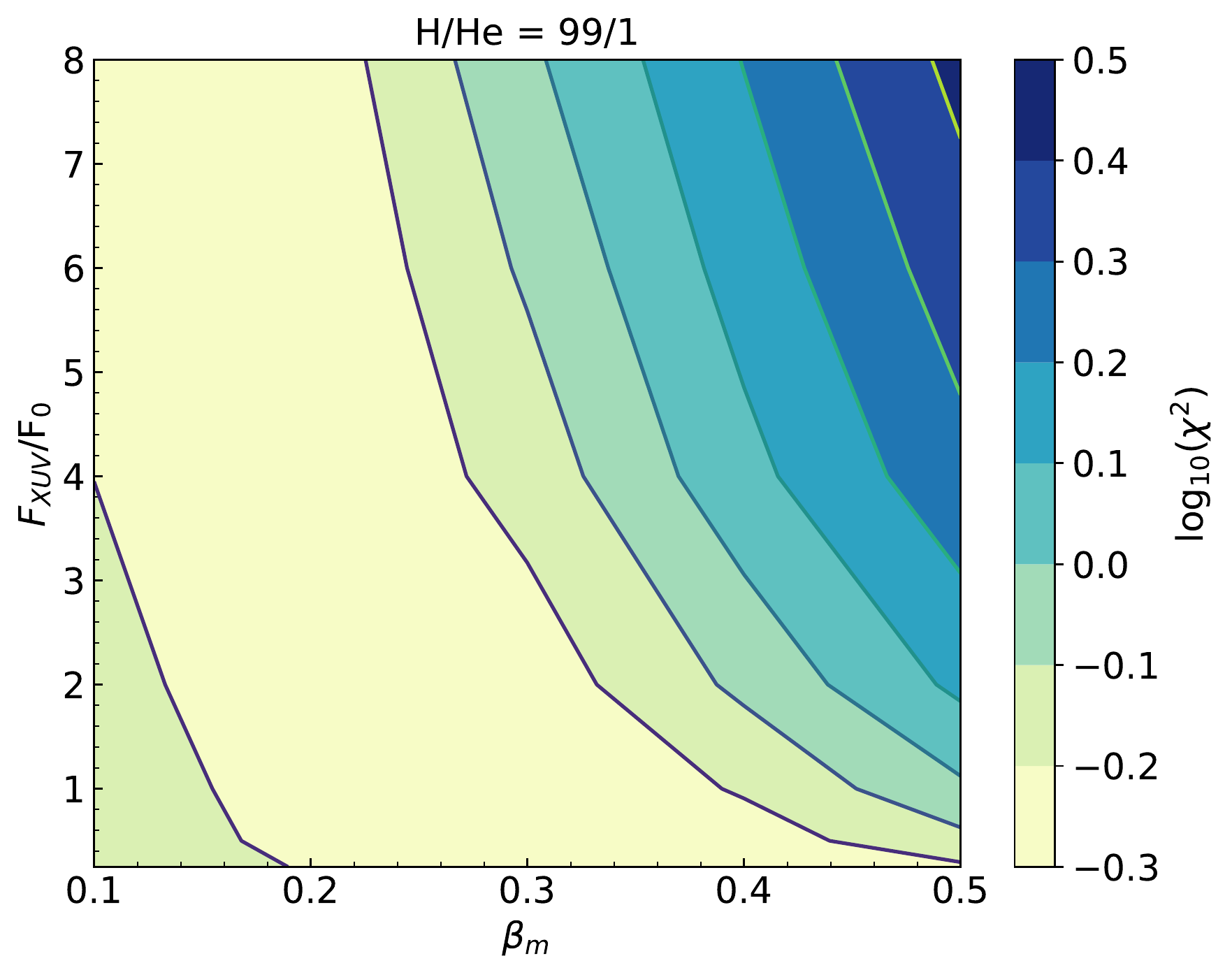}{0.5\textwidth}{(c)}
 \fig{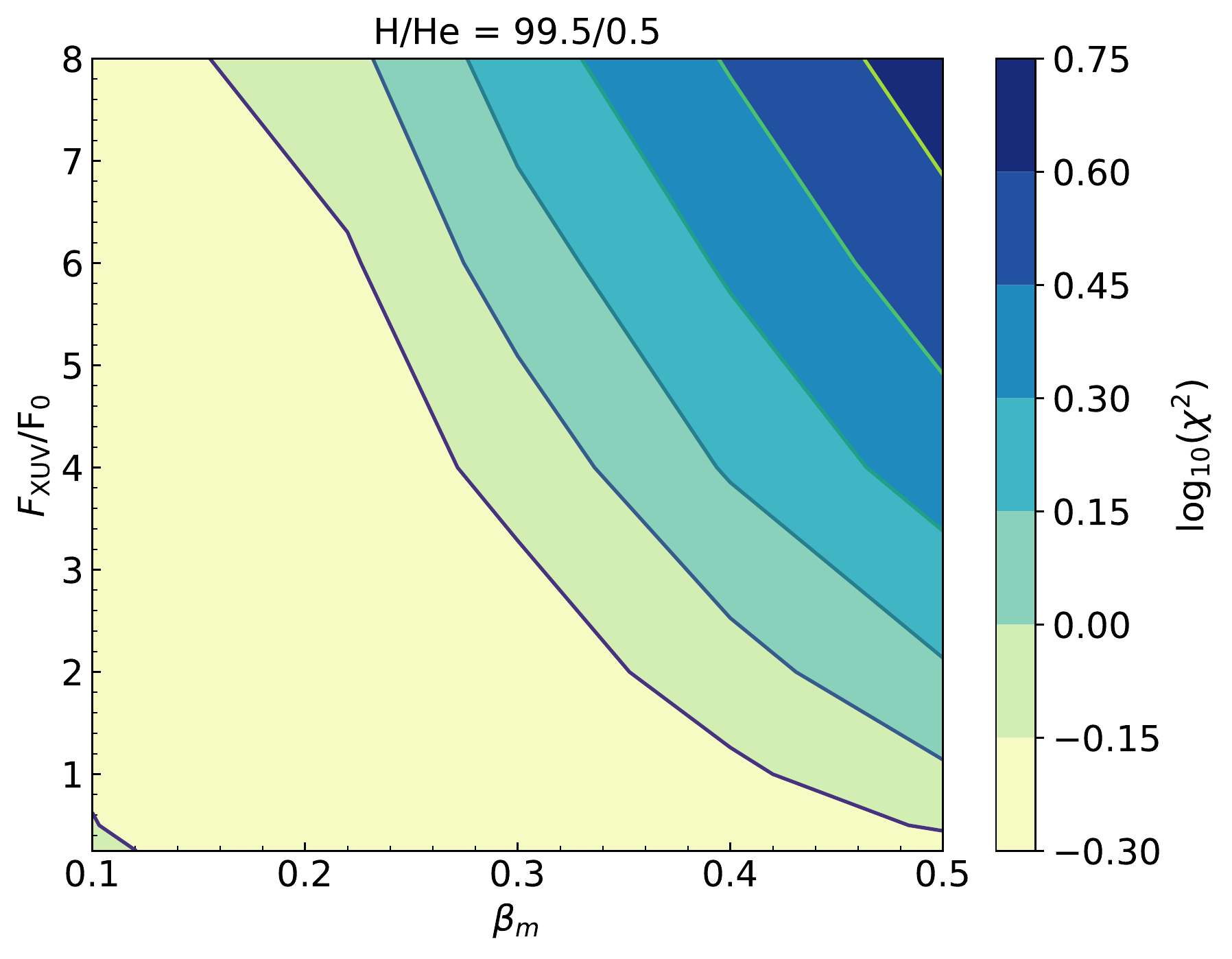}{0.5\textwidth}{(d)}
 }
\caption{$\chi^2$ contours for H$\alpha$ transmission spectrum, obtained by comparing the model spectrum with the observation data in the wavelength range of [6562.05, 6563.55]$\rm\AA$. In (a)-(d), the panels show the $\chi^2$ contours for various hydrogen to helium abundance ratios. Note the $\chi^2$ contour levels in color bars are shown in logarithm scale}. 
\label{fig_HaTS_chi_unbinned}
\end{figure*}

We modeled the H$\alpha$ transmission spectrum and compared it with the observations. 
To compare with the observations, we calculated $\chi^2$ in the passband [6562.05, 6563.55]$\rm\AA$, i.e., taking 0.75$\rm\AA$ from each side of the line center. This wavelength range was chosen because the absorption is almost negligible outside of it.

Figures \ref{fig_HaTS_chi_unbinned} (a)-(d) show the contours of $\chi^2$ as a function of $F_{\rm XUV}$ and $\beta_m$ for models with H/He = 92/8, 98/2, 99/1, and 99.5/0.5. If we regard the models with  $\chi^2\leq 1$ to be good fits, then we find that $\beta_m$ should be low when $F_{\rm XUV}$ is high. For small $F_{\rm XUV}$, $\chi^2$ is found to be very small over a wide range of $\beta_m$. The region with $\chi^2\leq 1$ spans a wide $F_{\rm XUV}$ and $\beta_m$ range for all the H/He ratio cases.
For example, in the case of H/He = 92/8 and $F_{\rm XUV}$ = F$_0$, the models with $0.2 \leq \beta_m \leq 0.5$ can explain the observation well. The models with higher H/He ratios (98/2, 91/1, and 99.5/0.5) and smaller $F_{\rm XUV}$ ($\textless F_0$) can fit the observation regardless of $\beta_m$. For high $F_{\rm XUV}$, the models with large $\beta_m$ do not give a good fit. For instance, when $F_{\rm XUV}$ = 2F$_0$, the model with $\beta_m$ = 0.5 gives $\chi^2\geq1$. For $F_{\rm XUV}$ = 8F$_0$, the models can fit the observation only when $\beta_m\leq 0.3$.

\begin{figure*}
\gridline{\fig{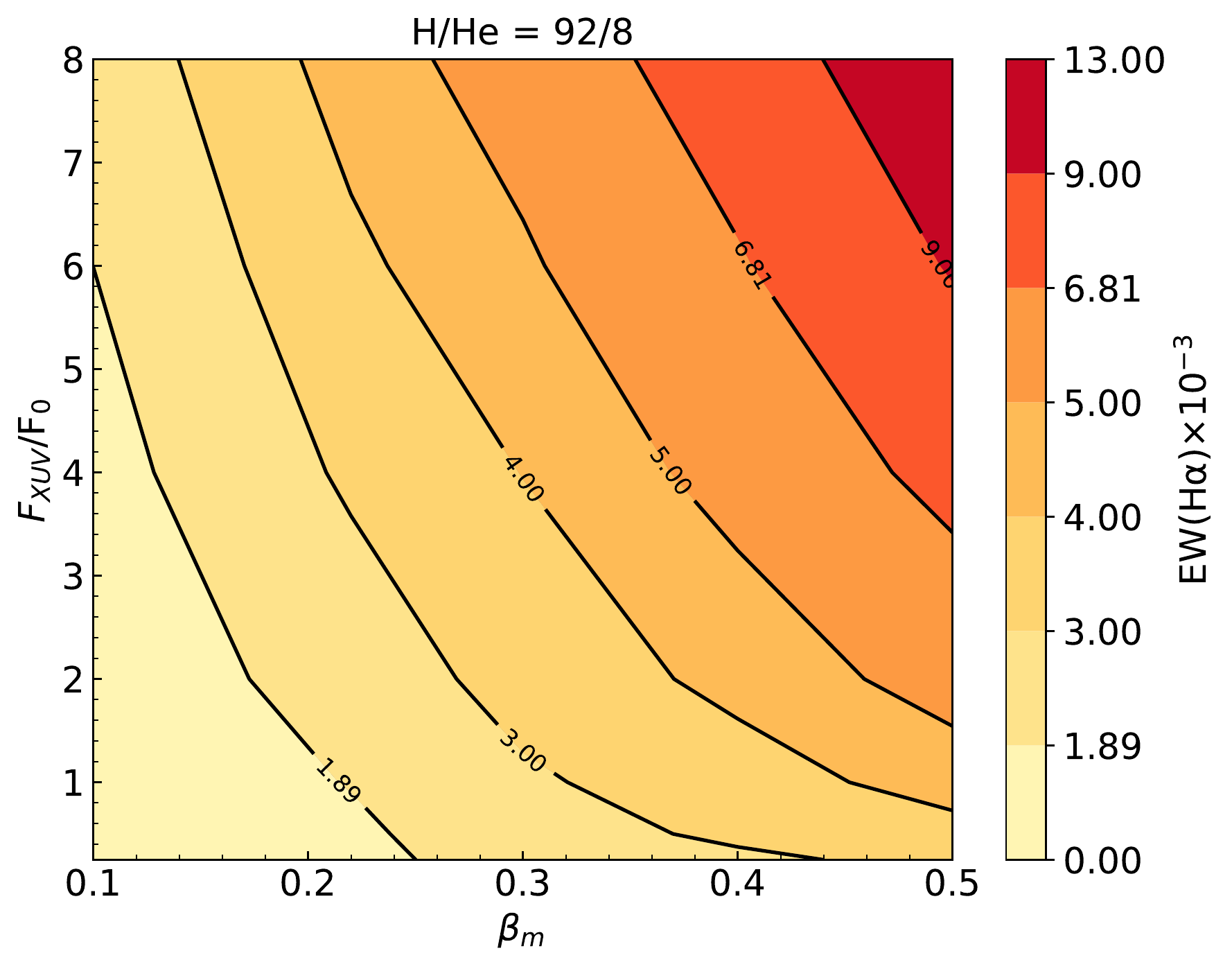}{0.5\textwidth}{(a)}
 \fig{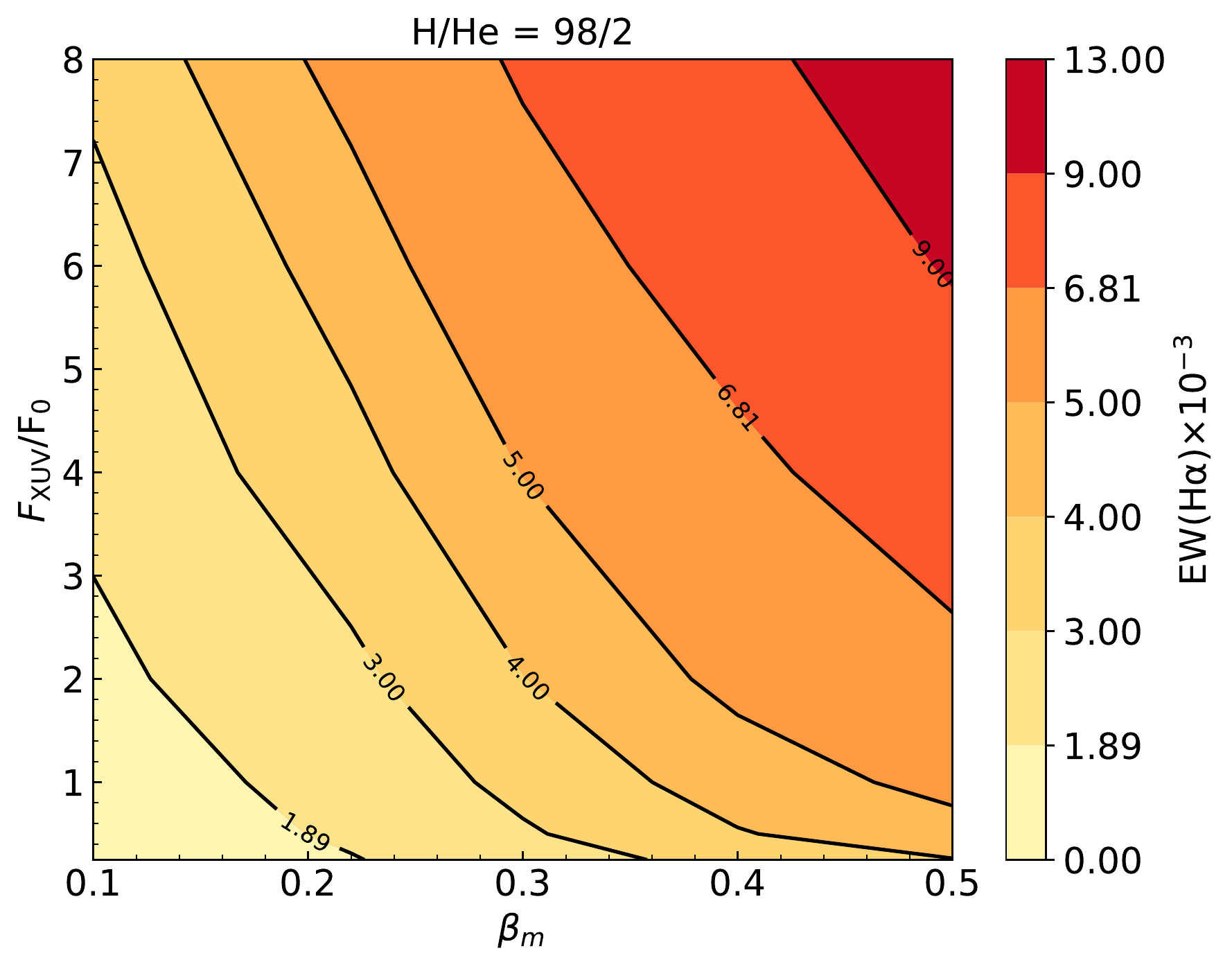}{0.5\textwidth}{(b)}
 }
\gridline{\fig{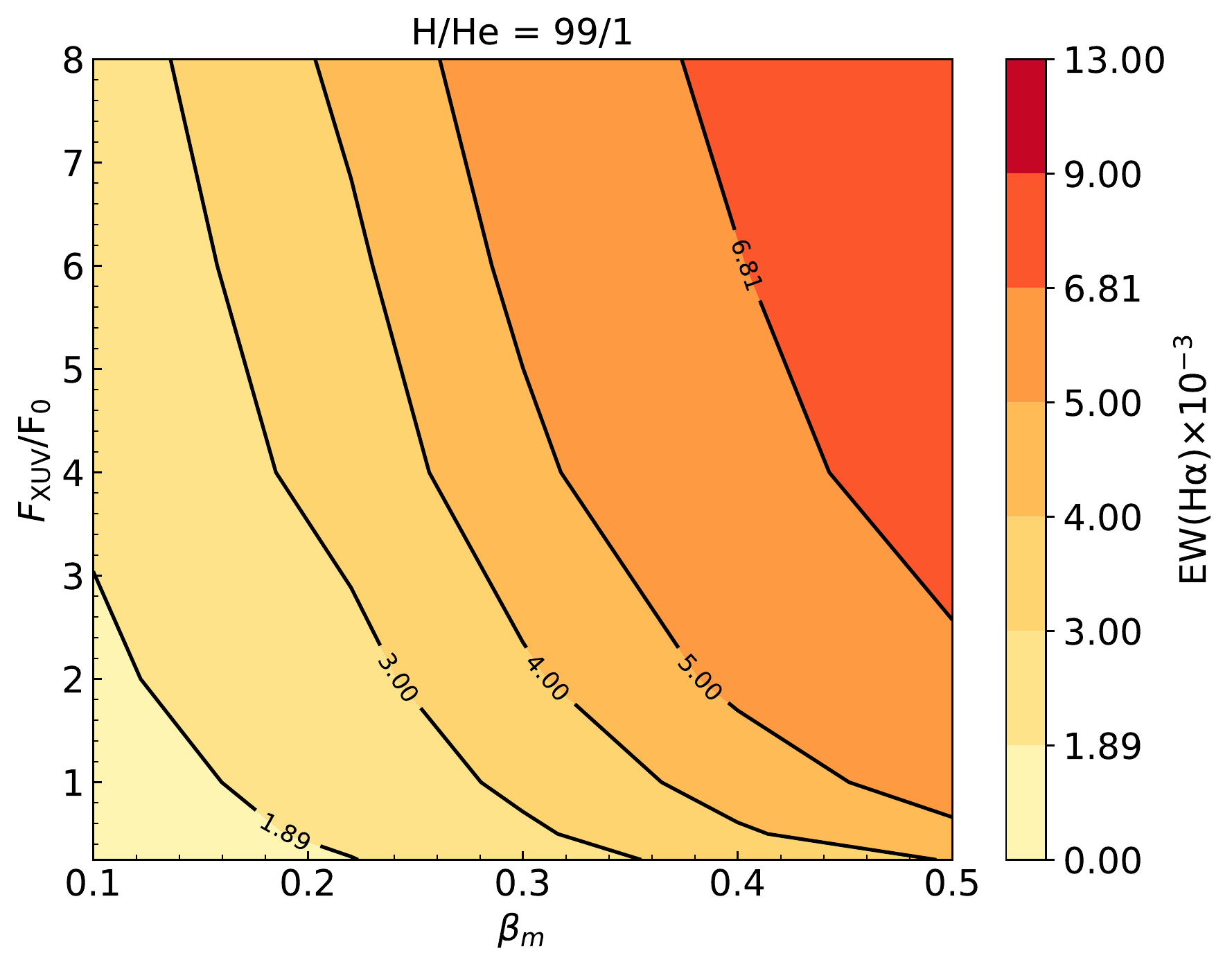}{0.5\textwidth}{(c)}
 \fig{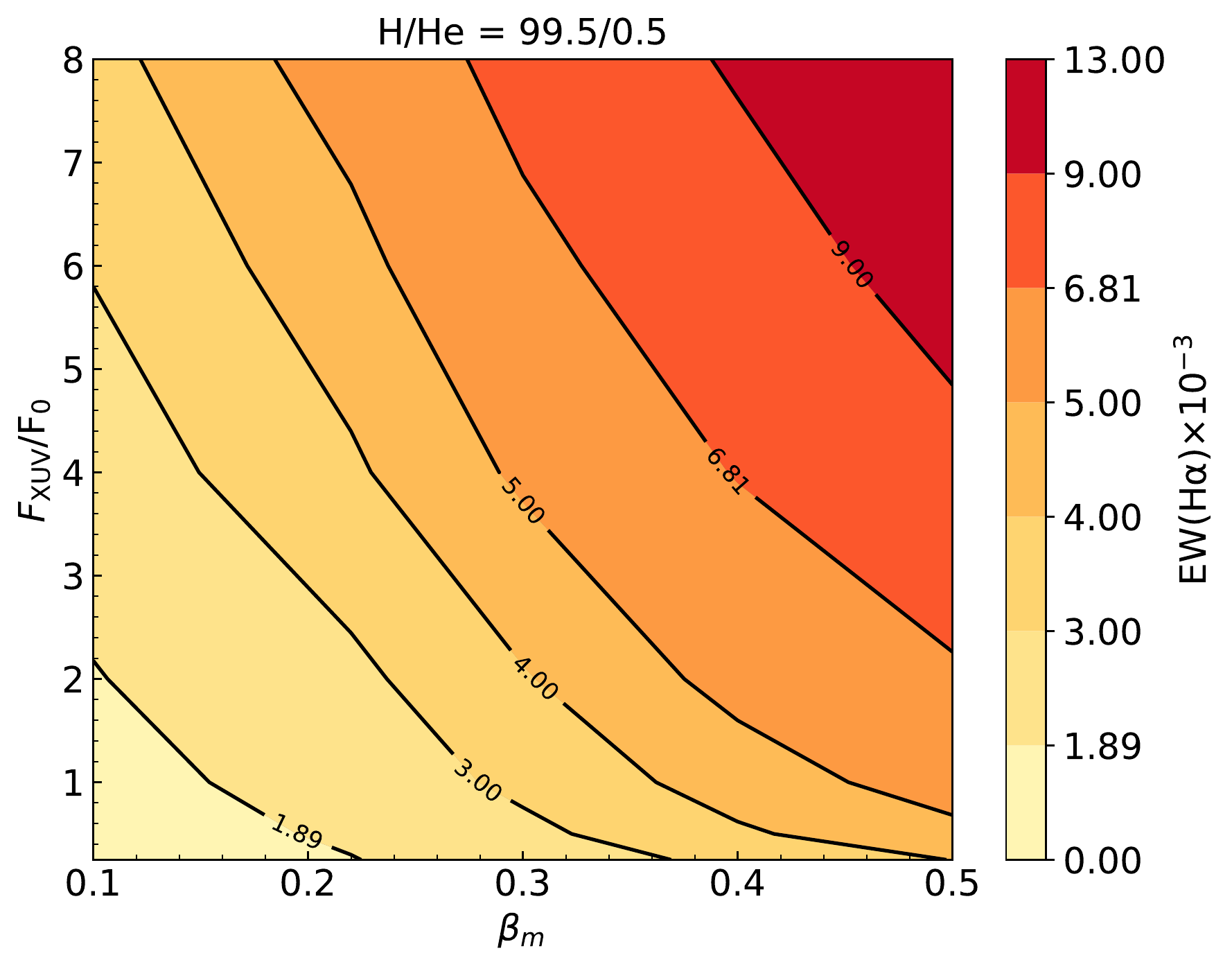}{0.5\textwidth}{(d)}
 }
 
\caption{Equivalent width contours of H$\alpha$ line, calculated in the passband of [6562.05, 6563.55]$\rm\AA$, as a function of $F_{\rm XUV}$ and $\beta_m$. The panels show the equivalent width contours for various hydrogen to helium abundance ratios.}
\label{EW_Ha_contour}

\end{figure*}

To compare the transmission spectra more quantitatively, we also compared the predicted equivalent width (EW) of H$\alpha$ line and the absorption depth at the H$\alpha$ line center (C(H$\alpha$)) with the observation. Figures \ref{EW_Ha_contour} (a)-(d) show the EW as a function of $F_{\rm XUV}$ and $\beta_m$ for models with H/He = 92/8, 98/2, 99/1, and 99.5/0.5. We find that the EW increases with the increase of $F_{\rm XUV}$ and $\beta_m$.
The mean EW value of the observed data is 1.89$\times 10^{-3}\rm\AA$, and the 1$\sigma$ confidence range is found to be 0$\sim$6.81$\times 10^{-3}\rm\AA$. 
From the figures, we find the regions in the space of $F_{\rm XUV}$ and $\beta_m$ encompassed by the contour lines 1.89$\times 10^{-3}\rm\AA$ and 6.81$\times 10^{-3}\rm\AA$ for models with various H/He ratios, and these regions are consistent with the regions defined by the $\chi^2$ constraints in Figure \ref{fig_HaTS_chi_unbinned}. This again demonstrates that the models with very high $F_{\rm XUV}$ and $\beta_m$ cannot fit the observation.

\begin{figure*}
\gridline{\fig{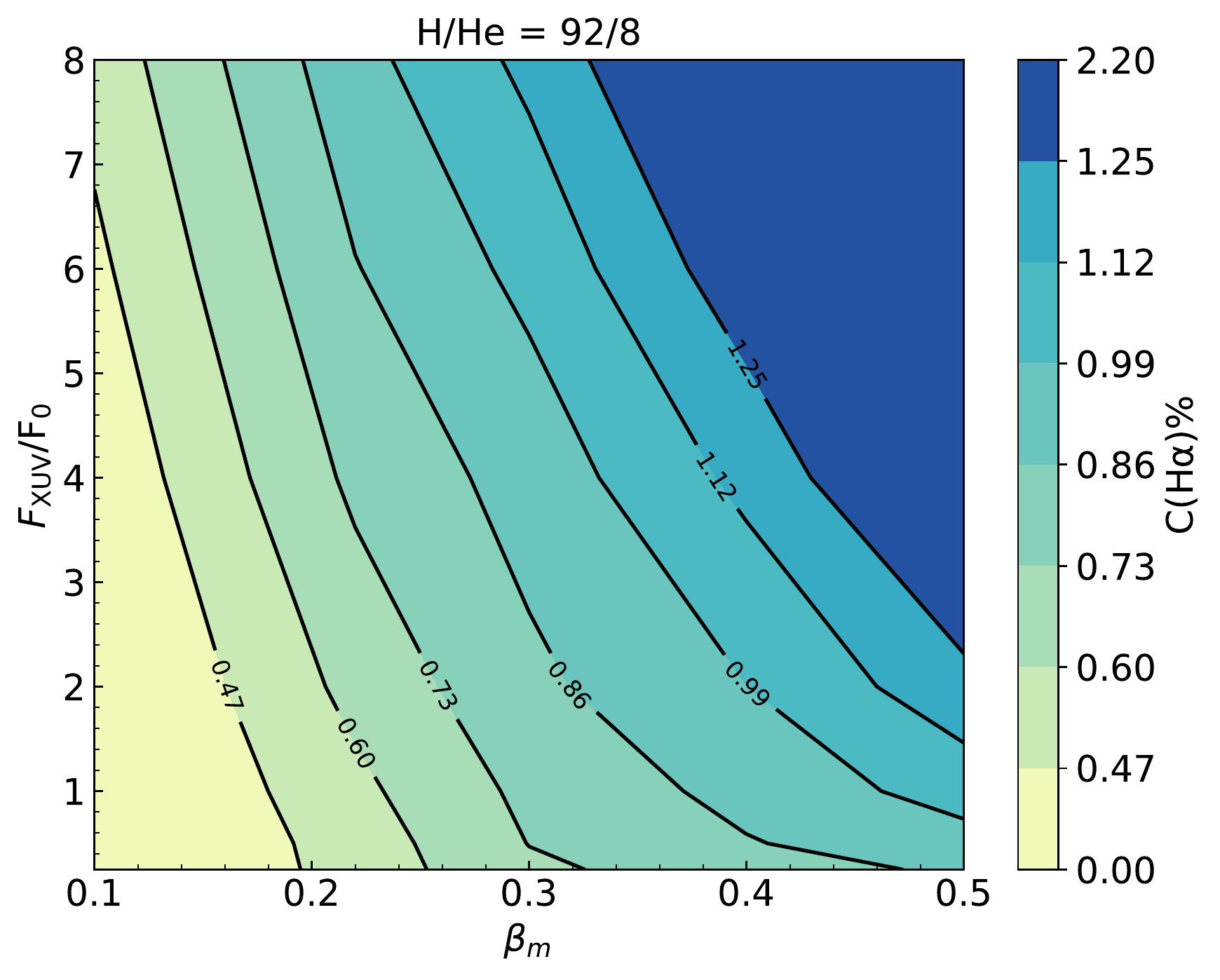}{0.5\textwidth}{(a)}
 \fig{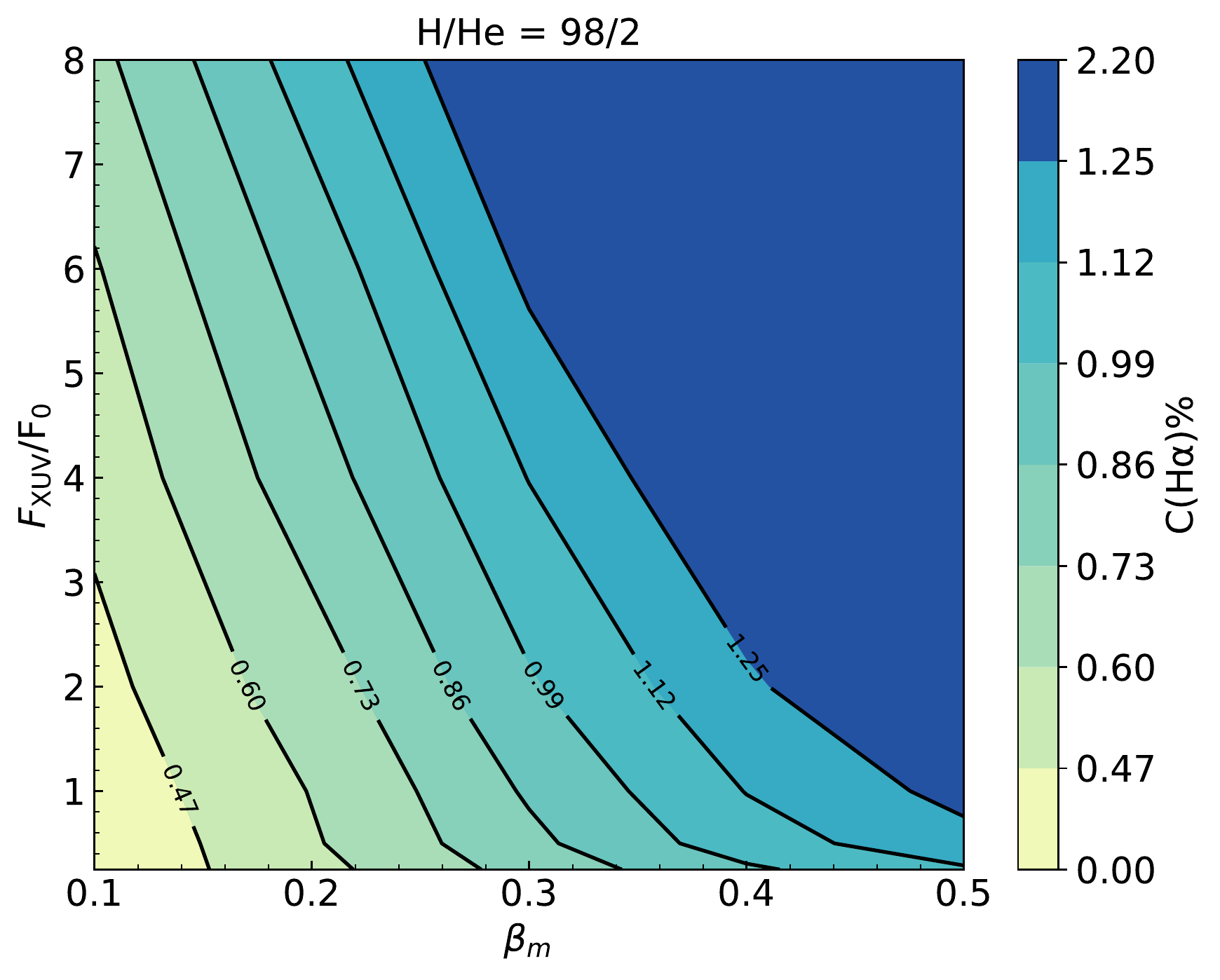}{0.5\textwidth}{(b)}
 }
\gridline{\fig{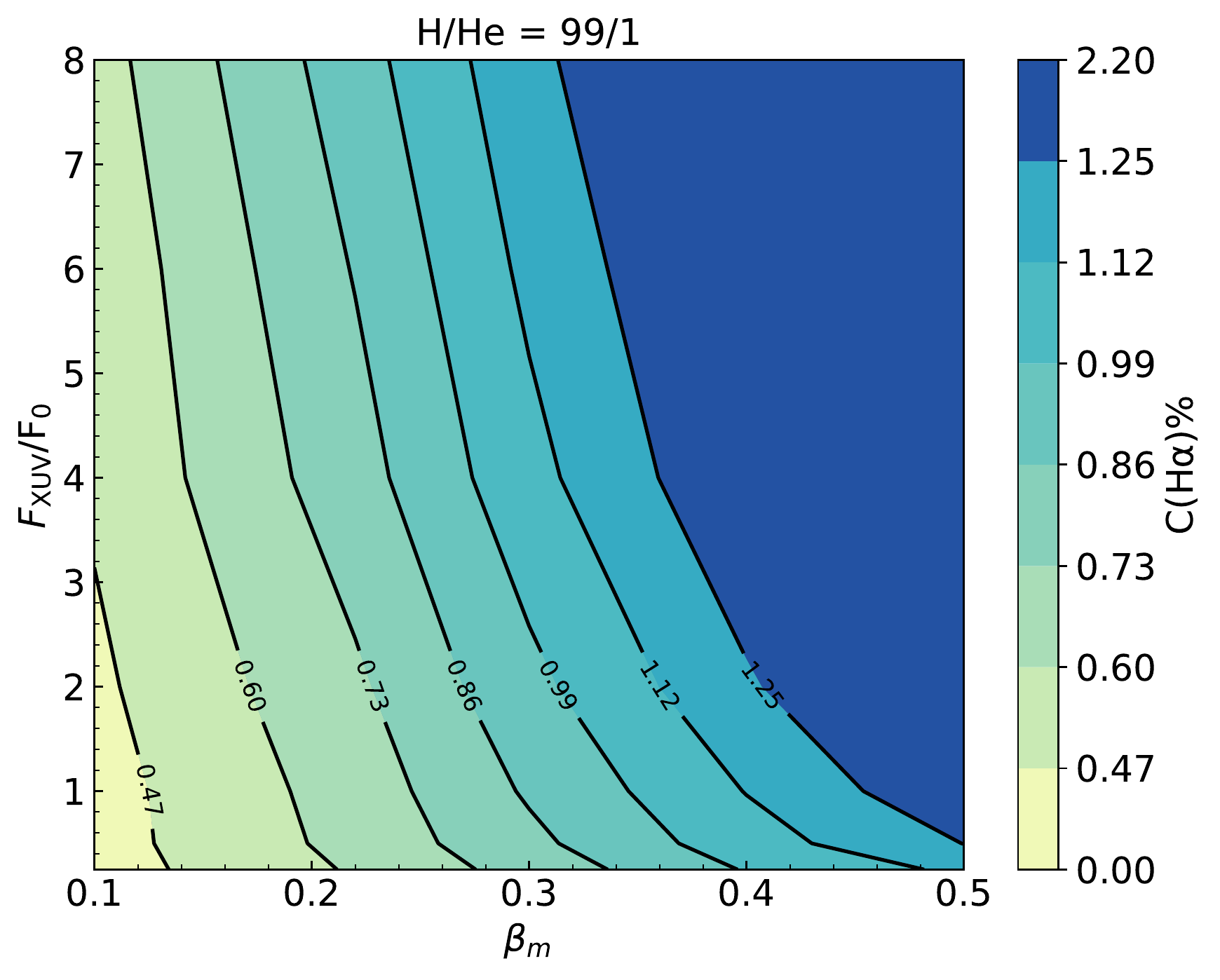}{0.5\textwidth}{(c)}
 \fig{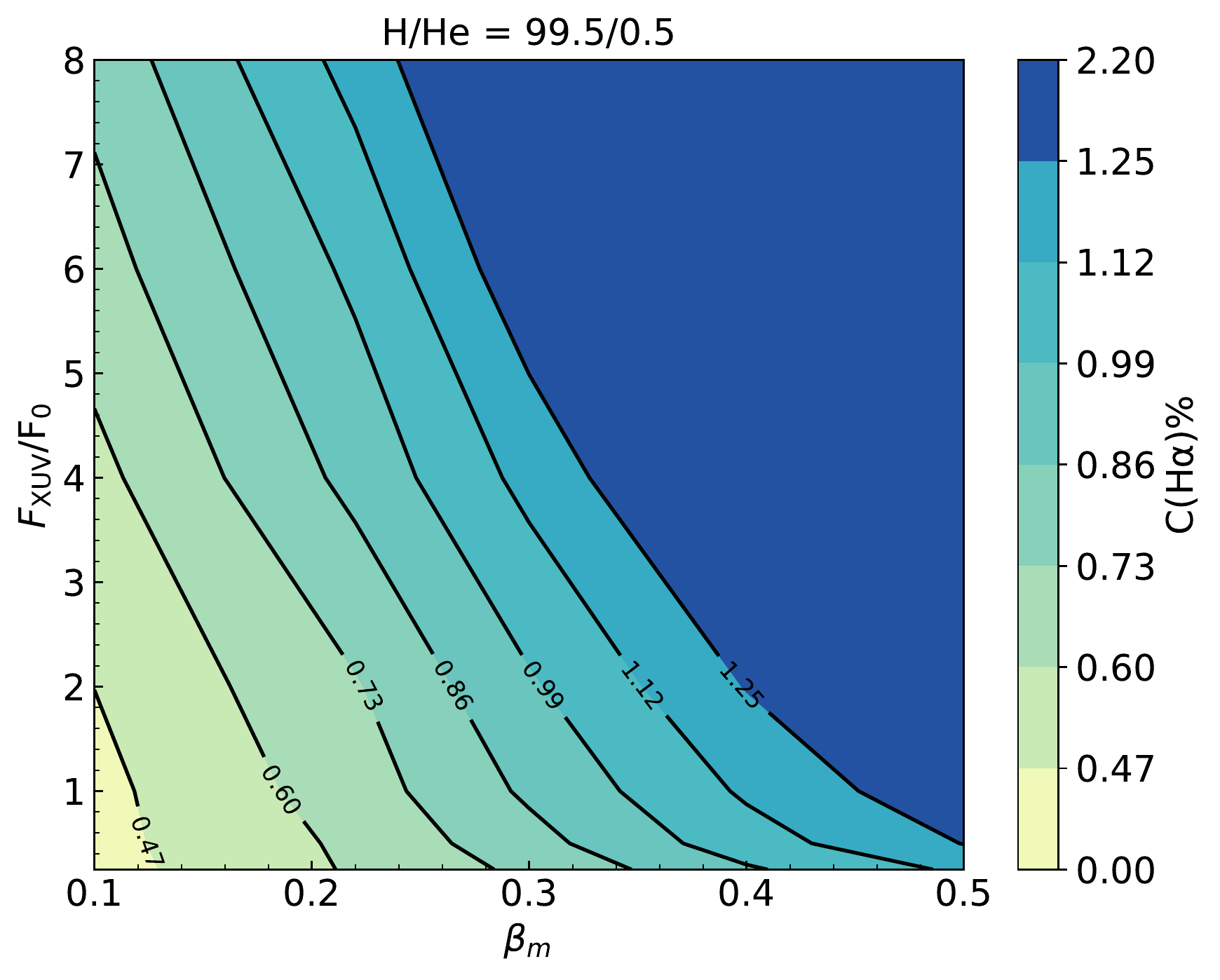}{0.5\textwidth}{(d)}
 }
 
\caption{Absorption depth at the H$\alpha$ line center as a function of $F_{\rm XUV}$ and $\beta_m$ for various H/He ratios.}
\label{CHa_contour}

\end{figure*}
According to \cite{2020A&A...635A.171C}, an absorption depth of 0.86\%$\pm$0.13\% was detected at the H$\alpha$ line center. Figures \ref{CHa_contour} (a)-(d) show the absorption depth at the H$\alpha$ line center as a function of $F_{\rm XUV}$ and $\beta_m$ for models with H/He = 92/8, 98/2, 99/1, and 99.5/0.5. In the figures, we show the contour lines of the lower (corresponding to C(H$\alpha$) = 0.73\%, 0.60\%, and 0.47\%) and upper limits at 1$\sigma$, 2$\sigma$, and 3$\sigma$ levels (corresponding to C(H$\alpha$) = 0.99\%, 1.12\%, and 1.25\%) of the observation. 
We find that the C(H$\alpha$) increases with the increase of $F_{\rm XUV}$ and $\beta_m$ in all models with four different H/He ratios. This trend also suggests that a high $\beta_m$ is required when $F_{\rm XUV}$ is low. For example, one can see that in Figure \ref{CHa_contour} (a) when H/He = 92/8 and $F_{\rm XUV}$ = 0.25F$_0$, only models with $\beta_m\geq 0.3$ can give C(H$\alpha$)$\geq$ 0.73\%. 
For a very small $\beta_m$, the models predict C(H$\alpha$) that are always lower than the lower limit of the observation. For instance, in the case of $\beta_m$=0.1, even with $F_{\rm XUV}$ = 8F$_0$, the models could not give C(H$\alpha$) reaching the lower limit of 1$\sigma$. However, as $\beta_m$ increases, the H$\alpha$ absorption depth increases, and thus a moderate $F_{\rm XUV}$ is found to be able to reproduce the observational data within 1$\sigma$ level. When $0.3 < \beta_m < 0.5$, a relatively small $F_{\rm XUV}$ can give an absorption level of $\sim 0.86$\%.
To constrain the physical parameters of the atmosphere, we define the models that yield 0.73\% $\leq$C(H$\alpha$)$\leq$ 0.99\% as acceptable ones. In Figures \ref{CHa_contour} (c-d), one can see that to reproduce C(H$\alpha$) within 1$\sigma$ level of the observation, a relatively larger $\beta_m$ (0.3 $\sim$ 0.4) is needed for models with a small $F_{\rm XUV}$ (0.25F$_0$ to F$_0$). On the other hand, for models with a larger $F_{\rm XUV}$, $\beta_m$ should be smaller. These models are also in the lowest $\chi^2$ regions in Figure \ref{fig_HaTS_chi_unbinned} as well as in the 1$\sigma$ EW regions in Figure \ref{EW_Ha_contour}.
We also notice that in a certain H/He range, a higher H/He ratio can lead to a higher C(H$\alpha$) and thus slightly reduced $F_{\rm XUV}$ or $\beta_m$ can fit the observation.
From Figure \ref{CHa_contour}, we can see that when H/He = 92/8 and $\beta_m$ = 0.1, the model even with the highest $F_{\rm XUV}$ = 8F$_0$ can not reach the lower limit of the observation. On the other hand, when H/He = 99.5/0.5 and $\beta_m$ = 0.1, a model with $F_{\rm XUV} \approx$ 7F$_0$ can fit the observation. 
%We can also see that when H/He $\gtrsim$ 99.5/0.5, changing the H/He ratio does not alter the resulting  H$\alpha$ absorption depth significantly.

\begin{figure*}
\gridline{\fig{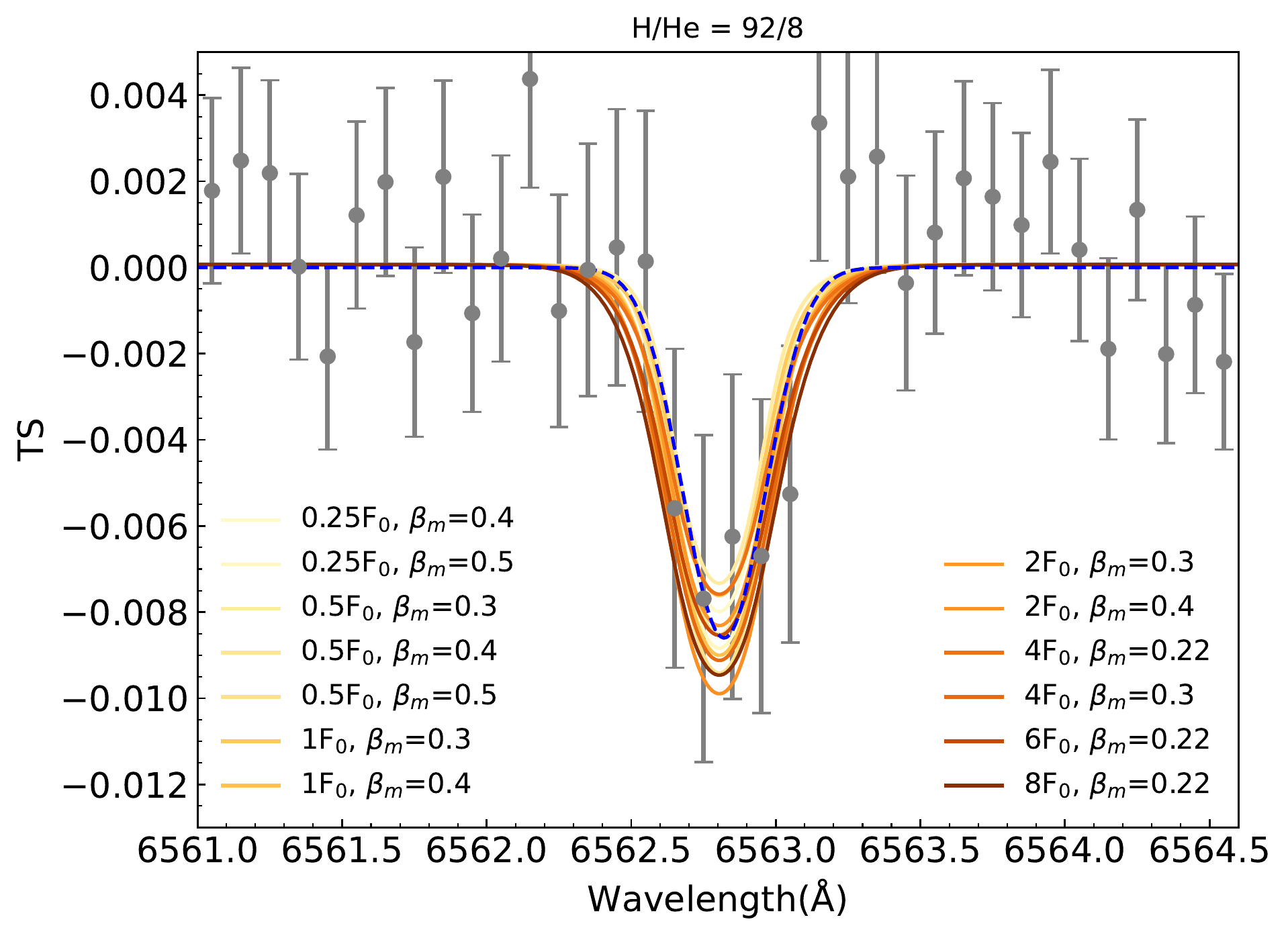}{0.5\textwidth}{(a)}
 \fig{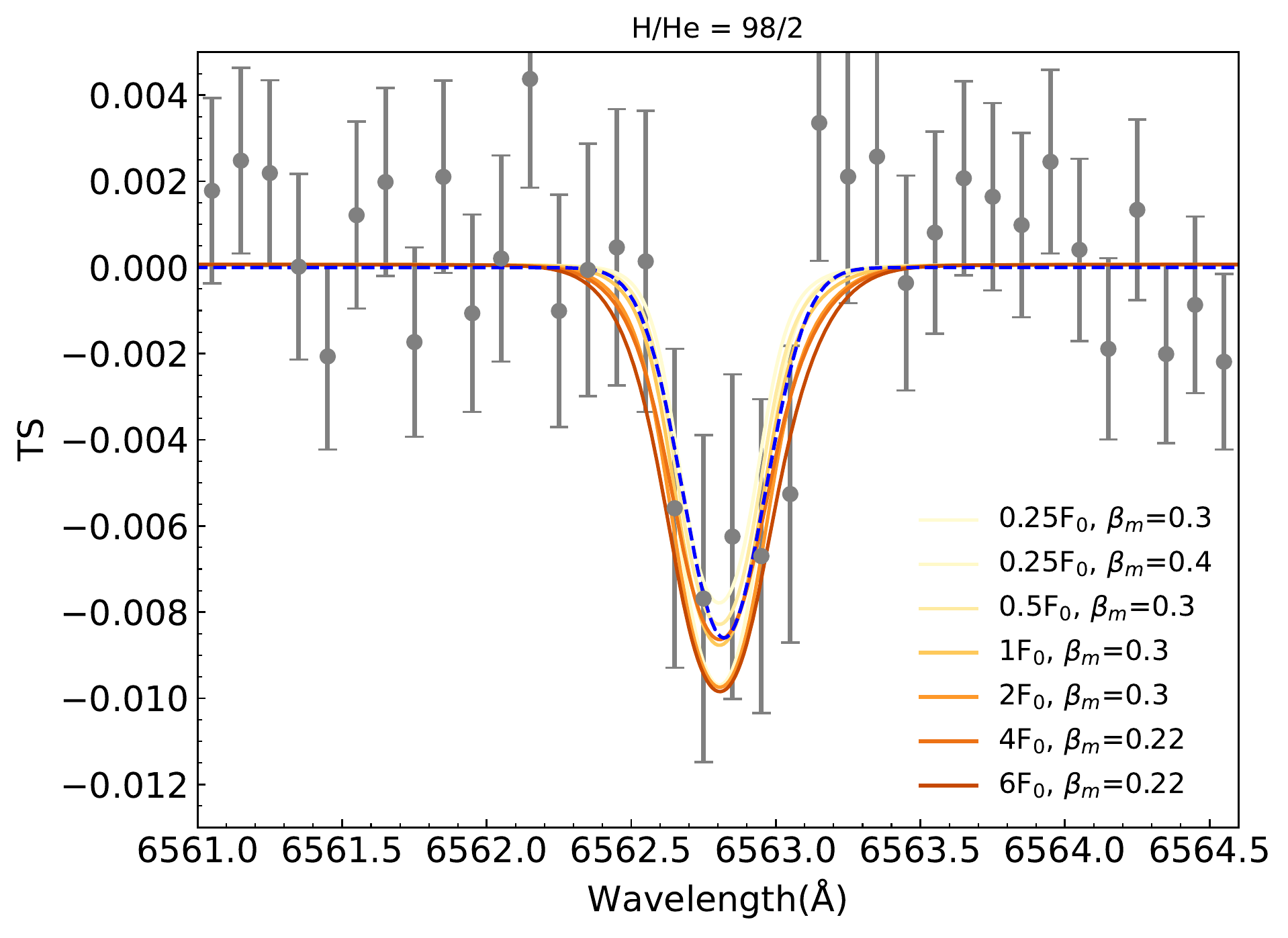}{0.5\textwidth}{(b)}
 }
\gridline{\fig{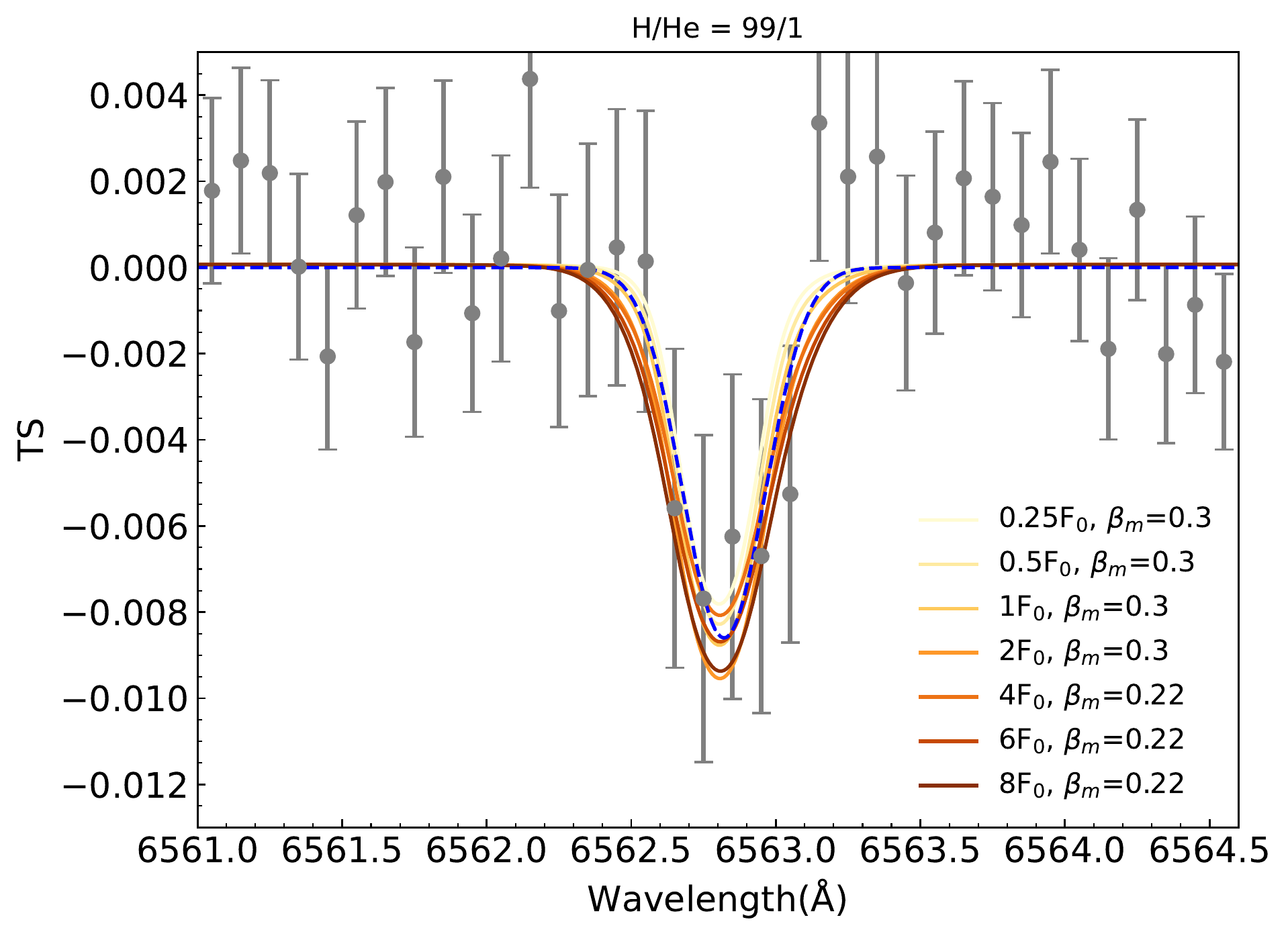}{0.5\textwidth}{(c)}
 \fig{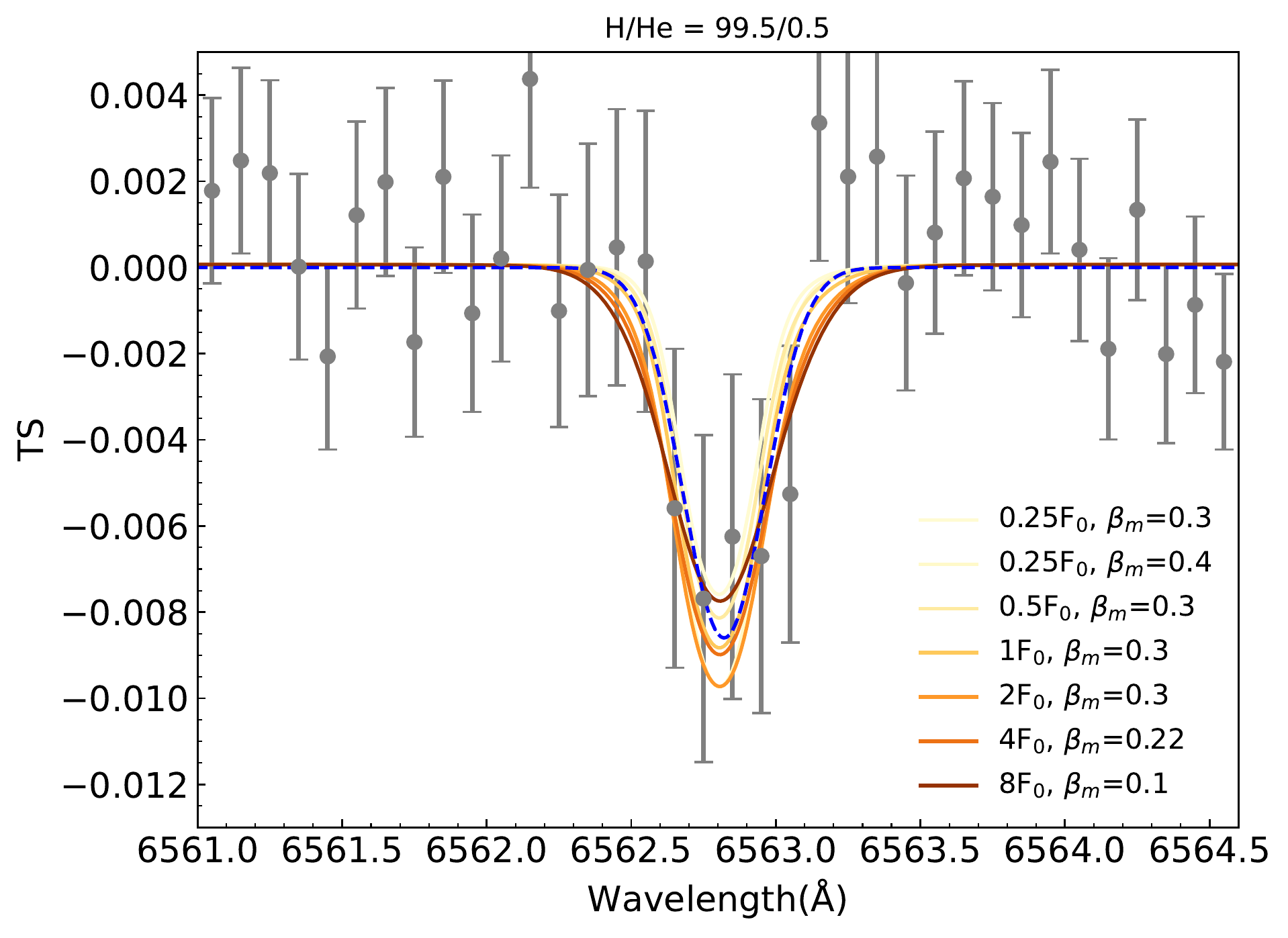}{0.5\textwidth}{(d)}
 }
 
\caption{H$\alpha$ transmission spectra of the models that predict C(H$\alpha$) between 0.73\% and 0.99$\%$. The gray scatters with error bars are the observed data from \cite{2020A&A...635A.171C}. The blue dashed lines are the best-fit Gaussian to the data. In (a)-(d), the H/He ratio is 92/8, 98/2, 99/1, and 99.5/0.5, respectively.} 
\label{Ha_TS_many}

\end{figure*}

In Figures \ref{Ha_TS_many} (a)-(d), we show the H$\alpha$ transmission spectra of the models that predict C(H$\alpha$) between 0.73\% and 0.99$\%$ when H/He = 92/8, 98/2, 99/1, and 99.5/0.5. The gray scatters with error bars are the observed data from \cite{2020A&A...635A.171C} and the blue dashed lines are the best-fit Gaussian to the data. In the case of H/He = 92/8, it shows that $\beta_m$ should be as high as 0.4 $\sim$ 0.5 for the models with $F_{\rm XUV}$ = 0.25F$_0$ and 0.5F$_0$. As $F_{\rm XUV}$ increases to (1-4) $\times$ F$_0$, a lower $\beta_m$ of 0.3 $\sim$ 0.4 can fit the absorption depth. For models with an even higher $F_{\rm XUV}$, for example, 6 and 8 $\times$ F$_0$, a value of $\beta_m$ = 0.22 provides a good fit.
Therefore, by comparing the models with the H$\alpha$ observation, we can only give a range of the XUV flux $F_{\rm XUV}$ and the spectral energy distribution index $\beta_m$. For models with H/He = 98/2, 99/1, and 99.5/0.5, there are also a wide range of combinations of $F_{\rm XUV}$ and $\beta_m$ that can fit the observation. In general, combinations of a small $F_{\rm XUV}$ and large $\beta_m$, or a large $F_{\rm XUV}$ and small $\beta_m$ can explain the observations. This means that there is a degeneracy in determining the physical parameters of the planet. To reduce the degeneracy, we need more restrictive conditions.
\subsection{Populations of He(2$^3$S)}\label{sub:He_pop}
\begin{figure*}
\gridline{\fig{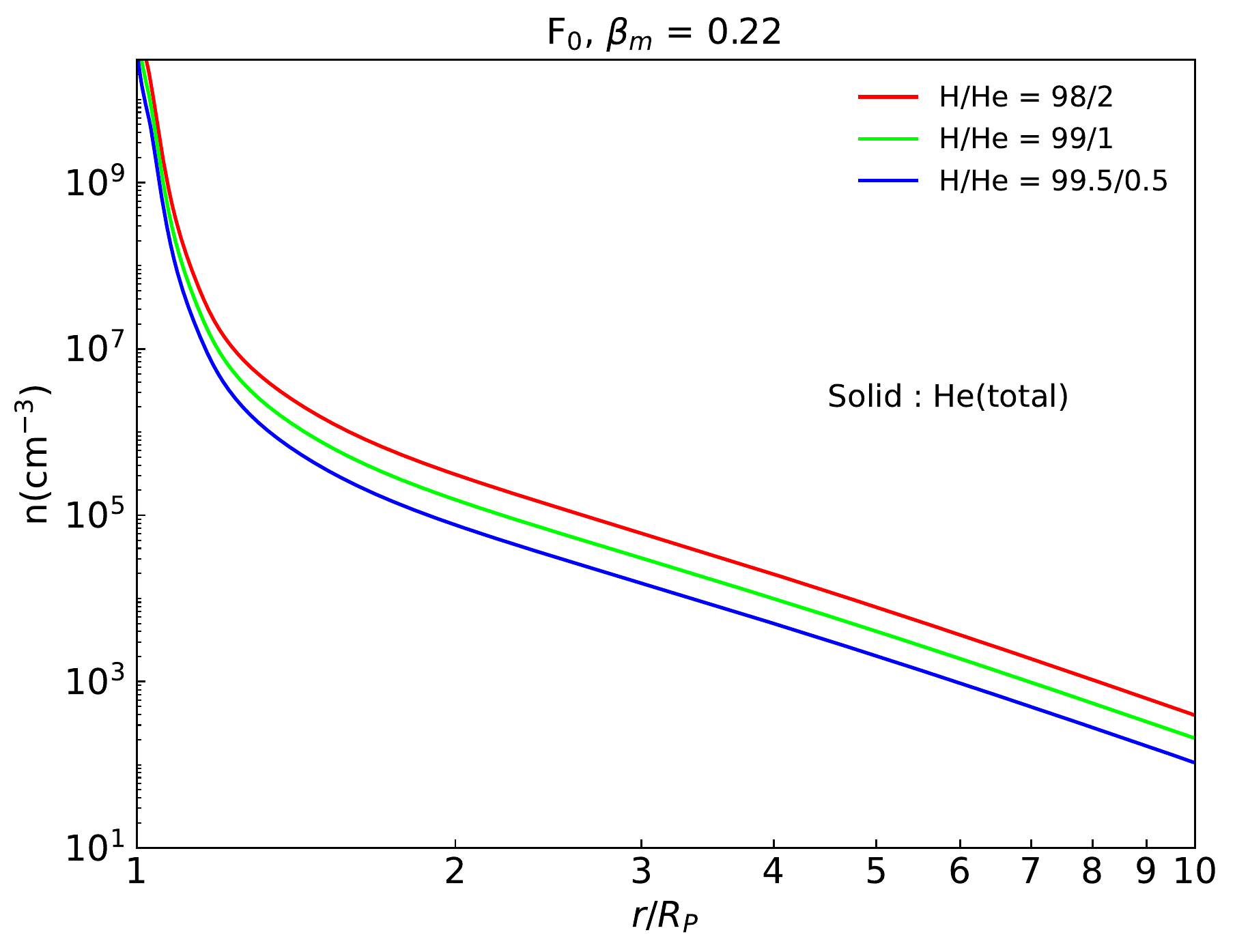}{0.5\textwidth}{(a)}
 \fig{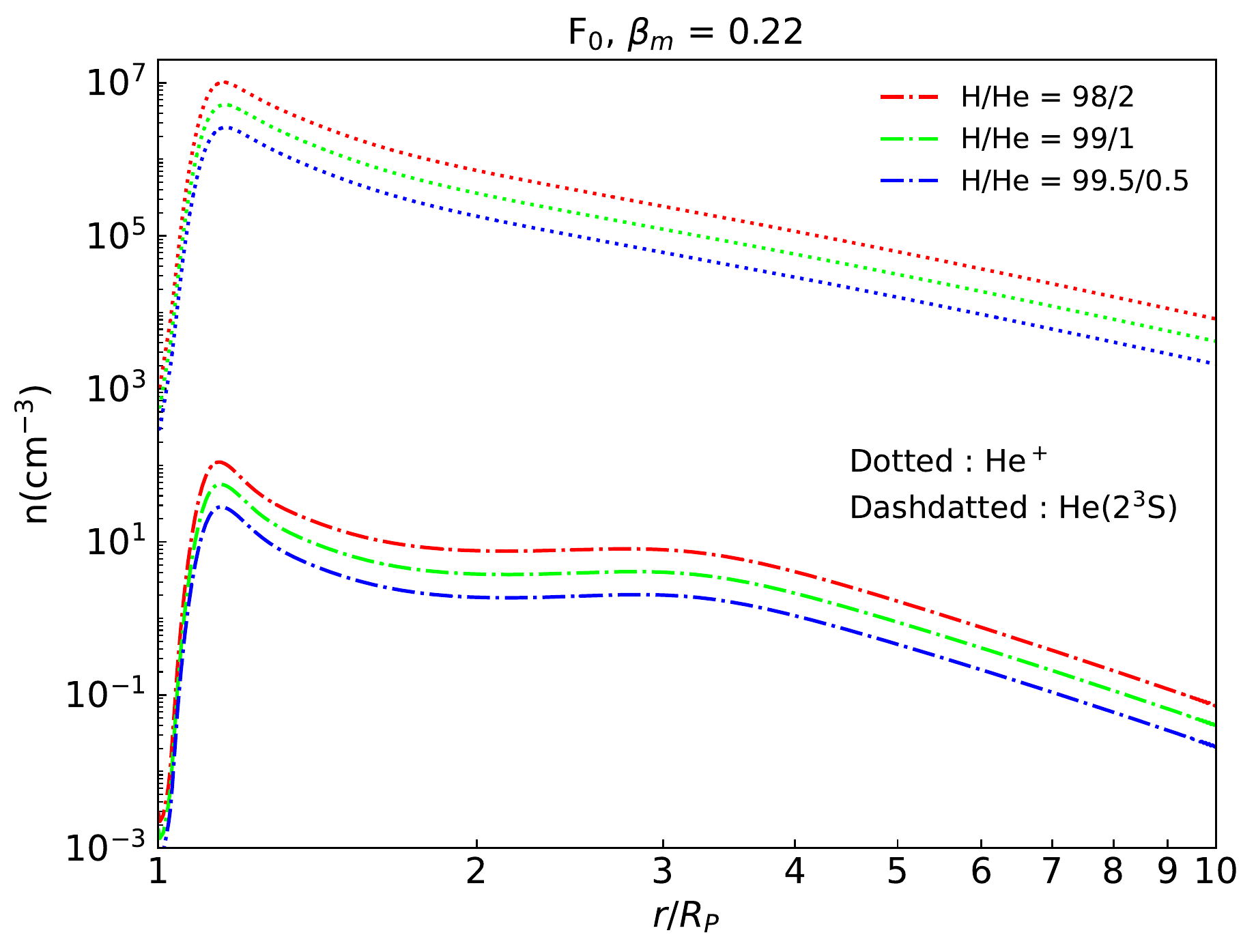}{0.5\textwidth}{(b)}
 }
\gridline{\fig{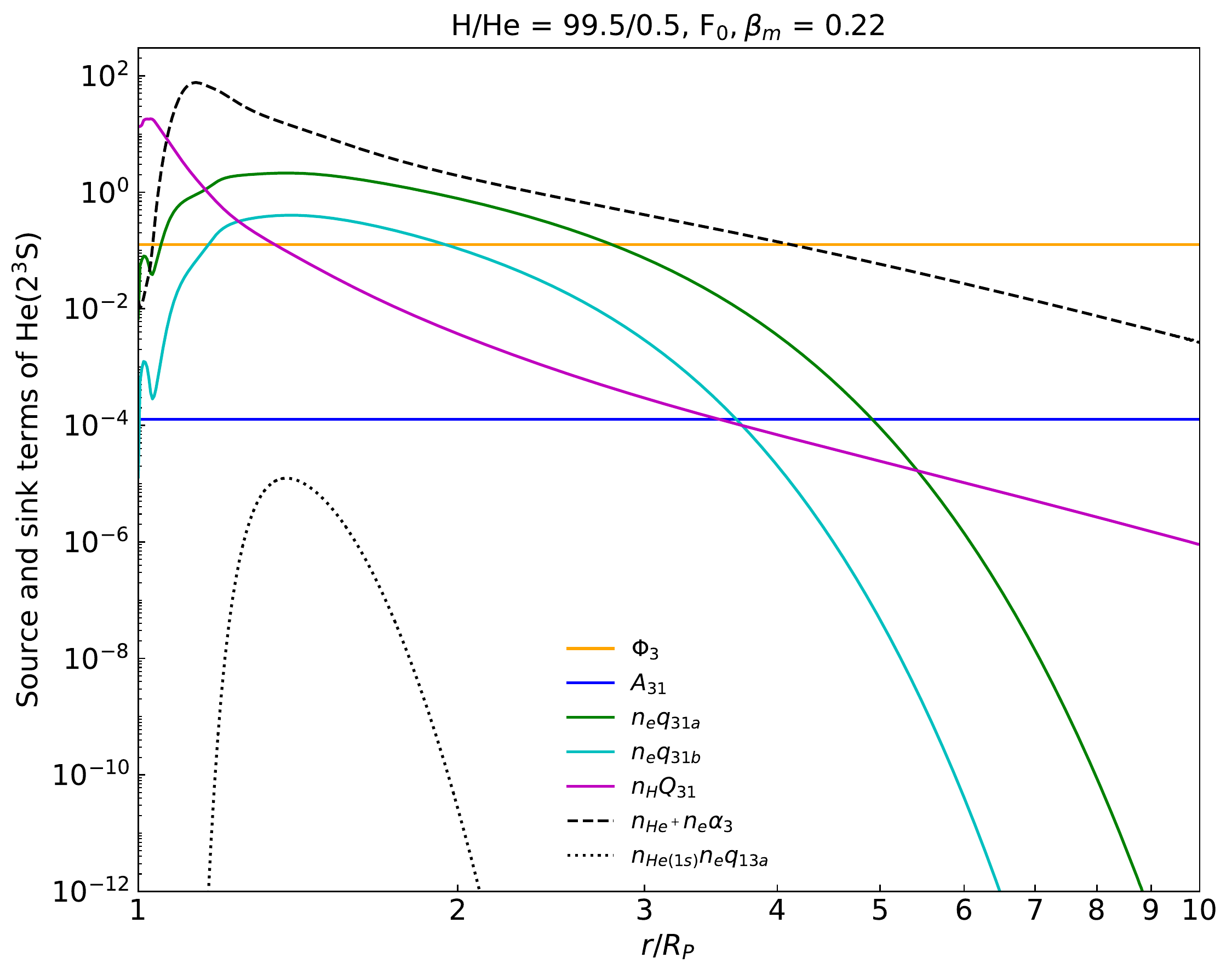}{0.5\textwidth}{(c)}
 \fig{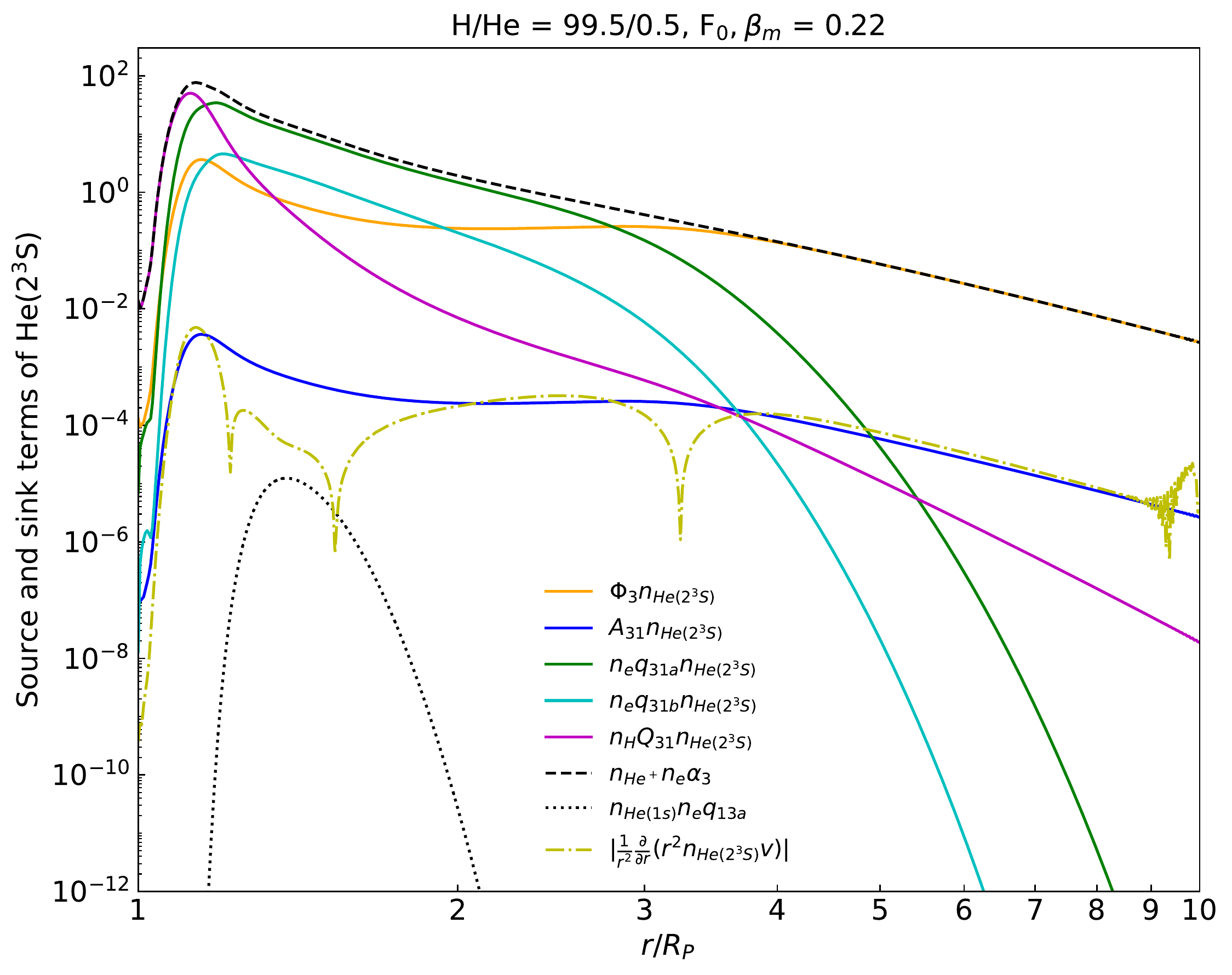}{0.5\textwidth}{(d)}
 }
\caption{(a)-(b) Density profiles of helium in models with different H/He ratios, but with the same $F_{\rm XUV}$ = F$_0$ and $\beta_m$ = 0.22; (c) Source and sink terms related to the population of He(2$^3$S) for the model with H/He = 99.5/0.5, $F_{\rm XUV}$ = F$_0$, and $\beta_m$ = 0.22. The source terms plotted in the black dashed and dotted lines are in units of cm$^{-3} \rm s^{-1}$, and the sink terms plotted in the solid lines are in units of s$^{-1}$; (d) is similar to (c), but additionally shows the advection term of He(2$^3$S) in the yellow dash-dotted line. Here, both the source and sink terms are shown in units of cm$^{-3} \rm s^{-1}$, unlike (c).}
\label{atm_He_fxuv_rates}

\end{figure*}

\begin{figure*}
\gridline{\fig{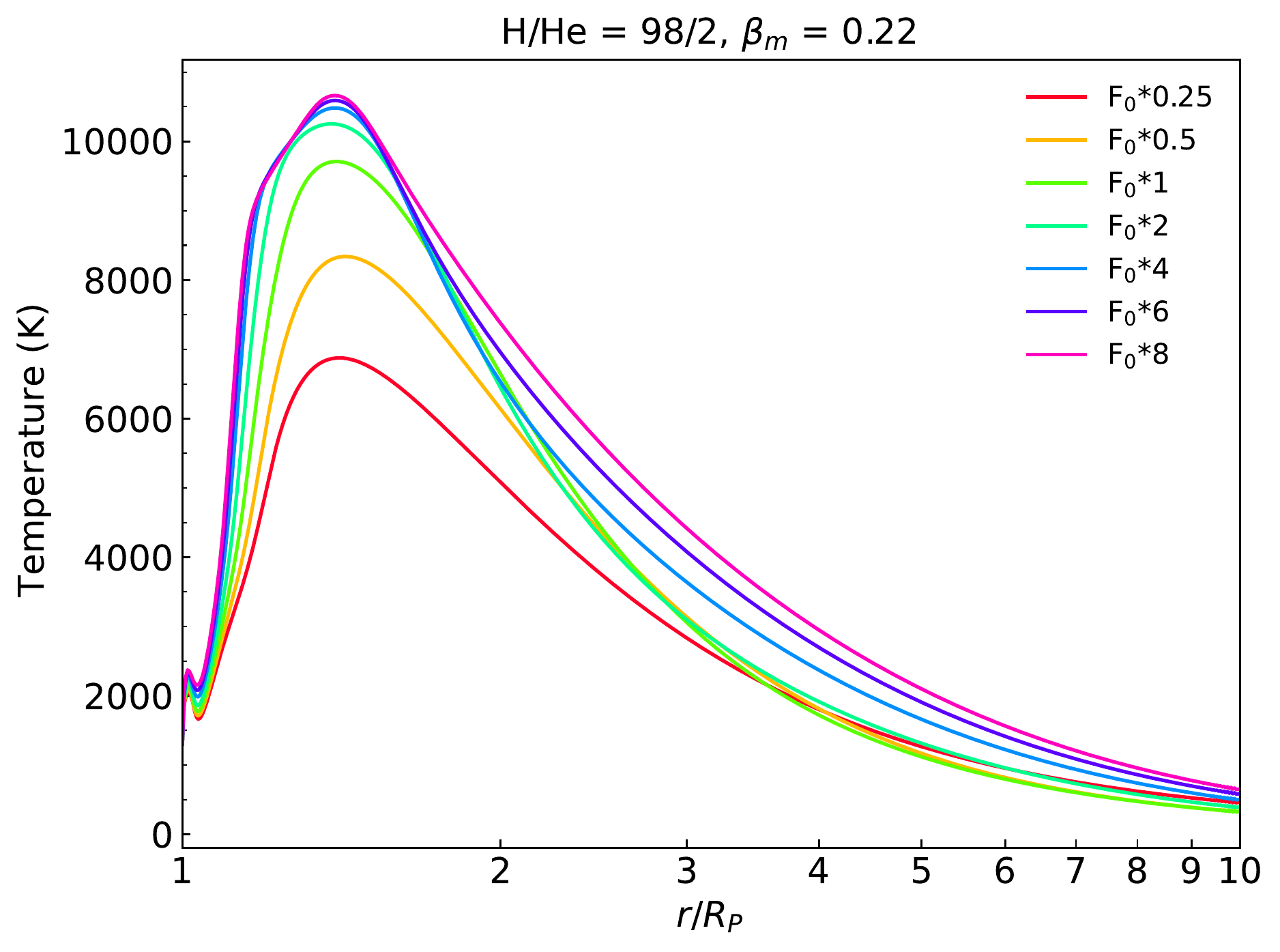}{0.53\textwidth}{(a)}
 \fig{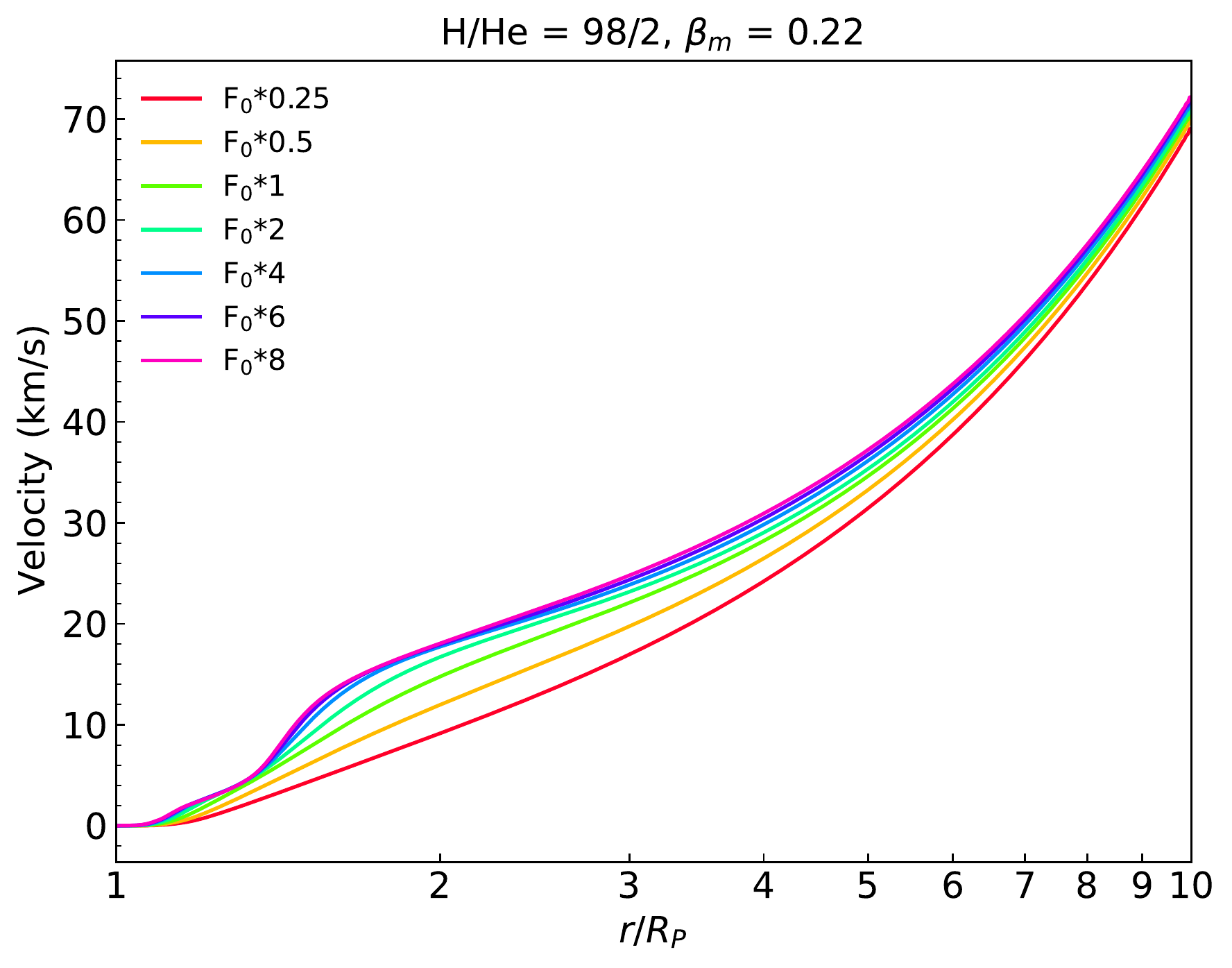}{0.5\textwidth}{(b)}
 }
\gridline{\fig{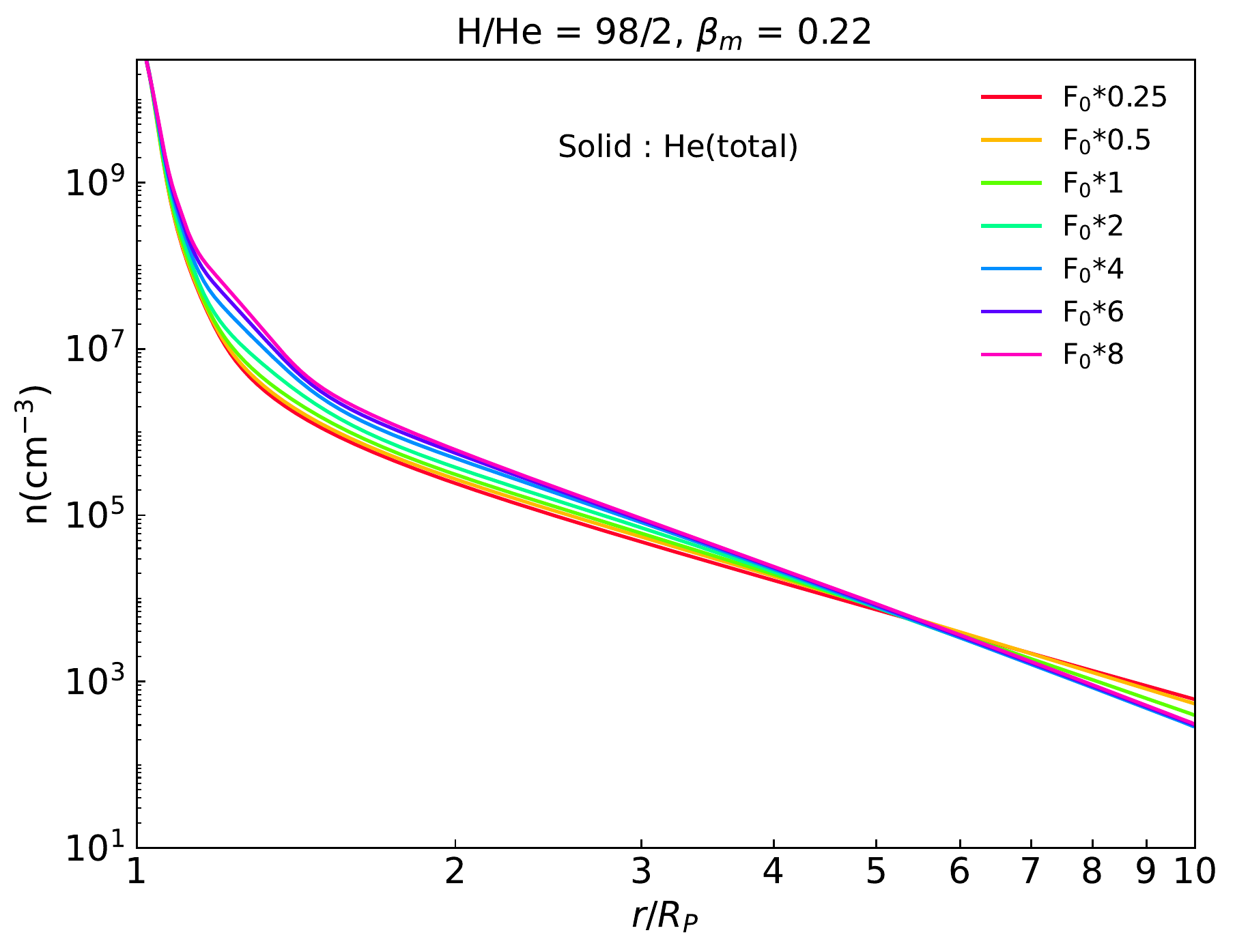}{0.5\textwidth}{(c)}
 \fig{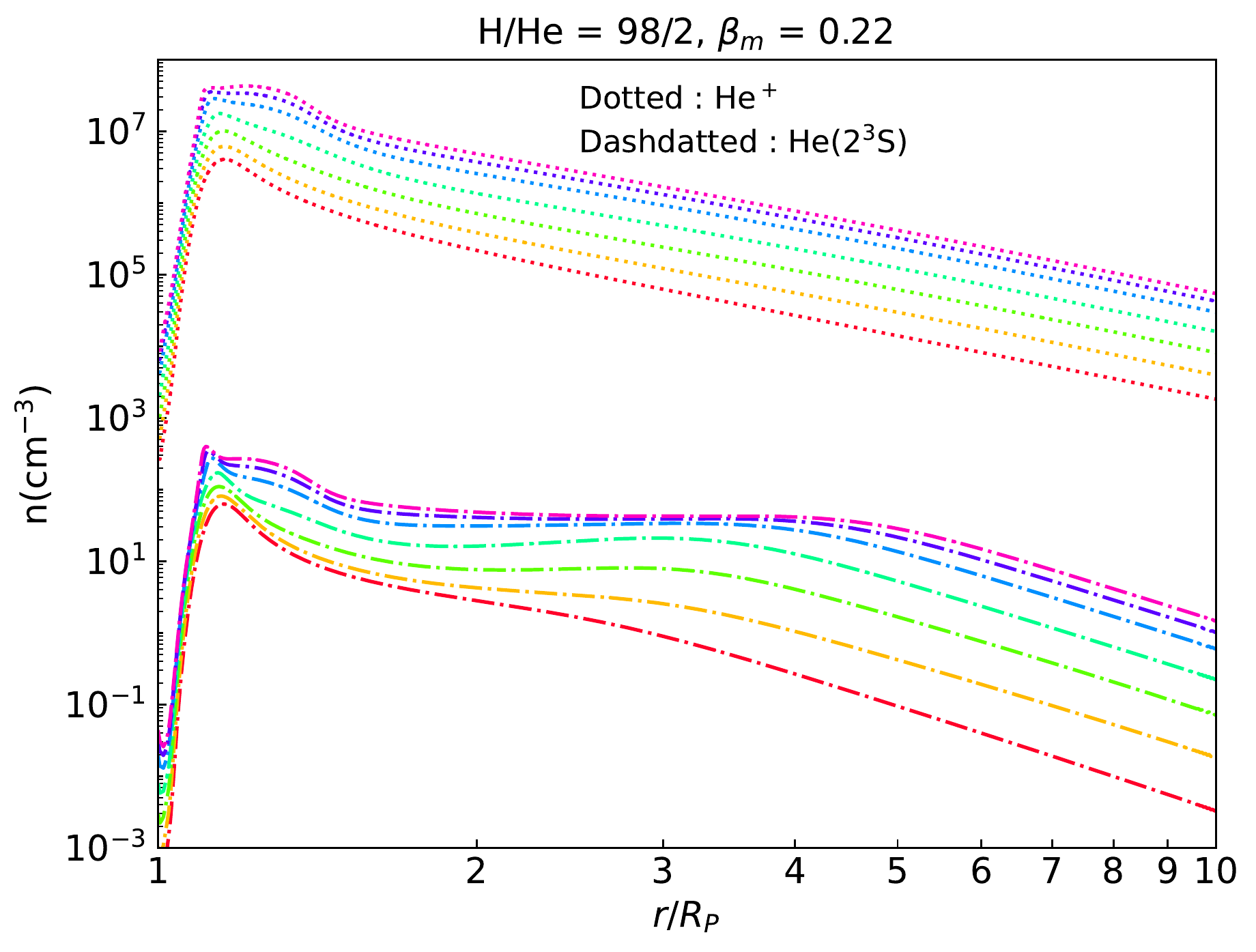}{0.5\textwidth}{(d)}
 }
\gridline{\fig{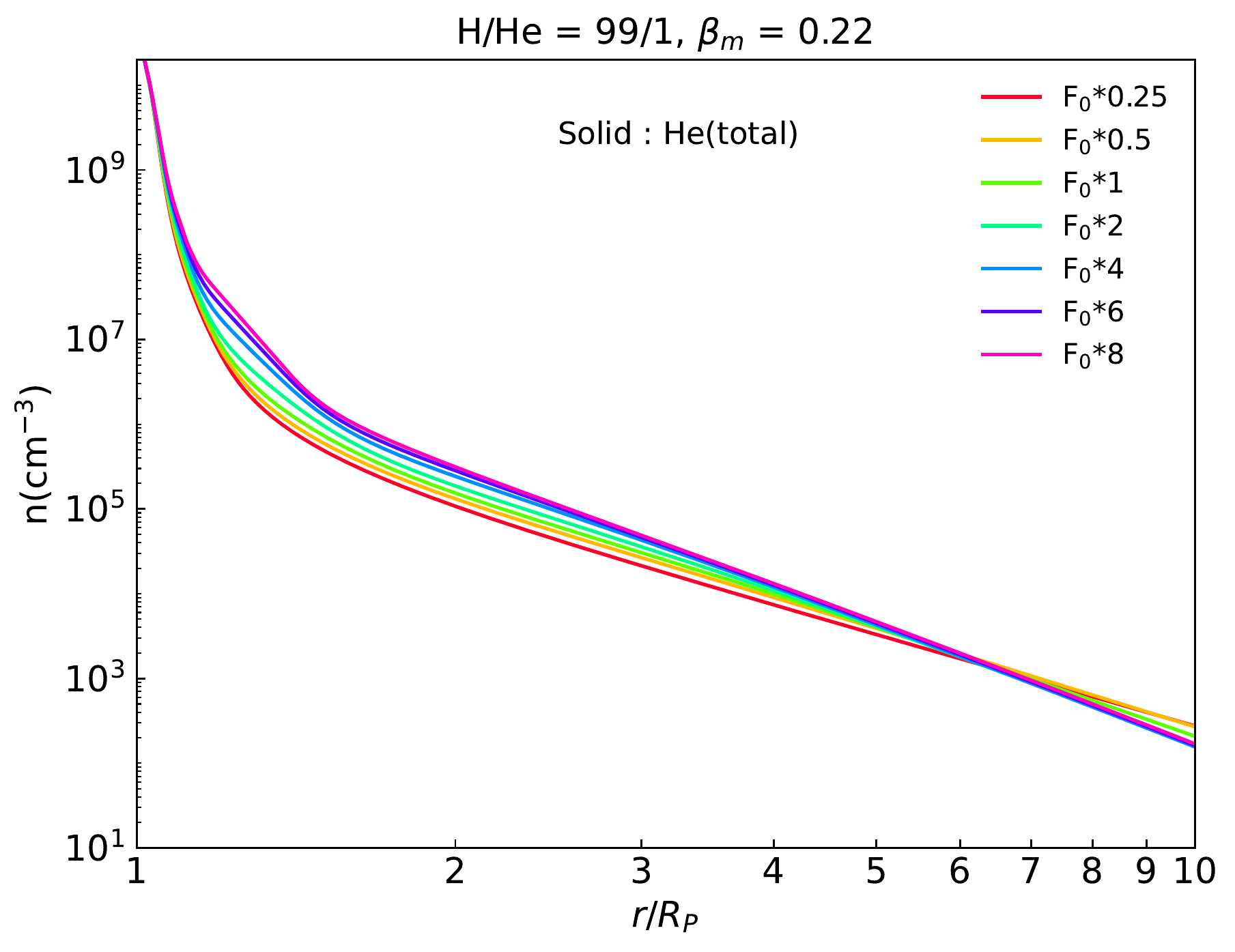}{0.5\textwidth}{(e)}
 \fig{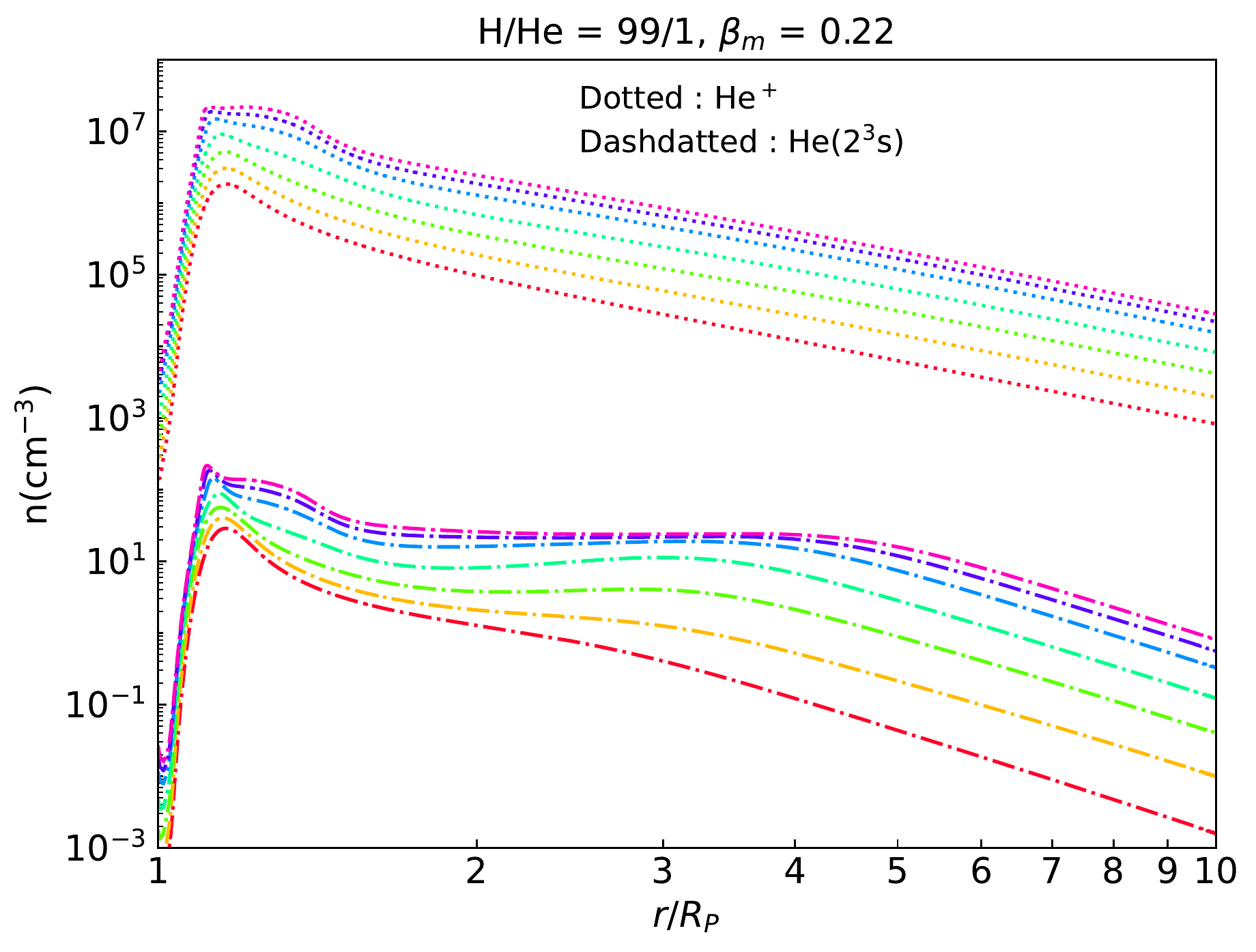}{0.5\textwidth}{(f)}
 }
 
\caption{(a)-(d) Atmosphere structures for helium of models with different $F_{\rm XUV}$, but with the same H/He = 98/2, $\beta_m$ = 0.22; (e)-(f) Atmosphere structures for helium of models with different $F_{\rm XUV}$, but with the same H/He = 99/1, $\beta_m$ = 0.22.}
\label{atm_He_b022_H999}

\end{figure*}

\begin{figure*}
\gridline{\fig{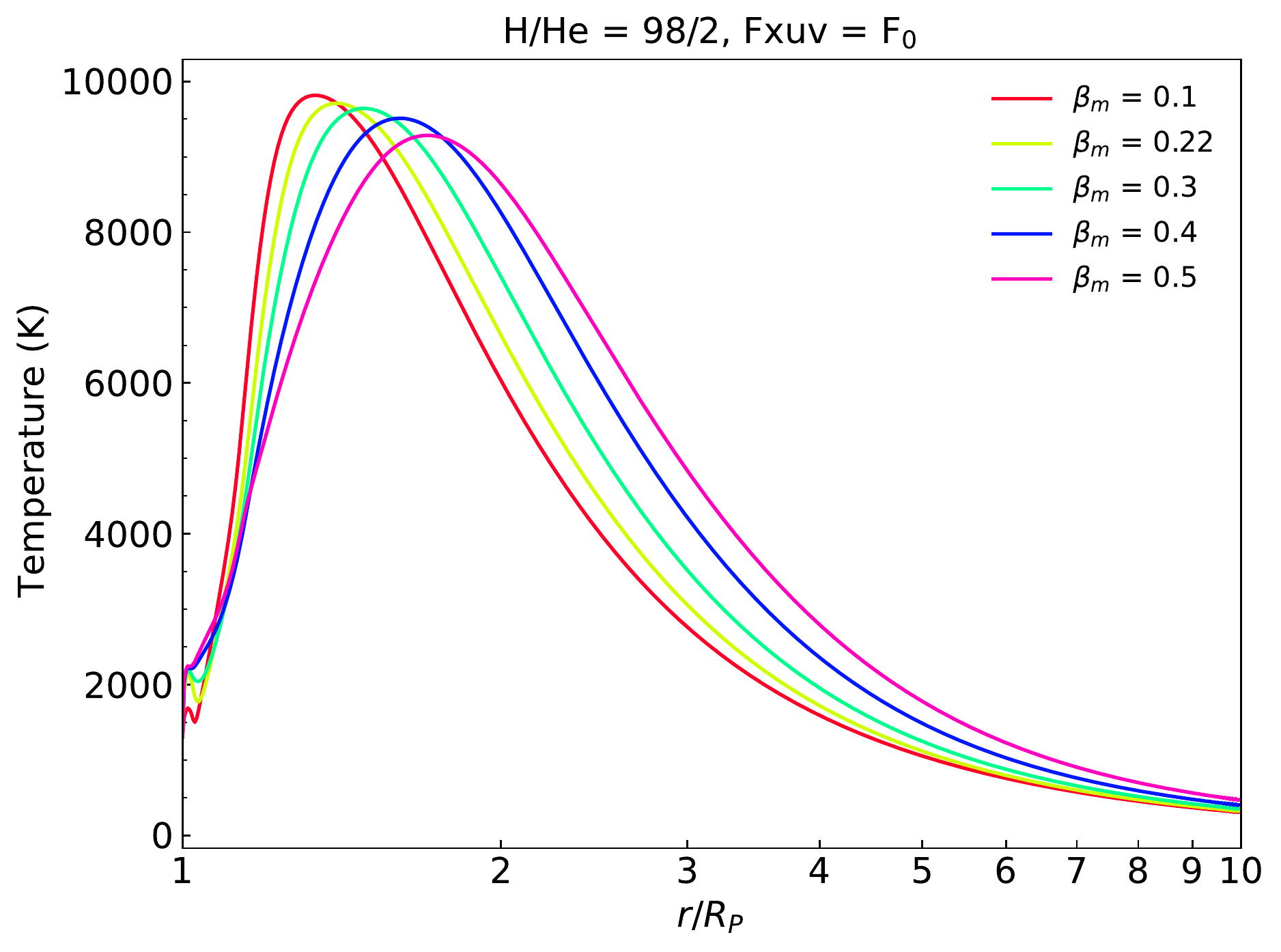}{0.53\textwidth}{(a)}
 \fig{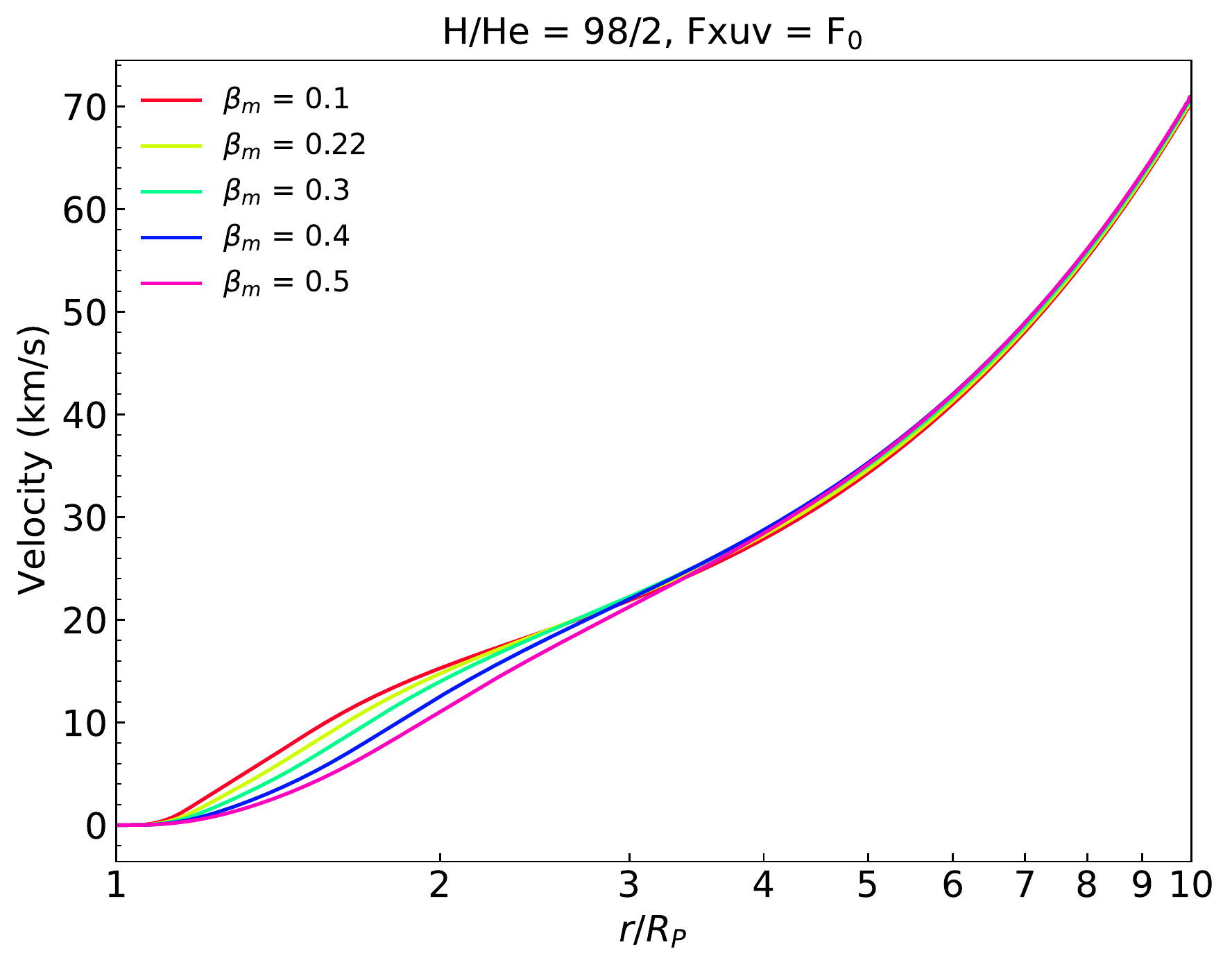}{0.5\textwidth}{(b)}
 }
\gridline{\fig{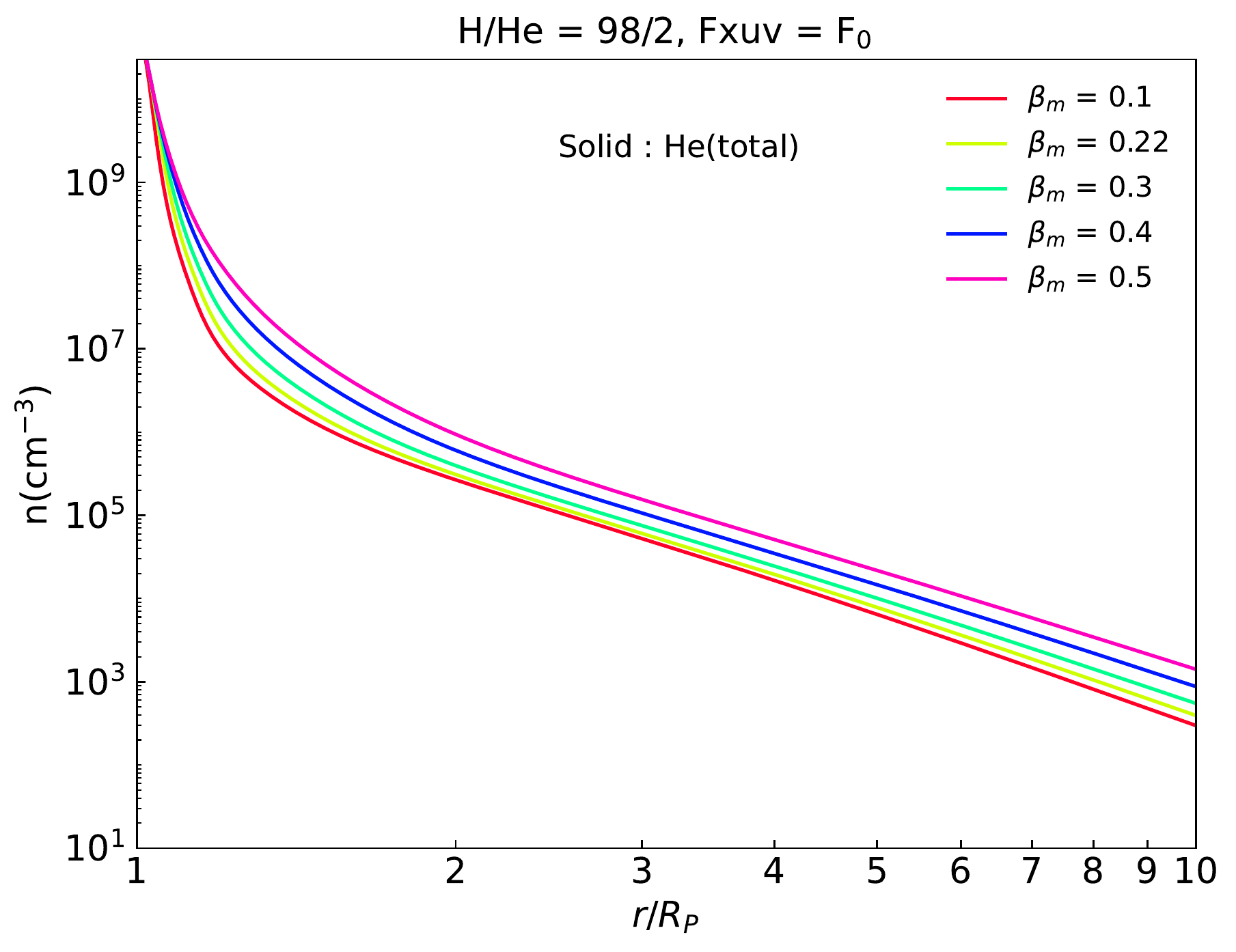}{0.5\textwidth}{(c)}
 \fig{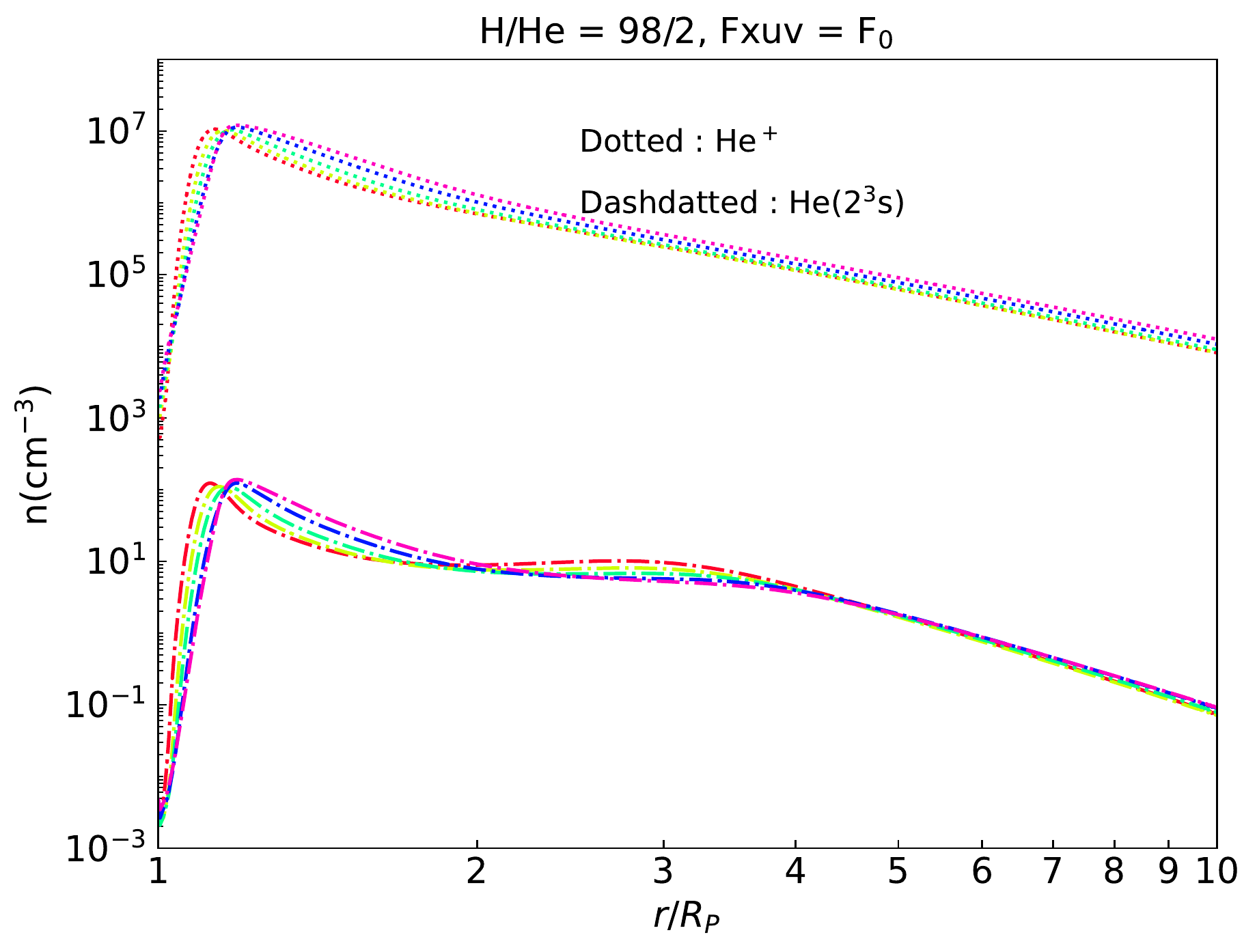}{0.5\textwidth}{(d)}
 }
 
\caption{Atmosphere structures for helium of models with different $\beta_m$, but with the same H/He = 98/2, $F_{\rm XUV}$ = F$_0$.}
\label{atm_He_fxuv1}

\end{figure*}

With ultranarrowband photometry, \cite{2020AJ....159..278V} placed an upper limit on the excess absorption of He 10830 in WASP-52b.
They found that assuming the spectral shape as is observed in WASP-69b \citep{2018Sci...362.1388N}, WASP-52b will have an amplitude of 1.31\% $\pm$ 0.94\% in the deepest line center of He 10830 triplet. The authors also noted that the line shapes can vary from planets to planets. This assumption, however, can give a sense of what one might expect from high-resolution spectroscopy. Recently, \cite{2022arXiv220511579K} measured the spectral shape of the absorption line of neutral helium at 10830 $\rm\AA$,
using high-resolution spectroscopy, in the atmosphere of WASP-52b, and estimated an excess absorption of He 10830 of $\sim 3.44\%\pm 0.31$\%. Our work is based on this spectroscopic observation.
In the present study, we find that the absorption depth at the He 10830 line center in all the models with H/He = 92/8 exceeds the 3$\sigma$ upper limit (4.37\%) of the observation. Therefore, we only show the results of models with H/He = 98/2, 99/1, and 99.5/0.5.
Figures \ref{atm_He_fxuv_rates} (a)-(b) show the number densities of neutral He, He$^+$, and He(2$^3$S) of models with different H/He ratios but with the same $F_{\rm XUV}$ = F$_0$ and $\beta_m$ = 0.22. We can see that increasing the H/He ratio decreases the number density of He(2$^3$S).
The number densities of He(2$^3$S) first increase then decrease with the increasing altitude.  At about 2-3 $R_P$, there seems to be a plateau where the number density of He(2$^3$S) almost keeps a constant. This is mainly because the destruction rate of He(2$^3$S) due to the collisional transition by electrons depends strongly on the atmospheric temperature. The collisional transition rates $q_{31a}$ and $q_{31b}$ increase with increasing temperature.
However, the He(2$^3$S) photoionization rate $\Phi_3$ almost remians constant throughout the atmosphere. Figure \ref{atm_He_fxuv_rates} (c) shows the source (black dashed and dotted lines, in units of cm$^{-3} \rm s^{-1}$) and sink (solid lines, in units of $\rm s^{-1}$) terms related to the population of He(2$^3$S) for the model of H/He = 99.5/0.5, $F_{\rm XUV} = F_0$, and $\beta_m$ = 0.22. We can see that the main production mechanism of He(2$^3$S) is the recombination of He$^+$. On the other hand, different sink terms play a dominant role depending on altitude. When $r < 1.1$ $R_P$, the destruction of He(2$^3$S) is dominated by the collisional transition by hydrogen atoms. When 1.1 $ R_P < r < 3$ $R_P$, the dominant terms are the collisional transitions by electrons as $q_{31a}$ and $q_{31b}$ are high in this radius range. Above 3 $R_P$, the destruction of He(2$^3$S) is mainly due to photoionization. The dominant term at a specific altitude would vary from model to model as the atmospheric structure changes.
Figure \ref{atm_He_fxuv_rates} (d) is similar to (c), except for expressing the sink terms in units of cm$^{-3} \rm s^{-1}$ and additionally showing the advection
term of He(2$^3$S), which is denoted by the yellow dash-dotted line. Here, the advection term was estimated by an ad-hoc
manner similar to those for H(2p) and H(2s).

Figures \ref{atm_He_b022_H999} (a)-(d) show the change of the atmosphere structure as
$F_{\rm XUV}$ varies, when H/He = 98/2 and $\beta_m = 0.22$.
Both the temperature and velocity increase with the increase of $F_{\rm XUV}$. The number density of helium atoms increases with $F_{\rm XUV}$ when the altitude is below a certain value (about 6 $R_P$ for the models with H/He = 98/2), but above this altitude the opposite trend occurs. The number densities of He$^+$ and He(2$^3$S) increase significantly as $F_{\rm XUV}$ increases.
To compare with models of other H/He ratios, we also show the number densities of He and He(2$^3$S) for the models with H/He = 99/1 in Figures \ref{atm_He_b022_H999} (e)-(f). We find the number densities He(2$^3$S) of these models also increase with $F_{\rm XUV}$.
In addition, we find in models of small $F_{\rm XUV}$ ($< \rm F_0$), the number densities of He(2$^3$S) first increase then decrease fast as the altitude increase, without a plateau appearing. This is mainly because the temperatures in such models are relatively low and photoionization of He(2$^3$S) starts to become the dominant sink term from a lower altitude.
Figures \ref{atm_He_fxuv1} (a)-(d) show the temperature, velocity, and number densities of He(1s), He$^+$, and He(2$^3$S) for the models with the same H/He = 98/2 and $F_{\rm XUV}$ = F$_0$ but with different $\beta_m$. In the models with H/He = 98/2, we find a similar trend to the models with H/He = 92/8 that the highest temperature in the atmosphere as well as the velocity decreases with the increase of $\beta_m$. 
The number density of He increases with $\beta_m$. The number density of He$^+$ decreases with $\beta_m$ at radii $\lesssim$ 1.2 $R_P$ before a maximum value reaches.
%On a whole, different $\beta_m$ does not cause significant difference to the He$^+$ population. As a result, the He(2$^3$S) population is not heavily dependent on $\beta_m$. 
However, the number density of He(2$^3$S) increases with $\beta_m$ when the altitude is between about 1.2 $R_P$ and 2 $R_P$. Between 2 $R_P$ and 4 $R_P$, the trend becomes opposite again. Above 4 $R_P$, the number density changes little with $\beta_m$. Overall, compared to the effect due to $F_{\rm XUV}$, the change in $\beta_m$ does not appear to cause a significant variation in the He(2$^3$S) population. As a result, the He(2$^3$S) population is relatively independent of $\beta_m$.

\subsection{Modeling He 10830 transmission spectrum}

\begin{figure*}
\gridline{\fig{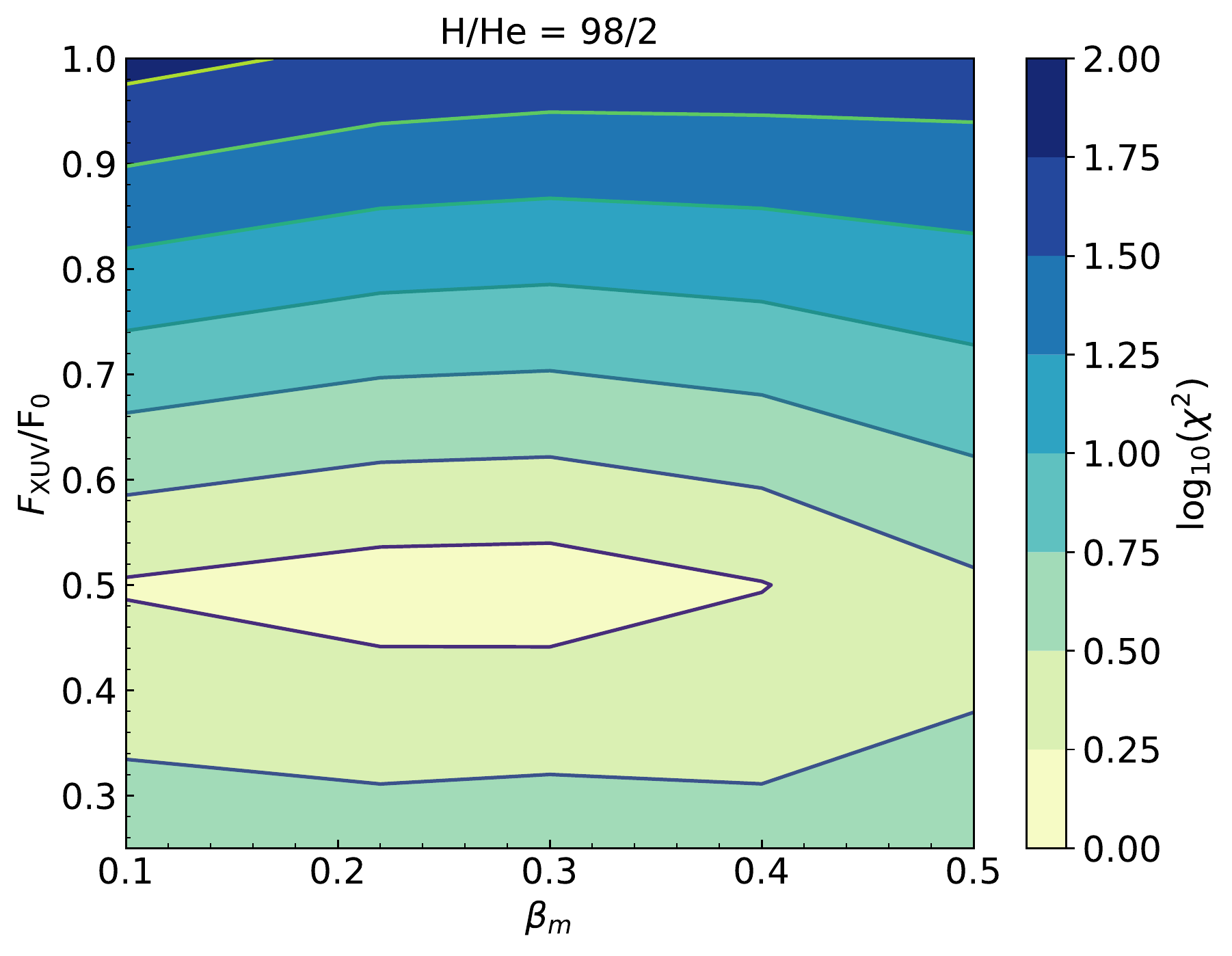}{0.5\textwidth}{(a)}
 \fig{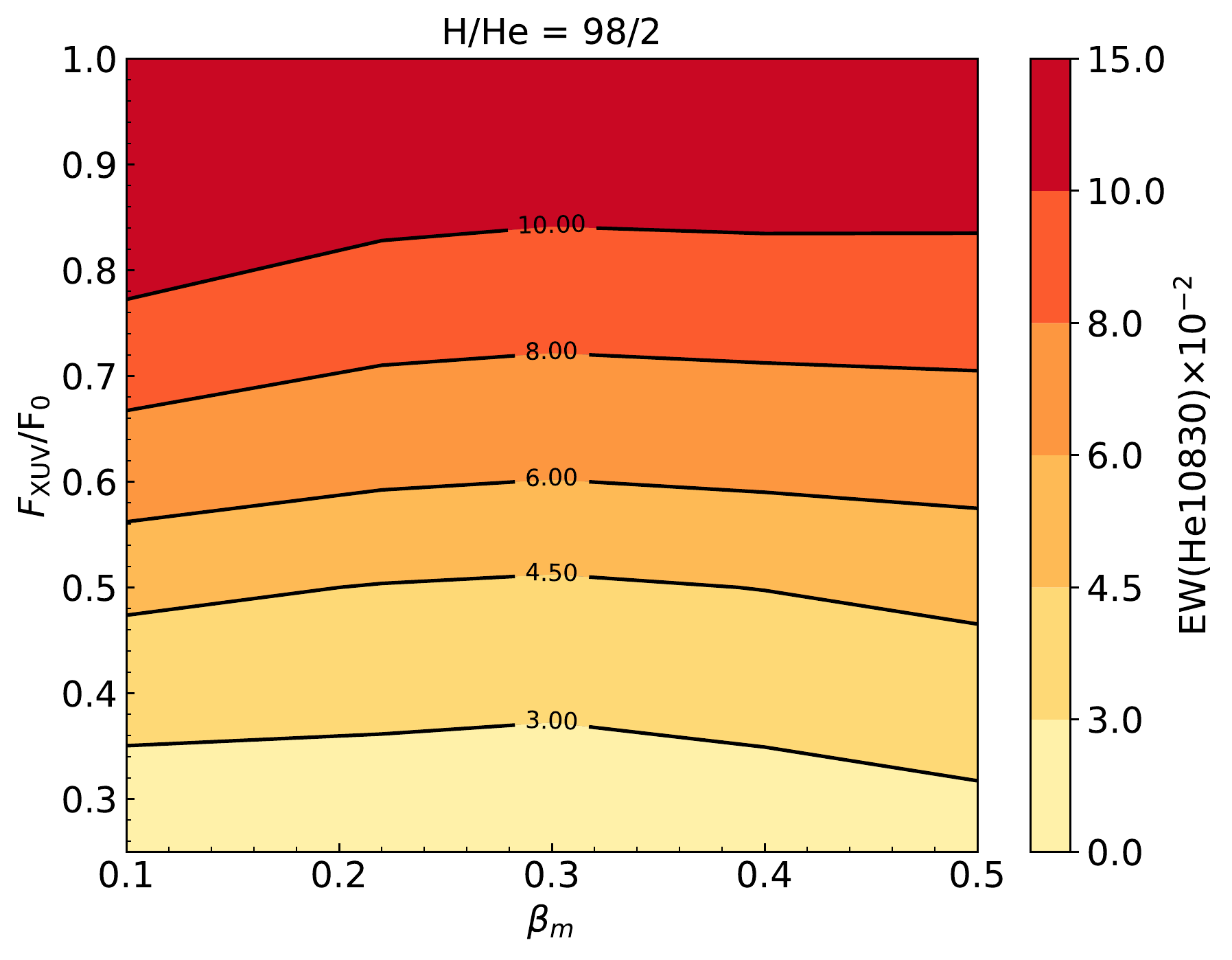}{0.5\textwidth}{(d)}
 }
\gridline{\fig{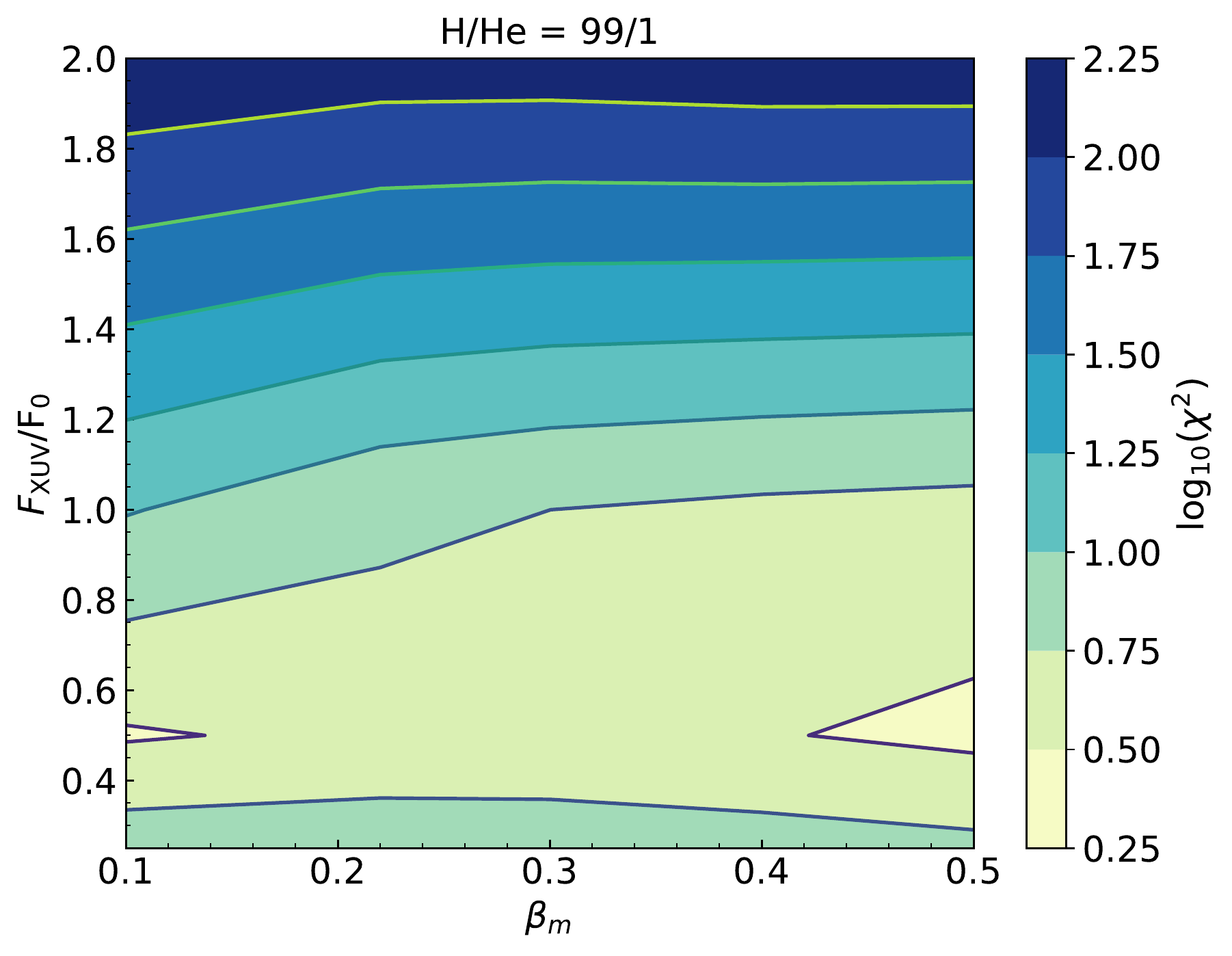}{0.5\textwidth}{(b)}
 \fig{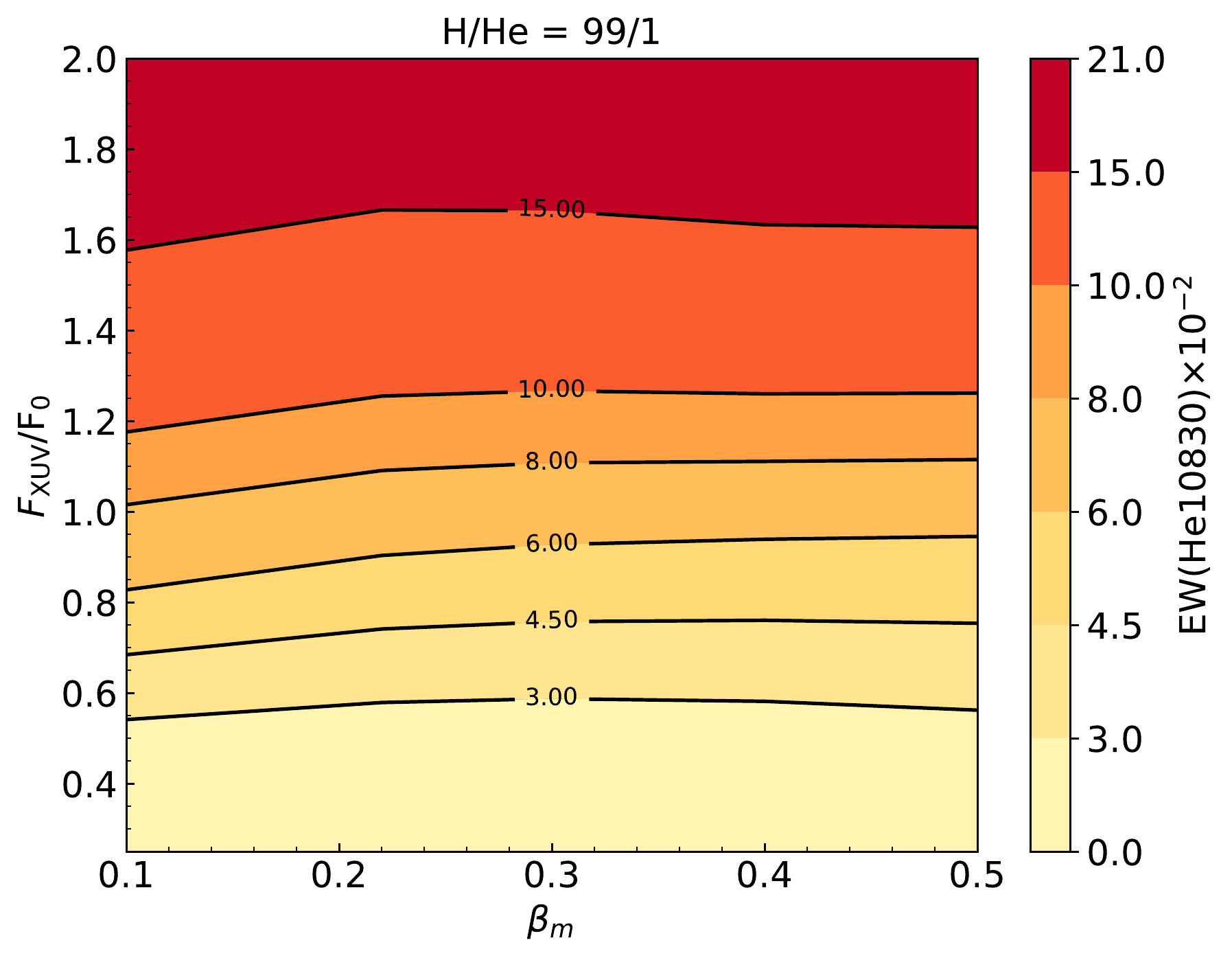}{0.5\textwidth}{(e)}
 }
\gridline{\fig{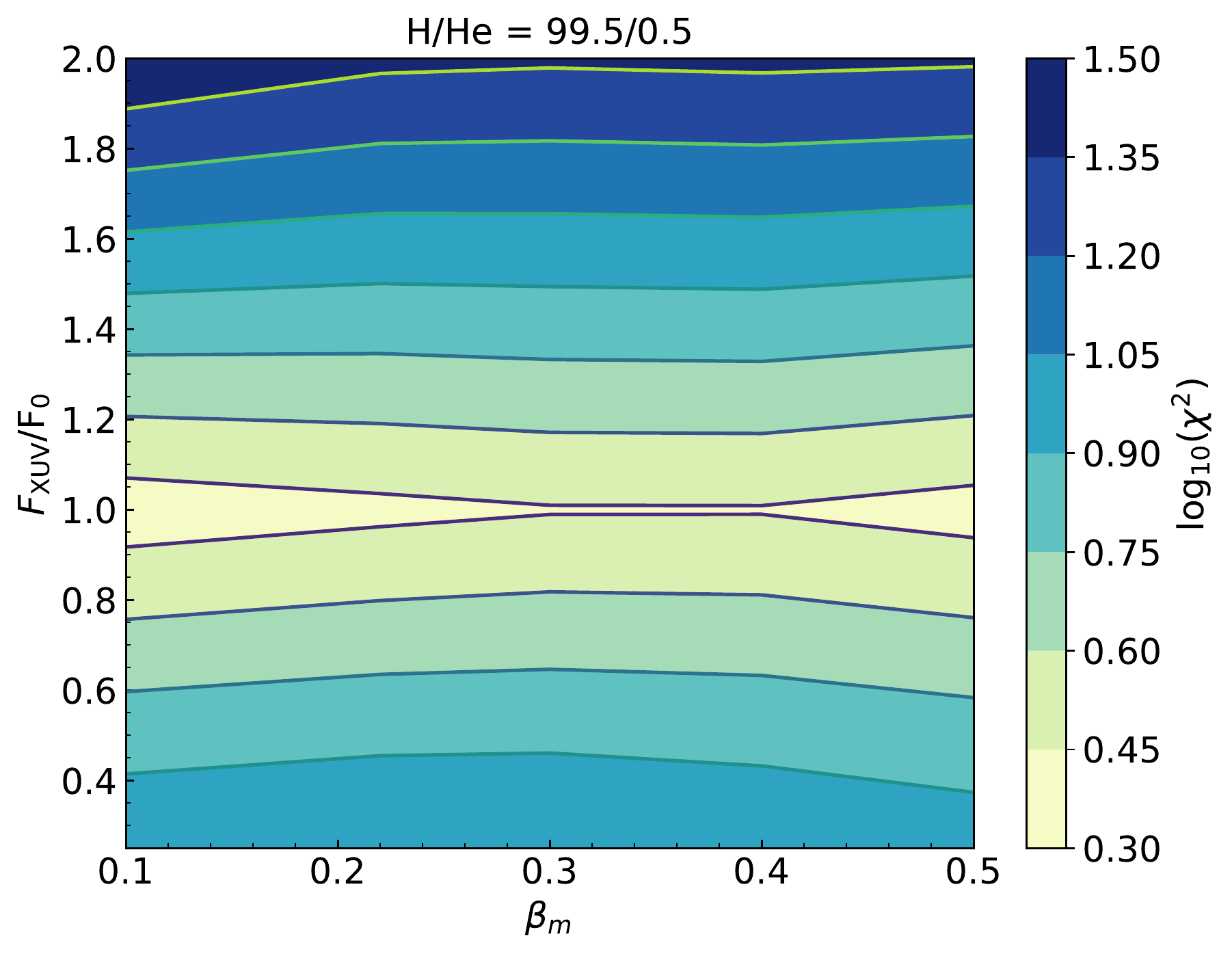}{0.5\textwidth}{(c)}
 \fig{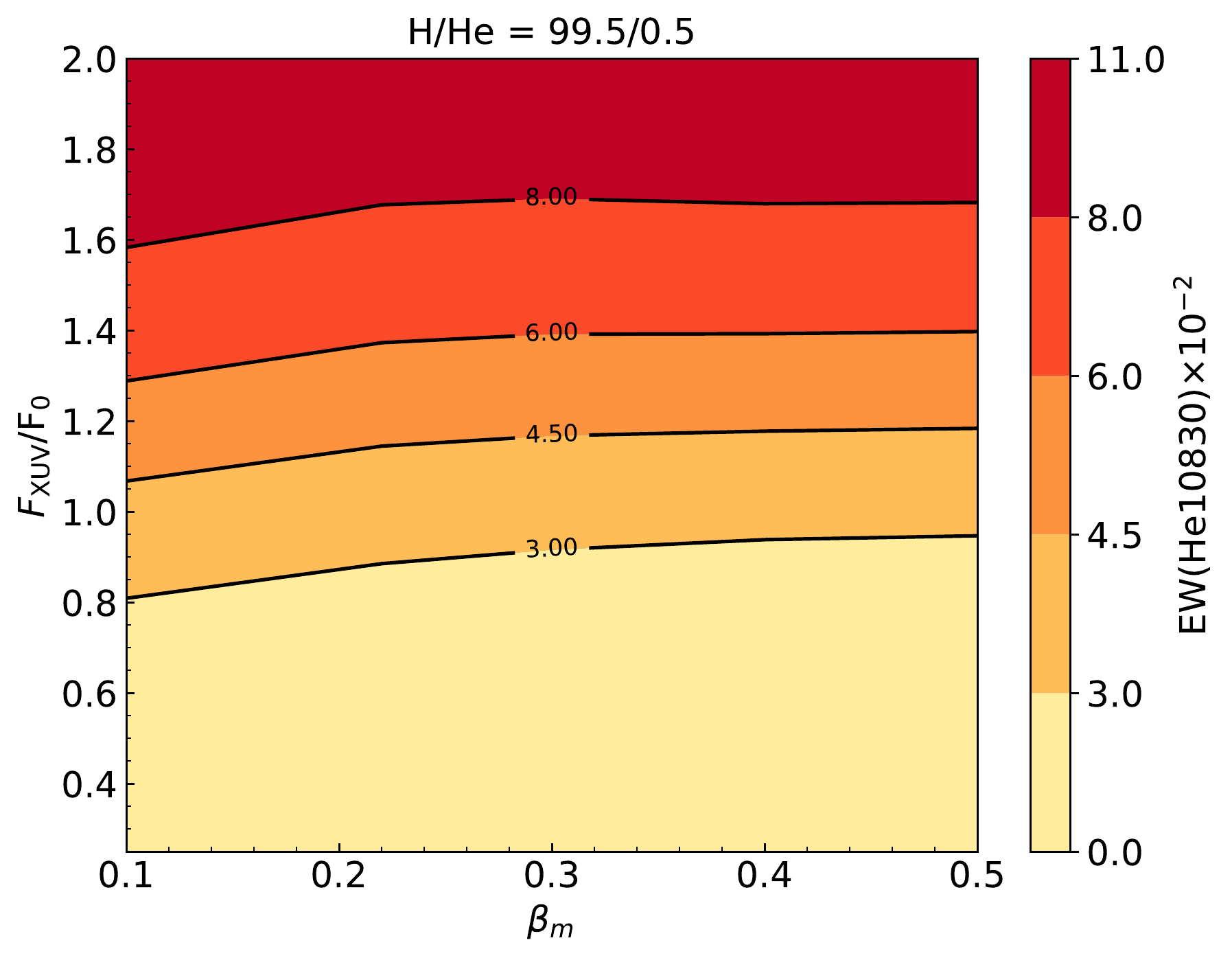}{0.5\textwidth}{(f)}
 } 
\caption{(a)-(c) $\chi^2$ contours for He10830 transmission spectrum, obtained by comparing the model spectrum with the observation data in the wavelength range of [10828.2, 10831.8]$\rm\AA$. (d)-(f) Equivalent width contours of He10830 line. The $\chi^2$ and EW contours were calculated for three hydrogen to helium abundance ratios.} 
\label{fig_HeTS_chi_unbinned}
\end{figure*}

\begin{figure*}
\gridline{\fig{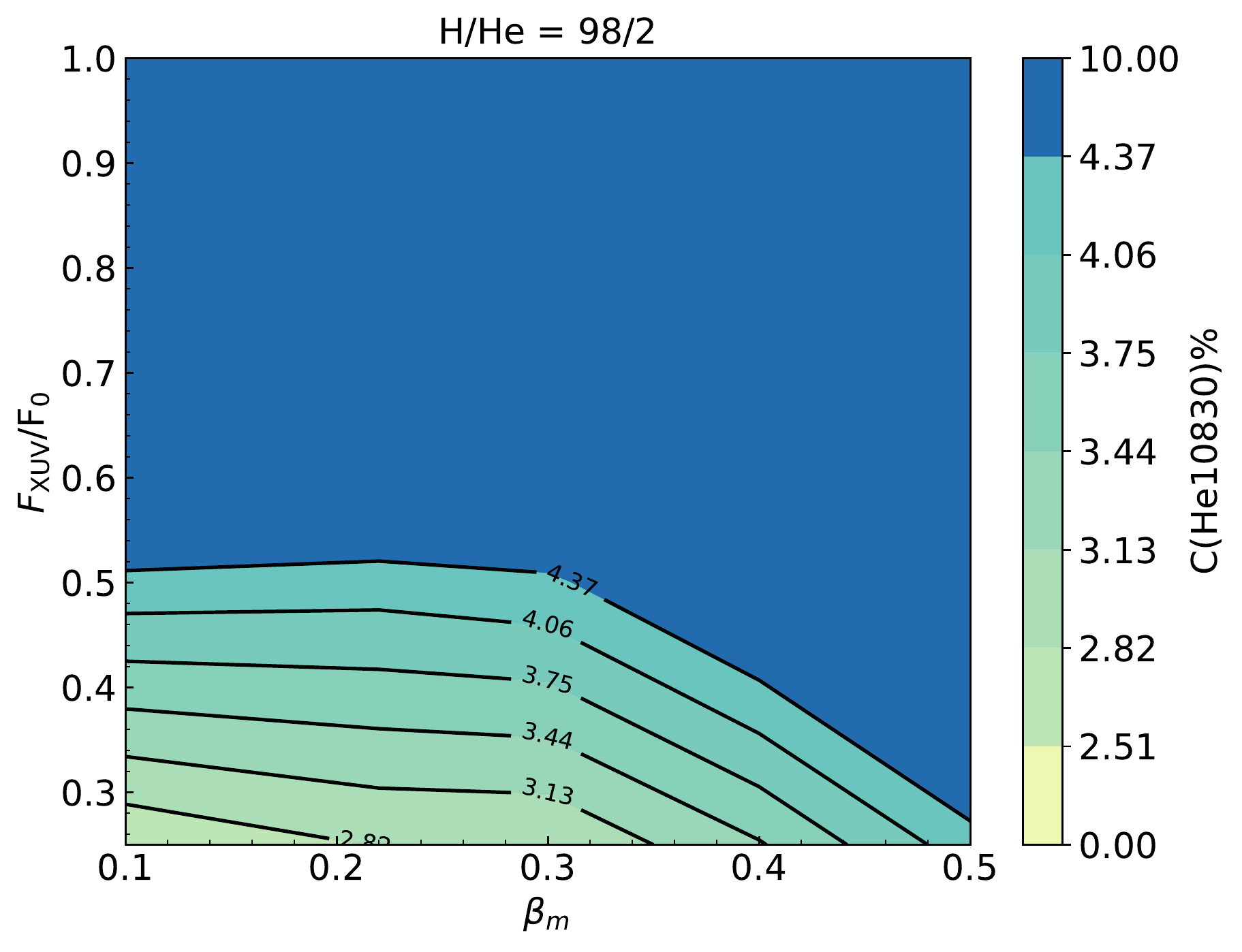}{0.5\textwidth}{(a)}
 \fig{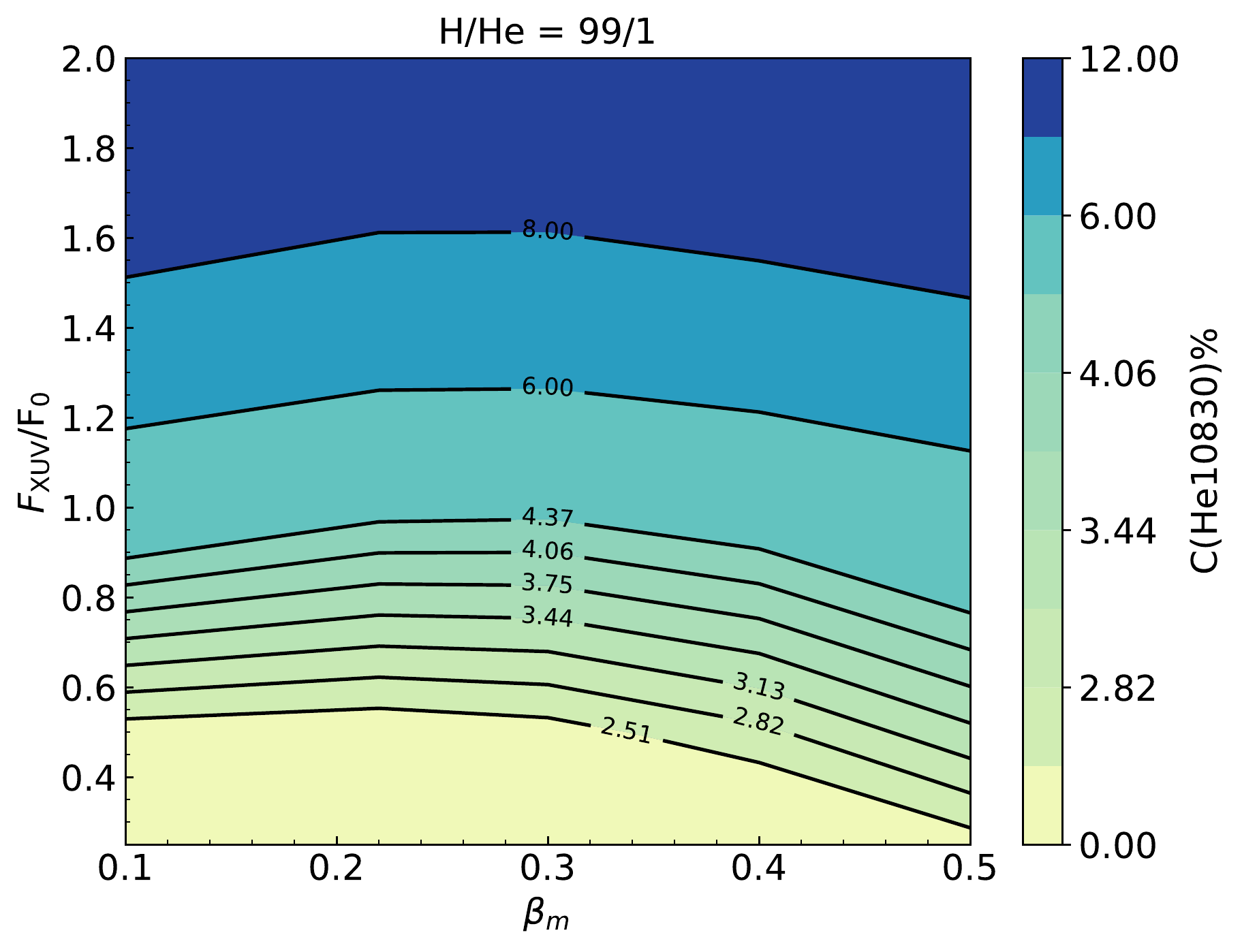}{0.5\textwidth}{(b)}
 }
\gridline{\fig{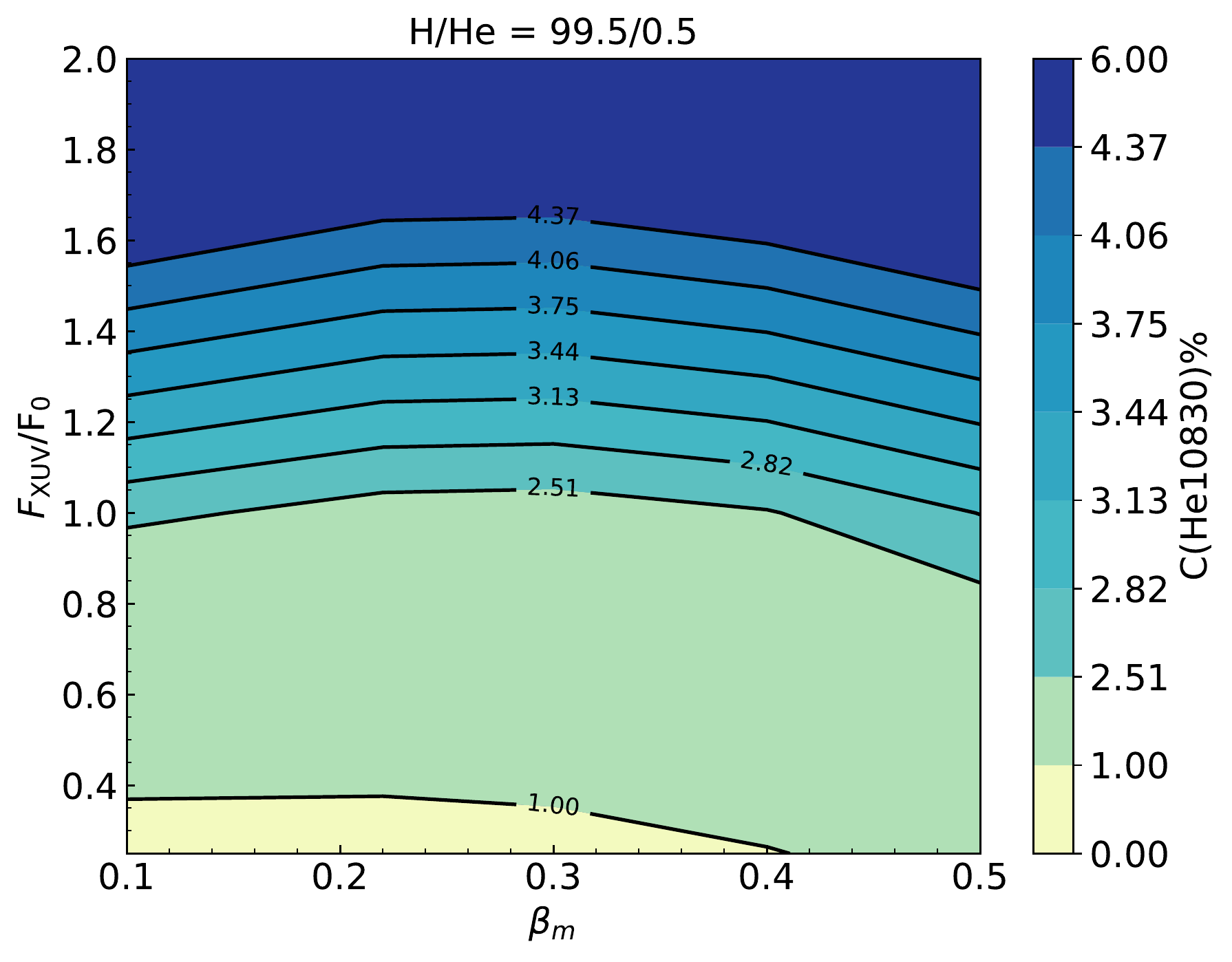}{0.5\textwidth}{(c)}
 \fig{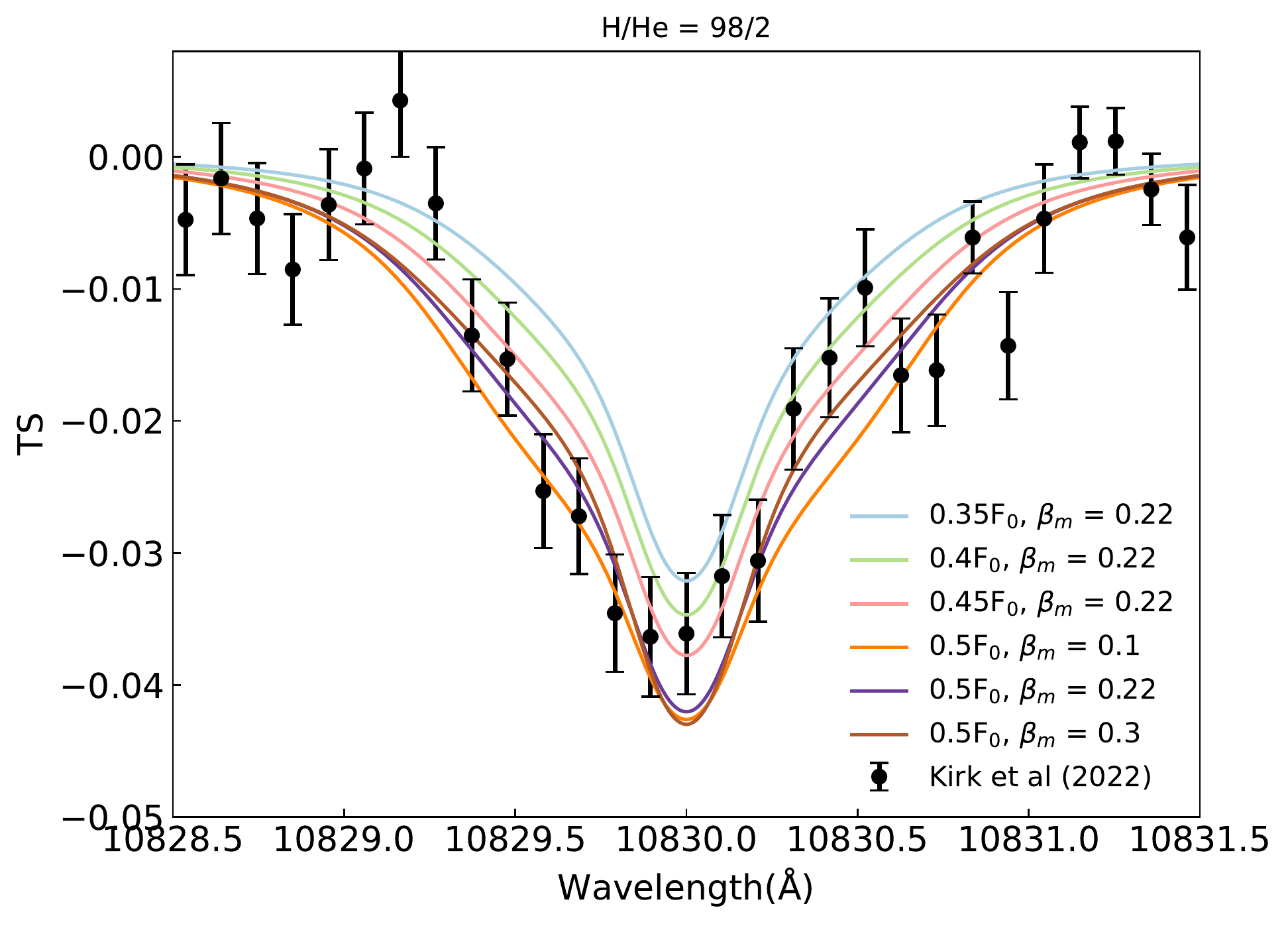}{0.5\textwidth}{(d)}
 }
\caption{(a)-(c) Absorption depth at the He10830 line center as a function of $F_{\rm XUV}$ and $\beta_m$ for the models of various H/He ratios. (d) He10830 transmission spectra of the models, which describe the observation well. Here, the models with $F_{\rm XUV}$ =
0.5F$_0$ are the best-fit models, and the others are those calculated for experiments.}
\label{CHe_contour}

\end{figure*}

\begin{figure*}
\gridline{\fig{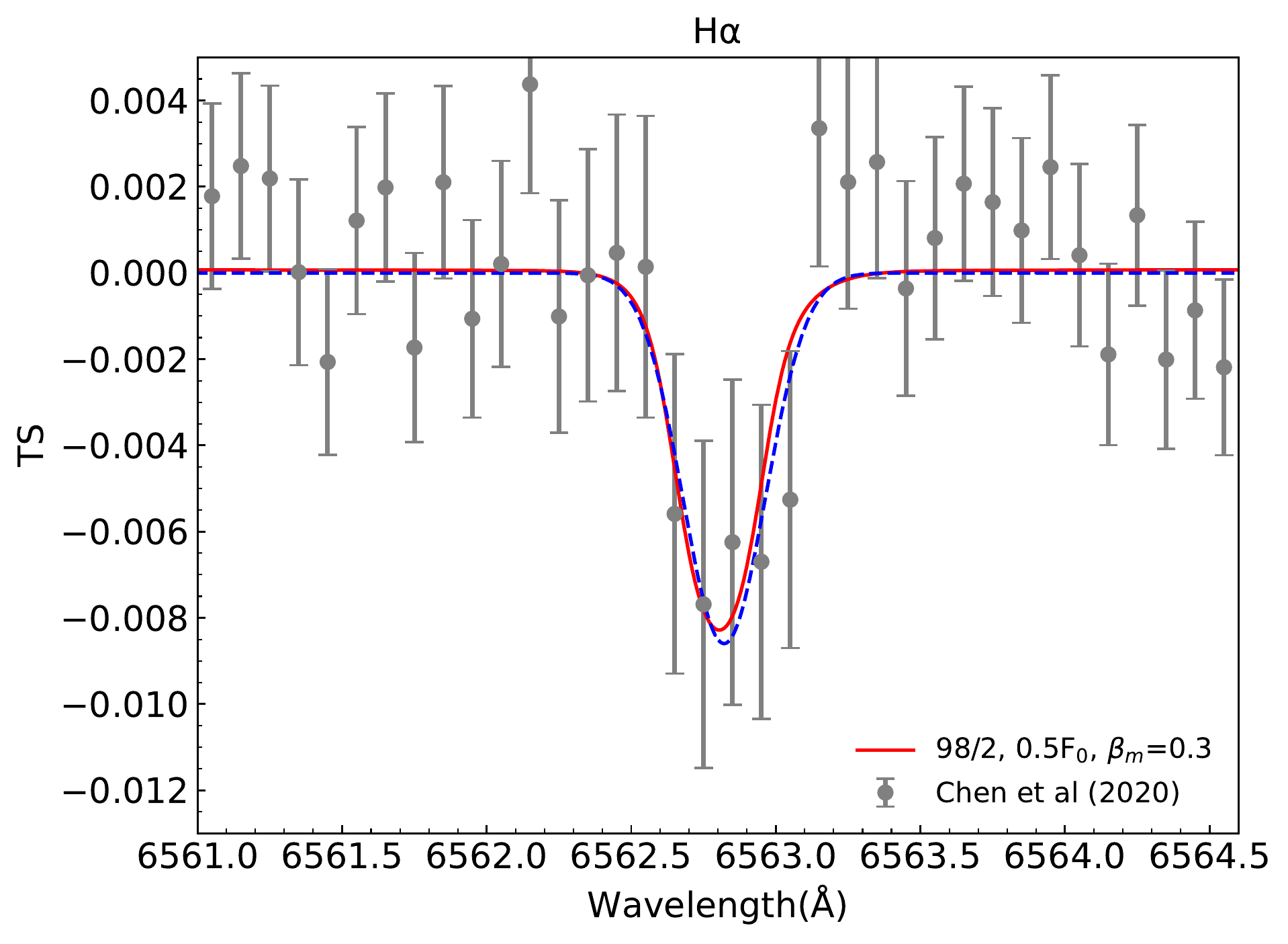}{0.5\textwidth}{(a)}
 \fig{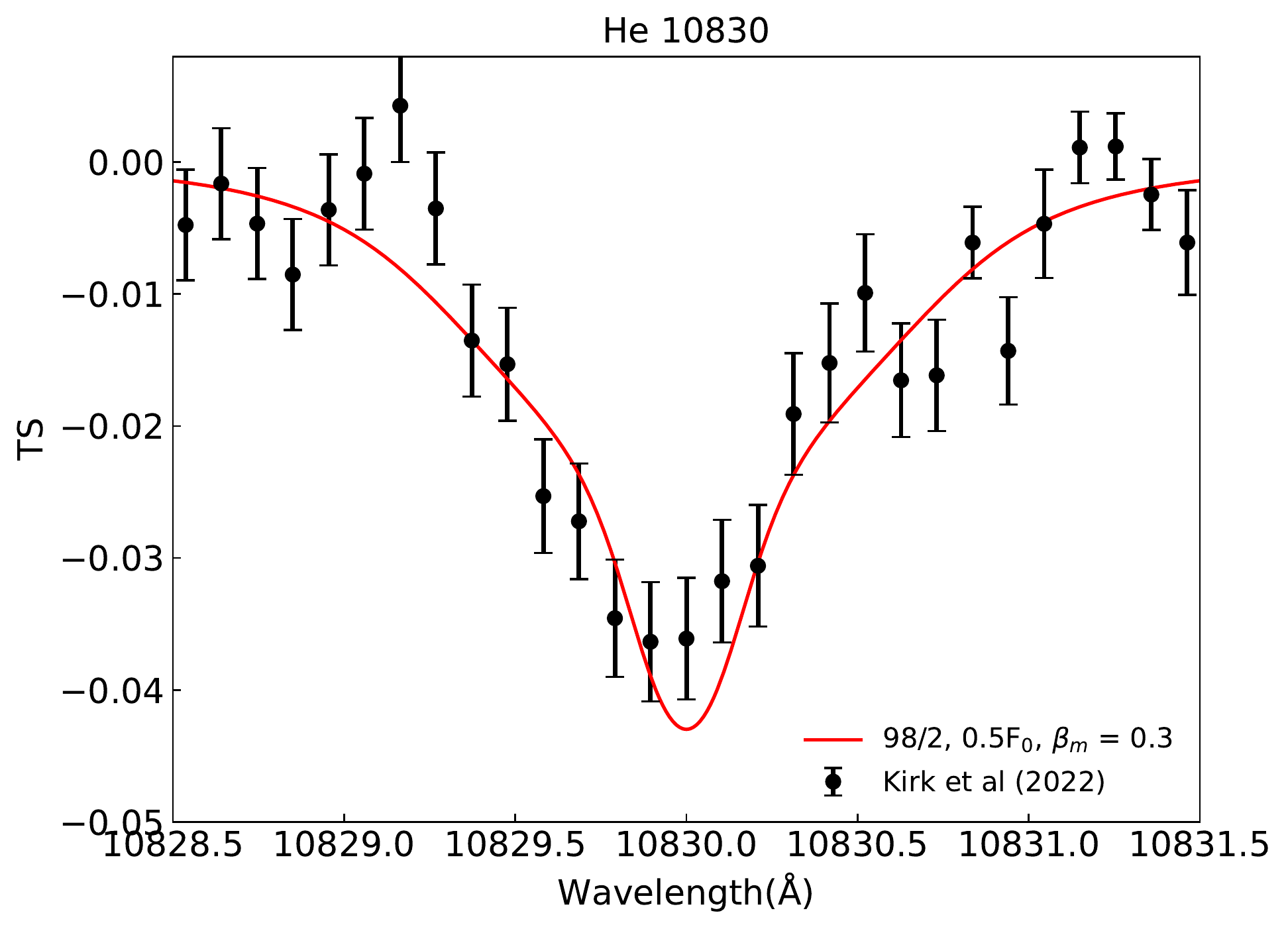}{0.5\textwidth}{(b)}
 }
 
\caption{H$\alpha$ and He10830 transmission spectra of the best-fit model.}
\label{Ha_He10830_TS_single}

\end{figure*}

Similar to H$\alpha$, we also calculate the He 10830 transmission spectra and compare them with the observation. To compare with the observation, we calculate the $\chi^2$ and equivalent width in the passband [10828.2, 10831.8]$\rm\AA$. We found that the results for very high $F_{\rm XUV}$ cannot fit the observation. Therefore, in this section, we only show the results for $F_{\rm XUV} \lesssim F_0$ in the case of H/He = 98/2, and $F_{\rm XUV} \lesssim 2F_0$ in the cases of H/He = 99/1 and 99.5/0.5.

Figures \ref{fig_HeTS_chi_unbinned} (a)-(c) show the contours of $\chi^2$ as a function of $F_{\rm XUV}$ and $\beta_m$ for models with H/He = 98/2, 99/1, and 99.5/0.5. We can see that the confidence region defined by the lowest $\chi^2$ contour differs from panel to panel. In the case of H/He = 98/2, the confidence region spans a parameter space of $F_{\rm XUV}\sim 0.5F_0$ and $\beta_m \lesssim 0.4$. In the case of H/He = 99/1, it covers the space of $F_{\rm XUV}\sim 0.5F_0$, $\beta_m\sim 0.1$ or $\beta_m\gtrsim 0.4$. And in the case of H/He = 99.5/0.5, the least $\chi^2$ is found when $F_{\rm XUV}\sim F_0$, regardless of $\beta_m$. Among these three cases, the smallest $\chi^2$ appears when H/He= 98/2. In other words, a better fit can be found in the models with H/He = 98/2.

Figures \ref{fig_HeTS_chi_unbinned} (d)-(f) show the EW as a function of $F_{\rm XUV}$ and $\beta_m$ for models with H/He = 98/2, 99/1, and 99.5/0.5. Here, we note that the 1$\sigma$ confidence range of the observed EW is found to be (3.0$\sim$6.0)$\times 10^{-2}\rm\AA$, and the mean EW is 4.5$\times 10^{-2}\rm\AA$.
The figure shows that the model EW increases as $F_{\rm XUV}$ increases, but is not very sensitive to the varation of $\beta_m$. By comparing the contours of $\chi^2$ and EW, we find that the confidence region defined by $\chi^2$ is well within the 1$\sigma$ confidence region of EW in the cases of H/He = 98/2 and 99.5/0.5. In the case of H/He = 99/1, however, the $\chi^2$ confidence region does not coincide with that constrained by EW. This implies that the models of H/He = 99/1 can not reproduce the observed He10830 line profile well.

In the above, we compared the overall spectral shape of the He 10830 line by examining $\chi^2$ and EW. We now compare the absorption depth at the He 10830 line center C(He10830). According to \cite{2022arXiv220511579K}, the absorption depth of 3.44\% $\pm$ 0.31\% (11$\sigma$) was detected at the He 10830 line center. In comparing the absorption depth, we define the models with C(He10830) located in the 3$\sigma$ confidence ranges  to be acceptable.
Figures \ref{CHe_contour} (a)-(c) show the absorption depth at the He 10830 line center as a function of $F_{\rm XUV}$ and $\beta_m$ for models with H/He = 98/2, 99/1, and 99.5/0.5. In the figures, we show the contour lines of the lower (C(He10830) = 3.13\%, 2.82\%, and 2.51\%) and upper limits at 1$\sigma$, 2$\sigma$, and 3$\sigma$ levels (3.75\%, 4.06\%, and 4.37\%) of the observation. 
We find C(He10830) in all cases increases with increaing $F_{\rm XUV}$, but its variation with $\beta_m$ is more complicated. In the case of H/He = 98/2, C(He10830) varies little with $\beta_m$ when $\beta_m\lesssim 0.3$ while increases with $\beta_m$
when $\beta_m\gtrsim 0.3$. In the cases of H/He = 99/1 and 99.5/0.5, C(He10830) seems to first decreases then increases with $\beta_m$.

The absorption depth of He 10830 is obviously directly related to the number density of He(2$^3$S). In this regard, we also
note that, as shown in Figure \ref{atm_He_fxuv1} (d), there are points (at about 2 $R_P$) after which the number density of H(2$^3$S) decreases with increasing $\beta_m$, but before the points the opposite trend occurs.
This behavior reflects the fact that the absorption of He 10830 depends not only on the properties of the planetary atmosphere, but also on the intensity and spectral shape of the incident stellar radiation field, as discussed by \cite{2019ApJ...881..133O}.

In addition, we find that increasing the H/He ratio (H/He $\geq$ 98/2) doesn't significantly alter the H$\alpha$ absorption, while it significantly reduce the He10830 absorption. 
From Figures \ref{CHe_contour} (a)-(c), we can see that for a samll $F_{\rm XUV}$ (for example, F$_0$), a reduction of the He abundance by a factor of 5 leads to a decrease of $\sim 2$\% in the He 10830 absorption depth. In the case of a larger $F_{\rm XUV}$, the decrease can be as high as 3\%-4\%. This confirms our idea that increasing the H/He will lead to a better fit to the observation.

Finally, the models simultaneously constrained by $\chi^2$, EW, and C(He10830) are the best-fit ones. They are the models with H/He = 98/2, $F_{\rm XUV} = 0.5$F$_0$, and $\beta_m$ = 0.1, 0.22, 0.3. Figure \ref{CHe_contour} (d) shows the transmission spectra of these models. We can see that the models fit both the absorption depth and the profile well. However, the absorption depth at the He 10830 line center of the models with $F_{\rm XUV} = 0.5$F$_0$ almost reaches the upper limit of the observation. To find models that give a relatively smaller absorption depth, we also examined models with $F_{\rm XUV}$ = 0.35, 0.4, and 0.45 F$_0$ when $\beta_m$ = 0.22. We find that the model with $F_{\rm XUV}$ = 0.45F$_0$ can yield a better
fit to the line center depth. We also note that these model parameters are very close to $F_{\rm XUV} = 0.5$ and $\beta_m = 0.3$. Therefore, we conclude that the H/He ratio ($\sim 98/2$) should be higher than the solar value ($\sim 92/8$), $F_{\rm XUV}$ is about 0.5 times the fiducial value, and $\beta_m \lesssim$ 0.3 to reproduce the He 10830 observation.

\subsection{Models that fit both the H$\alpha$ and He10830 lines}
Because both the H$\alpha$ and He 10830 absorption were detected from this planet, we attempt to find models that can fit both lines at the same time. According to the above analysis, it is clear that the H/He ratio should be 98/2. 
From Figure \ref{Ha_TS_many} (b) and Figure \ref{CHe_contour} (d), we find only the model with $F_{\rm XUV}$ = 0.5F$_0$ and $\beta_m$ = 0.3 can fit both lines simultaneously.
In Figures \ref{Ha_He10830_TS_single} (a) and (b), we show the H$\alpha$ and He 10830 transmission spectra of the best-fit model, respectively. We can see both model spectra agree well with the observed data.

\cite{2020AJ....159..278V} set an upper limit of the mass-loss rates ($\dot{M}$) of WASP-52b by applying an isothermal Parker wind. They obtained that $\dot{M}\textless1.25\times10^{10}$ g s$^{-1}$ ($\dot{M}\textless1.25\times10^{11}$ g s$^{-1}$) for an isothermal atmosphere with a temperature of 7000 K (12,000 K). In \cite{2022arXiv220511579K}, they used an isothermal Parker wind to model the He 10830 transmission spectrum of WASP-52b and found a mass-loss rate of about 1.4 $\times10^{11}$ g s$^{-1}$.
In our hydrodynamic simulations, the mass-loss rate increases with $F_{\rm XUV}$ and is slightly affected by $\beta_m$ and the H/He ratio.
We find the mass-loss rate of about 2.8 $\times10^{11} \rm g$ s$^{-1}$ from the best-fit model, which is twice as higher as theirs. The difference is mainly caused by the difference in the adopted atmosphere models.
In Section \ref{subsec:isothermal}, we will discuss the difference between an isothermal Parker wind and our hydrodynamic model.

\section{Discussion}\label{sec: Diss}
\begin{figure*}
\gridline{\fig{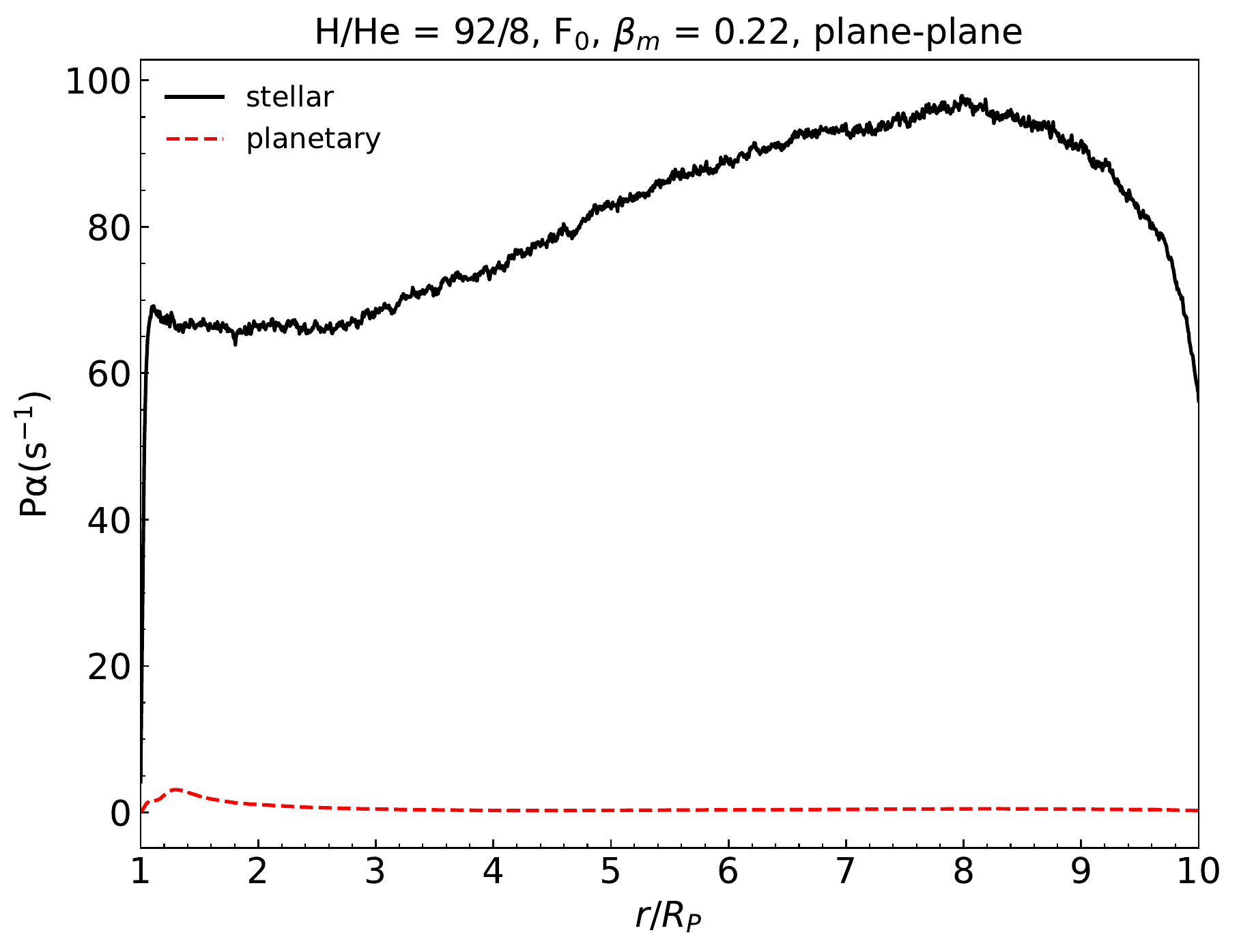}{0.35\textwidth}{(a)}
 \fig{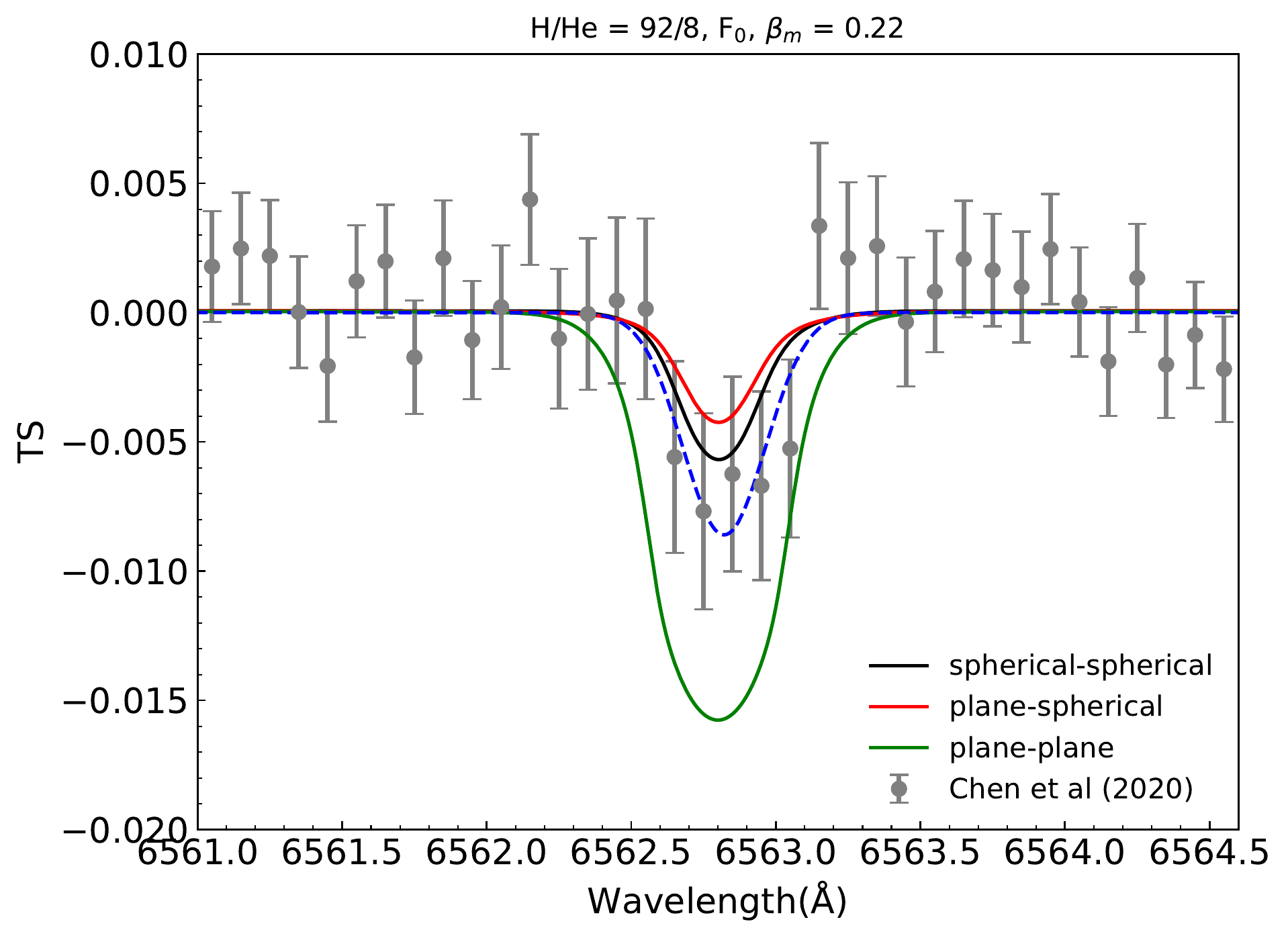}{0.35\textwidth}{(b)}
 \fig{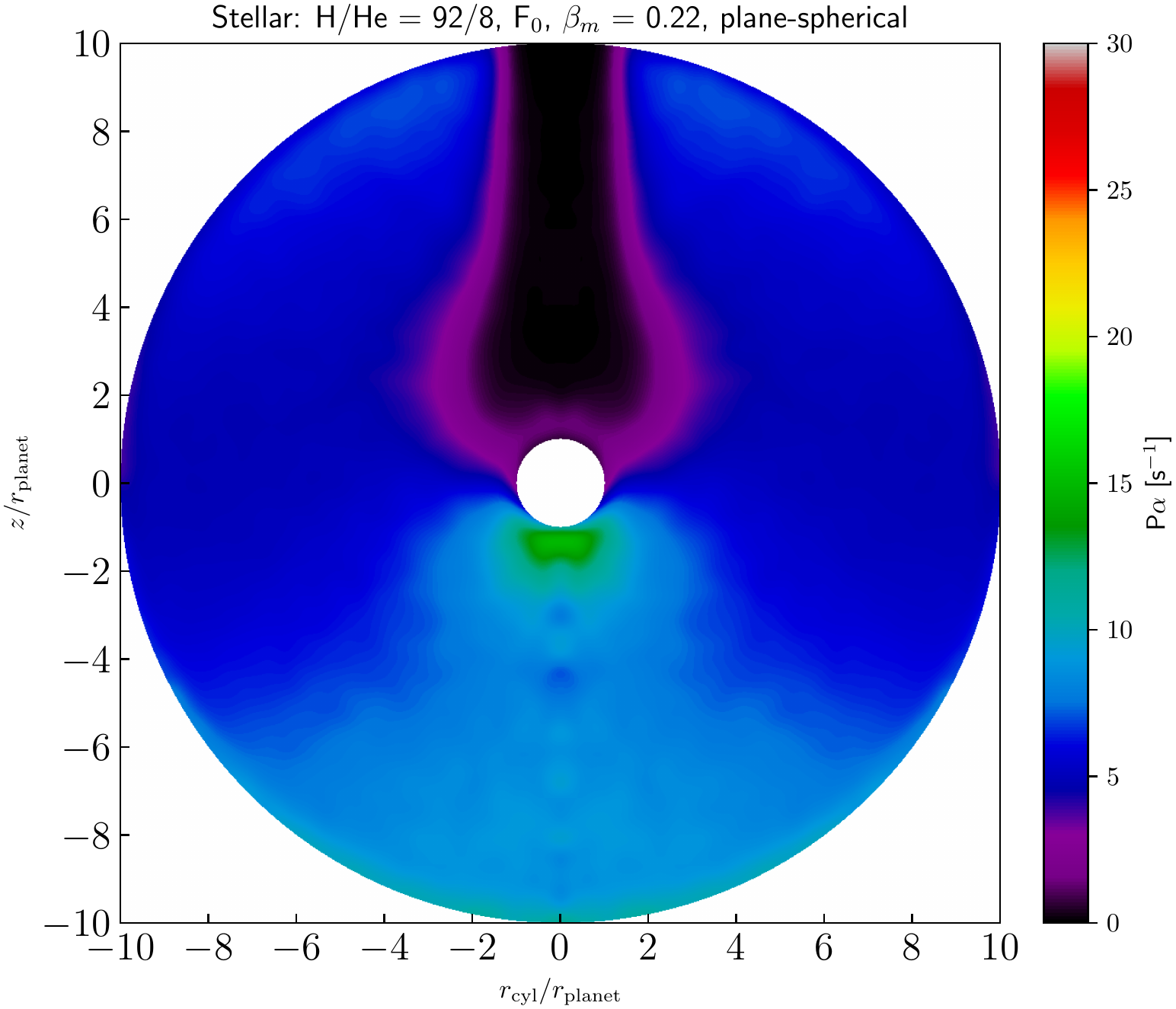}{0.35\textwidth}{(c)}
 }
\gridline{\fig{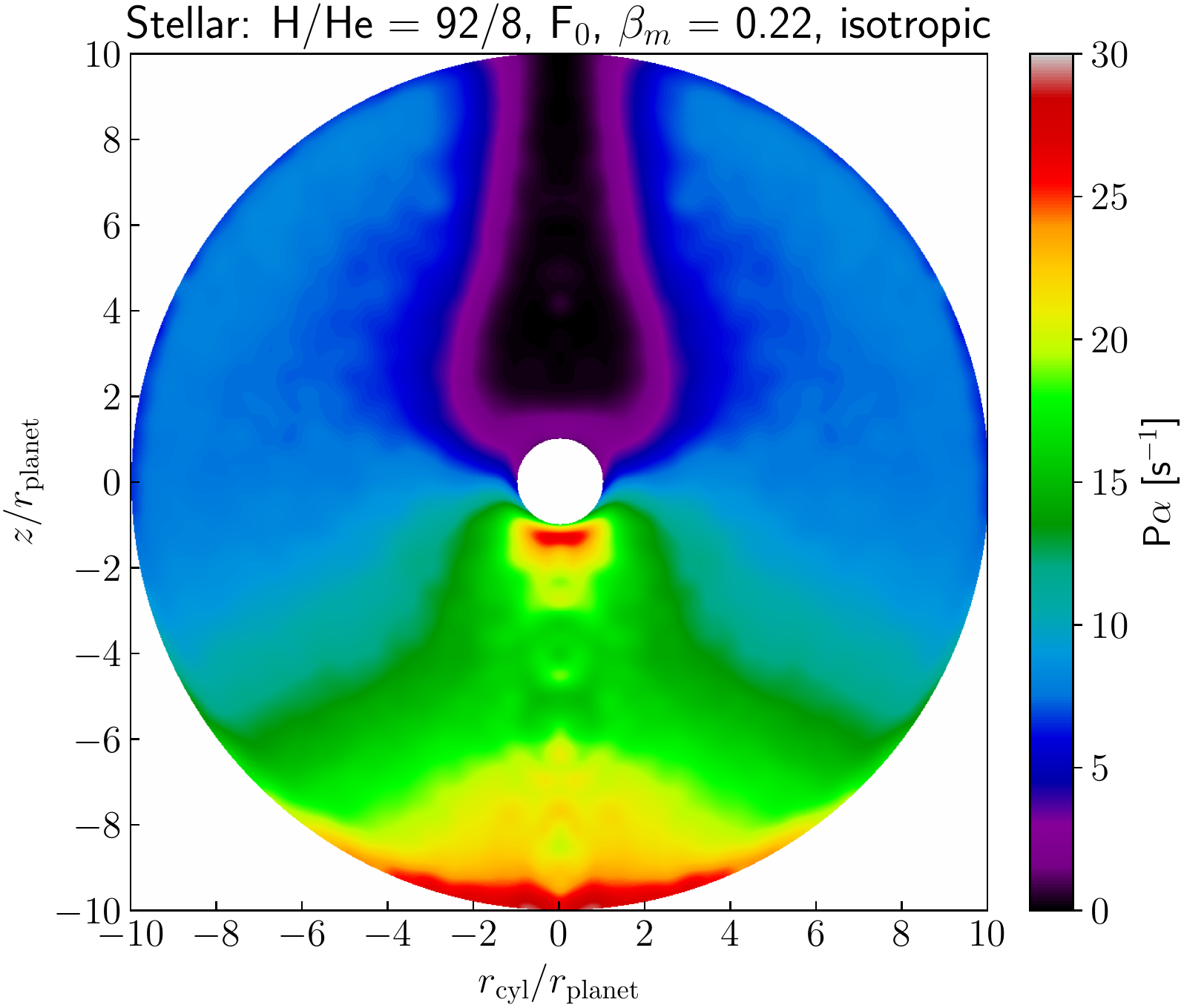}{0.35\textwidth}{(d)}
 \fig{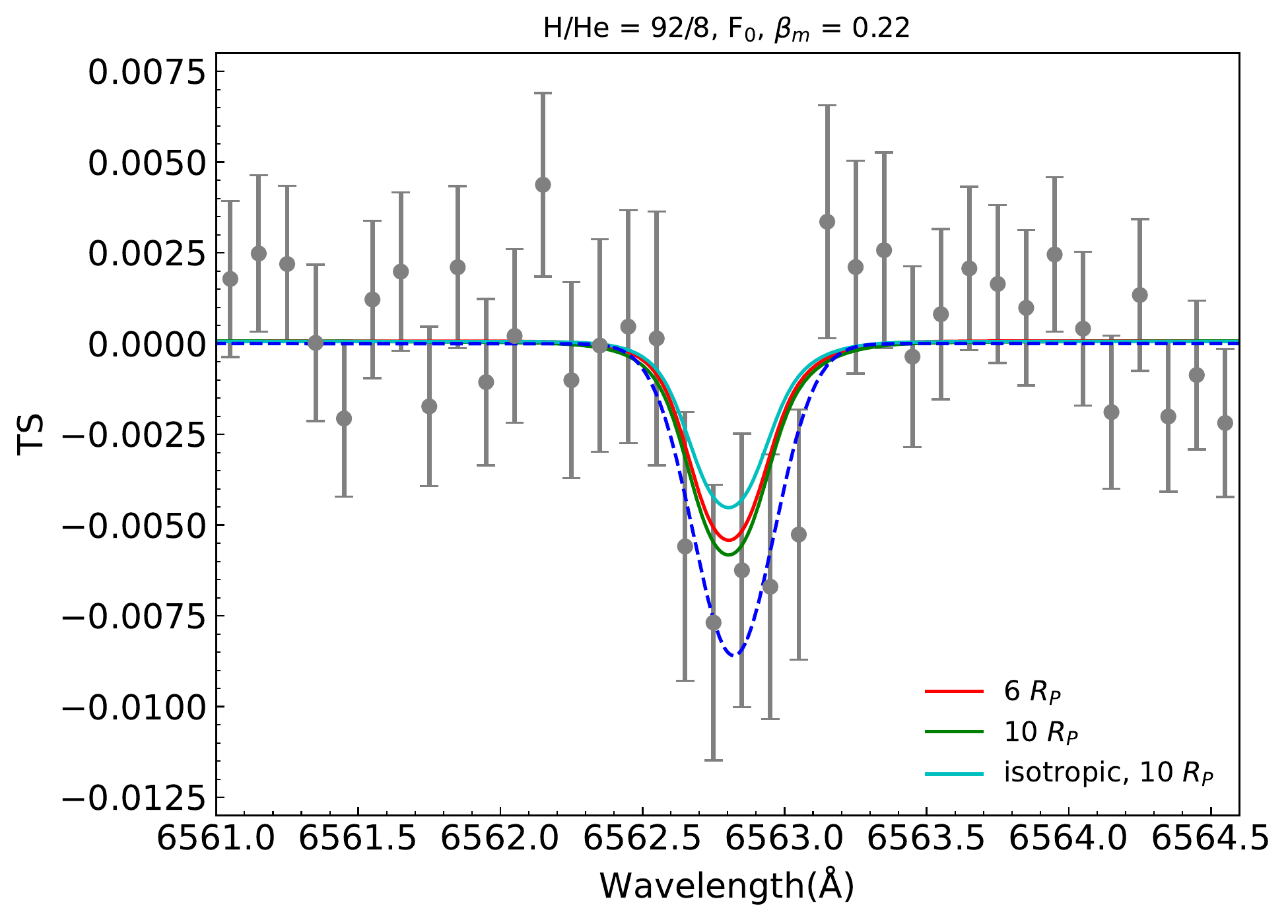}{0.35\textwidth}{(e)}
 \fig{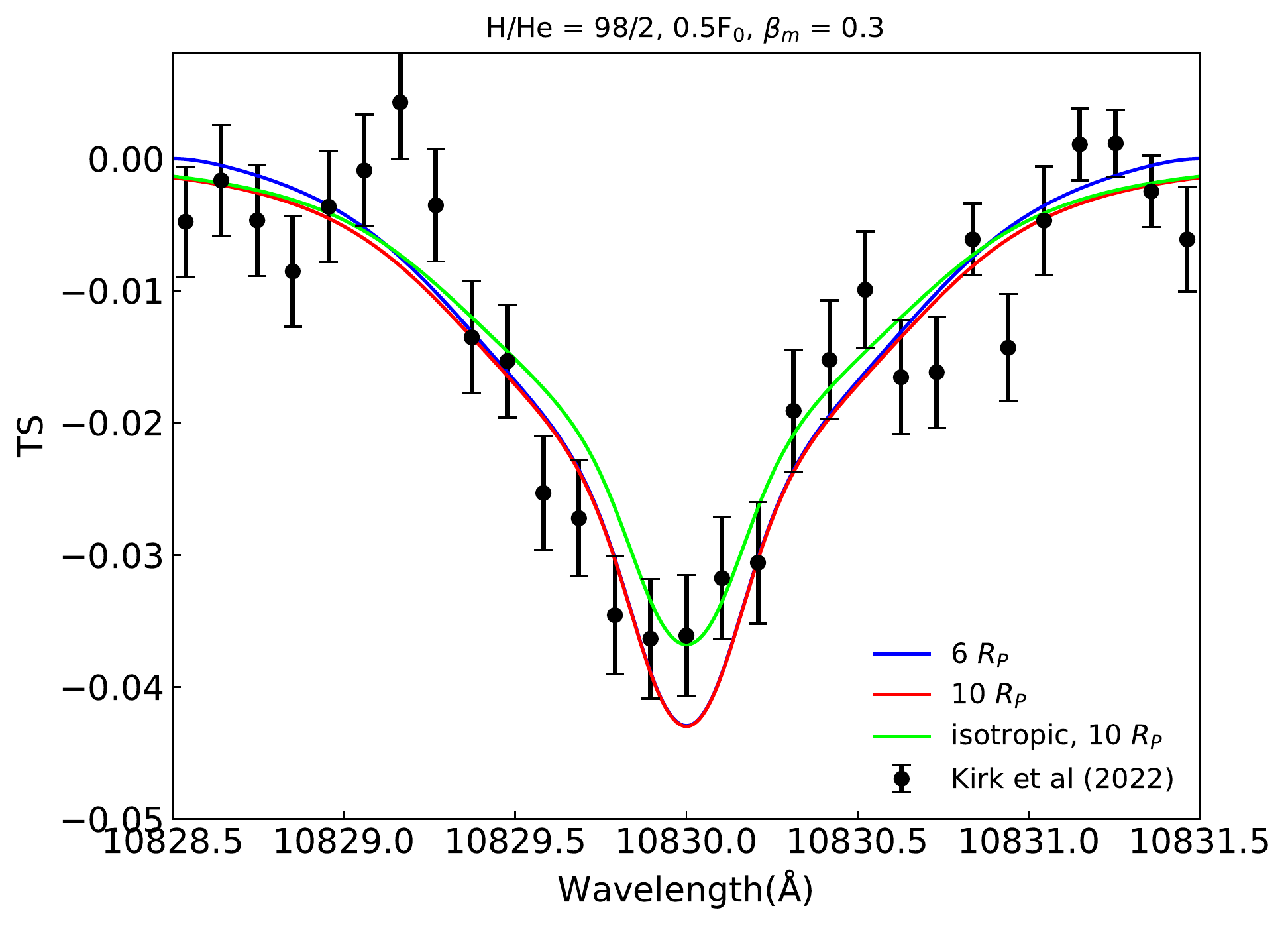}{0.35\textwidth}{(f)}
 }
 \gridline{\fig{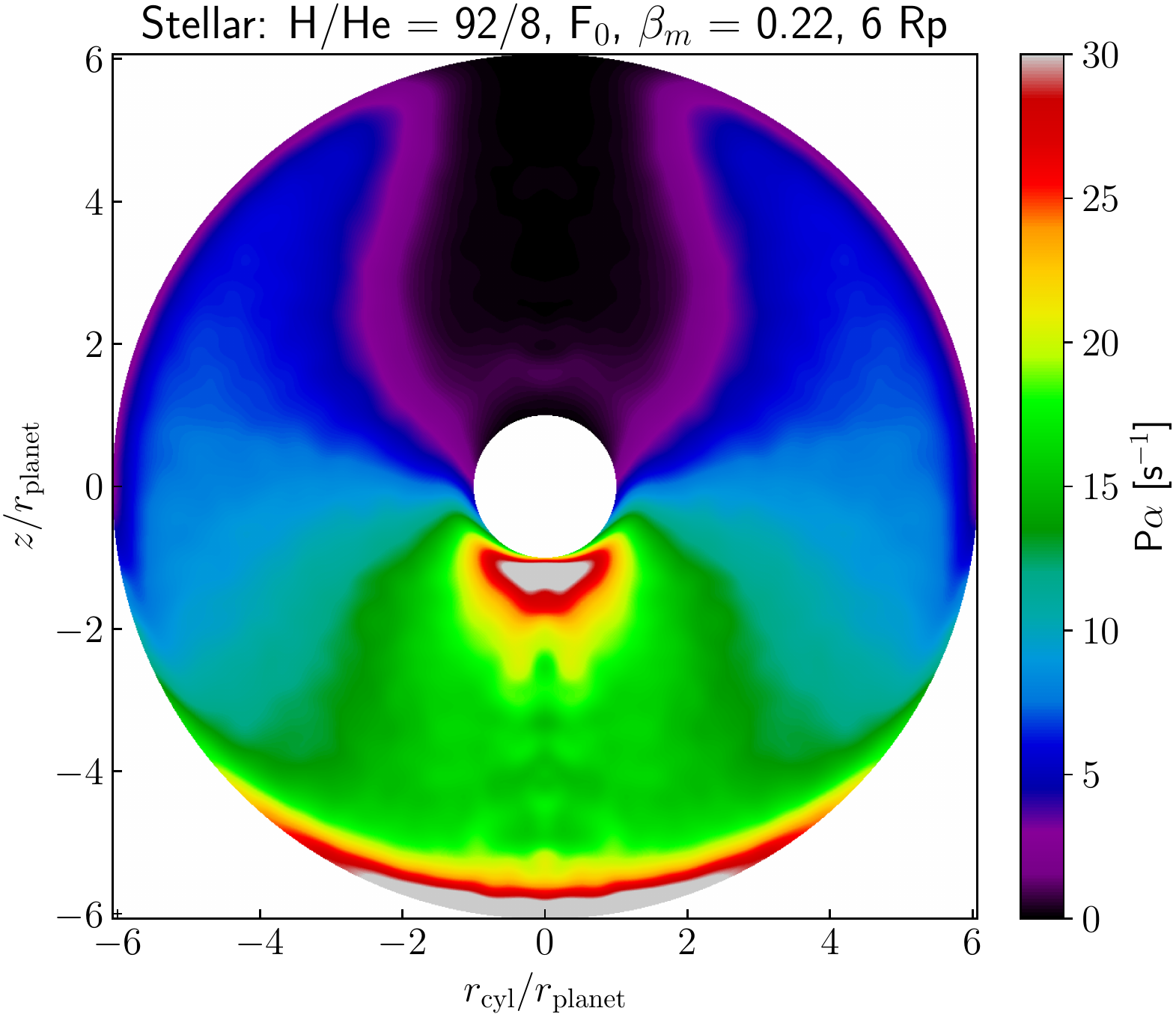}{0.35\textwidth}{(g)}
 }
\caption{(a) $P\alpha$ as a function of atmosphere altitude for the plane-plane model; (b) Comparison of H$\alpha$ transmission spectrum between the plane-plane, plane-spherical and spherical-spherical models; (c) $P\alpha$ of the plane-spherical model; (d) $P\alpha$ of the spherical-spherical model with isotropic illumination; (e) Comparison of H$\alpha$ transmission spectrum between three spherical-spherical models: a limb darkened model with an atmosphere outer boundary size of 6 $R_P$, a limb darkened model with an atmosphere size of 10 $R_P$, and an isotropic model with an atmosphere size of 10 $R_P$; (f) Similar to (b), but for the He 10830 transmission spectrum; (g) $P\alpha$ of the spherical-spherical model with an outer boundary size of 6 $R_P$. Note in panel (g), the value of $P\alpha$ is slightly larger than 30 in the gray regions.}
\label{fig:plane_illum}

\end{figure*}

\subsection{Plane-parallel stellar Ly$\alpha$ illumination}
Here, we discuss two plane-parallel illumination models, in which the stellar Ly$\alpha$ radiation is plane-parallel while the atmosphere can be either plane-parallel or spherical. These two models are the plane-plane model and plane-spherical model defined in Section \ref{sec:LaRT}.
First, Figure \ref{fig:plane_illum} (a) shows $P\alpha$ as a function of atmosphere altitude for the plane-plane model. The black solid and red dashed lines represent the contribution from the stellar and planetary Ly$\alpha$ source, respectively. We can see that the total $P\alpha$ is dominated by the stellar Ly$\alpha$. The total $P\alpha$ is much higher than in the spherical illumination model. In particular, in the plane-plane model, the $P\alpha$ in the nightside of the atmosphere is the same as the dayside.
This will dramatically increase the populations of H(2), and finally result in a much deeper absorption of H$\alpha$ compared to the spherical-spherical model. We can clearly see this difference in the H$\alpha$ transmission spectra in Figure \ref{fig:plane_illum} (b). In the spherical-spherical model, the absorption depth at the H$\alpha$ line center is only about 0.6\% when H/He = 92/8, $F_{\rm XUV}$ = F$_0$, and $\beta_m$ = 0.22 , but it increases to 1.6\% for the case of the plane-plane model.
Therefore, using a plane-plane assumption, the parameters that fit the observation must be altered. We can conclude that smaller $F_{\rm XUV}$ and $\beta_m$ would be required in this scenario.
Second, Figure \ref{fig:plane_illum} (c) shows $P\alpha$ as a function of atmosphere altitude for the plane-spherical model. Compared to the spherical-spherical (or spherical illumination) model, this model gives relatively lower $P\alpha$ values. This is because the planet would receive less Ly$\alpha$ photons from a plane-parallel illumination source, which is estimated to be about 60\% of the spherical illumination model. Consequently, it turns out that the H$\alpha$ absorption is the lowest in the plane-spherical model, as can be seen in Figure \ref{fig:plane_illum} (b).

\subsection{Isotropic and limb darkening illumination}
As mentioned in Section \ref{subsec:model-3}, the limb darkening effect of the stellar illumination was taken into account in this work. Because the H(2) population that leads to the H$\alpha$ absorption is determined by the Ly$\alpha$ mean intensity inside the atmosphere, we have used the same limb darkening law for Ly$\alpha$ and H$\alpha$ self-consistently. That is, the Eddington approximation is assumed for both the stellar Ly$\alpha$ and H$\alpha$ illumination when considering the limb darkening effect. Here, we discuss the isotropic radiation case of Ly$\alpha$ and H$\alpha$, as well as He 10830. 
Figure \ref{fig:plane_illum} (d) shows the $P\alpha$ distribution for the model with H/He = 92/8, $F_{\rm XUV}$ = F$_0$, and $\beta_m$ = 0.22 when the Ly$\alpha$ radiation on the stellar surface is isotropic (a flat disk). $P\alpha$ doesn't show much difference compared to the model with the limb darkening effect. However, the H$\alpha$ absorption appears to increase a little when 
the limb darkening effect is considered. Figure \ref{fig:plane_illum} (e) shows the H$\alpha$ transmission spectrum when the stellar H$\alpha$ source is isotropic (the cyan line) and limb darkened (the green line).
In Figures \ref{fig:plane_illum} (e) and (f), the limb darkening case is found to lead to a deeper absorption in both H$\alpha$ and He 10830, compared to the isotropic case.
\subsection{Size of the atmospheric outer boundary}\label{sec:size}

Note that in the above models, the atmospheric outer boundary is assumed to be 10 $R_P$ (larger than the stellar radis).
The Ly$\alpha$ intensity or $P\alpha$ distribution may be altered if the size of the outer boundary changes. In Figure \ref{fig:plane_illum} (g), we show the $P\alpha$ of the model with H/He = 92/8, $F_{\rm XUV}$ = F$_0$, $\beta_m$  = 0.22, and outer boundary of 6 $R_P$. The $P\alpha$ shows a similar trend to the model whose outer boundary is 10 $R_P$,
although its absolute value is relatively larger. This is reasonable, because a smaller atmosphere will recieve less stellar Ly$\alpha$ radiation at the surface of the atmosphere and thus have a smaller flux factor ($L_{A}/L_{S}$ as defined in Section \ref{sec:cal_RT}). However, compared to the case of 10 $R_P$, there is no medium (in other word, it is vacuum) in the region between 10 $R_P$ and 6 $R_P$, which scatters and reflects the Ly$\alpha$ photons. As a consequence, the absolute value of $P\alpha$ is slightly higher compared to the case of 10 $R_P$. 
However, the size change in atmosphere from 6 $R_P$ to 10 $R_P$ has no significant influence on the H(2) column density. As a result, the H$\alpha$ absorption will increase only a little when the outer boundary is increased to 10 $R_P$ as shown in Figure \ref{fig:plane_illum} (e). In other words, the contribution of the H(2) atoms in the region between 6 $R_P$ and 10 $R_P$ would be very small compared to that below 6 $R_P$. This is mainly because the density is very low and the velocity is high in the outer regions.
In addition, we also compared the He 10830 transmission spectrum for the models with outer boundary of  6 $R_P$ and 10 $R_P$. As shown in Figure \ref{fig:plane_illum} (f), the two model transmission spectra are almost identical (black and red lines), illustrating that the influence of atmosphere size, when changed from 6 $R_P$ to 10 $R_P$, is negligible also on the absorption of He 10830.

\begin{figure*}
\gridline{\fig{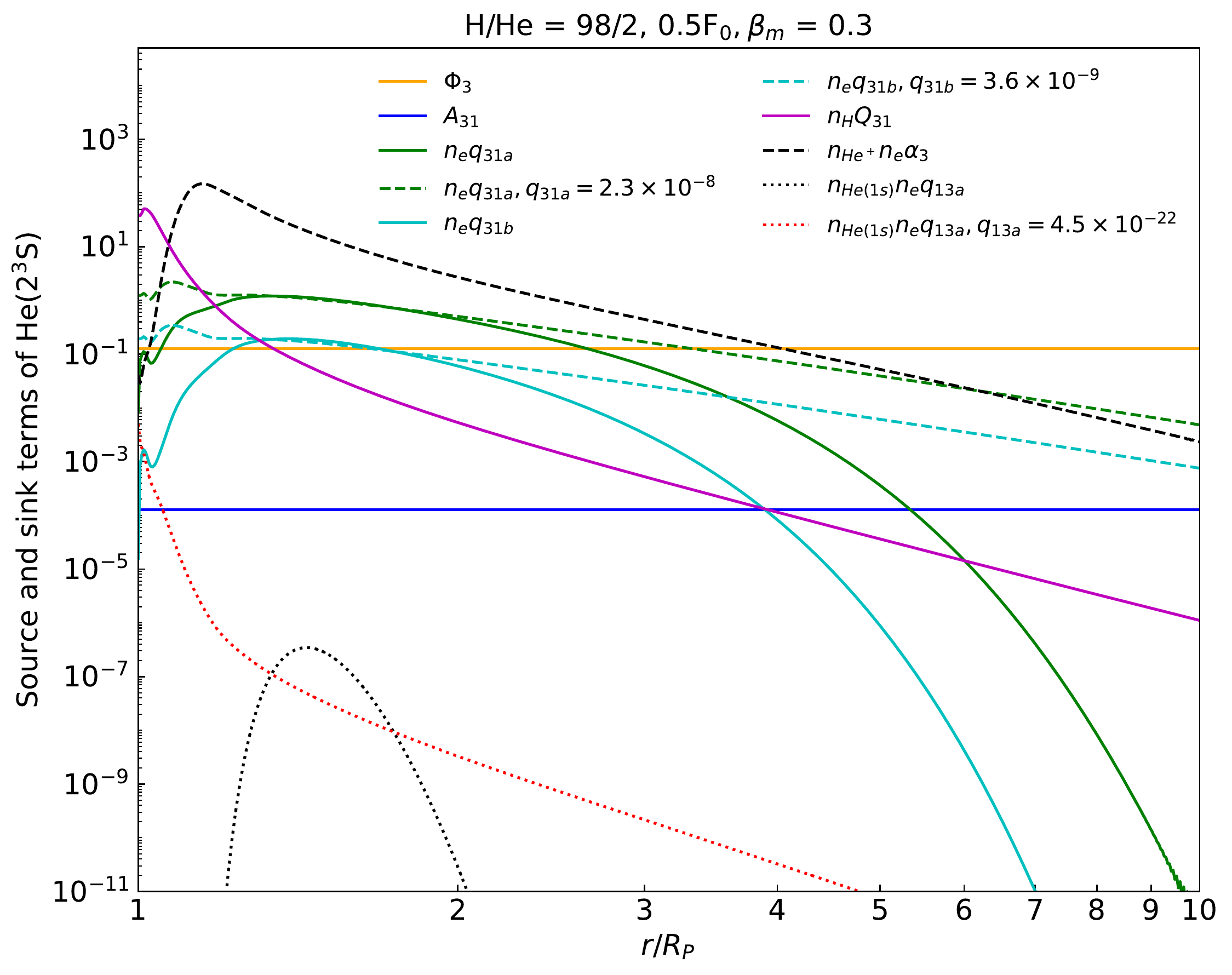}{0.47\textwidth}{(a)}
 } 
 \gridline{\fig{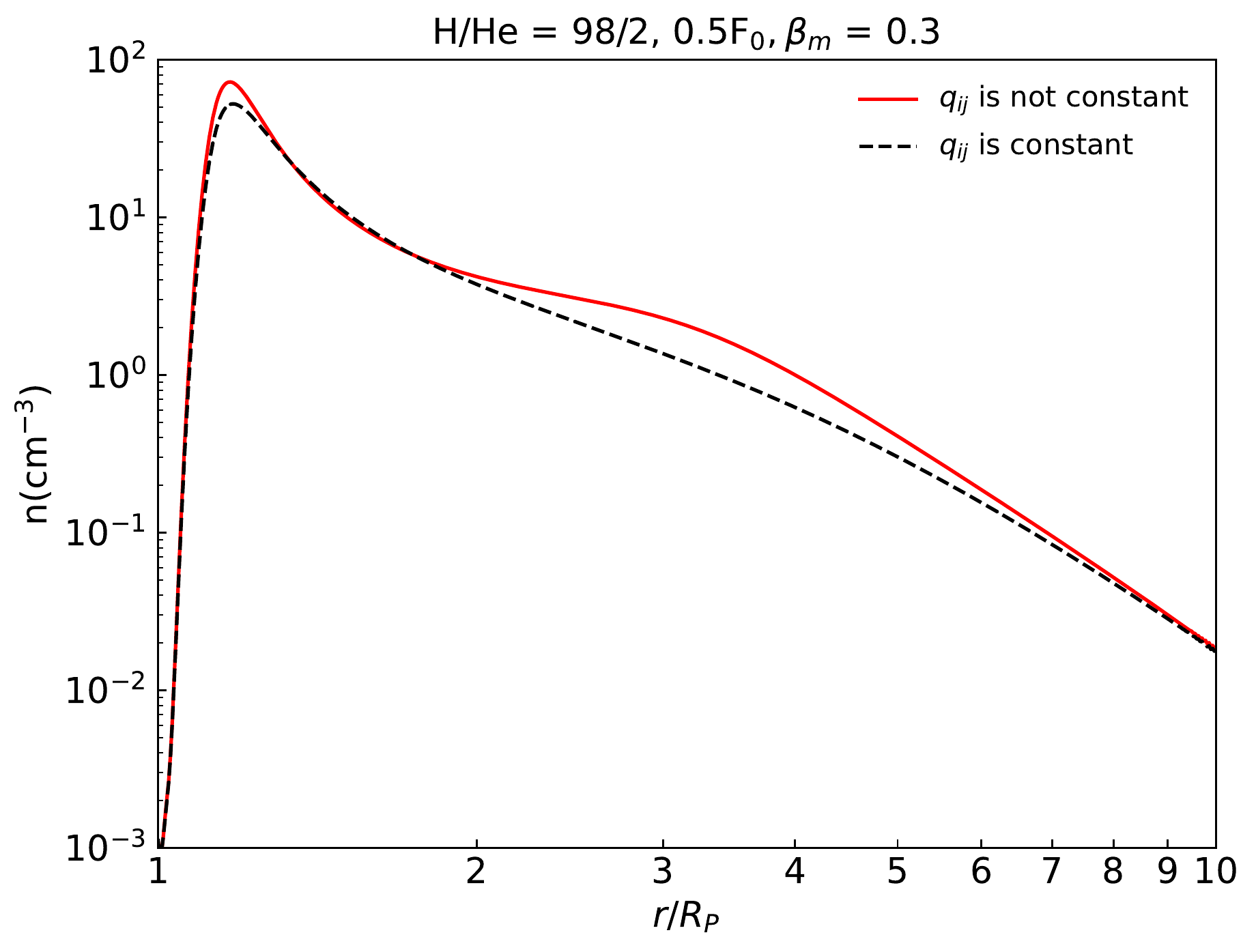}{0.47\textwidth}{(b)}
 \fig{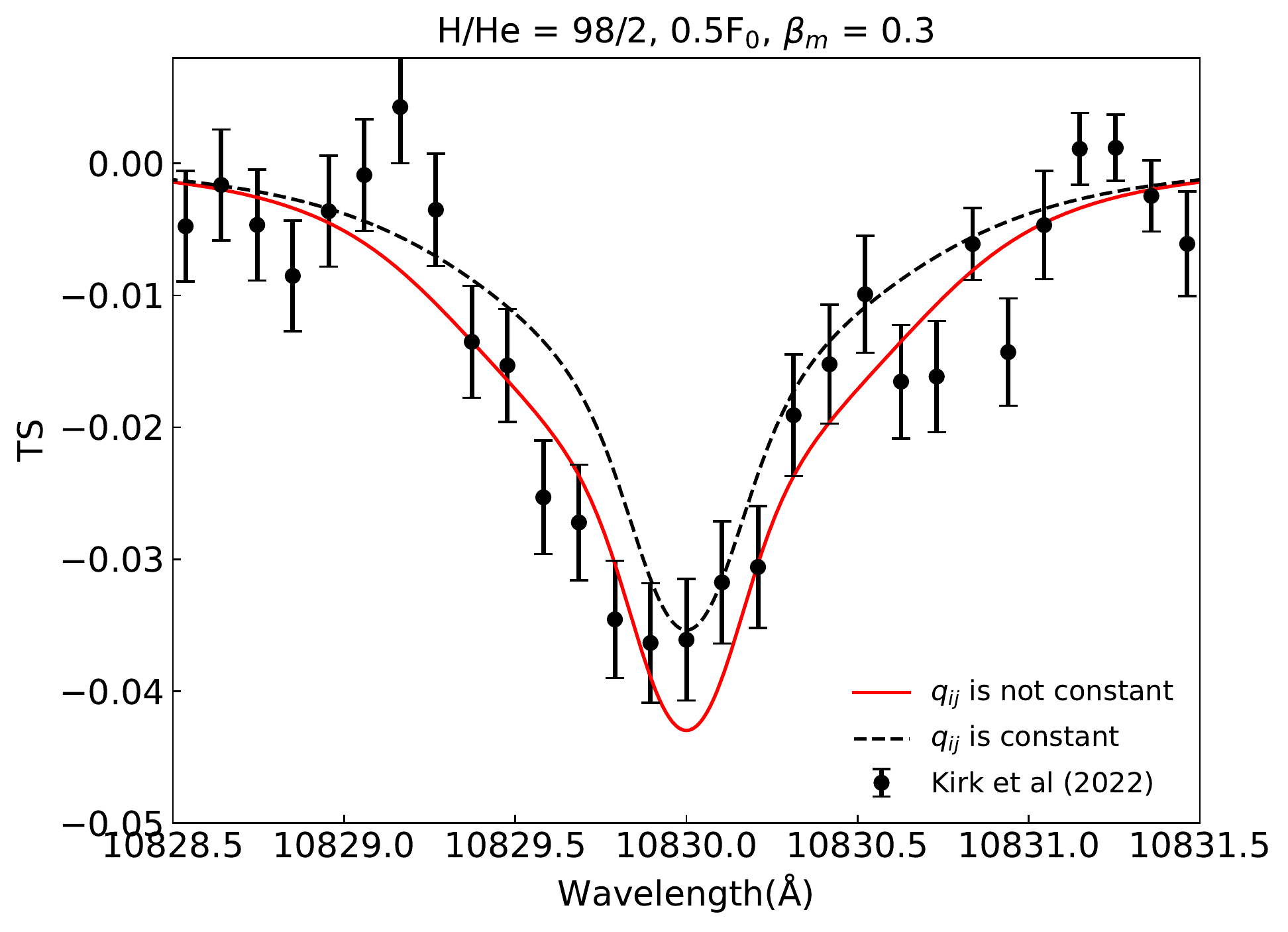}{0.5\textwidth}{(c)}
 }
\caption{(a) Source and sink terms related to the population of He(2$^3$S) for the model with H/He = 98/2, $F_{\rm XUV}$ = 0.5F$_0$, and $\beta_m$ = 0.3. The source terms plotted in the black dashed and dotted lines are in units of cm$^{-3} \rm s^{-1}$, and the sink terms plotted in the solid lines are in units of s$^{-1}$. The dotted red lines and dashed green and cyan lines represent $n_eq_{13a}$, $n_eq_{31a}$, and $n_eq_{31b}$ calculating using constant rate coefficients $q_{13a} = 4.5\times10^{-22}\rm cm^{3}s^{-1}$, $q_{31a}$ = 2.3$\times$10$^{-8} \rm cm^{3}s^{-1}$, and $q_{31b}$ = 3.6$\times$10$^{-9} \rm cm^{3}s^{-1}$. (b) The number density profile of He(2$^3$S). The red solid and black dashed lines are calculated by using a temperature-dependent $q_{ij}$ and constant q$_{ij}$, respectively. (c) The transmission spectra of He 10830. The red solid and black dashed lines are calculated by using a temperature-dependent $q_{ij}$ and constant $q_{ij}$, respectively.}
\label{fig:isothermal_com}

\end{figure*}
\subsection{Isothermal Parker wind and our hydrodynamic model}\label{subsec:isothermal}
\cite{2018ApJ...855L..11O} developed a 1D isothermal Parker wind model to simulate the escaping planetary atmosphere and used it to calculate the density structure of He(2$^3$S) and the transmission spectra of HD 209458b and GJ 436b. In the isothermal model, the atmospheric temperature and mass-loss rate are free parameters; by adjusting them one can obtain a density profile that results in a good fit to the He 10830 absorption. Similarly, \cite{2020A&A...636A..13L, 2021A&A...647A.129L} also used an isothermal Parker wind model to simulate the atmosphere of HD 209458b, HD 189733b, and GJ3470b, and modeled the He 10830 transmission spectra of these planets. In these models, the number densities of He(2$^3$S) first increase rapidly and then decrease with the atmospheric altitude, without showing a plateau as in our models. We find the main reason for the absence of the plateau could be that they used constant collisional transition rate coefficients ($q_{13a}$, $q_{31a}$ and $q_{31b}$). In the work of \cite{2018ApJ...855L..11O}, $q_{13a}$ = 4.5$\times$10$^{-20} \rm cm^{3}s^{-1}$, $q_{31a}$ = 2.6$\times$10$^{-8} \rm cm^{3}s^{-1}$, and $q_{31b}$ = 4.0$\times$10$^{-9} \rm cm^{3}s^{-1}$, evaluated at an `isothermal' temperature of 9000 K, were used. In a similar way, \cite{2022arXiv220511579K} reproduced the He 10830 line of WASP-52b by assuming an `isothermal' temperature of 7600 K. This temperature yields $q_{13a}$ = 4.5$\times$10$^{-22} \rm cm^{3}s^{-1}$, $q_{31a}$ = 2.3$\times$10$^{-8} \rm cm^{3}s^{-1}$, and $q_{31b}$ = 3.6$\times$10$^{-9} \rm cm^{3}s^{-1}$.
However, our models are not isothermal and $q_{13a}$, $q_{31a}$, and $q_{31b}$ depend on the temperature. When the temperature increases, these rate coefficients increase. As we have mentioned before, the collisions with electrons are the dominant destruction process when the atmospheric temperatures are high. The photoionization of He(2$^3$S) is the main destruction mechanism of He(2$^3$S) at high altitudes where the temperature is low. Thus, we speculate that assuming a constant $q_{31a}$ and $q_{31b}$ may lead some errors in the He(2$^3$S) population. To examine the effect of assuming constant collisional rate coefficients, we take the model with H/He = 98/2, $F_{\rm XUV}$ = 0.5F$_0$, and $\beta_m$ = 0.3. Figure \ref{fig:isothermal_com} (a) shows the source and sink terms related to He(2$^3$S). The dotted red line and dashed green and cyan lines represent $n_{He(1s)}n_eq_{13a}$, $n_eq_{31a}$, and $n_eq_{31b}$ for the model where the constant rate coefficients are assumed. Compared to the model using the temperature-dependent $q_{ij}$ ($ij=$ $13a$, $31a$, and $31b$), these quantities show higher values in most regions. The dominant sink terms in low altitudes ($r < 3R_P$) are the collisional transition by electrons as in the model using temperature-dependent coefficients. Unlike the case of using
temperature-dependent coefficients, the collisional transition still plays an essential
role in destructing He(2$^3$S) in higher altitudes ($r > 3 R_P$).
This difference decreases the number density of He(2$^3$S) significantly compared to the case of the temperature-dependent $q_{31a}$ and $q_{31b}$, as shown in Figure \ref{fig:isothermal_com} (b). As a result, the absorption of He 10830 appears to decrease obviously in this situation, as shown in Figure \ref{fig:isothermal_com} (c). Therefore, an isothermal Parker wind model is likely to give different conclusions compared to a non-isothermal hydrodynamic model, as adopted in this paper.

\subsection{The influence of the Ly$\alpha$ radiation pressure}\label{subsec:radiation_pressure}
The Ly$\alpha$ radiation pressure exerted upon atomic hydrogens may impact the atmospheric structures of exoplanets. It can accelerate the atmosphere to high velocities and has been used in some systems to explain the transmission spectra of Ly$\alpha$ \citep{2003Nature...422..143,2013A&A...557A.124B,2016A&A...591A.121B}. However, some hydrodynamic simulations have shown that the Ly$\alpha$ radiation pressure has little influence on the results \citep{2007P&SS...55.1426G,2007ApJ...671L..61B,2008ApJ...688.1352B,2010ApJ...723..116K,2016ApJ...832..173S,2017ApJ...847..126K}. In this work, the Ly$\alpha$ radiation pressure is not included in the 1D hydrodynamic simulations.

\cite{2016ApJ...832..173S} discussed that the Ly$\alpha$ radiation force would become noticeable only at distances of tens of $R_p$. In Section \ref{sec:size}, we examined the case where the size of the outer boundary of the atmosphere is reduced to $6R_p$ and demonstrated that the H$\alpha$ and He 10830 absorption lines are caused mainly by the atmosphere in a
relatively inner part of $R\lesssim 6R_p$. Therefore, it is highly likely that the Ly$\alpha$ radiation pressure would not significantly alter the present results. For detailed investigation, we
may need to calculate the radiation force directly while Ly$\alpha$ radiative transfer is performed, as done for $P_\alpha$, and iterate the hydrodynamic simulations and the radiative
transfer calculations until the result converges. We defer the study of the radiation pressure effect to future work.

\subsection{The influence of the Coriolis force}\label{subsec:coriolis_force}
In a rotating planet, the radially outflowing atmosphere will experience an inertial force, the Coriolis force, affecting the direction of the atmospheric flow. The tangential velocity component produced by this effect would give rise to a difference in the Doppler shift regarding the line absorption. Using 3D MHD models, \cite{2020MNRAS.491.3435S} showed that the spiraling planetary atmosphere due to the Coriolis force causes a lower absorption level in the line of interest compared to their 2D model.

In 1D hydrodynamic simulations, the Coriolis force is disregarded inevitably. Neglecting this effect may cause an overestimation of the absorption depth around the line center of H$\alpha$ and He 10830, since the Coriolis force can result in a decrease of the radial velocity.
For WASP-52b, the tangential velocity aquired per unit length is about $|\frac{dv_t}{dr}|\sim 2\omega\sim 8$ km$^{-1}R_P^{-1}$, where $v_t$ is the tangential velocity and $\omega$ is the angular rotation velocity of the planet. Assuming the planet is tidally-locked to the star, $\omega$ is the same as the angular velocity of the planet's revolution. The tangential velocity can be comparable to the radial velocity in some atmospheric regions.
However, calculating the tangential velocity component in detail along the line-of-sight
direction is complicated. Thus it is difficult to evaluate the effect of the Coriolis force
with our 1D hydrodynamic model.
Better models that include the effect of Coriolis force is beyond the scope of this paper and will be studied in our future work.

\section{Summary}\label{sec: summary}
In this work, we have modeled the transmission spectra of the H$\alpha$ and He 10830 lines simutaneously for the first time in hot Jupiter WASP-52b by using a 1D hydrodynamic model combined with a non-local thermodynamic model. 
To calculate the number density of H(2), which causes the absorption of H$\alpha$, we also calculated the Ly$\alpha$ mean intensity distribution inside the planetary atmosphere by using a Monte Carlo code and assuming a spherical stellar Ly$\alpha$ radiation source and a spherical planetary atmosphere.
Then we explored a parameter space of $F_{\rm XUV}$, $\beta_m$, and H/He and found that models with different H/He ratios can fit the H$\alpha$ observations well if the host star has a high XUV flux and the X-ray fraction in XUV radiation is low or the other way around. Further constraints on the atmosphere can be obtained by analyzing the He 10830 absorption line. The simulations of the He 10830 $\rm\AA$ triplet suggest that a higher H/He abundance ratio  ($\sim$ 98/2) than the solar ratio ($\sim$ 92/8) is required to fit the observation. Basically, the absorption of He 10830 increases with increasing $F_{\rm XUV}$, but its variation with $\beta_m$ is more complicated. Only the model with H/He = 98/2, $F_{\rm XUV}$ = 0.5F$_0$ and $\beta_m$ = 0.3 can reproduce both H$\alpha$ and He 10830 lines.
Our simulations also suggest that hydrogen and helium originate from the escaping atmosphere, and the mass-loss rate is about 2.8$\times 10^{11}$ g s$^{-1}$, which can influence the evolution of planetary atmosphere.
We also discussed the plane-parallel Ly$\alpha$ illuminaiton models. Under this assumption, in a plane-parallel atmosphere, the Ly$\alpha$ radiation strength in the atmosphere is overestimated. In contrast, the Ly$\alpha$ intensity calculated in a spherical atmosphere appears to be slightly lower than in the spherical illumination models. In this work, we also considered the limb darkening effect of the stellar Ly$\alpha$, H$\alpha$ and He 10830 illumination self-consistently. Compared to the models with an isotropic illumination, the models with the limb darkening effect produced stronger absorption in the H$\alpha$ and He 10830 transmission spectra.
Our work would be a benchmark in studying the exoplanetary hydrogen and helium atmosphere by modeling H$\alpha$ and He 10830$\rm\AA$ absorption simultaneously. 
The findings can help to constrain the physical paramters of the atmosphere and to better understand its composition and structure. Finally, we note that the stellar Ly$\alpha$ profile and flux adopted in this work are based on some assumptions due to the lack of observations. We believe that future observation and reconstruction of the Ly$\alpha$ line will bring more insights into the atmosphere study.

\vspace{12 pt}
\textbf{Acknowledgement.} We thank the anonymous reviewers for their constructive comments to improve the manuscript. We thank James Kirk for sharing the observed data of He 10830 transmission spectrum with us. We thank XiaoLin Yang for his fruitful discussion of the spherical illumination model in the LaRT code and for the discussion of the limb darkening effect in the transmission spectrum. We are also grateful to Zhi Xu for her helpful discussion of the calculation of He(2$^3$S) populations.
The authors acknowledge supports by the Strategic Priority Research Program of Chinese Academy of Sciences, Grant No. XDB 41000000 and the National Key R$\&$D Program of China (Grant No. 2021YFA1600400/2021YFA1600402).
The authors acknowledge the supports by the National Natural Science Foundation of China though grants through grants 11973082 to JG, 42075122 and 12122308 to GC. and by the Natural Science Foundation of Yunnan Province (No. 202201AT070158).
The authors gratefully acknowledge the “PHOENIX Supercomputing Platform” jointly operated by the Binary Population Synthesis Group and the Stellar Astrophysics Group at Yunnan Observatories, Chinese Academy of Sciences. K.-I. Seon was partly supported by a National Research Foundation of Korea (NRF) grant funded by the Korean government (MSIT; No. 2020R1A2C1005788) and by the Korea Astronomy and Space Science Institute grant funded by the Korea government (MSIT; No. 2022183005). This work has made use of the MUSCLES Treasury Survey High-Level Science Products; doi:10.17909/T9DG6F.

\end{document}